\documentclass[10pt,a4paper,psfig,epsfig,graphicx,psfrag,bibtotoc]{scrbook}
\usepackage[bindingoffset=15mm,paperwidth=180mm,paperheight=260mm,scale=0.807,heightrounded,mag=1142]{geometry}
\usepackage{graphicx}
\usepackage{amsfonts,amssymb,amsmath}
\usepackage{psfrag}
\usepackage{hyperref}
\usepackage{geometry}
\usepackage[all]{xy}
\usepackage{feynmf}
\usepackage{sidecap}
\usepackage[spanish,english]{babel}
\usepackage[latin1]{inputenc}
\newcommand{\be}{\begin{equation}}
\newcommand{\ee}{\end{equation}}
\newcommand{\ba}{\begin{eqnarray}}
\newcommand{\ea}{\end{eqnarray}}
\newcommand{\bea}{\begin{eqnarray}}
\newcommand{\eea}{\end{eqnarray}}
\newcommand{\tr}{{\mathrm{tr}}}
\newcommand{\arctanh}{\mathrm{arctanh}}
\def\L5{\tilde{\Lambda}}

\newcommand{\m}{\hat m}
\def\pd{\partial}
\def\a{\alpha}
\def\b{\beta}
\def\g{\gamma}
\def\di{\mathrm{d}}
\def\k{{\bf k}}
\def\m{\mu}
\def\n{\nu}

\def \M{\mathcal{M}}

\def\t{\tau}
\def\th{\theta}

\def\th{\theta}
\def\l{\lambda}
\def\r{\rho}
\def\s{\sigma}
\def\I{\mathrm{i}}
\def\e{\epsilon}
\def\H{\mathcal H}
\def\L{\Lambda}
\def\kf{\tilde \kappa_f}
\def\D{\rm Diff}
\def\TD{\rm TDiff}
\def\WTD{\rm WTDiff}
\def\CD{\rm CDiff}
\def\WRS{\rm WRS}
\def\RS{\rm RS}

\def\A{A}
\def\La{{\mathcal L}}
\def\pdi{/\hspace{-.2cm}\pd}

\renewcommand{\d}{{\delta}}

\def\1{\mathchoice{\rm 1\mskip-4.2mu l}{\rm 1\mskip-4.2mu l}%
{\rm 1\mskip-4.6mu l}{\rm 1\mskip-5.2mu l}}
\begin{document}

\frontmatter
\selectlanguage{english}

\begin{titlepage}
\vskip 2cm

\begin{center}
\textsf{Departament de F\'isica Fonamental\\[.3cm]
Grup de Gravitaci\'o i Cosmologia}\\[3cm]
{\LARGE {\bf Aspects of Infrared Modifications of  Gravity}}\\[2cm]
{\Large Diego Blas Temi\~no}\\[5.5cm]
{\Large Advisor: Dr. Jaume Garriga Torres.\\[.2cm]
April  2008}\\[1.5cm]
\begin{figure}[h]  \centering
\includegraphics[width=0.4\textwidth]{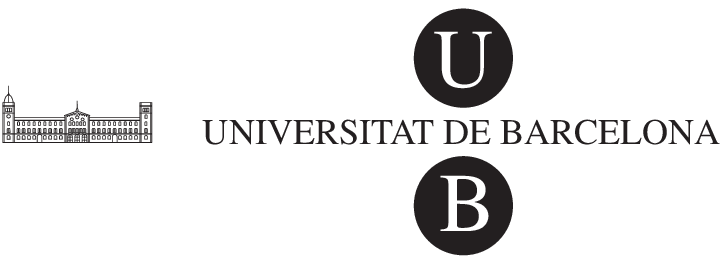}
\end{figure}
\end{center}

\cleardoublepage

\thispagestyle{empty}
\hspace{3cm}
\begin{flushright}
{\em A Gin\'es y Estefan{\'\i}a}.
\end{flushright}
\newpage

\end{titlepage}

\tableofcontents

\newpage

%%%%%%%%%%%%%%%%%%%%%%%%%%%%%%%%%%%%%%%%%%%%%%%%%%%%%%%%%%%%%%%
\chapter{Acknowledgements}
%%%%%%%%%%%%%%%%%%%%%%%%%%%%%%%%%%%%%%%%%%%%%%%%%%%%%%%%%%%%%%%%%%%%%%%%%%%%%%%%%%%%%%%%%%%%%%%%%%%%

\selectlanguage{english}
I am very grateful to the following people for their support, influence and
for sharing with me many ideas about Physics during my Ph.D.: Jaume Garriga,
Enrique Álvarez, Cedric Deffayet, Gia Dvali, Enric Verdaguer,
 Roberto Emparan, Jorge Russo,
José Ignacio Latorre, Joan Soto, Joaquim Gomis, Guillem Pérez-Nadal,
 Daniel Arteaga, Tasos Avgoustidis,
 Manuel Asorey, Oriol Pujolàs, Michele Redi, Javier Redondo, Carla Biggio,
Eduard Massó,
Antón Faedo, Albert Roura, Andi Ross,
Gregory Gabadadze Alberto Iglesias,  Gastón
Giribet, Zurab Berezhiani, Denis Comelli, Luigi Pilo and Frabizio Nesti.
%\selectlanguage{english}

%%%%%%%%%%%%%%%%%%%%%%%%%%%%%%%%%%%%%%%%%%%%%%%%%%%%%%%%%%%%%%%
\chapter{Conventions}\label{conventions}
%%%%%%%%%%%%%%%%%%%%%%%%%%%%%%%%%%%%%%%%%%%%%%%%%%%%%%%%%%%%%%%%%%%%%%%%%%%%%%%%%%%%%%%%%%%%%%%%%%%%

Throughout the dissertation we will follow the Landau-Lifshitz time-like
conventions; the $n$-dimensional flat metric in particular, reads
$\eta_{\mu\nu}=\mathrm{diag}\,(1,-1,\ldots,-1)$. $n$ is the space-time
dimension that will be taken to be 4 in some parts of the Thesis. We will also
use  $N=n-1$ as the space dimension.
Lagrangians are written in
momentum space as well as in configuration space, depending on the
context. It is usually trivial to shift from one language to the
other. For the totally antisymmetric tensor we choose
$\e^{0123}=1.$\\

We will define the Laplacian  operator as $\Delta=\sum_i\partial_i \partial_i=
-\partial^i\partial_i$ and $\Box=\eta^{\m\n}\pd_\m \pd_\n$.\\

Given a connection, the Riemann tensor will be defined as
\be
R^\a_{\phantom{\a}\m\b\n}\equiv \Gamma^\a_{\phantom{\a}\m\n,\b}-..., \quad R_{\m\n}\equiv R^\a_{\phantom{\a}\m\a\n}.
\ee
Similarly,  given a spin-connection $\omega_{\m ab}$,
\be
R_{\m\n ab}(\omega)=\pd_\m\omega_{\n ab}-\pd\omega_{\n ab}+\omega_{\m a}^{\phantom{\m a}c}\omega_{\n cb}-
\omega_{\n a}^{\phantom{\m a}c}\omega_{\m cb}.
\ee
The (anti)symmetrization is performed with a weight factor,
\be
\phi_{(ab)}=\frac{1}{2}\left(\phi_{ab}+\phi_{ba}\right)
, \quad \phi_{[ab]}=\frac{1}{2}\left(\phi_{ab}-\phi_{ba}\right).
\ee
The gamma matrices in 4-dimensions
will be (see also \cite{deWit:1985aq})
\ba
\g^0=\left(
 \begin{array}{cc}
 1&\ 0\\
 0&-1
 \end{array}\right), \quad \g^i=\left(
 \begin{array}{cc}
 0&\s^i\\
 -\s^i&0
 \end{array}\right),\quad \g_5=\I\g^0\g^1\g^2\g^3,\quad
\quad C=\I\g^2\g^0,
\ea
satisfying
\be
\{\g^\m,\g^\n\}=2\eta^{\m\n}.
\ee
We would also like to write a list of some abbreviations that appear throughout this
Thesis:

\begin{itemize}
\item Eq.: equation,
\item KK: Kaluza-Klein
\item PDoF: Propagating degree(s) of freedom,
\item EoM.: Equations of motion,
\item GR: General relativity,
\item CC: Cosmological constant,
\item RS: Rarita-Schwinger,
\item FP: Fierz-Pauli,
\item TDiff: Transverse diffeomorphisms,
\item Diff: Diffeomorphisms,
\item GCT: General coordinate transformations,
\item r.h.s.: Right hand side.
\end{itemize}

The references are sorted alphabetically.

\mainmatter
%%%%%%%%%%%%%%%%%%%%%%%%%%%%%%%%%%%%%%%%%%%%%%%%%%%%%%%%%%%%%%%
%\part{Preliminaries}
%%%%%%%%%%%%%%%%%%%%%%%%%%%%%%%%%%%%%%%%%%%%%%%%%%%%%%%%%%%%%%%%%%%%%%%%%%%%%%%%%%%%%%%%%%%%%%%%%%%%

\renewcommand*{\dictumwidth}{.50\textwidth}
\setchapterpreamble[o]{\dictum[Silvio Rodríguez]{Lo más terrible se aprende
en seguida y lo hermoso nos cuesta la vida.}}
%%%%%%%%%%%%%%%%%%%%%%%%%%%%%%%%%%%%%%%%%%%%%%%%%%%%%%%%%%%%%%%
\chapter{Introduction}\label{chapterintro}
%%%%%%%%%%%%%%%%%%%%%%%%%%%%%%%%%%%%%%%%%%%%%%%%%%%%%%%%%%%%%%%%%%%%%%%%%%%%%%%%%%%%%%%%%%%%%%%%%%%%

In this Chapter, we will first review some of the proposals for modifying gravity at large
distances, explaining the difficulties that appear in these models together with possible solutions.
In the second part of the chapter, we present an outline of the rest of the Thesis.

%%%%%%%%%%%%%%%%%%%%%%%%%%%%%%%%%%%%%%%%%%%%%%%%%%%%%%%%%
\section[Massive gravity and related models]{Massive gravity and related models of modifications of gravity}
%%%%%%%%%%%%%%%%%%%%%%%%%%%%%%%%%%%%%%%%%%%%%%%%%%%%%%%%%

The non-renormalizability of Einstein's theory of general relativity (GR) suggests that GR
will be superseded by a quantum theory of gravity at high enough energies with respect to
a certain mass scale $M_{QG}$. For dimensional reasons, it is customary to associate this scale with the
Planck mass\footnote{Source http://physics.nist.gov/.} $$M_P=\sqrt{\frac{\hbar c}{G}}=1.220 892(61)\cdot 10^{19} \ \mathrm{GeV \cdot c^{-2}},$$
or the corresponding
Planck length $l_P= G M_P c^{-2}=1.616252(81)\cdot 10^{-35}\ \mathrm{m}$. The standard
assumption is that GR is valid as an effective field theory (EFT\footnote{For reviews on EFT see {\em e.g.}
\cite{Burgess:2003jk,Donoghue:1995cz,Burgess:2007pt,Goldberger:2007hy,Polchinski:1992ed}
(see also \cite{Faller:2007sy}). Henceforth, we will take units such
that $\hbar=c=1$.}) for length scales much larger than $l_P$.
If this
is true the expectation of learning something about the actual theory of quantum gravity
from experiments to be performed within the near future
 is almost hopeless\footnote{It is true that there are some astrophysical
phenomena that involve very high energy events, and that may shed
some light at energies beyond the possibilities of accelerators (see {\em e.g.} \cite{Albert:2007qk}).}.

Yet, when the cosmological observational data is analyzed within the framework of GR, the most
successful models imply the existence of a vacuum energy $\Lambda$ whose magnitude is
{\em unnatural} from the EFT point of view\footnote{The value of a constant is \emph{
technically unnatural} if it is much smaller than the size of quantum corrections to it.}.
Hence,
a very fine-tuned  vacuum energy (or \emph{dark energy}) is needed to reconcile
 GR with the observations \cite{Weinberg:2000yb,Weinberg:1988cp} (see
 \cite{Nobbenhuis:2006yf} for a quite comprehensive review
 of the cosmological constant (CC) problem).
 This problem is rather pressing as
it corresponds to the explanation of actual data \cite{Spergel:2006hy,Adelman:2007wu,Astier:2005qq}.
In fact, the problem can be divided into
two: first why the vacuum energy is not as high as it should be (fine tuning problem)
and second why is it so small that becomes dominant precisely at the present time
(coincidence problem). For a modern review article see, \emph{e.g.}, \cite{Copeland:2006wr}.\\

To address the previous problems, GR can be modified  at short (ultraviolet, UV)
or long (infrared, IR) distances. This requires the introduction of new length scales $L$
in the theory which can be combined with $l_P$ to build new constants with dimensions
of length
\be
\label{L_q}
L_q=l_P \left(\frac{L}{l_P}\right)^q.
\ee
When $L$ and $l_P$ are very different, we find a hierarchy of length scales
larger than the Planck length where GR may be modified.
For instance, we may assume that the fundamental scale of quantum gravity is a certain $L_{q}$
in (\ref{L_q}),
and that $l_P$ is a derived quantity.
The energy scale at which quantum gravity effects are important, $L_{q}^{-1}$,
 may be  as low as TeV
in which case the phenomenology of LHC  could
 probe the true quantum theory of gravity and shed some light in the existing
hierarchy between the Planck energy and the electroweak energy
\cite{ArkaniHamed:1998rs,ArkaniHamed:1998nn,Antoniadis:1998ig}. Later on,
we will discuss some models where this possibility is realized.

 A related possibility
is that there exists a certain low energy scale $L_{ir}^{-1}$ below which GR may be modified.
 In particular, if this length scale $L_{ir}$ is
of the order of the present cosmological horizon, $L_{ir}\sim 10\ \mathrm{Gpc}$,
 we expect modifications of GR to be important
at current
cosmological scales. Thus, all the predictions of GR at these scales (including the existence
and amount of dark energy)
may be modified within this new framework of infrared modifications of gravity.\\

%%%%%%%%%%%%%%%%%%%%%%%%%%%%%%%%%%%%%%%%%%%%%%%%%%%%%%
\subsubsection*{{\em Linearized Massive Gravity}}
%%%%%%%%%%%%%%%%%%%%%%%%%%%%%%%%%%%%%%%%%%%%%%%%%%%%%

The appearance of the length scale $L$ can be motivated in several ways. One of the first
possibilities dates back to the work of Fierz and Pauli \cite{Fierz:1939ix} and consists of
adding a mass to  the graviton. More concretely,
if one considers a small gravitational field propagating in Minkowski space-time,
\be
\label{metricper}
g_{\m\n}=\eta_{\m\n}+h_{\m\n},
\ee
the Lagrangian for the perturbations $h_{\m\n}$ corresponds to that of a massless particle
of spin-2 \cite{Einstein:1916cc,Fierz:1939ix,WeinbergGC}. In the linear approximation, one
can solve the field equations for $h_{\m\n}$ in the presence of a conserved energy-momentum
tensor\footnote{In
 the massless case the energy-momentum tensor must be conserved from consistency reasons.}
and Newton's law  and the deflection of light  for weak gravitational fields are recovered
\cite{WeinbergGC,Ortin}. The interaction between two sources can be understood as due to
 the exchange of a massless particle so that, ignoring the tensor structure,
the corresponding potential between two test particles of mass $m_1$, $m_2$ can be written as
\be
\label{potential}
V(r)\sim \frac{m_1m_2}{M^2_P}\frac{1}{r}.
\ee
After the addition of a mass term to the mediator of gravity we expect that the potential will
acquire a Yukawa form for length scales larger that the inverse of the mass scale. Namely, we
expect it to behave as
\be
V(r)\sim \frac{m_1m_2}{M^2_P}\frac{e^{-mr}}{r}.
\ee
If the mass is as small as
$m\sim (10 \ \mathrm{Gpc})^{-1}\sim  10^{-33}\ \mathrm{eV}$,
 we expect that gravity fades away at cosmological distances and that at
 smaller distances the usual predictions of GR are recovered.
This would imply that sources of the scale of the Universe would
gravitate less than those smaller than this scale,
 which could alleviate the CC problem.

There are some obstacles in the way of
this naive expectation. Assuming the Diff invariant kinetic term,
there is only a possible mass term at the linear level which respects
Lorentz invariance and does not contain ghost
degrees of freedom\footnote{By a ghost we mean
a field with negative kinetic energy in the Lagrangian.} \cite{Fierz:1939ix},
\be
\label{PFmass}
\La_m\sim h_{\m\n}h^{\m\n}-h^2.
\ee
The interaction between two conserved sources computed from this
linearized  Lagrangian
 suffers from a discontinuity with respect to its massless counterpart, coming
from the different tensor structure of the propagator. As shown in \cite{vanDam:1970vg,Zakharov:1970cc},
when coupled to conserved sources,
the propagator of the massive theory reduces to
\be
\label{masslesspr}
P_{\m\n\r\s}=\frac{1}{k^2-m^2+\I \e}\left(\eta_{\m(\r}\eta_{\n)\s}-\frac{1}{a}\eta_{\m\n}\eta_{\r\s}\right),
\ee
with $a=(n-1)$  where $n$ is the dimension of the space-time.
In the massless case,
the propagator corresponds to the massless limit of  (\ref{masslesspr}),
but with $a=(n-2)$, which means that the propagator of the massless theory \emph{does not} agree
with the massless limit of the massive case. This fact, known as vDVZ discontinuity, has drastic consequences.
 From
measurements of the deflection of light by the Sun, the linear massive case can be
excluded completely for any value of $m$ \cite{vanDam:1970vg,Zakharov:1970cc}. Notice that the
difference between the massless and the massive case comes from the scalar part of the
propagator. One may think that  the massless case can be recovered
by adding a scalar field coupled to the trace of the
energy-momentum tensor. This is obviously true, but
the fact that $a(m=0)>a(m\neq 0)$ implies that the new field will be a ghost\footnote{If
 \emph{non-local} couplings are considered, the previous argument can be
circumvented by choosing a coupling of the scalar field to matter
that vanishes in the UV. Recently, a local model with a running $a$ has been discovered
in certain local brane models with two extra dimensions, but the vDVZ
discontinuity is still present \cite{deRham:2007xp}.}, \emph{i.e.} its propagator
will have a negative residue \cite{Zakharov:1970cc}. The existence of these states with
negative norm destroys unitarity, and it is usually understood that quantum theories with
ghosts are ill-defined. One can modify the quantization procedure
to get rid of the the negative norm states but in this case the vacuum is unstable.
In Lorentz-invariant theories its decay rate is in fact infinite\footnote{If a Lorentz breaking cut-off is introduced
in the theory, the decay rate can be regularized to be consistent with the observations. Similarly,
as the linearized theory is understood as an effective field theory valid to a certain scale, beyond
this scale new degrees of freedom can make the theory well-behaved \cite{Creminelli:2005qk}.} \cite{Cline:2003gs}.\\

As first noticed in \cite{Vainshtein:1972sx},
another way in which the discontinuity may disappear
is through the non-linear effects.
 The main idea is that there is a
source dependent scale $r_\star$
below which the three graviton vertex (\emph{i.e.} the operators involving three gravitons)
becomes of the same order as the quadratic terms and the classical
linearized approximation breaks down. In other words, in the presence of a source
the theory is strongly coupled
for distances smaller than $r_\star$.
If $r_\star$ is bigger than the length scales at which an experiment probing gravity is performed, one
must solve the whole non-linear system to give reliable predictions and there is a chance that
the nonlinear effects restore agreement with GR. For the massless case, given a source of mass $M$,
the non-linear
effects of GR become important at a scale $r_\star\sim  r_s \equiv M M_P^{-2} $, which for the Sun is
much smaller than the distance at which the light deflection is measured. Naively, we would think
that for length scales smaller than  $m^{-1}$, the dynamics of the massive case would be similar to the
massless one, and that non-linear effects will not show up for $r>r_s$
 also for the massive case. However, as found
in \cite{Vainshtein:1972sx}, this naive expectation is incorrect. It was shown in
\cite{ArkaniHamed:2002sp} (see also \cite{Deffayet:2001uk,Nicolis:2004qq})
that for the FP
massive case  the spin-0 polarization of the massive graviton interacts strongly
(in the presence of a source of mass $M$) at a scale
which can not be smaller than
\be
\label{strongcoupl}
r_\star\gtrsim(m^{-2} M M_P^{-2})^{1/3}.
\ee
This scale diverges for $m\rightarrow 0$. For a source of Solar mass $M\sim M_\odot$ and
 $m$ of the order of the
Hubble length,
 $r_\star$ is larger than the size of the Solar System ($r_\star\sim 10 \ \mathrm{pc}$).
 The tensor structure of the massive graviton
at distances $r\ll r_\star$ where non-linear
effects are important
is still an open issue. For a
related model that we will discuss later (DGP), it was argued that the
correct tensor structure is recovered
and the vDVZ discontinuity is not present \cite{Deffayet:2001uk,Dvali:2006su} (see also
\cite{Damour:2002gp}).

Even if this effect is welcome,  it is intimately related to another
potential disaster of the theories that modify GR in the infrared: \emph{strong coupling} at the
quantum level.
This pathology shows up
when one considers the scale at which sources of the scale of quantum gravity are strongly coupled
\cite{ArkaniHamed:2002sp} (see also \cite{Aubert:2003je} for an explicit calculation). From
(\ref{strongcoupl}) we see that this scale is $\Lambda\sim (m^2 M_P)^{1/3}$ which
for $m$ of the order of the present Hubble parameter is of the order of
$\Lambda\sim (1000 \ \mathrm{km})^{-1}$.
This energy scale is much lower than the Planck mass and also than the naive scale that
one would expect from the analogous calculation for spin-1, $\sqrt{m M_P}$. The reason why this happens
is that the strongly coupled polarization does not have a  standard kinetic term, but gets it from
its mixing with other polarizations \cite{ArkaniHamed:2002sp}.

In a non-renormalizable
theory like the one at hand, quantum corrections imply the presence of an infinite
tower of higher dimensional operators suppressed by inverse powers of the interaction scale $\Lambda$
and
a theory of quantum gravity would be needed to deal with calculations at distances smaller than
$\Lambda^{-1}\sim 1000\ \mathrm{km}$. These conclusions depend on the UV completion of the theory and,
as outlined in \cite{Nicolis:2004qq},
there may exist a non-generic prescription to choose the counterterms in such a way that the quantum
corrections are not important in all the astrophysical situations (see also \cite{Dvali:2004ph}).
In other words, the loop expansion may admit a resummation such that the scale $\Lambda^{-1}$ is unphysical
(indeed, this is what happens for the classical expansion \cite{Dvali:2004ph}).

To sum up,  let us state again that whenever a Lorentz invariant
theory has a massive
graviton as the mediator of gravity, it requires the presence of \emph{strong coupling}
to be phenomenologically acceptable, which generically requires a UV completion at very low energy scales.\\

A related aspect of massive gravity is that when propagating on a curved background,
it behaves differently than in flat space\footnote{Writing
the action for a spin-2 field in an arbitrary background is problematic
as the structure of the constraints is modified and a ghost mode may appear
or causal propagation can be lost (see, {\em e.g.} \cite{Aragone:1979bm}).
 These problems have been recently reconsidered in \cite{Buchbinder:1999ar} for
 the coupling of the spin-2 field to gravity
 (see also \cite{Porrati:2008an,Deser:2006sq,Aragone:1971kh} where
the coupling of spin-2 fields to  electromagnetism is studied).}. In particular, in anti-de Sitter (AdS) space there is no vDVZ
discontinuity \cite{Kogan:2000uy,Porrati:2000cp,Karch:2001jb} while in de Sitter (dS) a light
massive graviton becomes a ghost \cite{Higuchi:1986py}.  The reason why this happens is simply that the
mode that becomes strongly coupled in the flat case acquires a kinetic term proportional
to the curvature in the curved background case\footnote{As
 shown in  \cite{Dilkes:2001av}, the discontinuity reappears
at the quantum level, but then its effects happen at very short distances.} \cite{ArkaniHamed:2002sp}.\\

For the sake of completeness, we should mention that
there are some theories with massive gravitons
 which only involve the four dimensional metric and are invariant under diffeomorphisms.
An example of these theories is gravity with higher derivatives
\cite{Stelle:1977ry,Stelle:1976gc,Starobinsky:1980te,DeFelice:2007ez,Nojiri:2006ri}.
One can show that the spectrum of this theory can be decomposed into a massless graviton
and a massive graviton with a mass term different from (\ref{PFmass}) in general. In
this sense, these models  resemble  {\em bigravity} theories (see below).

However, these models
have a very serious drawback, namely the appearance of ghost states.
Only in certain instances where the massive states disappear
 this pathology may be absent. In these cases,  called Modified Gravity Models,
 the term in higher
 derivatives is simply $f(R)$ and the theory is equivalent
 to a scalar-tensor theory (cf. \cite{Wands:1993uu,Stelle:1977ry}). The gravitational interaction can be modified
both at long and short distances\footnote{The modification at large
distances occurs, {\em e.g.}, when one considers functions of the
form $R^a$, with $a<1$ \cite{Woodard:2006nt}.} but a successful model is still absent \cite{DeFelice:2007ez}.
 Yet another possibility
is provided by topological massive gravity in $2+1$ dimensions \cite{Deser:1981wh,Deser:1982vy}
or the possibility of mass generation through matter loops in AdS \cite{Porrati:2001db}.
Besides, we could also consider non-local modifications
of gravity \cite{Dvali:2007kt,ArkaniHamed:2002fu,Dvali:2006su}.\\

%%%%%%%%%%%%%%%%%%%%%%%%%%%%%%%%%%%%%%%%%%%%%%%%%%%%%%
\subsubsection*{{\em Non-linear Massive Gravity}}
%%%%%%%%%%%%%%%%%%%%%%%%%%%%%%%%%%%%%%%%%%%%%%%%%%%%%

From the discussion above, it seems clear that
it is essential for any theory of massive gravity to have a formulation beyond
the linear regime. In fact, this is also true for the massless case both from
observational (perihelion of Mercury) and theoretical (the {\em equivalence principle})
considerations. In the massless case, the gauge invariance
can be a guiding principle in this
extension and it is usually stated that the only consistent final result is GR in the usual
geometrical formulation (\emph{i.e.} having the whole group of diffeomorphism as a gauge group)
\cite{Kraichnan:1955,Ogiev:1965,Boulanger:2000rq,Wald:1986bj,Deser:1969wk,FeynmanGrav,Gupta:1957}.
The presence of the mass term (\ref{PFmass}) breaks the gauge invariance of the
linear theory and  it is not clear how to build a non-linear theory consistently.
One could consider adding a term to the full GR Einstein-Hilbert Lagrangian
that in the weak field limit reduces to (\ref{PFmass}). Since
no scalar can be built out of the metric alone without including derivatives,
 either one relaxes the invariance under
diffeomorphisms, or other dynamical fields
should be added  to the theory (see below). A  possibility in the first approach
 consist of adding a static background ({\em e.g.} Minkowski
space-time) and defining $h_{\m\n}$ and the mass term as in (\ref{metricper}) and (\ref{PFmass}).
However,
in this case, besides breaking of the background independence of the theory, the
Hamiltonian is not bounded from below. This can also be  understood through the appearance of a
mode  with a negative kinetic energy which propagates at the nonlinear level (Boulware and
Deser mode)
 \cite{Boulware:1973my,Creminelli:2005qk} (see also \cite{Gabadadze:2004iv}).
A related problem of this proposal is that the spherically symmetric solution with
 flat boundary conditions\footnote{Remind that, in general,
 the Birkhoff theorem does not hold
 in
 modified theories of gravity \cite{Stelle:1977ry,Dai:2007rv}.}
  presents a singularity at finite radius \cite{Damour:2002gp} (see also
  \cite{Jun:1986hg}).\\

An approach more similar to the massless case can be followed, based on the St\"uckelberg formalism of
compensators for massive gauge theories \cite{Stuck:1938} (see \cite{Ruegg:2003ps}
for a review).  Currently, this
approach has been developed until third order \cite{Zinoviev:2006im}.
Besides, a version of the Brout-Englert-Higgs mechanism to give mass
to vector fields can be applied to spin-2 \cite{Hooft:2007bf}. The idea in both cases
is to add new degrees of freedom coupled to the massive graviton
in a way that the theory has a gauge invariance which makes them spurious.
The presence of a gauge invariance at linear order may then be used to guess the non-linear terms
as the non-linear extensions of linear gauge invariance must satisfy certain consistency
conditions, such as the closure of the associated algebra \cite{Henneaux:1997bm}.
Both approaches encounter problems with unitarity, which may be
understood from the counting of the degrees of freedom. The number of new fields required
for a diffeomorphism invariant formulation of massive gravity is 4, whereas the massive and
the massless theories differ by just 3 degrees of freedom. This means that besides
the spin-2 degrees of freedom, the gauge invariant formulations generically
include a new scalar. This field must be a ghost in flat space since
the only ghost-free
possibility for Lorentz invariant massive gravity only has tensor
degrees of freedom
and this destroys the consistency of the theory (see, however, \cite{Porrati:2001db}
for a successful model in AdS).

 The Fierz-Pauli
mass term is singled out from the rest of Lorentz preserving mass terms because
at the {\em linear level}
this new degree of freedom disappears in Minkowski space. This
allows for a successful St\"uckelberg formulation of massive gravity at linear order \cite{ArkaniHamed:2002sp}.
 However, the dangerous ghost mode
 reappears once the non-linear effects are taken into account
\cite{Boulware:1973my}. Furthermore, around non-trivial
sources the ghost is also present at the linear level \cite{Creminelli:2005qk}.
In particular, this means that for the Fierz-Pauli mass term
 \emph{any} non-linear extension breaks down at length scales beyond the radius
where the non-linear effects can cure the vDVZ discontinuity.\\

The previous negative conclusions may change
if Lorentz invariance is broken \cite{Gabadadze:2004iv} (see \cite{Rubakov:2008nh}
for a review).
In that case, there are more possibilities for mass terms which are unitary and are
not affected by strong coupling \cite{Rubakov:2004eb,Dubovsky:2004sg}
(see also \cite{Dvali:2007ks,Bluhm:2007bd,Jackiw:2007br} for other aspects of Lorentz violation
and gravity).
As the mass term explicitly breaks Lorentz invariance, the massive polarizations
do not necessarily correspond to spin states. This kind of models
appears naturally when more fields are added to GR, and
 {\em bigravity} (to be discussed below) is perhaps the simplest possibility\footnote{Another
  possible generalization is to consider non-local extensions \cite{Dvali:2002pe,Dvali:2006su}.}.\\

%To sum up,  the guiding principles that one may use  to
%find a non-linear extension of the theory
%of massive spin-2 are the  equivalence principle, the conservation of the number
%of {\em gravitational}
%degrees of freedom, unitarity, absence of strong coupling
%at low energies, smooth massless limit and causality.

%%%%%%%%%%%%%%%%%%%%%%%%%%%%%%%%%%%%%%%%%%%%%%%%%%%%%%
\subsubsection*{{\em Large Extra Dimensions and Braneworlds}}
%%%%%%%%%%%%%%%%%%%%%%%%%%%%%%%%%%%%%%%%%%%%%%%%%%%%%

From the previous section, it seems clear that a covariant non-linear theory with massive gravitons
requires the presence of new fields coupled to the graviton.
The theories with extra spatial dimensions provide such fields as the
pure massless graviton in higher dimensions can be understood as a
  four dimensional field theory with an infinite tower of modes
  interacting with each other\footnote{Besides,
the presence of extra-dimensions is necessary for consistent string theory \cite{Polchinski}.} \cite{KKBook}.
This  provides a method to find
consistent coupling of massive gravitons in fixed backgrounds \cite{Argyres:1989cu,Nappi:1989ny}.
Nevertheless, it should be noted that those completions are not consistent
in general unless the infinite tower of modes is considered
\cite{Duff:1989ea}.\\

The simplest possibility  is that the
 extra dimensions are compact
 with a typical size $L$.   In this case, the extra dimensions can be
understood as a massless graviton coupled to a
\emph{discrete} tower of massive fields with masses depending on the size and
topology of the compact manifold \cite{ArkaniHamed:2001ca,KKBook}.
If two test masses  $m_1$, $m_2$ are placed within a distance
$r\gg L$ the gravitational flux lines can not spread in the extra compact dimensions.
Only the massless mode is excited at this energy scale
and the usual four dimensional potential potential (\ref{potential}) is obtained,
\be
V(r)\sim\frac{m_1m_2}{M^{2+d}_{Pd} L^d}\frac{1}{r},
\ee
where $d$ is the number of extra dimensions and $M_{Pd}$ is the
gravitational scale of the theory.
The effective four dimensional Plank mass in this set-up is easily read comparing the previous
expression with (\ref{potential}),
$$
M^2_P=M^{2+d}_{Pd} L^d.
$$
For distances of the order $L$ and below, the gravitational interaction is modified by the
tower of massive modes. The fact that Newton's law has not been probed at distances smaller than
than $10^{-2}$ millimeters \cite{Decca:2007jq,Geraci:2008hb,Kapner:2006si}
allows for a $L\sim 10\ \m\mathrm{m}$ and a fundamental Planck mass
$M_{Pd}\gtrsim 1 \ \mathrm{TeV}$ for $d\geq 2$ \cite{Kapner:2006si,ArkaniHamed:1998rs}.

 If the Standard Model fields live in the
 bulk, the Kaluza-Klein (KK) reduction affects
all the interactions. However, the Standard Model interactions
 have been accurately
measured at the weak scale $m_{EW}\sim 1 \ \mathrm{TeV}$
and this gives the constraint
$L<  m^{-1}_{EW}\sim \ 10^{-17}\ \mathrm{mm}$.

A way to circumvent the previous arguments is by localizing the Standard Model fields in a four dimensional
submanifold of a certain width $L_{D}$ (\emph{domain wall} or {\em brane})
\cite{Rubakov:1983bb,ArkaniHamed:1998rs,Dvali:1996xe}. This idea
introduces
two length parameters apart from the Planck length: the size of the extra dimensions $L$ and the
width\footnote{This length scale can be arbitrarily small.}
of the defect $L_{D}$. If gravity is not localized, these parameters can be chosen so that
 gravity is modified at the submillimeter scale and the compact
extra dimensions are large in comparison with the electroweak scale. In this scenario,
 gravity is  modified
at high energies and remains massless and four dimensional at large distances\footnote{Another
 way of localizing fields in submanifolds is provided by string theory and $D$-branes
\cite{Polchinski}.}.\\

An alternative to the existence of compact dimensions is provided by warped extra dimensions
(not necessarily compact but of finite volume and with the
Standard Model fields localized in a brane)
 \cite{Randall:1999vf,Randall:1999ee} (see \cite{Maartens:2003tw} for a review).
  In this scenario, known as Randall-Sundrum scenario, the extra dimensions are not
 factorized and solutions with nontrivial warped factors of typical curvature
 $L_{W}^{-1}$ exist and give rise to  massless zero modes and a continuous tower of massive states
 without a mass gap.
 Nonetheless, the gravitational interaction is again four dimensional for length scales
 larger than $L_{W}$. The effect of the warped factor can be understood as a potential
that makes the wave functions of massive states to be suppressed in the brane,
and the final effective non-relativistic potential for two sources
in the brane can be written as
\be
\label{RSpot}
V(r)\sim \frac{m_1 m_2}{M_P^2}\frac{1}{r}+\frac{m_1 m_2}{M_P^2}L_W^2\int_0^\infty  \di m
\ m \frac{e^{-m r}}{r}=
 \frac{m_1 m_2}{M_P^2}\frac{1}{r}\left(1+\frac{L_W^2}{r^2}\right).
\ee
From the previous expression we see that a mass gap in the spectrum in not required
to obtain a correct Newtonian limit because
 the coupling of massive modes to matter is suppressed by a factor $m L_W^2$.
Again, this model proposes modifications to the gravitational interaction only
at high energies.\\

There are many generalizations of the previous model, and we would like to focus on
those where GR is also modified in the infrared.
In \cite{Kogan:1999wc}, the number of branes is increased to three:
two of positive tension and laying in the fixed point of an orbifold and a third brane with
negative tension placed between those two. The final result is the existence of a mass gap
 between the first massive mode and the rest of the tower of KK states. That
  makes it possible to integrate out the heavy modes and consider a theory with only
two gravitons at intermediate distances (\emph{bigravity}).
Finally, for large distances, the massive mode
is frozen and only the massless mode remains. Thus, there are two scales in which
gravity is modified: one related to the first massive mode and the other one related
to the mass of the second massive mode. Unfortunately, the branes of negative
tension do not satisfy the
null energy condition. This has been related to
Hamiltonians which are unbounded from below, which makes the theory ill-defined \cite{Witten:2000zk}.
This problem is related to the stabilization
of the branes positions. In principle, the branes are dynamical objects whose
relative  distances fluctuate
and these fluctuations must be stabilized. For the case of branes with negative
tension, this degree of freedom (the relative
distance of the branes or \emph{radion})
 is a ghost and its stabilization is an important issue in brane
physics \cite{Goldberger:1999wh,Goldberger:1999uk,Garriga:2000jb,Garriga:2002vf}.

A ghost-free \emph{bigravity} scenario
was presented  in \cite{Kogan:2000vb}, where the addition of a non-trivial
background in the branes allows for a model with two light gravitational modes
without ghosts or vDVZ discontinuity. However, in this case the deviations from
GR occur at distances which are not observable.

In \cite{Padilla:2004mc},
the author considers two five-dimensional spacetimes separated by a domain wall and allows
for different Planck masses in the two separated regions. This setup
admits solutions with asymmetric warp factors and introduces
modifications of GR both at long and at short distances.
This model suffers from the vDVZ discontinuity which may be cured
through the non-linear interactions. As we discussed previously, this
implies that the theory has a low energy cut-off, although it was argued in
\cite{Padilla:2004mc} that this scale may be set to the Planck scale.

Other possible generalizations including
regularized (thick) branes
and intersecting branes can be found in \cite{Csaki:2000fc} and references therein.
Finally, we would like to mention a recent proposal of
 an asymmetric background with a induced gravity
term (see below) where some of the previous
problems are absent \cite{Charmousis:2007ji}.\\

Besides the linear approximation, it is interesting to study how some non-linear predictions
of GR are modified in the models with large extra dimensions. Many studies have been
devoted to cosmology in the presence of large extra dimensions
(see {\em e.g.} \cite{Brax:2003fv,Brax:2004xh,Langlois:2002bb}). In the models related to the Randall-Sundrum
scenario,
 the standard Friedmann equation is modified at high energies
  on the brane of positive tension,
which sets some phenomenological constraints in the parameters of the theory
 and there is also no-conservation
of energy on the brane as some matter can leak to the extra-dimensions
\cite{Binetruy:1999ut,Csaki:1999jh,Cline:1999ts}.
 The parameters in the models can be tuned so that
these modifications are phenomenologically acceptable.

Inflation is also modified in models with large extra dimensions and branes.
Apart from new mechanisms of inflation (such as collision
of branes) the modification of Friedmann equation implies that slow-roll inflation may be
possible for potentials that are too steep for ordinary cosmology \cite{Maartens:2003tw}.
Besides, some other aspects of cosmology, such as the growth of cosmological
perturbations and structure formation, may be modified
in the presence of large extra dimensions  (see \emph{e.g.}
\cite{Koyama:2006ef,Koyama:2007rx,Maartens:2003tw,Charmousis:2006pn,Gregory:2007xy} and references therein).

%%%%%%%%%%%%%%%%%%%%%%%%%%%%%%%%%%%%%%%%%%%%%%%%%%%%%%
\subsubsection*{{\em Metastable gravitons}}
%%%%%%%%%%%%%%%%%%%%%%%%%%%%%%%%%%%%%%%%%%%%%%%%%%%%%

Another way in which gravity is modified at large distances is provided by models
where the four dimensional graviton is not a normalizable eigenstate of the linearized
theory but a metastable resonance with a finite lifetime \cite{Csaki:2000pp,Dvali:2000rv}. The
basic idea is that if  the graviton is a resonance, its propagator for momentum close
to  the resonance mass $m_r$  can be
written as (neglecting the tensor
structure)
\be
P(k)\sim  \frac{1}{k^2-m_r^2+\I m_r \Gamma},
\ee
where $\Gamma$ is the width  of the resonance.
The previous expression admits a spectral representation
$$
\frac{1}{k^2-m_r^2+\I m_r \Gamma}=\int \di s \frac{\r(s)}{s-k^2+\I\e},
$$
where $s$ is the Mandelstam variable and $\r(s)$ is a spectral density
\cite{ArteagaBarriel:2007hk,Dvali:2000rv}.
Assuming that the resonance lifetime is very big the potential produced by exchanging of such
a particle between two static sources is
\be
V(r)\sim \int \di s \r(s)\frac{e^{-\sqrt{s}r}}{r},
\ee
which for a peaked spectral density $\r(s)$ around the resonance mass $s=m_r^2$ reduces to the standard
Newtonian interaction at distances $r\ll m_r^{-1}$ and is modified at large distances
(or late times) where the resonance decays into the eigenvalues of the theory.\\

This kind of behavior can be reproduced by higher dimensional set-ups.
A particular model where gravity opens up  at long distances due to the presence
of a metastable four dimensional graviton and which can have also a modified fundamental
scale of quantum gravity is provided by
the localization of gravitons on a brane, but not
completely \cite{Gregory:2000jc,Karch:2000ct}. In this set-up,  the relevant fact is that the
extra dimension is warped, asymptotically flat but with an infinite volume
which makes the zero mode non-normalizable. This background yields two length scales
related to the length at which the crossover to flat space occurs and to the curvature in the extra dimension.
 In this model, there is a
resonant mode at zero momentum in the extra dimension that can   be interpreted
 as a  metastable four dimensional graviton with a certain width $\Gamma$
and decaying into the eigenmodes of the theory which spread in
the extra dimensions \cite{Csaki:2000pp}.
This $\Gamma$ is thus related to the large scale at which
the four dimensional description breaks down.

Generalizations of these models
which connect them to the bigravity scenario are provided by the inclusion of more
3-branes in the model
\cite{Kogan:2000xc,Kogan:2000cv}. These scenarios
interpolate between a spectra with more than one ultralight massive graviton and
the appearance of resonances \cite{Kogan:2000xc}. Many aspects
of these models were summarized in \cite{Papazoglou:2001cc}.

The fact that the resonant mode is built out of massive modes (without a
massless zero mode) implies the presence
of the vDVZ discontinuity in these models \cite{Dvali:2000rv}. However,
the presence of matter in the brane produces a bending of the brane which restores the
right tensor structure of the propagator \cite{Garriga:1999yh,Gregory:2000iu}. As we have
argued, the only way in which the vDVZ discontinuity can be cured at the
linear level is through the introduction of ghost states and their presence in these models
 was shown explicitly in
 \cite{Pilo:2000et}. This makes them
quantum mechanically ill-defined at the linear level\footnote{Of course, at non-linear level or at
high energies, the theory can have a well defined UV completion, even if this possibility
has been questioned in \cite{Adams:2006sv}.}. It was argued in \cite{Kogan:2000cv} that,
in the brane models, the condition that
the energy-momentum tensor must satisfy to stabilize
 the brane configuration directly implies
the right tensor structure.
In this case the ghost state decouples from matter at the linear level \cite{Kogan:2000xc}.
Besides, the previous models involved branes with negative tension free to fluctuate
which implies the lack of energy-positivity in this scenario \cite{Witten:2000zk}.

%%%%%%%%%%%%%%%%%%%%%%%%%%%%%%%%%%%%%%%%%%%%%%%%%%%%%%
\subsubsection*{{\em Induced gravity: DGP}}
%%%%%%%%%%%%%%%%%%%%%%%%%%%%%%%%%%%%%%%%%%%%%%%%%%%%%

A related possibility, pointed out by Dvali, Gabadadze and Porrati (DGP henceforth),
 is provided by factorized non-compact extra dimensions of infinite
volume with induced
terms in a $3$-brane \cite{Dvali:2000hr}. In these models
 one includes a four dimensional action for gravity in the brane which is
compatible with
the symmetries of the set-up. Thus, even if it is absent classically,
it may be generated on a brane by the loops of the matter localized
in the brane. For simplicity let us consider the case of just
one extra dimension.

The gravitational interaction is five dimensional
except in the brane where the induced term produces modifications to this behavior at distances
smaller than $$l_{DGP}=\frac{L^3_5}{L^2_{4}}$$ where $L_5$ is the five dimensional Planck length
which sets the scale of quantum gravity effects
and $L_{4}$ is the length scale of the induced term. The propagator in this case evaluated on the brane
takes the form
\be
\label{dgpprop}
P(x)\sim \int \di^4 k \frac{e^{-\I kx}}{k^2+2 \sqrt{k^2}/l_{DGP}},
\ee
whose interpretation is the following. A graviton emitted by the
source localized on the brane propagates along the brane and
gradually dissipates into the bulk. The lower the frequency of the
signal, the faster it leaks in the extra dimension. This is similar
to what happened in the previous model of metastable gravitons (see
also \cite{Dvali:2002pe}).
The potential between two test particles in the
brane and separated by a distance $L_5\ll r\ll l_{DGP}$ is \cite{Dvali:2000hr}
\be
V(r)\sim  L_{4}^2\frac{m_1 m_2}{r}\left(\frac{\pi}{2}+\frac{r}{2l_{DGP}}\left[-1+\g+
\ln \left(\frac{r}{2l_{DGP}}\right)\right]+O(r^2)\right),
\ee
which implies the identification $L_{4}\sim l_P$. For $r\gg l_{DGP}$ the gravitational
interaction is five dimensional, \emph{i.e.}, the potential satisfies the five dimensional Laplace equation
whose solution is of the form $r^{-2}$. It is interesting to note that for similar setting with more
than one extra-dimensions the evaluation of the propagator is more involved (see {\em e.g.}
\cite{deRham:2007xp} and references therein).\\

The tensor structure of the propagator in DGP is that of a massive graviton (which may be related
to the infinite volume of the extra dimension) which means that it suffers
 from the vDVZ discontinuity
\cite{Dvali:2000hr,Luty:2003vm}. As argued in \cite{Deffayet:2001uk},
its resolution in this model may be related to the strong coupling
phenomenon. As happens for the Fierz-Pauli mass term of massive gravity, in DGP there is a mode (related to
the extrinsic curvature of the brane)
which gets strongly coupled at large distances as compared to the rest of modes \cite{Luty:2003vm}.
More concretely, the cross-over scale
at which there is a strongly coupled mode is \cite{Luty:2003vm,Rubakov:2003zb,Nicolis:2004qq}
$$\Lambda_{DGP}\sim ( L_4 l^2_{DGP})^{-1/3}.$$
In the presence of a source $M$, the non-linearities set in  at a distance
$r_c\sim (M L_4^{2}l_{DGP}^2)^{1/3}$ which for the Solar System is far bigger than the
distance where the deflection of light by the Sun has been measured. Even more,
it was shown in \cite{Deffayet:2001uk} that for certain sources,
at distances smaller than $r_c$ the full non-linear solution approaches
that of GR (see also \cite{Gruzinov:2001hp}). Unfortunately, the
exact solution for a static spherically symmetric source in the brane is not known even
if one expects that the non-linearities may also help to circumvent the
vDVZ discontinuity \cite{Gabadadze:2004iy} (see also \cite{Dvali:2006if} for the exact domain wall
solution).

As happens in \emph{massive gravity}, the strong coupling of a mode at
relatively small
energy scales can be quite problematic as it may introduce a rather low UV cut-off.
If the crossover scale to Newtonian gravity is of the order
of the Hubble length, the scale of strong coupling is  $\Lambda_{DGP}\sim (1000\ \mathrm{km})^{-1}$ \cite{Luty:2003vm}
and a theory of quantum gravity would be needed to deal with calculations at distances smaller than
$\Lambda^{-1}_{DGP}$. As for massive gravity,
 these conclusions depend on the UV completion of the theory. For DGP,  a non-generic prescription
to choose the counterterms was proposed in \cite{Nicolis:2004qq}, in  a way that the quantum
corrections are not important in all the astrophysical situations (see also \cite{Dvali:2004ph}).
As we already said, the loop expansion may admit a resummation such that the scale
$\Lambda_{DGP}$ is unphysical
(indeed this is what happens for the classical expansion \cite{Dvali:2004ph}).

Similarly to the case of massive gravity, the previous results change in the presence
of curvature. More concretely, positive curvature increases the scale of strong interaction
and yields a ghost for large curvatures (compared to $l_{DGP}^{-1}$)
whereas negative curvature decreases it \cite{Luty:2003vm}.\\

DGP models are phenomenologically very interesting because they not only modify
 the scale of quantum gravity (which
is now $L_5$) but they also predict a modification of
the gravitational interaction at long range which
may have interesting consequences in cosmology (see \cite{Lue:2005ya} for a review).
In the DGP
model, the Friedmann equation is modified and can mimic the behaviour of
a cosmological constant \cite{Deffayet:2000uy,Deffayet:2001pu,Koyama:2007rx}.
In particular, self-accelerating solutions are found in the brane without the
need of a cosmological constant, and they provide an alternative
to dark energy \cite{Deffayet:2000uy}.
Even if these solutions are
interesting it has been argued that they suffer from the presence of a ghost state
which makes them quantum mechanically unstable
\cite{Nicolis:2004qq,Luty:2003vm,Izumi:2006ca}.

Again, other aspects of cosmology, such as inflation
or the behaviour of perturbations
and structure formation in the DGP model may differ
from GR \cite{Koyama:2007rx,Koyama:2005kd,Lue:2004rj}.\\

%%%%%%%%%%%%%%%%%%%%%%%%%%%%%%%%%%%%%%%%%%
\subsubsection*{{\em Addition of Scalar or Vector Fields}}
%%%%%%%%%%%%%%%%%%%%%%%%%%%%%%%%%%%%%%%%%%

So far we have presented models of non-linear massive gravity
which involved only the metric (possibly in the presence of
extra dimensions). As we already stated, from the four dimensional point of view,
the introduction of extra dimensions can be understood as
the addition of an infinite number
of fields  in a precise way which allows
for general covariance in higher
dimensions \cite{ArkaniHamed:2001ca}\footnote{A related possibility is
considering higher dimensional QFT
where the presence of a four dimensional defect induces GR in it
 \cite{Dvali:2000xg,Adler:1982ri,Akama}}.
The reason why these modifications are considered natural nowadays is because of the need
of extra dimensions in some extensions of GR, such as string theory.
However, from a purely four dimensional point of view the addition of a
 finite collection
of new fields coupled to the graviton and/or to matter
 seems a much simpler possibility\footnote{Besides, as we have seen, there
are models with extra-dimensions with a spectrum with a mass
gap which yield these theories at low energies.}. Indeed, independently of the
 modern ideas of extra dimensions,
the phenomenology of the addition of new fields which couple to matter
 has been a subject of constant research \cite{Will}. The more conservative
possibility is adding relativistic fields of different spin. These fields
may condensate generically giving rise to Lorentz breaking mass terms for the
gravitons (or to a cosmological constant in  certain cases). Let us say a few
words about the most studied possibilities.

Before, it is fair to say that the possibility that a simple
 model gives rise to an adjustment mechanism yielding
 a small
cosmological constant does not seem possible \cite{Weinberg:1988cp}\footnote{It is also
true that none of the previously mentioned possibilities provides this mechanism.}
.
\\

Models where a scalar field is added
 to the gravitational interaction have been studied for many years
 \cite{Fujii,WeinbergGC,Will,Brans:1961sx}.
The standard approach consist of adding a  scalar field  to the
GR action with some free parameters which allow for interesting new phenomenology
\cite{Will:2001mx,Will}. For a recent review on some proposals of scalar fields models of dark energy see
\cite{Copeland:2006wr}.
The origin of the scalar field can be fundamental, as happens in string theory, or purely phenomenological.
This field can also couple to matter and, depending on parameters such as the mass of the field, the interaction
 is modified at a certain distance.

Recently, there has been some interest in models with   non-standard Lagrangians, such as the
case of the ghost condensate \cite{ArkaniHamed:2003uy} (earlier attempts to
apply non canonical kinetic terms to the CC problem can be found in \cite{ArmendarizPicon:2000dh}).
 In these models, the vacuum solution is a time
dependent configuration for the scalar field together with a flat metric. The fact that the vacuum
breaks some of the Lorentz symmetries gives rise to a consistent modification
of GR at large distances and the model can be generalized to obtain a Lorentz breaking mass term
for the graviton \cite{Dubovsky:2004sg,Rubakov:2008nh}.
The phenomenology of this scenario is very interesting and different from the standard approach
(see \emph{e.g.} \cite{Bebronne:2007qh,Rubakov:2008nh} and references therein). On the other hand,
the thermodynamic properties of black holes are problematic
when the Lorentz symmetry is violated \cite{Jacobson:2008yc}.\\

The next possibility to modify gravity in the infrared is by adding a vector field that condensates.
Some examples with
spontaneous breaking have also been considered in recent years (see {\em e.g.}
\cite{Tartaglia:2007mh,Libanov:2005nv,Gripaios:2004ms,Zlosnik:2006zu}). Again, those
models present some regions in the parameter space which are phenomenologically acceptable
and more non-trivial checks are necessary to rule them out or to accept them
as plausible models.\\

Recently, models which include a vector and a scalar field coupled to the graviton
have been considered in the context of dark matter.
Along with the cosmological observations, another motivation to modify GR at large distances
is that the total gravitational field of different astrophysical objects in the Universe surpasses
by far what we expect from the baryonic mass we can see. The standard solution of this
problem is to invoke the existence of a exotic form of matter which does not couple to light (dark
matter, DM) \cite{Navarro:1995iw}. However, one can take a different point of view and try to modify Newton's
law to avoid the introduction of exotic matter. A very successful possibility dubbed MOND (Modified
Newton Dynamics) consist of modifying Newton's law
not at a certain length scale but at a certain \emph{acceleration} scale \cite{Milgrom:1983ca}. Recently
a relativistic version of MOND has been proposed. It includes vector and scalar fields which couple
non trivially to the metric\footnote{The fact that
gravity is modified at a curvature scale seems to be related to the presence of
derivative couplings.}, and thus can be considered as a particular example of the general
scalar-vector-tensor theories (see {\em e.g.} \cite{Skordis:2005xk,Bruneton:2007si} for a recent review
and \cite{Moffat:2007nj} for the related MOG theory).

%%%%%%%%%%%%%%%%%%%%%%%%%%%%%%%%%%%%%%%%%%%%%
\subsubsection*{{\em Addition of a Tensor Field: Bigravity}}
%%%%%%%%%%%%%%%%%%%%%%%%%%%%%%%%%%%%%%%%%%%%%%

One of the possibilities we will  focus on in this dissertation
is \emph{bigravity}.
This theory consists
of two rank-2 tensor fields, \emph{i.e.}  two metrics. The first
thing we may notice is that there are some theorems
that forbid the interaction of massless gravitons (see {\em e.g.} \cite{Boulanger:2000rq}). This means that when
two metric fields interact non-trivially one of them will always acquire a mass.
The phenomenology
of theories with a fixed metric background (or aether),  known as
\emph{bimetric} theories, has been studied in   \cite{Will}.
A slight generalization consists of allowing for both metrics to by dynamical  (see
\cite{Damour:2002wu} and references therein). This possibility is known as \emph{bigravity}.
One of the key ingredients of the theories with more than one field is the \emph{physical}
metric, \emph{i.e.} the field that produces the gravitational interaction between the matter of
the Standard Model.
Having two metrics at our disposal, any combination of them can be considered
as the \emph{physical} metric while the interaction between both metrics will
produce a massive and a massless graviton.\\

The main motivation to focus on {\em bigravity} is that
it offers a simple modification of GR where the
gravitons {\em can} be massive and where there are known {\em non-linear} exact solutions.
This may help to clarify some of the difficulties that we have outlined. Besides,
the Lorentz breaking mass terms appear quite naturally in these theories,
which means that some of the difficulties of the linear analysis encountered
in the Lorentz invariant case may be absent.

%%%%%%%%%%%%%%%%%%%%%%%%%%%%%%%%%%%%%
\subsubsection*{{\em Unimodular Gravity}}
%%%%%%%%%%%%%%%%%%%%%%%%%%%%%%%%%%%%%%%%%

Hitherto we have presented modifications to GR which appear at a certain length scale related
to  some parameters with dimension of length which are present in the model\footnote{Besides,
there may be a source dependent scale.}. As we have seen,
they are sometimes related to the
appearance of a preferred frame which breaks the diffeomorphism invariance of the theory. One may wonder about
the mildest way of introducing this modification, \emph{i.e.} about
the possibility of sending the length scale
to infinity or about keeping a large
subgroup of the diffeomorphisms as a gauge invariance of the theory. It turns out that
both possibilities are related and this modification of GR is dubbed \emph{unimodular} gravity
\cite{vanderBij:1982,Unruh:1988in}. Unimodular gravity dates back to the work of Einstein himself
who discovered that the Einstein's equations are equivalent to their traceless part except for the
appearance of an integration constant which plays the role of a cosmological constant. Thus,
both equations of motion coincide except for a zero mode. The interesting thing is that
the traceless part of the Einstein's equations can be derived from Lagrangians which have a fixed volume
element. In a sense, this is the minimal way in which a background can be added:
we just include a privileged volume form, whose presence breaks the group of diffeomorphisms
to its transverse part. As we just said, this is enough to modify the problem of the cosmological
constant, even if it does not quite solve it \cite{Weinberg:1988cp}.
As we shall see, the transverse part of the diffeomorphisms (TDiff) appears
naturally in the theories of spin-2.\\

Finally, a common feature of the different scenarios that modify gravity is that
they must admit the embedding in a complete
theory of quantum gravity (UV completion). This issue has been addressed recently
in \cite{Adams:2006sv} but the results are controversial. It is fair to say that
there are some models whose embedding in string theory seems possible (as {\em e.g.} the Randall-Sundrum
 model \cite{Verlinde:1999fy})
whereas for other models such as DGP or the ghost condensate it is not clear how to find them
in UV complete theories (see also \cite{Gregory:2007xy} for a list of
other problems that may appear in DGP at the quantum level).

%%%%%%%%%%%%%%%%%%%%%%%%%%%%%%%%%%%%%%%
\section{Outline and Summary of the Thesis}
%%%%%%%%%%%%%%%%%%%%%%%%%%%%%%%%%%%%%%%%%%

The body of the Thesis is divided into three parts.
The first part (Chapters \ref{chapterLorentz} and \ref{chapterRS}) is devoted
to the analysis at the linear level of certain gauge theories related
to gravity, whereas the non-linear
extensions are presented in the second part (Chapters \ref{chapternl},
\ref{chapterbigra} and \ref{chapterperturbbigrav}).
The third part contains the conclusions (Chapter \ref{chapterconclu}) and
three appendices which contain aspects related
to the Thesis but which are not essential to it. Every Chapter begins
with a summary of the contents and main results.\\

In  Chapter  \ref{chapterLorentz}
 we will study the most  general quadratic
Lagrangian  of second order in derivatives
for rank-2 symmetric tensors which preserves Lorentz invariance,
in order to see which possibilities yield
a consistent modification of the usual Lagrangian coming from the linearization
of GR (with the possibility
of a mass term). The Chapter is based on \cite{Alvarez:2006uu}.
As it is well known, a symmetric rank-2 tensor has more degrees of
freedom than those required for the propagation of a massless particle, and the presence
of a gauge
invariance is required if we want to match both counts. This
is the reason why we will first focus on the characterization of the
different gauge invariances which the previous Lagrangians can enjoy.
Out of them, two possibilities are singled out as involving a larger
number of free parameters: the linearized diffeomorphisms (Diff)
of GR and its transverse part (TDiff) enlarged with a Weyl transformation (WTDiff).
Even if both possibilities correspond to inequivalent Lagrangians, we will
show that the equations of motion (EoM) coincide in both cases
except for the appearance of an integration constant.

We will then analyze the general Lagrangians and find the constraints in the
parameters that prevent the appearance of ghosts and tachyons. As
expected, the consistency of the theory will imply the presence of a gauge invariance
which can be smaller than the Diff or WTDiff. The consistent
theories are equivalent to {\em scalar-tensor} theories except in
those two cases.

The next step will be to study the consistency of the general Lagrangian once a
Lorentz preserving mass term is included. Contrary to what happens in the massless
case, we will find just \emph{one} possibility which is free of ghosts and tachyons and that
gives mass to the tensor modes, which corresponds
to the Fierz-Pauli (FP) choice \cite{Fierz:1939ix}.

After a comment on an alternative derivation of the WTDiff and Diff Lagrangians, we will
devote the rest of the Chapter to study  the propagators that mediate the
 interaction between conserved sources in the consistent cases.
We will discuss in some detail the gauge fixing of the TDiff theories, which is not trivial
as the gauge invariance is \emph{reducible} (\emph{i.e.}, there is a condition between the gauge parameters),
and the issue of the consistent coupling to matter, as the TDiff subgroup allows
 the graviton  to be coupled to a source which is conserved except for
a divergence. We will finally set some phenomenological bounds on the mass and coupling
constant of the extra scalar field present in the TDiff invariant case. This mode
disappears in the theory invariant under the WTDiff group, whose propagator
coincides {\em on-shell} with that of linearized GR.\\

Chapter \ref{chapterRS} is devoted to the extensions of the ideas of Chapter \ref{chapterLorentz}
to the fermionic counterpart of spin-2: the spin-$3/2$ field. The Chapter is partially
 based on \cite{Blas:2008ce,Blas08}. We will first study the most
 general first order Lorentz invariant Lagrangian for the vector-spinor field
$\psi_\m$. As happens for any massless field of spin higher than $1/2$, the
description in terms of a covariant field includes more degrees of freedom than
the physical polarizations of the massless particle.
We will find that there are just two possible Lagrangians which enjoy a gauge invariance
that may render the extra degrees of freedom spurious: the Rarita-Schwinger (RS)
Lagrangian \cite{Rarita:1941mf} and another possibility endowed with a $S$-symmetry (WRS).
We will study the equations of motion for both possibilities and find
that the WRS Lagrangian has an extra spin-$1/2$ PDoF. To study whether
this new degree of freedom yields different physical predictions,
we will couple the field $\psi_\m$ to a conserved fermionic current
and study the propagator that mediates the interaction between the conserved currents
in the WRS case. As we will show, the propagator coincides with that of RS.

After making some remarks on the consistent coupling of the WRS Lagrangian to
$U(1)$ gauge fields, we will study the possibility of finding
a supersymmetric Lagrangian built out of the WTDiff Lagrangian for spin-2
 and a certain Lagrangian for the
spin-$3/2$ field. We will show in the last part of the Chapter that, unless more
ingredients are included in the set-up, this does not seem to be possible.\\

After the linearized study, in the second part of the Thesis
we embark on the non-linear extensions of the spin-2 Lagrangians. If the spin-2 particle is related
to the actual graviton, it must account for the \emph{equivalence principle}. In other words,
it must be coupled universally to any kind of energy including its own. This paves the way to
the addition of non-linearities to the Lagrangian to get a consistent self-interacting theory
of gravity.

In Chapter \ref{chapternl} we will study non-linear extensions of the TDiff Lagrangians
of Chapter \ref{chapterLorentz}.
This Chapter is based on \cite{Alvarez:2006uu,Blas:2007pp}.
We will first address the issue constructively following the approach developed in \cite{Deser:1969wk}
for the Diff case and we will find that the analogous construction is not successful for WTDiff. It
is however easy to construct a consistent extension based on the intuitive non-linear extension
of the TDiff group, which will be the transverse subgroup of the non-linear diffeomorphisms.
We will show the equivalence between these theories and {\em scalar-tensor} theories.
Concerning the WTDiff linear Lagrangian, we will find a {\em unique} non-linear
Lagrangian of second order in the derivatives of the metric whose
 equations of motion  are equivalent to the Einstein's equations even in the presence of matter except
for the appearance of an integration constant which acts as a cosmological constant
(they are equivalent to those of \emph{unimodular} gravity,
namely the traceless part of Einstein's equations \cite{Weinberg:1988cp}).

Finally, we will consider the first order formulation of the WTDiff non-linear Lagrangian
and comment on the possibility of coupling the metric consistently to a spin-$3/2$ field.\\

Chapter \ref{chapterbigra} is concerned with \emph{bigravity}. It is based on the work that appeared in
\cite{Blas:2005sz,Blas:2005yk,Blas:2007zz,Blas:2007ep}. The framework in which we will
be interested consists of two metrics  interacting through a
a non-derivative term which can be considered as a mass term in the linear approximation.
 We will choose a minimal possibility for the coupling to matter in which there
are two kinds of matter each of which is coupled to one of the metrics ({\em weakly
interacting worlds}).

After finding the conditions for the interaction term to admit
maximally symmetric metrics as solutions
of the equations of motion, we will
focus on spherically symmetric static
solutions and a certain subclass of them with both metrics being  Schwarzschild-(anti)de Sitter
in different coordinates.
 It is interesting
to notice that any potential admits this kind of solutions. Similarly, we will show that
the system of two maximally symmetric and proportional metrics is a general solution of
bigravity and the interaction term reduces to a cosmological constant term.

The rest of Chapter \ref{chapterbigra} is devoted
to the global structure analysis of certain bigravity solutions.
 We will focus on  geodesic completeness and global hyperbolicity
of the solutions.
 One might think that the presence of two causal structures could give
rise to new pathologies, but we will find that this is not necessarily the case.
We will study the behaviour of
the null geodesics for one metric in the conformal compactification of the other metric.
This will lead us to propose a  prescription to construct geodesically complete manifolds
even in the case where one the metrics is geodesically complete whereas
  the companion  metric of the solution is not. We will
illustrate the procedure with some examples.

We will see that, in general, this maximal extension implies the loss of
the global hyperbolicity of the solution.
This problem is not as catastrophic as it may seem and it also appears
in GR. Besides, as we will argue, one expects this  solution to be  unstable
near the analogous of the Cauchy horizon.

Another related issue that we will study is the possibility of building closed timelike curves (CTC)
by using both metrics to propagate signals.
We will prove that this is not possible for {\em all} the solutions of \emph{bigravity}
that we studied in the Thesis.
The coexistence of two causal structures can also have very important consequences in black hole physics
and in the homogeneity problem, but we will not elaborate on them.\\

The next Chapter of the second part, Chapter \ref{chapterperturbbigrav}, deals with the stability
of certain bigravity solutions and is based on \cite{Blas:2007ep}.
 We will first focus on a solution with two flat metrics which breaks
the Lorentz invariance to a common $SO(3)$ invariance. The linearized analysis
will include a Lorentz breaking mass term for one of the gravitons
 and  the PDoF  will be
a spin-2 massless graviton and a spin-2 massive graviton with two polarizations.
We will proceed by coupling the system to matter and show that the corrections to Newton's
law scale with the coupling constant of the metrics (related to the mass
of the graviton). In the limit where this coupling constant
goes to zero (massless limit) we recover the predictions of linearized GR,
 which means that the  vDVZ discontinuity
is absent. We will comment on the apparent contradiction
of this correction with the fact that the non-linear
theories accept Schwarzschild as a solution (where Newton's law is not modified).

The next section is devoted to the analysis of perturbations around two de Sitter metrics which
are proportional to each other. The PDoF will be a massless graviton and a massive graviton with a mass term
which in general will differ from the FP form.
The appearance of a new mass scale in the Lagrangian makes the analysis of the PDoF quite different
from the similar analysis in Minkowski and
one could think that the new mass scale would allow for a hierarchy of scales where
 deviation from FP could be well defined as an EFT till a certain cut-off scale built out of the curvature
 scale
and the mass. We will show that this expectation is not fulfilled in the Lorentz invariant case
and only FP survives as a stable possibility. After a brief comment
on a possible mechanism to {\em offload} the cosmological constant in bigravity,
we will devote the last
section of Chapter \ref{chapterperturbbigrav} to
study the degrees of freedom for  non-covariant mass term in de Sitter and find
that this hierarchy can be realized. This constitute the last section of the body of the
Thesis.\\

The third part of the dissertation contains some general conclusions and the outlook of possible
future directions (Chapter \ref{chapterconclu}) and is supplemented with three appendices.

Appendix \ref{AppendixQ} is devoted
to the study of some quantum aspects of TDiff theories and is based on unpublished
results \cite{Blas08}. The final aim of this approach is to tell whether the TDiff invariant
theories which are
classically equivalent to GR are still equivalent to GR at the quantum level.
We will first comment on the possible differences at the {\em semiclassical} level
and present regularization schemes compatible with TDiff, WTDiff
and Diff invariant theories. The counterterms associated to the different regularizations
may yield observable differences between them.

We will
 then  present a BRST construction that may allow for a covariant quantization
of the theories. The fact of dealing with a {\em reducible}
gauge theory means that new ghosts besides the usual Fadeev-Popov ghosts are required
and we will find a minimal set of fields that makes the BRST transformation
nilpotent.

The Chapter ends with a section devoted to the Euclidean Quantum Gravity formalism
for WTDiff theories where we will show that the convergence of the path integral
in this case
seems to be as problematic as for the Diff invariant case.\\

The second appendix, Appendix \ref{AppendixBig}, has some extra information
on unimodular gravity and bigravity. The first section is devoted
to the integration of
tensor densities on manifolds and some comments on the gauge invariance of
the WTDiff theories.
Finally,
in the last section
we will prove the uniqueness of
the solutions dubbed Type II (see Chapter \ref{chapterbigra}) for a
specific form of the potential.

In Appendix \ref{chaptercastella} we present a summary of the Thesis in Spanish.

%%%%%%%%%%%%%%%%%%%%%%%%%%%%%%%%%%%%%%%%%%%%%%%%%%%%%%%%%%%%%%%%%
\part{Linearized Theories}
%%%%%%%%%%%%%%%%%%%%%%%%%%%%%%%%%%%%%%%%%%%%%%%%%%%%%%%

%%%%%%%%%%%%%%%%%%%%%%%%%%%%%%%%%%%%%%%%%%%%%%%%%%%%%%%%%%%%%%%%%%
\chapter{Lorentz Invariant Healthy Lagrangians} \label{chapterLorentz}
%%%%%%%%%%%%%%%%%%%%%%%%%%%%%%%%%%%%%%%%%%%%%%%%%%%%%%%%

As stated in Chapter \ref{chapterintro},
 it is important to study how gravity can be modified
to obtain a consistent theory of gravitation which differs from GR
in the infrared. This Chapter is motivated by the possible modifications
at the linear level where GR can be understood as
a theory of a massless particle of spin-2 represented by a symmetric rank-2
tensor\footnote{We will restrict to this possibility even if it is also possible to represent
the gravitational field by a \emph{vielbein} $e_{\phantom{a}\m}^a$, whose linearized
limit does not necessarily coincide with that of $g_{\m\n}$,  see {\em e.g.} \cite{Nibbelink:2006sz}.}  $h_{\m\n}$. More precisely,
we will study the most general quadratic Lorentz invariant
 Lagrangians for the  tensor $h_{\m\n}$
and will characterize those which are free from tachyon or ghost instabilities (which will
be dubbed \emph{healthy}).

For the case where the tensor modes are massless, we will show that there is a whole family of Lagrangians
which are phenomenologically viable and which are equivalent to the usual scalar-tensor
theories. Besides, we will find two inequivalent possibilities where the degrees of freedom
are purely tensor modes and which share the same equations of motion (EoM). For the massive
case, we will see that the only healthy possibility is the Pauli-Fierz mass term.
Besides the study of the degrees of freedom, we
will provide the propagator for the \emph{healthy}
theories from which we can read the interaction between
conserved sources and set the first phenomenological constraints.
 As expected, we find a whole family of scalar-tensor possibilities
together with two massless tensor possibilities.
This Chapter is based on \cite{Alvarez:2006uu} (see also \cite{VanNieuwenhuizen:1973fi}
for related previous work and \cite{Kuhfuss:1986rb,Sezgin:1981xs} for a extension including propagating
torsion and higher derivatives).

%%%%%%%%%%%%%%%%%%%%%%%%%%%%%%%%%%%%%%%%%%%%%%%%%%%%%%%%%%%%%%%%%%%%%%%%%%%%%%%%%%%%%%%%%%%%%%%%%%
\section{Massless theory}
%%%%%%%%%%%%%%%%%%%%%%%%%%%%%%%%%%%%%%%%%%%%%%%%%%%%%%%%%%%%%%%%%%%%%%%%%%%%%%%%%%%%%%%%%%%%%

Let us begin our discussion with the most general Lorentz invariant
local Lagrangian for a free massless symmetric tensor field
$h_{\m\n}$ involving just two derivatives,
\be \label{MA}
{\mathcal{L}}= {\mathcal{L}^{I}}+\beta\ {\mathcal{L}^{II}}+ a\
{\mathcal{L}^{III}}+b\ {\mathcal{L}^{IV}},\ee
 where we have
introduced
\bea &&{\mathcal{L}}^I={1\over 4}\ \partial_\mu
h^{\nu\rho}\partial^\mu h_{\nu\rho}, \quad
{\mathcal{L}}^{II}=-{1\over 2}\
\partial_\mu h^{\mu\rho}\partial_\nu h^\nu_\rho, \nonumber\\
&&{\mathcal{L}}^{III}={1\over 2}\ \partial^\mu h\partial^\rho
h_{\mu\rho}, \quad {\mathcal{L}}^{IV}=-{1\over 4}\ \partial_\mu
h\partial^\mu h.
\eea
The first term is strictly necessary for the propagation of spin-2
particles, and we give it the conventional normalization. Before
proceeding to the dynamical analysis it will be useful to consider the possible
symmetries of (\ref{MA}) according to the values of $\beta$, $a$ and
$b$.

%%%%%%%%%%%%%%%%%%%%%%%%%%%%%%%%%%%%%%%%%%%%%%%%%%%%%%%%%%%%%%%
\subsection{$\TD$ and enhanced symmetries}\label{subseTD}
%%%%%%%%%%%%%%%%%%%%%%%%%%%%%%%%%%%%%%%%%%%%%%%%%%%%%%%%%%%%%%%%%%%%%%%

Under a general transformation of
the fields $h_{\mu\nu}\mapsto h_{\mu\nu}+\delta h_{\mu\nu}$, and up
to total derivatives, we have\footnote{Notice that we keep the coordinates
fixed under this transformation. By construction, the Lagrangians are also invariant
under Lorentz transformations. In the standard GR case,
 both kind of transformations blend at the non-linear level
to give rise to the non-linear diffeomorphism \cite{Ortin}.}
\bea
\delta {\mathcal{L}}^I&=&-{1\over 2} \delta h_{\mu\nu} \Box h^{\mu\nu}, \nonumber\\
\delta {\mathcal{L}}^{II}&=& \delta h_{\mu\nu}
\partial^\rho\partial^{(\mu} h_\rho^{\nu)},\nonumber\\
\delta {\mathcal{L}}^{III}&=&-{1\over 2}\Big(\delta h
\partial^\mu\partial^\nu h_{\mu\nu}+\delta h_{\mu\nu}
\partial^\mu\partial^\nu h\Big),\nonumber\\
\delta {\mathcal{L}}^{IV}&=& {1\over 2} \delta h \Box
h.\label{variIV}
\eea
 It follows that the combination
\be
{\mathcal{L}}_{\TD} \equiv {\mathcal{L}}^I+{\mathcal{L}}^{II} + a\
{\mathcal{L}}^{III} + b\ {\mathcal{L}}^{IV}, \label{tdl}
\ee
with arbitrary $a$ and $b$ is invariant under restricted
gauge transformations
\be
 \delta h_{\mu\nu}
=2\partial_{(\mu} \xi_{\nu)}, \label{gauge} \ee with \be
\partial_\mu\xi^\mu=0.\label{ttra}
\ee
These restricted (or more correctly \emph{reducible} \cite{HenneauxTeit}) gauge
transformations have been claimed to pay the crucial role for the
propagation of massless spin-particles \cite{vanderBij:1982,Alvarez:2005iy}.
Indeed, as shown in \cite{vanderBij:1982}, this reducible gauge invariance
is enough to get rid of the extra polarizations introduced by applying
the little group generators of the massless spin-2 particle to the usual
polarizations of spin-2
$$ h^{+}\equiv e^+\otimes e^+-e^-\otimes e^-
,\quad h^\times\equiv e^+\otimes e^-+e^-\otimes e^+,$$
where $e^{\pm}$ are the standard polarizations of spin $s=\pm 1$.
This can be understood from the fact that the transformations (\ref{gauge}-\ref{ttra})
are characterized by the Lorentz invariant condition of leaving the trace $h$ invariant
and the trace does not belong to the irreducible representation of the Lorentz
group which contains $h^{\pm}$ .
From now on we will call the transformations (\ref{gauge}-\ref{ttra}) transverse diffeomorphisms
(TDiff).\\

An enhanced symmetry can be obtained by adjusting the parameters
$a$ and $b$
appropriately. For instance, $a=b=1$ corresponds to the Fierz-Pauli (FP)
Lagrangian \cite{Fierz:1939ix}, which is invariant under the full group of
linear
diffeomorphisms ($\D$ henceforth), where the condition (\ref{ttra}) is dropped.
In fact, a one parameter family of Lagrangians can be obtained from
the FP one through the non-derivative field redefinitions
\be
\label{cov} h_{\m\n}\mapsto h_{\m\n}+\lambda h \eta_{\m\n},
\quad\quad (\lambda\neq -1/n)
\ee
where $n$ is the space-time
dimension and the condition $\lambda\neq -1/n$ is necessary for the
transformation to be invertible. Notice that the new variables are tensor
{\em densities} with respect to the transformation (\ref{gauge}).
Under this redefinition, the
parameters in the Lagrangian (\ref{tdl}) change as
\be
 a\mapsto
a+\lambda\left(an-2\right), \quad b\mapsto
b+2\lambda(nb-a-1)+\lambda^2(b n^2-n(2a+1)+2). \label{mas}
\ee
Starting from $a=b=1$, the new parameters are related by
\bea
\label{conftrd} b&=&\frac{1-2 a+(n-1) a^2}{(n-2)}.
\eea
It follows
that Lagrangians where this relation is satisfied are equivalent to
FP, with the exception of the case $a=2/n$, which cannot be
reached from $a=1$ with $\lambda\neq -1/n$ (cf. (\ref{mas})).

A second possibility is to enhance $\TD$ with an additional Weyl
symmetry,
\be
\delta h_{\mu\nu} = {2\over n} \phi
\eta_{\mu\nu},\label{wesu}
\ee
by which the action becomes
independent of the trace. This possibility is accomplished if
 in the generic transverse Lagrangian
${\mathcal{L}}_{\TD}[h_ {\mu\nu}]$ of Eq. (\ref{tdl}), one replaces $h_
{\mu\nu}$ with the traceless combination
\be h_{\mu\nu}\mapsto\hat
h_{\mu\nu}\equiv h_{\mu\nu}-(h/n) \eta_{\mu\nu}\label{repl}.
\ee
This
is formally analogous to the transformation (\ref{cov}) with $\lambda=-1/n$, but cannot
be interpreted as a field redefinition. As such, it would be
singular, because the trace $h$ cannot be recovered from $\hat
h_{\mu\nu}$. The resulting Lagrangian
\be
{\mathcal{L}}_{\WTD}[h_{\mu\nu}]\equiv{\mathcal{L}}_{\TD}[\hat
h_{\mu\nu}],
\label{wtddef}
\ee
is still invariant under $\TD$ (the
replacement (\ref{repl}) does not change the coefficients in front
of the terms ${\mathcal{L}}^I$ and ${\mathcal{L}}^{II}$). Moreover,
it is invariant under (\ref{wesu}), since $\hat h_{\mu\nu}$ is so.
Using (\ref{mas}) with $\lambda=-1/n$, we immediately find that this
``$\WTD$" symmetry corresponds to Lagrangian parameters \be
a={2\over n},\quad \quad
 b = {n+2\over n^2}. \ee
This is the exceptional case mentioned at the end of the previous
paragraph. Even if we will not deal with non-linearities till
Chapter \ref{chapternl}, we just want to remark that the metric density $\hat g_{\mu\nu} =
g^{-1/n} g_{\mu\nu}$ with $\hat g=1$ can be written at the linear level as
$$\hat g_{\m\n} = \eta_{\mu\nu} + \hat h_{\mu\nu}+O(h^2).$$
This is the starting point for the
non-linear generalization of the $\WTD$ invariant theory, which is
discussed in the second part of this Thesis. Notice also that the WTDiff
Lagrangian cannot be related to the Diff Lagrangian by gauge fixing.
To show it, it is enough to realize that the most general covariant gauge fixing term which breaks
Diff to TDiff and has two derivatives is simply
\be
{\mathcal{L}}_{gf}=
\l \pd_\m h\pd^\m h,
\ee
which cannot change the coefficient of the term ${\mathcal L}^{III}$.

Let us now show that $\D$ and $\WTD$ exhaust all possible
enhancements of $\TD$ for a Lagrangian of the form (\ref{MA}) (and
that, in fact, these are its largest possible gauge invariance
groups\footnote{Lagrangians for $h_{\m\n}$ with a larger WDiff gauge invariance can
be constructed by adding terms with higher derivatives to (\ref{MA}). However those
Lagrangians are problematic as the presence of higher derivative generically
implies the existence of ghosts \cite{Stelle:1977ry}.}). Note first, that the variation of ${\mathcal{L}}^I$
involves a term $\Box h^{\m\n}$.
 For arbitrary $h_{\mu\nu}$, the previous variation will only cancel against other
 terms in (\ref{variIV}) provided that the transformation is of the form
\be \delta h_{\mu\nu} = 2 \partial_{(\mu} \xi_{\nu)}+{2\phi\over
n}\eta_{\mu\nu}, \ee for some $\xi^{\mu}$ and $\phi$, \emph{i.e.}, the transformation
does not touch the spin-2 polarizations. The vector field $\xi_\m$ can
be decomposed as \be \xi_{\mu}=\eta_{\mu} +\pd_{\mu} \psi \ee where
$\partial_{\mu}\eta^{\mu}=0$. Using (\ref{variIV}) we readily find
\bea
\delta{\mathcal L}&=&
\eta_\nu (\beta-1)\Box(\partial_\mu h^{\mu\nu})\nonumber\\
&+&{\psi\over 2}\left[(b-a)\Box h+( 2\beta -a-1)\Box(\partial_\mu\partial_\nu h^{\mu\nu})\right]\nonumber\\
&+&{\phi\over n}\left[(bn-a-1)\Box h + (2\beta-na)
\partial_\mu\partial_\nu h^{\mu\nu}\right].
\eea
$\TD$ corresponds
to taking $\beta=1$, with arbitrary transverse $\eta^\mu$ and with
$\phi=\psi=0$. This symmetry can be enhanced with nonvanishing
$\phi$ and $\psi$ satisfying the relation \be n (a-1) \Box \psi = 2
(2-an) \phi,\label{relat1} \ee provided that \be
 b=\frac{1-2 a+(n-1) a^2}{(n-2)}.\label{relat2}
\ee Eq. (\ref{relat1}) ensures the cancellation of the terms with
$\partial_\mu\partial_\nu h^{\mu\nu}$, and Eq. (\ref{relat2})
eliminates  terms containing the trace $h$. Eq. (\ref{relat2})
agrees with (\ref{conftrd}), and therefore the Lagrangian with the
enhanced symmetry is equivalent to Fierz-Pauli, unless $a=2/n$,
which corresponds to the Lagrangian invariant under $\WTD$\footnote{Incidentally, it may be noted
that for $n=2$ both possibilities coincide, since in this case the
symmetry of the Fierz-Pauli Lagrangian is full diffeomorphisms plus
Weyl transformations.}.

It is worth noticing that the  Weyl symmetry of equation  (\ref{wesu})
is an internal symmetry  in contrast with the
 conformal symmetry which includes transformations of coordinates
 which are not transverse \cite{Isham:1970gz}. A conformal  covariant
 Lagrangian for spin-2 can be found in \cite{Barut:1982nj}. This Lagrangian
 has $\b=2/3$, which, as we will see, implies the existence
 of vector ghost states.

%%%%%%%%%%%%%%%%%%%%%%%%%%%%%%%%%%%%%%%%%%%%%%%%%%%%%%%%%%%%%%%%%%%%%%%%%%%
\subsection{Comparing $\D$ and $\WTD$}
%%%%%%%%%%%%%%%%%%%%%%%%%%%%%%%%%%%%%%%%%%%%%%%%%%%%%%%%%%%%%%%%%%%%%%%%%%%%%%

Let us briefly consider the  differences between the two enhanced
symmetry groups. A first question is whether the Fierz-Pauli theory
${\cal L}_{\D}$ is classically equivalent to ${\cal L}_{\WTD}$.
Since $\D$ includes $\TD$, we can use (\ref{wtddef}) to obtain
\be
{\delta {\cal S}_{\WTD}[h]\over \delta h_{\mu\nu}} = {\delta {\cal
S}_{\D}[\hat h]\over \delta \hat h_{\rho\sigma}}\
\left(\delta^\mu_{(\rho}\delta^\nu_{\sigma)} -{1\over
n}\eta_{\rho\sigma}\eta^{\mu\nu}\right).\label{eomrel}
\ee
Hence, the
$\WTD$ EoM are traceless
$$ {\delta {\cal
S}_{\WTD}[h]\over \delta h_{\mu\nu}}\eta_{\mu\nu}\equiv 0.$$ In the
$\WTD$ theory, the trace of $h$ can be changed arbitrarily by a Weyl
transformation, and we can always go to the gauge where $h=0$.
Likewise, in the familiar $\D$ theory we can choose a gauge where
$h=0$. Then, $h_{\mu\nu}=\hat h_{\mu\nu}$, and the $\WTD$ EoM
 are just the traceless part of the Fierz-Pauli EoM. Differentiating Eq. (\ref{eomrel}) with respect to $x^\m$ and
using the Bianchi identity
$$\partial_\rho\left({\delta {\cal S}_{\D}[h]\over \delta
h_{\rho\sigma}}\right)=0,$$ one easily finds that ${\delta {\cal
S}_{\WTD}[h]/ \delta h_{\mu\nu}}=0$ implies
$${\delta {\cal S}_{\D}[\hat h]\over \delta
h_{\rho\sigma}}\ \eta_{\rho\sigma}= \Lambda.
$$
Hence, the trace of the Fierz-Pauli EoM is also recovered from
the $\WTD$ EoM (in the gauge $h=0$), up to an arbitrary
integration constant $\Lambda$ which plays the role of a
cosmological constant\footnote{Consistency of the linear theory
implies $\Lambda = O(h)$.}.
Thus, the two theories are closely related, but they are not quite
the same. Another conclusion that stems from the previous analysis is
that the traceless part of the linearized Einstein's equations in the
gauge $h=0$ are equivalent to the full Einstein's equations except
for an integration constant. This statement is nothing but the linear
version of the well known result that the full Einstein's equations
are equivalent to its traceless part up to an integration
 constant \cite{Einstein:1916cc,Alvarez:2005iy}. As we will see in the
 next section and in Chapter \ref{chapternl}, there is also a TDiff
 invariant Lagrangian
which shares this property: the Lagrangian with a Diff invariant kinetic
term and a TDiff invariant mass term.

Let us now consider the relation between the corresponding symmetry
groups. Acting infinitesimally on $h_{\m\n}$ they give
\bea
\delta^{\D} h_{\m\n}&=&2
\partial_{(\m}\xi_{\n)}=2\partial_{(\m}\eta_{\n)}+\pd_{\m}
\pd_{\n}\psi\label{feq}\\
\delta^{\WTD} h_{\m\n}&=&2
\partial_{(\m}\bar\eta_{\n)}+\frac{2}{n}\phi\eta_{\m\n}\label{seq} \eea where
$\partial_\m \eta^\m=\partial_\m \bar\eta^\m=0$. In (\ref{feq}) we
have
 decomposed $\xi_{\n}=\eta_\nu + \pd_\nu\psi$ into transverse and longitudinal part.
 The intersection of $\D$ and $\WTD$ can be found by equating (\ref{feq}) and (\ref{seq})
\be 2\partial_{(\m}\eta_{\n)}+\pd_{\m} \pd_{\n}\psi=2
\partial_{(\m}\bar\eta_{\n)}+\frac{2}{n}\phi\eta_{\m\n}.\label{mac}
\ee Taking the trace, we have \be \Box \psi=2\phi.\label{mac2} \ee
The divergence of (\ref{mac}) now yields \be \Box
(\bar\eta_\mu-\eta_\mu) = {n-1\over n} \Box \pd_\mu\psi. \label{wac}
\ee Taking the divergence once more, we have \be \Box \phi=0. \ee
Taking the derivative of (\ref{wac}) with respect to $\nu$,
symmetrizing with respect to $\mu$ and $\nu$, and using (\ref{mac})
and (\ref{mac2}), we have $(n-2)\pd_{\m} \pd_{\n}\Box \psi=0$. For
$n\neq 2$ this implies $\pd_{\m} \pd_{\n}\phi=0$, {\em i.e.}
$$
\phi=b_\mu x^{\mu} + c,
$$
where $b_\mu$ and $c$ are constants. Hence, not every Weyl
transformation belongs to $\D$, since only the $\phi$'s which are
linear in $x^{\mu}$ qualify as such. Conversely, the subset of $\D$
which can be expressed as Weyl transformations are the solutions of
the conformal Killing equation for the Minkowski metric
\cite{Wald}, \be \pd_{(\m}\xi^{CD}_{\n)}=\frac{1}{n}\phi
\eta_{\m\n},\label{confd} \ee where $\phi= \pd^\r \xi^{CD}_\r$ (and,
as shown above, $\phi$ has to be a linear function of $x^{\mu}$).
These solutions generate the so called conformal group, which we may
denote by $\CD$. In conclusion,
 the enhanced symmetry groups $\D$ and $\WTD$ are not subsets of each other.
 Rather, their intersection is the set of $\TD$ plus $\CD$. As we have already mentioned,
the implementation of
 this conformal transformation differs from the
one of \cite{Isham:1970gz} which also involves  transformations in the coordinates.

Finally, for theories invariant under Weyl and Diff transformations, one can show that the
covariant group of the theory contains the conformal group as a subgroup
(see {\em e.g.} \cite{Fradkin:1985am}).
For the TDiff case, one can easily see that this is not the case, as the equation
\be
e^{-2\l(x)}\frac{\pd x^\m}{\pd y^\a}\frac{\pd x^\n}{\pd y^\b}\eta_{\m\n}=\eta_{\a\b},
\ee
which determines the covariant group of the theory in the Minkowski vacuum,
 implies $\l(x)=0$ for a TDiff change
of variables. This yields just the Poincar\'e group as the covariant group of symmetry of the
WTDiff theories.

%%%%%%%%%%%%%%%%%%%%%%%%%%%%%%%%%%%%%%%%%%%%%%%%%%%%%%%%%%%%%%%%%%%%%%%%%%%%
\subsection{Dynamical analysis of the general massless Lagrangian}
%%%%%%%%%%%%%%%%%%%%%%%%%%%%%%%%%%%%%%%%%%%%%%%%%%%%%%%%%%%%%%%%%%%%%%%%%%%%%%%

The little group argument mentioned above indicates
that if the quantum theory describes massless spin-2 particles
it is not unitary unless the Lagrangian is
invariant under $\TD$ \cite{vanderBij:1982}. In fact, as we will see, in the absence of $\TD$ symmetry
 the Hamiltonian is unbounded from
 below. This leads to pathologies such as classical
 instabilities or the existence of ghosts.

To show this, as well as to analyze the physical degrees of freedom
of the general massless theory (\ref{MA}),
 it is very convenient to use the
``cosmological" decomposition in terms of scalar, vector, and
tensor modes under spatial rotations $SO(3)$ (see {\em e.g.}
\cite{Mukhanov:1990me}),
\bea
h_{00}&=&\A, \nonumber\\
h_{0i}&=& \pd_i B+V_i,\nonumber\\
h_{ij}&=&\psi\delta_{ij}+\pd_i \pd_j E+2\pd_{(i}F_{j)}+t_{ij},
\label{cosdec}
\eea
where $\pd^iF_i= \pd^iV_i=\pd^it_{ij}=t^i_i=0$.
Two important features of this decomposition are that it is  local in time
and that in the linearized theory the
scalar ($\A,B,\psi,E$), vector ($V_i,F_i$) and tensor modes ($t_{ij}$)
decouple from each other. Also, we can easily identify the physical
degrees of freedom without having to fix a gauge by directly substituting
the constraints in the Lagrangian \cite{Jackiw:1993in}.

The tensor modes $t_{ij}$ only contribute to ${\mathcal L}^I$, and one
readily finds that their Lagrangian is
\be
{}^{(t)}{\mathcal L}=-\frac{1}{4}t^{ij} \Box
t_{ij}.\label{tensors}
\ee
The vector modes contribute both to ${\mathcal
L}^I$ and ${\mathcal L}^{II}$. Working in Fourier space for the
spatial coordinates and after some straightforward algebra, we have
\be   {}^{(v)}{\mathcal{L}}=\frac{1}{2}
\k^2\left(V^i-\dot F^i\right)^2 + {1\over
2}(\beta-1)\left(\k^2F^i+\dot V^i\right)^2.\label{vecs} \ee For
$\beta=1$, corresponding to $\TD$ symmetry, there are no derivatives
of $V^i$ in the Lagrangian. Variation with respect to $V^i$ leads to
the constraint $V^i-\dot F^i=0$, which upon substitution in
(\ref{vecs}) shows that there is no vector dynamics.

Other values of $\beta$ lead to pathologies.  The Hamiltonian is
given by \be {}^{(v)}{\mathcal H}={(\Pi_F+\k^2 V)^2 \over
2\k^2}-{[\Pi_V +(1-\beta)\k^2 F]^2\over 2(1-\beta)}  +{(1-\beta)\k^4
F^2\over 2}- {\k^2 V^2\over 2},\label{ham} \ee where the momenta are
given by $\Pi_F=\k^2\left(\dot F-V\right)$ and
$\Pi_V=(\beta-1)\left(\k^2 F+\dot V\right)$,
 and
we have suppressed the index $i$ in the vector modes $F$ and $V$. Because
of the alternating signs in Eq. (\ref{ham}), the Hamiltonian is not
bounded from below. Generically this leads to a classical instability.
The momenta satisfy the equations $\dot\Pi_F = \k^2 \Pi_V$ and
$\dot\Pi_V = -\Pi_F$. These have the general oscillatory solution
$$
|\k| \Pi_V+\I\ \Pi_F= C \exp{\I(|\k| t +\phi_0)},
$$
where $C$ and $\phi_0$ are real integration constants. On the other
hand, $V$ and $F$ satisfy \bea
\ddot V+\k^2 V= {-\beta\over (\beta-1)} \Pi_F,\\
\ddot F+\k^2 F= {\beta \over (\beta-1)} \Pi_V. \eea For $\beta\neq
0$ these are equations for forced oscillators. For large times, the
homogeneous solution becomes irrelevant and we have
$$
V + \I|\k| F \sim \left({\beta C t\over (\beta-1) |\k|}\right)
\exp{\I(|\k| t + \phi_0)},
$$
whose amplitude grows without bound, linearly with time. This
classical instability is not present for $\beta=0$. However, in this
case $F$ and $V$ decouple and we have
$$
 {}^{(v)}{\mathcal{L}}_{\beta=0}=\frac{1}{2} \k^2(\partial_\mu
F^i)^2 - {1\over 2}(\partial_\mu V^i)^2,
$$
so $V_i$ are ghosts. One may argue that these ghosts do not couple
to conserved matter at the linear level, and thus Lagrangians
with ghosts in the vector sector are stable. Even if this is true,
 these modes are coupled to matter and to the
other polarizations of the graviton through the non-linear terms and thus the theory is
quantum mechanically unstable
at the scales where those terms are important. By considering this criterium of stability, we
 are going one step beyond other analysis which restrict the theories
to be ghost free at the linear level  once the
propagator is coupled to conserved sources, as \cite{VanNieuwenhuizen:1973fi}.

Hence, the only case where the vector Lagrangian is not problematic
is $\beta=1$, corresponding to invariance under $\TD$. The scalar
Lagrangian is then given by
\bea
{}^{(s)}{\mathcal{L}}_{\TD}&=&{1\over 4}\left[ (\partial_\mu \A)^2-2\k^2(\partial_\mu B)^2+N(\partial_\mu\psi)^2-2\k^2\partial_\mu\psi\partial^\mu E+\k^4(\partial_\mu E)^2 \right]\nonumber\\
&-&{1\over 2}\left[(\dot \A + \k^2 B)^2-\k^2 \dot B^2-\k^2\psi^2+2\k^4 E\psi-\k^6E^2+2\k^2\dot B(\psi-\k^2 E)\right]\nonumber\\
&+&{a\over 2}\left[(\dot \A-N\dot\psi+\k^2 \dot E)(\dot \A+\k^2 B)-\k^2(\A-N\psi+\k^2 E)(\dot B-\psi+\k^2 E)\right]\nonumber\\
&-&{b\over 4}\left[\partial_\mu(\A-N \psi +\k^2
E)\right]^2,\label{scalax}
\eea
where $N=n-1$ is the dimension of
space. It is easy to check that $B$ is a Lagrange multiplier,
leading to the constraint \be (N-1)\psi=(a-1) h, \ee where
$h=\A-N\psi+\k^2E$ is the trace of the metric perturbation.
Substituting this back into the scalar action (\ref{scalax}) we
readily find
\be
\label{chLscalar}
{}^{(s)}{\mathcal{L}}_{\TD}= -{Z \over
4}(\partial_\mu h)^2,
\ee
where
\be
 Z\equiv b-\frac{1-2a+(n-1)a^2}{n-2}. \label{deltab0}
\ee
Hence, the scalar sector contains a single physical degree of
freedom, proportional to the trace. Whether this scalar is a ghost
or not is determined by the parameters $a$ and $b$
and we see that there is a whole family of Lagrangians
with a positive definite energy (\emph{i.e.} with $Z<0$). For
$b=(1-2a+(n-1)a^2)/(n-2)$, corresponding to the enhanced symmetries
which we studied in the previous subsection, the scalar sector
disappears completely, and we are just left with the tensor modes\footnote{Whenever (\ref{conftrd})
holds, we
find always the same Lagrangian for the physical degrees of freedom  without the appearance of
an integration constant
because we have assumed it to be zero when we solved the constraints.}.\\

The fact that we have found a Lagrangian with the WTDiff gauge invariance
that has the same degrees of freedom as the the usual Lagrangian invariant
under Diff is surprising. Indeed, a naive counting of the degrees
of freedom (see {\em e.g.} \cite{Skvortsov:2007kz}) implies that the number
of propagating degrees of freedom (PDoF) is three and not two for this Lagrangian. However, after
a canonical analysis of the Hamiltonian for the WTDiff theory
one readily sees that there is a tertiary constrain which appears in WTDiff
and which is not present in the Diff theory which kills the extra expected degree of
freedom \cite{Skvortsov:2007kz}. Indeed, something similar happens also for
higher spin Lagrangians \cite{Skvortsov:2007kz}.

%%%%%%%%%%%%%%%%%%%%%%%%%%%%%%%%%%%%%%%%%%%%%%%%%%%%%%%%%%%%%%%%%%%%%%%%%%%
\subsection{TDiff Lagrangians in terms of gauge invariant quantities}
%%%%%%%%%%%%%%%%%%%%%%%%%%%%%%%%%%%%%%%%%%%%%%%%%%%%%%%%%%%%%%%%%%%%%%%%%%%%%%

As the Lagrangian of (\ref{tdl}), ${\mathcal{L}}_{\TD}$,
 is invariant under TDiff, one should be able to write it in terms
of quantities invariant under these transformations (for the $\D$ case see {\em e.g.}
\cite{Mukhanov:1990me}). It is easy to
see that under a general transformation $h_{\mu\nu} \mapsto
h_{\mu\nu}+2\partial_{(\mu}\xi_{\nu)}$ the fields of the
cosmological decomposition transform as
\bea t_{ij}&\mapsto&
t_{ij},\quad V_i\mapsto V_i+\partial_0 \xi^T_i,\quad F_i\mapsto F_i+
\xi^T_i,\quad
A\mapsto A+ 2\partial_0 \xi_0, \nonumber\\
B&\mapsto& B+\partial_0 \eta +\xi_0,\quad E\mapsto E+2\eta,\quad
\psi\mapsto\psi,\nonumber
\eea
where $\xi_i=\xi_i^T+\partial_i\eta$,
with $\partial^i\xi^T_i=0$. Whereas for a Weyl transformation
$h_{\mu\nu} \mapsto h_{\mu\nu}+ \frac{1}{n}\phi \eta_{\mu\nu}$ only
$A$ and $\psi$ change as
\bea
A\mapsto A+\frac{\phi}{n},\quad
\psi\mapsto \psi -\frac{\phi}{n}.\nonumber
\eea
 For general transverse transformations
the only gauge invariant combinations are
\be
 t_{ij}, \quad w_i=V_i-\partial_0 F_i,
\ee
in the tensor and vector sectors respectively and \be
\Phi=A-2\partial_0 B+ \partial^2_0 E, \quad \psi ,\quad \Theta=(A-\Delta E),
\ee
for the scalar modes. In terms of these combinations, the tensor,
vector and scalar part of the Lagrangian (\ref{tdl}) can be written
as (we write also the TDiff invariant mass term ${\mathcal L}^V=-m^2 h^2$) \bea
{}^{(t)}{\mathcal{L}}_{\TD}&=&-\frac{1}{4}t^{ij} \Box t_{ij}, \quad
{}^{(v)}{\mathcal{L}}_{\TD}=-\frac{1}{2}
w^i\triangle w^i,\nonumber\\
{}^{(s)}{\mathcal{L}}^I+{}^{(s)}{\mathcal{L}}^{II}&=&\frac{1}{4}\left(-\dot\Theta^2-\Theta\Delta
(\Theta-2\Phi)
-2\Delta \psi(\Phi-\Theta)+(n-3)\psi\Delta\psi+(n-1)\dot\psi^2\right),\nonumber\\
{}^{(s)}{\mathcal{L}}^{III}&=&\frac{a}{4}\Big((\Theta-(n-1)\psi)(\Delta(\Theta-\psi-\Phi)-\ddot{\Theta})\Big)
,\nonumber\\
{}^{(s)}{\mathcal{L}}^{IV}&=&-\frac{b}{4}\Big((\dot\Theta -(n-1)\dot
\psi)^2+(\Theta-(n-1) \psi)\Delta (\Theta-(n-1)\psi)\Big)
,\nonumber\\
{}^{(s)}{\mathcal{L}}^V&=&-\frac{m^2}{4}(\Theta-(n-1)\psi)^2.\nonumber
\eea
From this decomposition we easily see that
$\Phi$ is always a Lagrange multiplier whose variation yields the
constraint \be \label{constr}
\triangle\left((1-(n-1)a)\psi-(1-a)\Theta\right)=0. \ee

In the Diff invariant case ($a=b=1$), only two scalar
combinations are gauge invariant, namely $\Phi$ and $\psi$. Thus,
the lagrangian for the scalar part can be expressed as \be
{}^{(s)}{\mathcal{L}}_{\D}= \frac{(2-n)}{4}\left(-2\Phi
\Delta\psi+(n-1)\dot \psi^2 +(n-3)\psi\Delta \psi \right). \ee

Concerning the Weyl transformations, we can write only two scalar
invariants which are also scalars for TDiff,
\bea \Xi=\Phi+\psi, \quad \Upsilon=\Theta+\psi.
\eea
Thus, for the
Weyl invariant choice $a=\frac{2}{n}$, $b=\frac{n+2}{n^2}$, we can write the
Lagrangian as \bea \label{kineticconf}
{}^{(s)}{\mathcal{L}}_{\WTD}&=&
\frac{1}{4n^2}\left((n-2)(2n\Xi-(n-1)\Upsilon)\triangle \Upsilon
-(2-3n+n^2)\dot \Upsilon^2\right). \eea
Varying the Lagrangian
with respect to $\Xi$ we find the constraint \be \Delta \Upsilon=0.
\ee Besides, the mass term can be written as \be
{}^{(s)}{\mathcal{L}}^V=-\frac{m^2}{4}(\Upsilon-n\psi)^2. \ee

%%%%%%%%%%%%%%%%%%%%%%%%%%%%%%%%%%%%%%%%%%%%%%%%%%%%%%%%%%%%%%%%%%%%%%%%%%%%%%%%%%%%%%%%%%%%%%%%%%
\section{Massive fields}\label{Massivefields}
%%%%%%%%%%%%%%%%%%%%%%%%%%%%%%%%%%%%%%%%%%%%%%%%%%%%%%%%%%%%%%%%%%%%%%%%%%%%%%%%%%%%%%%%%%%%%%%%%%%

Let us now turn our attention to the massive case. The most general
mass term takes the form\footnote{Here, we are disregarding the
possibility of Lorentz breaking mass terms, which has been recently
considered in \cite{Rubakov:2004eb}. We will say more about these massive
terms in the next part of the Thesis (see Chapter \ref{chapterperturbbigrav}).}
$$
{\mathcal L}_m = -{1\over 4} m_1^2 h_{\mu\nu}h^{\mu\nu} +{1\over 4}
m_2^2 h^2.
$$
First of all, let us note that for $m_1=0$, this mass term is still
invariant under $\TD$. The term $m_2^2 h^2$ gives a mass to the
scalar $h$, but not to the tensor or vector modes.
Hence, the analysis of the previous section remains
basically unchanged. At energy scales below the mass $m$,  the
extra scalar effectively decouples and we are back to the situation
where only the standard helicity polarizations of the graviton are
allowed to propagate\footnote{Note also that the addition of the
term $m_2^2 h^2$ to both the $\D$ or the $\WTD$ Lagrangian does not
change the propagating degrees of freedom of the theory. The
analogous statement in a non-linear context is illustrated by the
 addition of a ``potential" $f(g)$ to the non-linear extensions
 of these Lagrangians (something {\em does} change, though, by
the addition of the potential, since the new theory does have the
arbitrary integration constant $\Lambda$). Hence, one may in
principle construct classical Lagrangians which propagate only
massless spin-2 particles, and whose symmetry is only $\TD$,
although in this case radiative stability is not guaranteed ({\em
i.e.} we may expect other terms, such as kinetic terms for the
determinant $g$, which are not protected by the symmetry, to be
generated by quantum corrections).}. For a tachyon free situation
we require $-m_2^2
> 0$.

When $m_1\neq 0$, we must repeat the analysis\footnote{For a similar
analysis in terms of spin projectors see \cite{VanNieuwenhuizen:1973fi}.}. With the
decomposition (\ref{cosdec}), the Lagrangian for the tensor modes
becomes \be {}^{(t)}{\mathcal{L}}=-\frac{1}{4}t^{ij}\left( \Box+
m_1^2\right) t_{ij},\label{mtensors}\ee and in order to avoid
tachyonic instabilities we need $m_1^2>0$. For the vector modes, and
for $\beta\neq 1$, the potential term
$$
\Delta {\cal H}_v = {m_1^2\over 2}[\k^2 (F^i)^2 - (V^i)^2],
$$ is added to (\ref{ham}). The contribution proportional to $V^2$
is negative definite. Hence, to avoid ghosts or tachyons we must
take $\beta=1$. In this case, $\dot V^i$ does not appear in the
Lagrangian and $V^i$ can be eliminated in favor of $\dot F^i$. This
leads to
\be ^{(v)}{\cal L} = -{1\over 2}\left({\k^2 m_1^2\over \k^2
+m_1^2}\right)\ F^i \left(\Box + m_1^2\right)F^i. \label{av}\ee
Out of the $(N+2)(N-1)/2$ polarizations of the massive graviton in $n=N+1$
dimensions, $(N-2)(N+1)/2$ of these are expressed as transverse and
traceless tensor modes $t_{ij}$, and $N-1$ are expressed as transverse
vector modes $F^i$, whose dispersion relation must coincide.
 The remaining one (also with the same dispersion relation) must be contained in the scalar
sector. The scalar Lagrangian can be written as
\be
^{(s)}{\cal L}=
^{(s)}{\cal L}_{\TD} +^{(s)}{\cal L}_m,
\ee
where the first term is
given by (\ref{scalax}) and the second is given by \be ^{(s)}{\cal
L}_m = -{m_1^2\over 4} (A^2 -2 \k^2 B^2 + N \psi^2 - 2 \k^2 \psi E +
\k^4 E^2) +{m_2^2\over 4}(A-N \psi + \k^2 E)^2. \ee Variation with
respect to $B$ leads to the constraint
$$
 m_1^2\ B= {(1-a)(\dot A +\k^2\dot E)-(1-aN) \dot\psi}.
$$
To proceed, it is convenient to eliminate $E$ in favor of the trace
$h$,
$$
\k^2 E = h+ N\psi -A,
$$
and to further express $A$ and $\psi$ in terms of new variables $U$
and $V$, \bea (N-1)\ A &=& (aN-1)\ h+[2 (N-1) \k^2- N m_1^2]\ U,
\nonumber \\ (N-1)\ \psi &=& (a-1)\ h - m_1^2\ (U-V).\label{nv}\eea
With these substitutions, and after some algebra, we find \be
^{(s)}{\cal L}= -{Z\over 4} \dot h^2 + {[N
m_1^2-2(N-1)\k^2]m_1^2 \over 4 (N-1)}\left(\dot V^2- \dot
U^2\right)+{W(h,U,V)\over 4(N-1)^2}, \label{umv}\ee where $Z$
is given by (\ref{deltab0}) and \bea W \equiv&&
\left\{(N-1)^2(\k^2Z+m_2^2)
-[1+(1-4a+a^2)N+a^2N^2] m_1^2\right\} h^2\nonumber\\
&& +{(N-1)m_1^4\ [(N-2)\k^2-Nm_1^2]}\ V^2\nonumber \\
&& -{m_1^2\ [4(N-1)^2 \k^4 +  (2+N-3N^2) m_1^2 \k^2 + N(N+1)
m_1^4]}\ U^2 \nonumber\\
&&+4(N-1){m_1^2\k^2[N m_1^2 -(N-1) \k^2]}\ UV \nonumber \\
&& +{2 m_1^2\ [(N+1)a-2]}\ [(Nm_1^2-(N-1)\k^2)\ U- (N-1) \k^2\ V]\
h.\label{horta}\eea For $2(N-1)\k^2<Nm_1^2$ the variable $U$ has
negative kinetic energy, whereas for $2(N-1)\k^2> Nm_1^2$ the same
is true of $V$. Thus, the Hamiltonian is unbounded below, unless
\be Z=0.\label{uud}\ee In this case, $h$ is non-dynamical, and
it will implement a constraint between $U$ and $V$ provided that the
coefficient of $h^2$ in $W$ vanishes identically. This requires \be
m_2^2 =\left({1+(1-4a+a^2)N+a^2N^2 \over (N-1)^2}\right)m_1^2.
\label{mrel}\ee
As discussed in section \ref{subseTD}, as long as $a \neq
2/(N+1)$, all kinetic Lagrangians with $Z=0$ are related to
the Fierz-Pauli kinetic term by the field redefinition (\ref{cov}).
Thus, there are only two possibilities for eliminating the ghost\footnote{As
we already mentioned, the presence of ghosts is not problematic
as long as they are not coupled to ordinary matter at energies below a certain
cut-off. This allows to consider
 TDiff invariant Lagrangians with massive gravitons which are
 stable at energy scales larger than the interaction scale.
 Contrary to the Diff invariant case, the interaction scale for the ghost
 modes can be made arbitrarily small by a convenient choose of the coefficients
$a$ and $b$ \cite{Porrati:2004mz}, but
this is not a real progress since then the vDVZ discontinuity
is present till these scales, and those models are ruled
out phenomenologically. Besides, this result only holds at the linear
level.}:
either the kinetic term is invariant under $\D$ or it is invariant
under $\WTD$.

%%%%%%%%%%%%%%%%%%%%%%%%%%%%%%%%%%%%%%%%%%%%%
\subsection{$\D$ invariant kinetic term}
%%%%%%%%%%%%%%%%%%%%%%%%%%%%%%%%%%%%%%%%%%%%%%%%%

Without loss of generality, we can take $a=b=1$, and from
(\ref{mrel}) we have the usual Fierz-Pauli relation
$$
m_1^2=m_2^2.
$$
Variation with respect to $h$ leads to the constraint \be (N-1)\k^2
V=[Nm_1^2-(N-1)\k^2] U. \label{babel}\ee In combination with
(\ref{nv}), this yields \be (N-1)\k^2 \psi= m_1^2[N
m_1^2-2(N-1)\k^2]\ U. \label{simp}\ee Substituting (\ref{babel}) in
the Lagrangian, and using (\ref{simp}) we obtain
 \be
^{(s)}{\cal L}= -{N \over 4 (N-1)}\ \psi(\Box + m_1^2)\
\psi,\label{as} \ee which is the remaining scalar degree of freedom
of the graviton.

The tensor, vector and scalar Lagrangians
(\ref{mtensors}),(\ref{av}) and (\ref{as}) are not in a manifestly
Lorentz invariant form, and the actual form of the propagating
polarizations is obscured by the fact that the components of the
metric must be found from $F^i$ and $\psi$ with the help of the
constraint equations. Nevertheless, once we know that the system has
no ghosts and all polarizations have the same dispersion relation,
it is trivial to repeat the analysis in the rest frame of the
graviton, $\k=0$. In this frame, the metric is homogeneous
$\partial_i h_{\mu\nu}=0$ and we may write
$$h_{00}={A}, \quad h_{0i}={V}_i, \quad h_{ij}= \psi \delta_{ij} +
t_{ij},$$ where $t^i_i=0$. The Lagrangian for tensor modes becomes\be
{}^{(t)}{\mathcal{L}}=-\frac{1}{4}t^{ij}\left( \Box+ m_1^2\right)
t_{ij},\label{mtensorss}\ee Vectors contribute to ${\mathcal{L}}^I$
and ${\mathcal{L}}^{II}$, giving \be {}^{({ v})}{\mathcal{L}}=
{1\over 2}(\beta - 1) \dot { V}_i^2 + {1\over 2} m_1^2 {V}_i^2, \ee
which is non-dynamical in the present case because $\beta=1$.
Likewise, it can easily be shown that the scalar fields ${A}$ an $\psi$
are non-dynamical. Therefore, in the graviton rest frame the
propagating polarizations are represented by the $[N(N+1)/2]-1$
independent components of the symmetric traceless tensor $t_{ij}$.

%%%%%%%%%%%%%%%%%%%%%%%%%%%%%%%%%%%%%%%%%%%%%%
\subsection{$\WTD$ invariant kinetic term}
%%%%%%%%%%%%%%%%%%%%%%%%%%%%%%%%%%%%%%%%%%%%%%%

For $a=2/n=2/(N+1)$, the last term in Eq. (\ref{horta}) disappears,
and $U$ and $V$ do not mix with $h$. Because of that, there are no
further constraints amongst these variables and the ghost in the
kinetic term in (\ref{umv}) is always present for $m_1^2\neq 0$.
This means that the $\WTD$ theory cannot be deformed with the
addition of a mass term for the graviton without provoking the
appearance of a ghost.

Note that this is so even in the case of a mass term compatible with
the Weyl symmetry, \emph{i.e.} $m_1^2=n m_2^2$. This relation causes $h$ to
disappear from the Lagrangian, but of course it does nothing to
eliminate the ghost. Thus, we have found that from the Lagrangians
that describe the propagation of massless spin-2 particles only one, the Diff invariant one,
can be deformed to describe pure massive spin-2 particles.
Again, the ghost mode may be decoupled
from matter at the linear level, but we expect it
to reappear in the interactions. Concerning the
strong coupling phenomenon for these Lagrangians, we expect it to be absent  but
an explicit calculation has not been performed. \\

In the previous analysis we have restricted to Lorentz invariant mass
terms. However, if one lifts this restriction, one expects to find
mass terms for the WTDiff kinetic term which are free of ghosts or
tachyons as happens in the Diff invariant case \cite{Rubakov:2004eb}. An interesting
possibility would be to consider situations where even if Lorentz invariance
is broken a $SIM(2)$ subgroup of the Lorentz group is preserved \cite{Cohen:2006ky}.
Mass terms compatible with the gauge invariance and with the $SIM(2)$  symmetry
are known for spin-1 \cite{Lindstrom:2006xh} but the search for equivalent
terms for spin-2 is still in progress \cite{Blas08}. Besides, the mass terms may be non-local
operators that come from the integration of high-energy degrees of freedom as in
\cite{Dvali:2006su}.

%%%%%%%%%%%%%%%%%%%%%%%%%%%%%%%%%%%%%%%%%%%%%%%%%%%%%%%%%%%%%%%%
\section{Lagrangians from Tracelessness and from Unitarity}
%%%%%%%%%%%%%%%%%%%%%%%%%%%%%%%%%%%%%%%%%%%%%%%%%%%%%%

An alternative route to the $\WTD$ invariant theory is to try and
construct a Lagrangian which will yield the traceless part of
Einstein's equations. As we have shown, these field equations are equivalent
to the Einstein's equations except for an integration constant and
finding Lagrangians which yield these EoM is  interesting
by itself.

It is clear, however, that we can only obtain traceless equations of
motion from
 a Lagrangian which is invariant under Weyl transformations. If the EoM are traceless,
 then $\delta S=0$ for variations of the form for $\delta h_{\mu\nu} \propto \eta_{\mu\nu}$.
 This symmetry is not included in $\D$, and therefore the traceless part of Einstein's equations
 cannot be recovered from the $\D$ invariant Lagrangian in any gauge. Rather, we should look for
 a Lagrangian which will yield the traceless part of Einstein's equations in {\em some} gauge.

Let us consider the EoM of the $\D$ invariant theory
in momentum space \be
 {\delta {\cal S}_{\D}[h]\over \delta  h_{\rho\sigma}}=K^{\r\s\m\n}_{\D} h_{\m\n},
\ee where \bea &&8
K_{\D}^{\m\n\r\s}=k^2\left(\eta^{\m\rho}\eta^{\n\sigma}+\eta^{\m\sigma}\eta^{\n\rho}-
2\eta^{\m\n}\eta^{\rho\sigma}\right)-\nonumber\\
&&\left(k^{\m}k^{\rho}\eta^{\n\sigma}+
k^{\n}k^{\sigma}\eta^{\m\rho}+ k^{\m}k^{\sigma}\eta^{\n\rho}+
k^{\n}k^{\rho}\eta^{\m\sigma}-2 k^{\m}k^{\n}\eta^{\rho\sigma} -2
k^{\rho}k^{\sigma}\eta^{\m\n}\right). \eea We can also define the
traces \bea \tr K^{\m\n}_{\D}&=&\eta_{\r\s} K^{\r\s\m\n}_{\D}=
\frac{n-2}{4}\left(
k_{\rho}k_{\sigma}-k^2 \eta_{\rho\sigma}\right), \nonumber\\
 \tr \
\tr K_{\D}&=&\eta_{\m\n}\eta_{\r\s}
K^{\r\s\m\n}_{\D}=-\frac{(n-1)(n-2)}{4}k^2. \eea The traceless part
of the $K^{\r\s\m\n}_{\D}$, \bea
&&8K_{\D}^t=8\left(K_{\D}-\frac{1}{n}\eta^{\m\n}\tr\,K_{\D}^{\rho\sigma}\right),
\eea cannot be derived from a Lagrangian as it is not symmetric in
the indices $(\r\s)$ vs. $(\m\n)$. Nevertheless, we can still define
traceless symmetric Lagrangians. One might think of substituting
$\eta^{\m\n}$ in the previous expression by $\tr\,K_{\D}^{\m\n}$,
and dividing by its trace. However, this would yield nonlocal terms.

For a local Lagrangian which is still invariant under $\TD$, we must
restrict to deformations which correspond to changes in the
parameters $a$ and $b$ in (\ref{MA}). The most general symmetric
Lagrangian with these properties is of the form \be
K_{t\D}^{\m\n\rho\sigma}\equiv
K_{\D}^{\m\n\rho\sigma}-\eta^{\m\n}M^{\rho\sigma}-M^{\m\n}
\eta^{\rho\sigma}, \ee with $M_{\r\s}$ a symmetric operator at most
quadratic in the momentum. Asking that the result be traceless leads
to: \be M^{\m\n}=\frac{1}{n}\left(\tr\, K_{\D}^{\m\n}-(\tr\,
M)\eta^{\m\n}\right), \ee which implies \be \tr\,M =\frac{1}{2n}
\tr\,\tr\,K_{\D}. \ee Therefore \be
M^{\m\n}=\frac{1}{n}\left(\tr\,K_{\D}^{\m\n}-\frac{1}{2n}(\tr\,
\tr\,K_{\D})\eta^{\m\n}\right), \ee and we can write \bea 8
K_{t\D}^{\m\n\rho\sigma}=
&&k^2\left(\eta_{\m\rho}\eta_{\n\sigma}+\eta_{\m\sigma}\eta_{\n\rho}\right)-
\left(k_{\m}k_{\rho}\eta_{\n\sigma}+ k_{\n}k_{\sigma}\eta_{\m\rho}+
k_{\m}k_{\sigma}\eta_{\n\rho}+ k_{\n}k_{\rho}\eta_{\m\sigma}\right)\nonumber\\
&&\hspace{1cm}-\frac{2(n+2)}{n^2}k^2\eta_{\m\n}\eta_{\rho\sigma}+\frac{4}{n}(
k_{\m}k_{\n}\eta_{\rho\sigma}+ k_{\rho}k_{\sigma}\eta_{\m\n}). \eea
Moving back to the position space, this corresponds to the $\WTD$
Lagrangian, \emph{i.e.} the case $a=\frac{2}{n}$ and $b=\frac{n+2}{n^2}$ in
(\ref{tdl}). As shown before, this yields the traceless part of the
Fierz-Pauli EoM in the gauge $h=0$.

A similar analysis could be done for the massive case. However, as
we have seen  in the previous section, the corresponding Lagrangian has a
ghost.\\

We would also like to comment on a technique
 to obtain  the free
Lagrangian for a {\em massive} field of spin-2 based on unitarity \cite{Alvarez:2005iy,Veltman:1975vx}.
 The basic requirement
is that the propagator be transverse and traceless on shell, so
that it does not mix with scalar or vector modes at the tree level. One can show (cf. \cite{Veltman:1975vx})
that there is only one propagator transverse and traceless on the mass shell such that
the imaginary part of the tree level diagram corresponding to the interaction
of two identical sources is positive (as  unitarity demands because,
from the usual cut rules, the imaginary part of this diagram corresponds
to the emission of a spin-2 particle). Obviously this Lagrangian is the FP Lagrangian
that we found in the previous section.  Notice also that in the previous section
we showed that the vector and scalar parts, if included, would give rise to a
non-unitary Lagrangian, and thus asking for unitarity is indeed enough to get a unique Lagrangian
for massive spin-2 particles.

%%%%%%%%%%%%%%%%%%%%%%%%%%%%%%%%%%%%%%%%%%%%%%%%%%%%%%%%%%%%%%%%%%%%%%%%%%%%%%%%%%%%%%%%%%%%
\section{Propagators and coupling to matter}
%%%%%%%%%%%%%%%%%%%%%%%%%%%%%%%%%%%%%%%%%%%%%%%%%%%%%%%%%%%%%%%%%%%%%%%%%%%%%%%%%%%%%%%%%%%%

In this section we shall consider the propagators and the coupling
to external matter sources for the different \emph{healthy}
Lagrangians which we have identified in the previous sections.

On one hand, we have the standard massless and massive Fierz-Pauli
theories, which have been thoroughly studied in the literature.
There are also the generic ghost-free $\TD$ theories, which satisfy
the condition
\be
 Z\equiv b-\frac{1-2a+(n-1)a^2}{n-2} < 0. \label{deltab}
\ee
These may include a mass term of the form $m^2h^2$, which
affects the scalar mode
 but does not give a mass to the tensor modes.
The $\WTD$ invariant theory completes the list of possibilities.

Throughout this section, we will make use of the spin-2 projector
formalism of \cite{Rivers:1964}, which is very useful to
invert the equations of motion.
We can expand the
momentum space projector of the propagator as a sum over non-local
projectors
in the space of symmetric tensors of two indexes. These are known
as Barnes and Rivers projectors \cite{VanNieuwenhuizen:1973fi,Rivers:1964}.
 We start with the usual
transverse and longitudinal projectors
\bea
&&\theta_{\a\b}\equiv \eta_{\a\b}-\frac{k_{\a}k_{\b}}{k^2},\nonumber\\
&&\omega_{\a\b}\equiv \frac{k_{\a}k_{\b}}{k^2}.
\eea
and then define
projectors on the subspaces of spin-2, spin-1, and the two
different spin zero components, labeled by $(s)$ and $(w)$. We
introduce also the convenient operators that map between these two
subspaces,
\bea
&&P_2\equiv\frac{1}{2}\left(\theta_{\m\rho}\theta_{\n\sigma}+
\theta_{\m\sigma}\theta_{\n\rho}\right)-\frac{1}{(n-1)}\theta_{\m\n}\theta_{\rho\sigma},\nonumber\\
&&P_0^s\equiv\frac{1}{(n-1)}\th_{\m\n}\th_{\rho\sigma},\nonumber\\
&&P_0^w\equiv\omega_{\m\n}\omega_{\rho\sigma},\nonumber\\
&&P_1\equiv\frac{1}{2}\left(\th_{\m\rho}\omega_{\n\sigma}+\th_{\m\sigma}\omega_{\n\rho}
+\th_{\n\rho}\omega_{\m\sigma}+\th_{\n\sigma}\omega_{\m\rho}\right),\nonumber\\
&&
P_0^{sw}\equiv\frac{1}{\sqrt{(n-1)}}\th_{\m\n}\omega_{\rho\sigma},\quad
P_0^{ws}\equiv\frac{1}{\sqrt{(n-1)}}\omega_{\m\n}\th_{\rho\sigma}.\nonumber\\
\eea
These projectors obey
\bea \label{propproj}
&&P_i^a P_j^b=\d_{ij}\d^{ab} P_i^b,\nonumber\\
&&P_i^{ab} P_j^{cd}=\d_{ij} \d^{bc} \d^{ad}P_j^{a},\nonumber\\
&&P_i^a P_j^{bc}=\d_{ij}\d^{ab}P_j^{ac},\nonumber\\
&&P_i^{ab} P_j^c = \d_{ij}\d^{bc} P_j^{ac}.
\eea
And the traces:
\bea
&&\tr\, P_2\equiv \eta^{\m\n}(P_2)_{\m\n\rho\sigma}=0, \ \tr\, P_0^s
=\theta_{\rho\sigma}, \
\tr\, P_0^w=\omega_{\rho\sigma},\nonumber\\
&&\tr\, P_1=0, \ \tr\,P_0^{sw}=\sqrt{n-1}\omega_{\rho\sigma},\
 \tr\, P_0^{ws}=\frac{1}{\sqrt{n-1}}\theta_{\rho\sigma}.
\eea

Apart from the previous expressions, these projectors satisfy
\be
P_2+P_1+P_0^w+P_0^s=\frac{1}{2}\left(\d_{\m\n}\d_{\r\s}+\d_{\r\s}\d_{\m\n}\right),
\ee
and any symmetric operator can be written as \be K=a_2
P_2+a_1P_1+a_wP_0^w+a_sP_0^s+a_{sw}P_0^\times ,
\ee
where
$P_0^\times=P_0^{sw}+P_0^{ws}$. The inverse of the previous operator
 is easily found from (\ref{propproj}) to be
\be K^{-1}=\frac{1}{a_2}
P_2+\frac{1}{a_1}P_1+\frac{a_s}{a_sa_w-a^2_{sw}}P_0^w+
\frac{a_w}{a_sa_w-a^2_{sw}}P_0^s-\frac{a_{sw}}{a_sa_w-a^2_{sw}}\left(P_0^{ws}+P_0^{sw}\right),
\ee
provided that the discriminant $a_sa_w-a^2_{sw}$ never vanishes.

%%%%%%%%%%%%%%%%%%%%%%%%%%%%%%%%%%%%%%%%%%%%%%%%%%%%%%%%%%%%%%%%%%%%%%%%
\subsection{Gauge Fixing}
%%%%%%%%%%%%%%%%%%%%%%%%%%%%%%%%%%%%%%%%%%%%%%%%%%%%%%%%%%%%%%%%%%%%%%%%%%%%

As noted in \cite{Alvarez:2005iy}, for the $\TD$ gauge invariance there is
no linear covariant gauge fixing condition which is at most
quadratic in the momenta. This is in contrast with the Fierz-Pauli
case, where the harmonic condition contains first derivatives only.
The basic problem is that a covariant gauge-fixing carries a free
index, which leads to $n$ independent conditions. This is more than
what transverse diffeomorphisms can handle, since these have only
$(n-1)$ independent arbitrary functions. To be specific, let us
consider the most general possibility linear in $k$, \be M_{\a\b\g}
h^{\b\g}=0,
\ee
where
\be M_{\a\b\g}=a_1 \eta_{\a(\b}k_{\g)}+a_2
\eta_{\b\g}k_{\a}. \ee In order to bring a generic metric
$h_{\mu\nu}$ to this gauge by means of a $\TD$, we have \be
M_{\a\b\g} h^{\b\g}=M_{\a\b\g} \pd^\b\xi^\g. \ee However, deriving
the r.h.s. of the previous expression with respect to $x^\a$ and
summing in $\a$, this terms cancels, which implies that the
integrability condition \be \pd^\a M_{\a\b\g} h^{\b\g}=0, \ee must
be satisfied. This simply means that the gauge condition cannot be
enforced on generic metrics.

It is plain, however, that the transverse part of the harmonic gauge
(which contains only $n-1$
 independent conditions) can be reached by a transverse gauge transformation. The corresponding gauge
fixing piece is  obtained by projecting the harmonic condition with
$k^2 \eta_{\mu\nu} - k_\mu k_\nu\equiv k^2\theta_{\m\n}$:
\be
{\mathcal L}_{gf}=\frac{1}{2 M^4}(\pd_\a\pd^\m\pd^\n
h_{\m\n}-\Box\pd^\m h_{\a\m})^2.\label{gfterm}
\ee
The gauge fixing
parameter is now dimensionful, and this has been explicitly
indicated by denoting it by $M^4$. A study of this kind of term and
its associated FP ghosts and BRST transformations can be found in
\cite{Alvarez:2006bx} (see also Appendix \ref{AppendixQ}).
We would like to remind that when projector operators are present
in the gauge fixing term, there may appear ghosts of ghosts
in the  quantization process \cite{HenneauxTeit}
and also Kallosh-Nielsen ghosts \cite{Kallosh:1978de,Nielsen:1978mp}.

By contrast, in the case of $\WTD$, the additional Weyl symmetry
allows for the use of gauge fixing terms which are linear in the
derivatives (such as the standard harmonic gauge).

%%%%%%%%%%%%%%%%%%%%%%%%%%%%%%%%%%%%%%%%%%%%%%%%%%%%%%%%%%%%%%%%%%%
\subsection{Propagators}
%%%%%%%%%%%%%%%%%%%%%%%%%%%%%%%%%%%%%%%%%%%%%%%%%%%%%%%%%%%%%%%%%%%%%%%%%%%%%%%%%%%%%

The generic Lagrangian including a mass term can be written in Fourier space as
\bea
\label{tutto}
{\mathcal L}&=&{\mathcal{L}^{I}}+\beta\
{\mathcal{L}^{II}}+ a\
{\mathcal{L}^{III}}+b\ {\mathcal{L}^{IV}}+{\mathcal L}_m+{\cal L}_{gf}={1\over 4}h_{\m\n}K^{\m\n\r\s}h_{\r\s}=\nonumber\\
&&{1\over 4}h_{\m\n}\Big\{\left(k^2-m_1^2 \right)P_2 +\left[(1-\b)\
k^2-m_1^2+\lambda^2(k)\right]P_1\nonumber\\
&&\hspace{4cm} + a_s P_0^s+ a_w P_0^w+a_\times
P_0^\times\Big\}^{\m\n\r\s}h_{\r\s},
\eea
where $P_1$ and
$P_2$ are the projectors onto the subspaces of spin-1 and spin-0
respectively, while the operators $P_0^s$, $P_0^w$ and
$P_0^\times\equiv P_0^{sw} + P_0^{ws}$ project onto and mix the
different spin-0 components. The coefficients in front of
the spin-0 projectors are given by \bea
a_s &=& [1-(n-1)b]k^2-m_1^2+ (n-1) m_2^2,\nonumber\\
a_w &=& (1-2\b+2a-b)k^2 - m_1^2 +  m_2^2,\nonumber\\
a_\times&=&\sqrt{n-1}\left[(a-b)k^2+m_2^2\right]. \eea In
(\ref{tutto}), we have included the term $\lambda^2(k) P_1$ which
can be used to gauge fix the $\TD$ symmetry whenever it is present.
Indeed, (\ref{gfterm}) can be written as \be {\cal L}_{gf}=
\lambda^2(k) h_{\mu\nu} P_1^{\mu\nu\rho\sigma}h_{\rho\sigma}.
\label{gf2} \ee where $\lambda^2(k)=(1/4M^4)k^6$. Even though we are
primarily
 interested in the $\TD$ Lagrangian (which corresponds to $\beta=1$),
we have kept generic $\beta$ throughout this subsection. This can be
useful to handle the cases with enhanced symmetry, since a generic
$\beta$ arises, for instance, from the conventional harmonic gauge
fixing term (as we shall see below). When invertible, the previous
Lagrangian yields a propagator $\Delta \equiv K^{-1}$, \bea
&&\Delta=\frac{P_2}{k^2-m_1^2}+ \frac{P_1}{(1-\b)\
k^2-m_1^2+\lambda^2(k)}+\frac{1}{g(k)}\Big(a_w P_0^s+a_s P_0^w
-a_\times P_0^\times\Big),\nonumber \eea where, \bea g(k)&=& a_s a_w
- a_\times^2.\label{geico} \eea Consider a generic coupling of the
form
\be {\cal L}_{int}(x)={1\over 2}(\kappa_1 T^{\mu\nu}+\kappa_2 T
\eta^{\mu\nu}) h_{\m\n}\equiv {1\over 2}{\cal
T}_{tot}^{\mu\nu}h_{\mu\nu}.
\ee
 For conserved external sources\footnote{For the theories which
 are not invariant under the whole Diff, the external source is
 not necessarily conserved. Nevertheless, the coupling to
 a non-conserved source may imply the loss of unitarity.
 See also \cite{Ford:1980up} for the study of the FP Lagrangian coupled to
 non-conserved sources.}
\be
\pd_\m  T^{\m\n}=0,\label{const}
\ee
this coupling is invariant under
$\TD$ for all values of $\kappa_1$ and $\kappa_2$. Moreover, it is
$\D$ invariant when $\kappa_2=0$, and $\WTD$ invariant for the
special case $\kappa_1=-n\kappa_2$. The interaction between sources
is completely characterized by \cite{Boulware:1973my}
\be {\cal
S}_{int}\equiv {1\over 2}\int \di^n k {\cal L}_{int}(k)={1\over 2}\int
d^n k \ {\cal T}_{tot}(k)^*_{\mu\nu} \Delta^{\mu\nu\rho\sigma} {\cal
T}_{tot}(k)_{\rho\sigma}. \ee

From the properties of the projectors $P_i$, it
is straightforward to show that \be {\cal L}_{int}(k) =\kappa_1^2\
T^*_{\mu\nu}\ \left( \frac{P_2^{\mu\nu\rho\sigma}}{k^2-m_1^2}\right)
\ T_{\rho\sigma}+{\cal P}_0\ |T|^2,\label{flint} \ee where the
operator
\ba
{\cal P}_0={1\over g(k)}\biggl[ {\kappa_1^2 a_w\over
(n-1)} &+&2\kappa_1\kappa_2 \left(a_w
-{a_\times\over\sqrt{n-1}}\right)\nonumber\\
&&+\kappa_2^2 \left[ (n-1) a_w + a_s
-2 \sqrt{n-1}a_\times\right]\bigg]\label{bigpro}
\ea
encodes the
contribution of the spin-0 part. We are now ready to consider the
different particular cases, which we  present by order of increasing
symmetry.

%%%%%%%%%%%%%%%%%%%%%%%%%%%%%%%%%%%%%%%%%%%%%%%%%%%%%%%%%%%%
\subsection{Massive Fierz-Pauli}
%%%%%%%%%%%%%%%%%%%%%%%%%%%%%%%%%%%%%%%%%%%%%%%%%%%%%%%%%%%%

In this case the parameters in the Lagrangian are given by
$\beta=a=b=1$ and $m_1^2=m_2^2$. From (\ref{geico}), we have
$$
g(k)=-(n-1)\ m_2^4,
$$
which does not depend on $k$. Because of that, the denominator of
the operator ${\cal P}_0$ does not
 contain any derivatives. Its contribution to Eq. (\ref{flint}) corresponds only to contact terms,
 which do not contribute to the interaction between separate sources. We are thus left with the
 spin-2 interaction, which ignoring all contact terms, can be written as
\be {\cal L}_{int}=\kappa_1^2\ T^*_{\mu\nu}\
\left(\frac{P_2^{\mu\nu\rho\sigma}}{k^2-m_1^2}\right)\
T_{\rho\sigma} = {\kappa_1^2 \over k^2 -
m_1^2}\left[T^*_{\mu\nu}T^{\mu\nu}-{1\over
(n-1)}|T|^2\right].\label{lintfp} \ee The factor $1/(n-1)$ is
different from the familiar $1/(n-2)$ which is encountered in
linearized GR, and produces
 the well known vDVZ discontinuity in the massless limit \cite{vanDam:1970vg,Zakharov:1970cc}.

%%%%%%%%%%%%%%%%%%%%%%%%%%%%%%%%%%%%%%%%%%%%%%%%%%%%%%%%%%%%
\subsection{$\TD$ invariant theory}
%%%%%%%%%%%%%%%%%%%%%%%%%%%%%%%%%%%%%%%%%%%%%%%%%%%%%%%%%%%%

In this case, we set $m_1^2=0$ and $\beta=1$. Note that the gauge
fixing term (\ref{gf2}) will not play a role, since the term
proportional to $P_1$ does not contribute to the interaction between
conserved sources. With these values of the parameters we have \be
g(k) = (n-2) (Z\ k^2 - m_2^2)\ k^2, \ee which is quartic in
the momenta. The terms proportional to $\kappa_2$ in the numerator
of Eq. (\ref{bigpro}) are also proportional to $k^2$, so this factor
drops out and we obtain the propagators for an ordinary massive
scalar particle (provided that $Z < 0$, in agreement with our
earlier dynamical analysis).

However, for the first term in Eq. (\ref{bigpro}) (the one
proportional to $\kappa_1^2$) there is no global factor of $k^2$ in
the numerator, and we must use the decomposition \be {1\over g(k)}
={-1\over (n-2)m_2^2}\left({1\over k^2} - {1\over k^2 - {m_2^2\over
Z}}\right). \ee Substituting in (\ref{bigpro}), and
disregarding contact terms, we obtain \be {\cal
P}_0=-\left({\kappa_1^2 \over (n-1)(n-2)}\right)\ {1\over
k^2}-\left(\kappa_2 +{1-a\over n-2} \kappa_1\right)^2{1\over Z k^2 -m_2^2}.\label{wdy} \ee Substituting in (\ref{flint}) and
adding the contribution of $P_2$ for $m_1^2=0$, which can be read
off form (\ref{lintfp}), we have \be {\cal L}_{int} =
{\kappa_1^2}\left[T^*_{\mu\nu}T^{\mu\nu}-{1\over
(n-2)}|T|^2\right]{1 \over k^2} -\left(\kappa_2 +{1-a\over n-2}
\kappa_1\right)^2{|T|^2\over  Z\ k^2 -\
m_2^2}.\label{lintfp2} \ee Note that the massless propagator in
(\ref{wdy}) combines with the second term in the spin-2 part to give
the factor $1/(n-2)$ in front of $|T|^2$. Eq. (\ref{lintfp2}) shows
that the massless interaction between conserved sources is the same
as in standard linearized General Relativity.

In addition, there is a massive scalar interaction, with effective
mass squared \be m^2_{eff}= { m_2^2\over Z} >0. \ee (note
that both parameters $m_2^2$ and $Z$ must be negative to yield a \emph{healthy}
interaction,
according to our earlier analysis), and effective coupling given by
\be \kappa^2_{eff} = {-1\over Z}\left(\kappa_2 +{1-a\over
n-2} \kappa_1\right)^2.\label{kappaeff} \ee These are subject to the
standard observational constraints on scalar tensor theories. If the
scalar field is long range, then the strength of the new interaction
has to be very small $\kappa_{eff} \lesssim 10^{-5} \kappa_1$
\cite{Will:2005va,Will:2001mx}. Alternatively, the interaction could
be rather strong, but short range, shielded by a sufficiently large
mass $m_{eff}\gtrsim( 30 \ \mu \mathrm{m})^{-1}$
\cite{Kapner:2006si,Will:2005va,Will:2001mx,Adelberger:2003zx}. In fact,
this mass term is not protected by any symmetry which
makes it sensitive to radiative corrections that will push it
till the cut-off scale of the theory. This way, the previous limit in the
mass is easily achieved. If the mass for the scalar field
is raised to the cut-off then any
value for $Z$ is possible (as long as  tachyons are not present), as the ghost
states only propagate at the cut-off scale and the propagation of
 new degrees of freedom is expected
at this scale which can render the theory unitary.

%%%%%%%%%%%%%%%%%%%%%%%%%%%%%%%%%%%%%%%
\subsection{Enhanced symmetry: WTDiff and Diff invariant theories}
%%%%%%%%%%%%%%%%%%%%%%%%%%%%%%%%%%%%%%%%

From general arguments, the interaction between sources in the
$\WTD$ theory is expected to be the same as in standard massless
gravity, since both theories only differ by an integration constant
but have the same propagating degrees of freedom.

In fact the result for  $\WTD$ can be obtained from the analysis of
the previous section by setting $Z=0$. In this case, the term
$m_2^2 h^2$ can be thought of as the additional gauge fixing which
removes the redundancy under the additional Weyl symmetry. With
$Z=0$ the second term in (\ref{lintfp2}) becomes a contact
term, and we recover the same result as in the standard massless
Fierz-Pauli theory \cite{Boulware:1973my}\footnote{Note
 also that the $\WTD$ invariant coupling to conserved sources
 requires $\kappa_1=-n\kappa_2$.  Using this and $a=2/n$ in (\ref{kappaeff})
  we have $\kappa_{eff}=0$, which again eliminates the scalar contribution.},
\be {\cal L}_{int} =
{\kappa_1^2}\left[T^*_{\mu\nu}T^{\mu\nu}-{1\over
(n-2)}|T|^2\right]{1 \over k^2},\label{stdfp} \ee as expected.

Note that in the $\D$ and $\WTD$ invariant theories, there is a
different possibility for gauge fixing. Rather than using the term
(\ref{gf2}) in order to take care of the $\TD$ part of the symmetry,
and then the $m_2^2h^2$
 to take care of the Weyl part, we can gauge fix the entire symmetry group with a standard term of the form
\be
{\cal L}_{gf}={\alpha\over 4}\left (\partial_\b
h^{\b\mu}+\gamma \pd^\mu h\right)^2,
\ee
where $\alpha$ and
$\gamma$ are arbitrary constants. This can be absorbed in a shift of
the parameters $a$, $b$ and $\beta$
$$
a\mapsto a+\alpha \gamma, \quad b\mapsto b-{\alpha\gamma^2\over
2},\quad \beta\mapsto \beta-{\alpha\over 2}.
$$
With these substitutions, the propagator becomes invertible, even if
it is not for the original values of $a,b$ and $\beta$ which
correspond to $\D$ or to $\WTD$. Needless to say, the result
calculated in this gauge coincides with (\ref{stdfp}).\\

Before ending this Chapter we would like to emphasize that
even if both theories give the same predictions at tree
level, this behaviour can change once interaction terms
are considered. First,
the vertices for the non-linear extensions may be different. Besides,
even if the vertices coincide, the fact that the off-shell
propagators for WTDiff and Diff are not related by a gauge-fixing
term makes it possible that the contributions from loops differ in both
cases \cite{Gabadadze:2005uq}.

%%%%%%%%%%%%%%%%%%%%%%%%%%%%%%%%%%%%%%%%%%%%%%%%%%%%%%%%%%%%%%%%%%%
\chapter{TDiff and Higher Spin: The Spin $3/2$ Case}\label{chapterRS}
%%%%%%%%%%%%%%%%%%%%%%%%%%%%%%%%%%%%%%%%%%%%%%%%%%%%%%%

In the previous Chapter we have shown that the free massless
spin-2 field can be consistently described by a traceless
tensor field with transverse gauge invariance.
This analysis has been extended to bosonic fields of higher
spin in \cite{Skvortsov:2007kz} and a similar result has been
found\footnote{This formulation
is in some sense opposite to the standard approach of higher
spin which resorts to the introduction of auxiliary fields to
build a covariant Lagrangian which yields the correct
equations of motion \cite{Fierz:1939ix,Fronsdal:1978rb,Fang:1978wz,deWit:1979pe} (see
also \cite{Singh:1974qz,Singh:1974rc} for the massive case).}.
Again, in the higher spin case,
although the new Lagrangian can be obtained from the
Fronsdal Lagrangian of \cite{Fronsdal:1978rb} by restricting to the traceless
part of the field, the equivalence
between both Lagrangians
is not trivial. In fact,  as shown in \cite{Skvortsov:2007kz} and similarly to spin-2,
the equivalence of the EoM is due to the appearance of a tertiary
constraint in the trace-free case
that kills the extra degree of freedom and makes
both theories equivalent at the classical level.

The covariant description of fermionic fields of spin $s>1/2$ also
needs the introduction of auxiliary fields which are rendered spurious
by an associated gauge invariance \cite{Fang:1978wz}.
A natural question one may ask is whether,
as happens in the bosonic case, there exists more than one Lagrangian
that describes the propagation of just the degrees of freedom of the spin under
consideration. In this Chapter we will  restrict to the
$s=3/2$ case. Again, we will
find that there are two possible Lagrangians
which satisfy the previous requirement:
the standard Lagrangian for spin-$3/2$ (the Rarita-Schwinger Lagrangian \cite{Rarita:1941mf})
and a traceless version of it which  enjoys a $S$-symmetry.
 We will also comment on the possibility of consistently
coupling the field $\psi_\m$ to the electromagnetic field in the last case.

Besides,
 the interacting spin-$3/2$ field appears very naturally in supergravity (SUGRA) \cite{VanNieuwenhuizen:1981ae}.
 At the linear level, the action built out of the addition of the Diff invariant spin-2 action
 and the Rarita-Schwinger (RS) action for the massless spin-$3/2$ constitute a supersymmetric action
\cite{VanNieuwenhuizen:1981ae}.
We will devote the last section of the Chapter to
prove that for the WTDiff Lagrangian
there is no minimal supersymmetric counterpart in the spin-$3/2$ sector.

We will follow the
conventions of \cite{deWit:1985aq}  and work
with a Majorana vector-spinor $\psi_\m$ (see also the Chapter
on Conventions). This Chapter is based on \cite{Blas:2008ce}
and  work in progress \cite{Blas08}.

%%%%%%%%%%%%%%%%%%%%%%%%%%%%%%%%%%%%%%%%%%%%%%%%%%%%%%%%%%%%%%%%%
\section{Lagrangians for Pure Massless Spin-$3/2$}\label{RSsection}
%%%%%%%%%%%%%%%%%%%%%%%%%%%%%%%%%%%%%%%%%%%%%%%%%%%%%%%

The most general local Lorentz invariant action for a Majorana
 vector $\psi_\m$ and first order in derivatives is given by\footnote{For
  a Dirac spinor, the coefficients in front of the first and second terms do not
necessarily coincide.}
\be
\label{gaction}
{\mathcal S}^{(3/2)}=\int \di^4 x \ \bar \psi_\m \left(\lambda (\g^\m \pd^\n
+ \g^\n \pd^\m)+\vartheta \g^\m \pdi \g^\n+\zeta
\eta^{\m\n}\pdi\right)\psi_\n.
\ee
After a transformation of the form
\be
\label{conftr}
\psi_\m\mapsto \psi_\m- \frac{a}{4} \g_\m\g^\r\psi_\r,
\ee
the coefficients are transformed as
\be
\label{coefftr}
\l\mapsto \l\left(1-a\right)-\frac{a}{2}\zeta, \quad \vartheta\mapsto
\vartheta(1-a)^2-\frac{a(1-a)}{2}\l+\frac{a}{2}\left(1-\frac{a}{4}
\right)\zeta.
\ee
This transformation is a field redefinition
which makes one of the coefficients spurious  except for the case $a=1$.
In this pathological case, the transformation is not invertible (see the
comment after (\ref{gamm_traceless})).

The Majorana field $\psi_\m$ has 16 real independent components, all of which
will be dynamical for a general action of the form (\ref{gaction}).
However, if the action is to describe a massless particle, only
the $\pm3/2$ polarizations should be dynamical, which implies
the need for a gauge invariance to render the remaining
polarizations non-dynamical\footnote{Recall also that fermions
have half as many PDoF as components as the other half are canonical momenta.}.
 The RS action, characterized
by $\l=-\vartheta=-\zeta$ (and the coefficients
related to it by a transformation (\ref{coefftr}) for $a\neq 1$)
is invariant under the transformation
\be
\psi_\m\mapsto \psi_\m+\pd_\m \e.
\ee
Let us consider now the transformation
\be
\label{generalgauge3/2}
\psi_\m\mapsto \psi_\m+\pd_\m \e+\g_\m \varphi,
\ee
which is the most general covariant gauge invariance
for the field $\psi_\m$ which does not involve the
spin-$3/2$ components of the field. Under the previous transformation,
the action changes as
\ba
\d S^{(3/2)}=-2\int \di^4 x \Big(\{(\l+\vartheta)\Box \bar\e&+&(\l+4\vartheta-\zeta)
\pd^\a \bar \varphi \g_\a\}\g^\m \psi_\m\nonumber\\
&& -\{(\l+ \zeta)\pd^\a \bar\e\g_\a+2(2 \l+\zeta)\bar\varphi\}\pd^\m\psi_\m\Big).\nonumber
\ea
For $2\l+\zeta\neq0$, the previous variation cancels for
\be
\bar \varphi=-\frac{(\l+\zeta)\pd^\a\bar\e\g_\a}{2(2\l+\zeta)},\quad (3\l^2+2\zeta\l+\zeta^2-
2\vartheta\zeta)\Box\bar \e=0.
\ee
In other words, for
\be
\vartheta=\frac{\zeta^2+2\zeta\l+3\l^2}{2\zeta}, \quad 2\l+\zeta\neq0,
\ee
the action (\ref{gaction})
is invariant under (\ref{generalgauge3/2}) with
$$\bar \varphi=-\frac{(\l+\zeta)\pd^\a\bar\e\g_\a}{2(2\l+\zeta)},$$
and $\e$ remains a free parameter. As it is clear from (\ref{coefftr}),
all these possibilities correspond to the RS action and field redefinitions
of the form (\ref{conftr}) with
$$a=\frac{2(\l+\zeta)}{\zeta}.$$
For the singular case $2\l+\zeta=0$ the variation cancels provided that
\be
\pdi \e=0, \quad (\l+4\vartheta-\zeta)\pdi\varphi=0.
\ee
In this case, the condition for a free gauge parameter $\varphi$ requires
the condition
\be
\l=\zeta-4\vartheta,
\ee
which, together with $2\l+\zeta=0$, imply that
\be
\l=-\frac{1}{2}\zeta, \quad \vartheta=\frac{3}{8}\zeta.
\ee
Substituting the previous values in (\ref{gaction})
(and fixing $\zeta$), one finds the action
\be
\label{WRSaction}
{\mathcal S}^{(3/2)}_{\WRS}={\mathcal S}_{\RS}(\hat \psi_\m)=
-\frac{1}{2}\int \di^4 x\ \bar{\hat \psi}_\m\e^{\m\n\r\s}\g_5 \g_\n \pd_\r \hat \psi_\s,
\ee
where $\hat \psi_\m\equiv\psi_\m-\frac{1}{4}\g_\m\g^\a\psi_\a$.
This action corresponds to the singular transformation  of the RS action, (\ref{conftr})
with $a=1$.
The WRS label
stands for the analogy of the transformation in (\ref{generalgauge3/2})
involving the field $\varphi$ (known as special supersymmetry, or simply, $S$-symmetry \cite{Fradkin:1985am})
with the Weyl gauge invariance.
 Notice that, as happens for the WTDiff case, the WRS action
is written in terms of a traceless field with {\em fewer} components
than the original field. In particular,
\be
\label{gamm_traceless}
\g^\m\hat\psi_\m=0,
\ee
which means that $\hat\psi_\m$ has just 12 independent real components. Besides, in complete
analogy with the WTDiff case, even if the action is invariant
under the Lorentz and the $S$-symmetries, the rigid \emph{superconformal} group
is not a symmetry of the Lagrangian (which happens when $\e$ and $\varphi$
are arbitrary \cite{Fradkin:1985am}).
As with in the spin-2 case, there is no action in (\ref{gaction}) invariant
under the general transformation (\ref{generalgauge3/2})\footnote{As happens for the Weyl and Diff symmetries,
 an action with this gauge group is possible once
higher derivatives terms are included (see \cite{Fradkin:1985am}), but the theory
is not unitary.}. Hence some of the low spin components
of the field $\psi_\m$ may be dynamical, as they are not automatically killed by the
gauge invariance.

It is important to note that this action is not related to the RS action
by a gauge fixing term, as the only covariant
gauge fixing term just involves the $\vartheta$ term
$$\bar \psi_\m\g^\m \pdi \g^\n\psi_\n$$
in (\ref{gaction}).
To our knowledge, the WRS action has not been studied
in the past\footnote{For the Lagrangians equivalent to
 RS see  \cite{VanNieuwenhuizen:1981ae,deWit:1985aq}.}.
The remaining possibilities will include both spin-$1/2$ polarizations,
one of which will be a ghost \cite{VanNieuwenhuizen:1981ae} (see also below).\\

The analysis of the
degrees of freedom can be performed in a covariant way
after introducing a system of projectors
as in \cite{Deser:1977ur,VanNieuwenhuizen:1981ae} or performing the decomposition
\be
\psi_0=A, \quad \psi_i=t_i+\g_i \chi+\pd_i E,
\ee
with $\g_i t^i=\pd_i t^i=0$. Notice that the
presence of the $\g_i$ matrices in the definition of
$\chi$ implies that it is an anti-Majorana
fermion\footnote{We could have defined $\chi=\g_0\eta$ with $\eta$ being a
Majorana spinor.}
$$
\bar\chi=-\chi^TC.
$$
This decomposition breaks the Lorentz
invariance, but this allows to identify the
actual PDoF and the constraints of the theory.
It is also very useful to show that
the RS and the WRS are the only possibilities
out of the general action (\ref{gaction}) endowed
with a gauge invariance. To prove it, it suffices
to show that these are the only possibilities
 where the kinetic term of the
associated EoM
is singular   \cite{HenneauxTeit}.

In terms of the previous fields, the general Lagrangian
(\ref{gaction}) can be written as
\be
\label{lagran_3/2}
{\mathcal L}={\mathcal L}^{(3/2)}+{\mathcal L}^{(1/2)},
\ee
where ${\mathcal L}^{(3/2)}\equiv-\zeta\bar t_i \pdi\ t_i$ and
\ba
&&{\mathcal L}^{(1/2)}\equiv\bar E\left\{(\zeta-\vartheta)\g_0\pd_0-
(2\l+\vartheta+\zeta)\g_i\pd_i\right\}\Delta E\nonumber\\
&&+\bar A\left\{
(2\l+\vartheta+\zeta)\g_0\pd_0-\g_i\pd_i(\zeta-\vartheta)\right\}A
+\bar \chi\left\{3(3\vartheta-\zeta)\g_0\pd_0-\g_i\pd_i(6\l+9\vartheta-
\zeta)\right\}\chi
\nonumber\\
&&\hspace{2cm}+2\bar\chi\left\{-(4\l+3\vartheta+\zeta) \Delta E-(3\vartheta-\zeta)\g_0\g_i
\pd_0\pd_i E\right\}\nonumber\\
&&\hspace{2cm}\quad +
2\bar A\left\{-(\l+3\vartheta)\g_0\g_i\pd_i \chi+(\l+\vartheta)
[\pd_0(3\chi-\g_i\pd_i E)-\g_0\Delta E]\right\}.\nonumber
\ea
The kinetic
part  can be written as,
\begin{displaymath}
(\bar E, \bar \chi , \bar A)\left(\begin{array}{lll}
(\zeta-\vartheta)\g_0\Delta & \ (\zeta-3\vartheta)\g_0\g_i\pd_i&\ (\l+\vartheta)\g_i\pd_i\\
(\zeta-3\vartheta)\g_0\g_i\pd_i&\ 3(3\vartheta-\zeta)\g_0&\ 3(\l+\vartheta)\\
-(\l+\vartheta)\g_i\pd_i&\ 3(\l+\vartheta)&\ (2\l+\vartheta+\zeta)\g_0
\end{array}\right)
\left(\begin{array}{c}
\dot E\\ \dot \chi \\ \dot A
\end{array}\right),
\end{displaymath}
and the determinant of the matrix multiplying the time derivative of the fields
is
\be
16\zeta^4(-2\vartheta \zeta+\zeta^2+2\zeta\l+3\l^2)^4\Delta^4.
\ee
Thus, we find that the theory will include constraints whenever (we take $\zeta\neq0$ as otherwise
the spin-$3/2$ degrees of freedom are not present)
\be
\label{cond_3/2}
\vartheta=\frac{\zeta^2+2\zeta\l+3\l^2}{2\zeta}.
\ee
As we found previously, this condition correspond to the existence of a gauge invariance
 of the form (\ref{generalgauge3/2}). In the singular case, the
kinetic term will be non-singular once the constraints are introduced back
in the Lagrangian. Besides, notice that for the general case, the determinant
has a definite positive sign, to be contrasted with the negative sign of the determinant
of kinetic part of the spin-$3/2$ case. Thus, the kinetic
term of the total Lagrangian (\ref{lagran_3/2}) has not a definite sign unless
(\ref{cond_3/2}) is satisfied. This means that if (\ref{cond_3/2})
does not hold, the
action (\ref{gaction}) has propagating
ghosts in its spectrum, as claimed in \cite{VanNieuwenhuizen:1981ae}.

%%%%%%%%%%%%%%%%%%%%%%%%%%%%%%%%%%%%%%%%%%%%%%%%%%%%%%%%%%%%%%%%%
\section{Propagator and Coupling of the WRS action}
%%%%%%%%%%%%%%%%%%%%%%%%%%%%%%%%%%%%%%%%%%%%%%%%%%%%%%%

For the RS Lagrangian, the propagator, spin content and
unitarity properties can be found in \cite{Das:1976ct,Sterman:1977ds,VanNieuwenhuizen:1981ae}.
In this case, the gauge invariance including a derivative
allows to kill all the low-spin states, leaving just the $\pm3/2$
polarizations as physical.

For the  WRS action (\ref{WRSaction}), the naive
counting of PDoF implies the existence of spin-$1/2$ components. To show
that this is the case, we analyze the EoM derived from the
action (\ref{WRSaction}).
One readily finds that they correspond to the
 $\g$-traceless part
of the RS case in the gauge $\g^\m\psi_\m=0$, which can be reached
by a $S$-transformation in the WRS case and by a gauge
 transformation in the RS case \cite{VanNieuwenhuizen:1981ae},
\be
{\mathcal R}_{\WRS}^\m\equiv\frac{\d {\mathcal L}_{\WRS}}{\d \bar\psi_\m}=
\left(\d^\m_\a-\frac{1}{4}\g^\m\g_\a\right)
\frac{\d {\mathcal L}_{\RS}(\hat\psi_\m)}{\d \bar{\hat \psi}_\m}\equiv
\left(\d^\m_\a-\frac{1}{4}\g^\m\g_\a\right){\mathcal R}_{\RS}^\a(\hat\psi_\m)=0,
\ee
with $\g_\a {\mathcal R}^\a_{\WRS}=0$, which is the Bianchi identity associated
to the fermionic  $S$-symmetry.
Contracting the EoM with the derivative operator, one finds
\be
\pd_\m {\mathcal R}^\m_{\WRS}=-\frac{1}{4}\pdi\left(\g_\a{\mathcal R}^\a_{\RS}(\hat\psi_\m)\right)=0.
\ee
Thus, contrary to what
happens in the bosonic case, we do not recover the missing equations of the RS
Lagrangian (in this case
the $\g$-trace of the RS EoM)\footnote{This result was expected as there is no
gauge invariance left in the  WRS action
written in terms of $\hat \psi_\m$, which means that no new constraints
can appear in the EoM.}.
From the identity
$$
\g_\a{\mathcal R}^\a_{\RS}(\hat\psi_\m)=-2\pd^\a\hat\psi_\a,
$$
we see that there is a spin-$1/2$ PDoF as the equation of motion for
$\pd^\a\hat\psi_\a$ is
\be
\label{newspin}
\pdi\pd^\a\hat\psi_\a=0,
\ee
in contrast to the RS case where $\pd^\a\hat\psi_\a$ cancels on shell\footnote{Similar
 equations of motion are also obtained if we add a term
\be
\l \bar \psi_\m\g^\m \g^\n \psi_\n
\ee
to the RS action. This is reminiscent to what happens
in unimodular gravity \cite{Henneaux:1989zc}.}.
Besides,  the residual gauge transformation satisfies $\pdi \e=0$,
which leaves this combination
invariant as
\be
\d \pd^\a\hat\psi_\a=\Box \e=0.
\ee
This implies that, in principle, the WRS case is not classically equivalent to the RS case
as there is one more spin-1/2 PDoF. However, from the fact that
this new PDoF does not mix with the spin-$3/2$ part, we can consistently
fix it to cancel by  the initial condition
$$\pd^\a\hat \psi_\a\big|_0=0.$$
In this case,  equation (\ref{newspin}) implies that
the missing equation also holds and that both systems are equivalent.
This situation is analogous to
what happens in ordinary gauge theory when one fixes the gauge through a covariant
quadratic gauge fixing term (see {\em e.g.} \cite{Das:1976ct,Itzykson:1980rh}).

The previous result is trivial in the case of {\em free} theories
but it may change in the presence of sources. Let
us see that for {\em conserved} sources this is not the case, {\em i.e.}
both theories yield the same physical results in this case.
To show this, we will
 consider the  coupling of the free spin-$3/2$ field to a conserved source $J_\a$,
 $\pd^\a J_\a=0$.
 The most general non-derivative
 covariant coupling will be of the form
$$
{\mathcal S}_{int}=\int \di^4x \bar\psi_\m\left(J^\m-\frac{b}{4}\g^\m\g_\a J^\a\right)+h.c.
$$
The consistency of the equations of motion implies that for the RS case $b=0$ whereas for
WRS $b=1$. The equations of motion for the WRS case are
\be
\label{eom}
\left(\d^\m_\a-\frac{1}{4}\g^\m\g_\a\right)\left({\mathcal R}_{\RS}^\a(\hat\psi_\m)
-J^\a\right)=0.
\ee
Again, from the conservation of the current and the Bianchi identity for ${\mathcal R}_{\RS}^\m$,
contracting the EoM with the derivative operator,
we obtain
\be
\pdi\left(\g_\a{\mathcal R}_{\RS}^\a(\hat\psi_\m)-\g_\a J^\a\right)=0.
\ee
After the imposition of the initial condition
$$
\left(\g_\a{\mathcal R}_{\RS}^\a(\hat\psi_\m)-\g_\a J^\a\right)\big|_0=0,$$
this is  equivalent to the missing equation of (\ref{eom}) compared to the RS case.
Thus the propagator that mediates the interaction between two conserved sources
is the same in both cases. In particular we find
\be
\pdi \hat \psi_{\WRS}^\m=J^\m-\frac{1}{2}\g^\m \g_\a J^\a+\g^\m\xi.
\ee
with  $\pdi \xi=0$.
The interaction between sources can be read from the quantity
\be
\bar J^\m \hat \psi_\m=
\bar J^\m\frac{1}{\Box}\left(\eta_{\m\n}\pdi+\frac{1}{2}\g_\m\pdi\g_\n\right)J^\n,
\ee
which coincide with that of the RS (see {\em e.g.} \cite{Deser:1977ur}). In particular,
this form guarantees the unitarity of the theory.
Thus, even if we have found an additional field
 $\xi$ in the WRS case, given that it is a free field it can be projected out consistently.

It is interesting to note that, as happens for the spin-2 Lagrangian, the WRS massive case
is completely different from the RS and the propagation involves new degrees of freedom.

%%%%%%%%%%%%%%%%%%%%%%%%%%%%%%%%%%%%%%%%%%%%%%%%%%%%%%%%%%%%%%%%%
\subsection{Remarks on Quantization and Consistent Coupling}
%%%%%%%%%%%%%%%%%%%%%%%%%%%%%%%%%%%%%%%%%%%%%%%%%%%%%%%

In the previous section we showed that apart from the Rarita-Schwinger (RS)
action and the actions related to it by a gauge fixing term or by a
field redefinition, there is another Lorentz invariant action
for the spin-$3/2$ field (the WRS action) with the
same physical predictions once coupled to a conserved
source. This equivalence needs the
imposition of initial conditions which may not be compatible
with the canonical (anti)commutators as happens for electromagnetism
in the Lorentz gauge. For the electromagnetic case, this problem
is solved by imposing the condition as a restriction in the physical
Hilbert space where the theory turns out to be unitary (Gupta-Bleuler formalism).
Even if we have not applied this formalism to the WRS theory, the similarities
with the standard case in the presence of a covariant
gauge fixing term, whose correspondence with the
canonical treatment in the gauge $\g^i\psi_i=0$ can be found in
 \cite{Das:1976ct}, makes one think that
it may also be valid in this case.
Besides, no Fadeev-Popov or Nielsen-Kallosh ghosts present
in the RS case (cf. \cite{VanNieuwenhuizen:1981ae})
will appear in the quantization of the WRS action, as it
has no gauge invariance.

The previous conclusions may change in the presence of interaction
where the extra spin-$1/2$ may become dynamical. Besides, the  proof
of unitarity of interacting massless theories resorts on gauge invariance
(see {\em e.g.} \cite{Das:1976ct} for supergravity)
and its absence in the WRS theory casts some doubts in the consistency
of any interacting theory.\\

Even more, the {\em interacting} theories
of higher spin, both massive and massless, may be problematic already at the
classical level.  For the
massive spin-$3/2$ field
 there are problems with unitarity and causal propagation once the
field is coupled to an external electromagnetic source \cite{Johnson:1960vt,Velo:1969bt}.
 For the massless case,
the inconsistency  occurs already at an algebraic level.

 Namely, if we substitute the
ordinary derivative by a covariant derivative in the RS action, differentiating
with the covariant derivative  $D_\m=\pd_\m-\I e A_\m$  and after using
 the Bianchi identity of the RS action
$$\pd_\m{\mathcal R}^\m_{\RS}=0,$$
 we find \cite{VanNieuwenhuizen:1981ae}
$$
F_{\m\n}\g^\m\psi^\n=0.
$$
The previous expression  means that either $\psi_\m=0$ or that the photon is a gauge excitation.
A similar problem occurs for every massless higher spin theory, as the
Bianchi identities of the free theory always imply some condition in the
background field. It was suggested in \cite{Skvortsov:2007kz} that
the description in terms of traceless fields may alleviate this problem as the
Bianchi identities are less stringent in this case.

 For the WRS case, coupling minimally the action to the electromagnetic field, one finds
  the equations of motion
\be
\label{chargedeom}
(\d_\m^\a-\frac{1}{4}\g^\a\g_\m)\e^{\m\n\r\s}\g_5\g_\n D_\r \hat \psi_\s=\I\left(\g^\m D_\m \hat \psi^\a
-\frac{1}{2}\g^\a D^\m \hat\psi_\m\right)=0.
\ee
After applying the covariant derivative, the equations of motion read
\be
eF_{\m\n}\g^\m\hat\psi^\m=\frac{\I}{2}\g^\b D_\b
(D_\a \hat\psi^\a),
\ee
which is not a constraint but a field equation\footnote{The same happens if
one considers the coupling of the gauge-fixed RS action.}.
 The hyperbolic structure of this equation
is independent of the connection, and  due to Lorentz invariance
there are just two possibilities:
either the determinant associated to this equation cancels identically (as happens for RS) or
 the characteristic surfaces have null normals \cite{Velo:1969bt}. The first possibility can not
be realized as it would indicate the presence of a gauge invariance, thus in the WRS case the
signals propagate in the null-cone. More explicitly, the symbol
of the system of differential equations is
\be
\label{det}
\sigma=\left((\g^\m)^{ab} \eta^{\a\s}-\frac{1}{2}(\g^\a)^{ab}\eta^{\m\s}\right)n_\m,
\ee
where $n_\m$ is an arbitrary vector. The determinant of this operator is
\be
\det \sigma=\frac{1}{16}(n^2)^8.
\ee
The main concern about  the previous coupling is that the states of low spin
corresponding to $\pd_\a\hat\psi^\a$ are turned on by the interaction, and this may spoil the
unitarity of the theory.

The absence of a gauge invariance implies that  Slavnov-Taylor identities
can not be derived in the standard fashion and unitarity may be violated even
at tree level. We leave
the study of these issues for future research\footnote{Even if unitarity
 is not preserved, one could try to introduce new fields of spin-$1/2$
to obtain a consistent theory.} \cite{Blas08}.

%%%%%%%%%%%%%%%%%%%%%%%%%%%%%%%%%%%%%%%%%%%%%%%%%%%%%%%%%%%
\section{Supersymmetric Extensions of WTDiff}
%%%%%%%%%%%%%%%%%%%%%%%%%%%%%%%%%%%%%%%%%%%%%%%%%%%%%%%%%%%%%%

A natural question concerning the possible extensions of the WTDiff Lagrangian of
the previous Chapter and its relation to the spin-$3/2$ field
is whether a minimal supersymmetric extension exists. In other words, as
 the number of \emph{off-shell} and \emph{on-shell}  degrees of freedom of the
massless WTDiff case
coincides with that of Diff (and RS) actions (see {\em e.g.} \cite{VanProeyen:2003zj}),
 we may
wonder about the existence of an action for the spin-$3/2$ field such that
the total action of WTDiff graviton plus gravitino has a certain
 global supersymmetry .
A first sign that this may  not be possible unless  more fields are added
to the
theory is that, as we showed in section \ref{RSsection}, the only Lagrangian
for the field $\psi_\m$ that describes purely spin-$3/2$ \emph{on-shell}
is the RS Lagrangian whose
supersymmetric counterpart is the usual linearized Einstein-Hilbert action\footnote{We
 could consider actions for the bosonic sector with more
degrees of freedom \emph{e.g.} allowing
for a propagating torsion or non-metricity, but this goes beyond the present work.}.
One may still think that the supersymmetric transformations can be
deformed so that the WTDiff action is also supersymmetric with the RS action.
We will study this possibility in a completely general way.

Let us first consider the variation of the WTDiff at linear level (\ref{wtddef})
under a variation $\d h_{\m\n}$ in four dimensions,
\ba
\d{\mathcal S}_{\WTD}^{(2)}&=&\int \di^4 x\ \delta \hat h_{\m\n}\left(R_{\m\n}^L(\hat h)
-\frac{1}{2}\eta_{\m\n} R^L(\hat h)\right)
\nonumber\\
&=&\frac{1}{4}\int \di^4 x\ \delta h_{\m\n}
\Big(4\eta^{\a b}\eta^{\b(\m}\eta^{\n)a} -2\eta^{\a\b}\eta^{a\m}\eta^{b\n}- \eta^{ab}\eta^{\m\a}
\eta^{\n\b}\nonumber\\
&&\hspace{2cm}-\eta^{\m\n}\left\{\eta^{\a a}\eta^{\b b}-\frac{3}{4}\eta^{a b}\eta^{\a\b}\right\}
\Big)\pd_\a \pd_\b h_{ab}.
\ea
For the spin-$3/2$  Majorana field $\psi_\m$ we will take the general action (\ref{gaction}).
 The most general supersymmetric transformation for Majorana spinors and gravitons
 can be written as\footnote{The supersymmetric transformation should preserve
the traceless condition of the WTDiff field $\hat h_{\m\n}$,
 which for the usual supersymmetric transformation of the graviton implies
\be
\d h=\bar \e \g^\m \psi_\m=0.
\ee
This seems to imply that the supersymmetric partner of the field $\hat h_{\m\n}$ should be
 the field $\hat \psi_\m$ but, as we will see, this is not so.}
\ba
\delta h_{\m\n}&=&\bar\epsilon \g_{(\m}\psi_{\n)}+A \eta_{\m\n}\bar\epsilon\g^\r\psi_\r,\nonumber\\
\delta \psi_\m&=&\left(B\pd_\m h+C\pd_a h^a_\m+D\g_\m \g^\n \pd_\n h
+E \g_\m \g^\a\pd_b h^b_\a+ F\s^{ab}\pd_a h_{\m b}\right)\e,
\ea
where $\s^{ab}\equiv \frac{1}{4}[\g^a,\g^b]$.
Some of the previous transformations are simply field redefinitions or gauge transformations
for certain Lagrangians but we will  consider all the coefficients as independent.\\

The variation of the bosonic Lagrangian can be written as
\ba
\label{susygr}
\delta{\mathcal S}_{\WTD}^{(2)}=\frac{1}{4}
\int \di^4 x\ \bar \e\Big(&&-\eta^{ab}\g^\a\psi^\b+2\eta^{\a a}\g^b\psi^\b+2\eta^{\a a}\g^\b\psi^b
-2\eta^{\a\b}\g^b\psi^a\nonumber\\
&&-\eta^{\a a}\eta^{\b b}\g^\r\psi_\r+\frac{3}{4}\eta^{ab}\eta^{\a\b}\g^\r\psi_\r
\Big)\pd_\a \pd_\b h_{ab}.
\ea
For the variation of the fermionic part  we find
\ba
\label{susyrs}
\d {\mathcal S}^{(3/2)}&&=-\int\di^4 x \ \bar \e
\Big\{(2B(\l+\zeta)+4D(2\l+\zeta)-F\l)\eta^{ab}\g^\a\psi^\b\nonumber\\
&&\hspace{-.5cm}
+(2C\l+4E(2\l+\zeta)+F\l)\eta^{\a a}\g^b\psi^\b+\zeta(2C-F)\eta^{\a a}\g^\b\psi^b
+F\zeta\eta^{\a\b}\g^b\psi^a\nonumber\\
&&\hspace{-.5cm}
+(2B(\l+\vartheta)+2D(\l+4\vartheta-\zeta)-F\vartheta)\eta^{ab}\eta^{\b\a}\g^\r\psi_\r
+\l(2C-F)\eta^{\a a}\eta^{\b b}\g^\r\psi_\r\nonumber\\
&&\hspace{-.5cm}+(2C\vartheta+2E(\l+4\vartheta-\zeta)+F(\l+\vartheta))\eta^{\a a}\g^b\g^\b\g^\r\psi_\r
\Big\}\pd_\a \pd_\b h_{ab}.
\ea
Comparing the third and forth coefficients of (\ref{susygr})
and (\ref{susyrs}), we find $C=0$. From the relation
between the last but one coefficient  and the forth one of (\ref{susygr}), we find $\zeta=-2\l$. Finally,
comparing
the second and forth coefficient we arrive at $F\zeta=0$. The condition
 $\zeta\neq 0$ is necessary if we want the fermionic action to describe
 spin-$3/2$ fields. This means that $F=0$, which, together with
$C=0$ and $(2\l+\zeta)=0$, implies that the third term
 of (\ref{susyrs}) cancels and  there is
no way in which both variations can cancel each other. Thus, we
conclude that there is not a minimal supersymmetric system
including the WTDiff Lagrangian.\\

One could try to add more fields to the theory to find
a supersymmetric action. In \cite{Nishino:2001gd} a supersymmetric
extension for unimodular gravity was found by the addition of Lagrange multipliers
to enforce a traceless conditions on the spin-$2$ and spin-$3/2$ fields.
 It was shown that the system
has a local {\em constrained} supersymmetry for any cosmological constant while the gravitino remains
massless. As we said, the addition of these Lagrange multipliers goes beyond
the minimal coupling considered in this section and can be problematic \cite{Gabadadze:2005uq}.

Finally, notice that the addition of a mass term or putting the gravitino in an anti-de Sitter background
can not help to build a supersymmetric action as the previous incompatibility will still be present.

%%%%%%%%%%%%%%%%%%%%%%%%%%%%%%%%%%%%%%%%%%%%%%%%%%%%%%%%%%%%%%%%%%
\part{Non-linear extensions: from Unimodular gravity to   Bigravity}
%%%%%%%%%%%%%%%%%%%%%%%%%%%%%%%%%%%%%%%%%%%%%%%%%%%%%%%%%%%%%%

%%%%%%%%%%%%%%%%%%%%%%%%%%%%%%%%%%%%%%%%%%%%%%%%%%%%%%%%%%%%%%%%%%
\chapter{Non-linear Extensions of TDiff Lagrangians}\label{chapternl}
%%%%%%%%%%%%%%%%%%%%%%%%%%%%%%%%%%%%%%%%%%%%%%%%%%%%%%%%

In  Chapter \ref{chapterLorentz}, we have studied different Lagrangians which are
phenomenologically  equivalent to  GR in the linearized approximation. In particular,
the TDiff invariant Lagrangians are admissible as long as the mass term compatible with
the TDiff symmetry is  set to  an energy scale beyond the
scales at which GR has been studied ($m \gtrsim (10\ \m\mathrm{m})^{-1}\sim 10^{-14}\ \mathrm{TeV}$).
Besides, we have found two
inequivalent possibilities which describe pure spin-2 massless propagation at any scale: the
usual Diff invariant Lagrangian and the WTDiff Lagrangian. As it is well known, the
linear theory of GR is not enough to describe the gravitational interaction. First, it
fails observationally as it does not predict the nonlinear effects of GR as
the right perihelion of Mercury \cite{Ortin}. Besides,  from the
\emph{strong equivalence principle}, gravity must couple to any kind
of energy including its own \cite{Will:2001mx}. If the gravitational interaction
 is described by a spin-2  particle, this particle must be coupled to its own energy-momentum tensor.
Both arguments imply the inclusion of interaction terms in the Lagrangian. As we are dealing
with a theory with a gauge invariance, the new terms must be compatible with this gauge invariance
as otherwise they generically impose new constraints in the propagating fields. This requirement
uniquely determines the nonlinear terms for the Diff case
\cite{Ogiev:1965,Deser:1969wk,Wald:1986bj,Boulanger:2000rq} (see also
\cite{Grishchuk:1984,FeynmanGrav,Gupta:1957}).
The Noether trick can  also be considered to constructively build the nonlinear theory.
However, for GR it is not very useful as it requires the knowledge of the deformation of the linear
algebra to be applied \cite{Ortin}.
For an argument based on quantum gravity for the nonlinear extension see \cite{Boulware:1974sr}.

For the TDiff and WTDiff cases much less is known about the possible nonlinear extensions. Transverse
diffeomorphisms form a group also at the nonlinear level, providing a first possibility for the nonlinear
gauge invariance \cite{vanderBij:1982,Buchmuller:1988wx} (see also \cite{Pitts:2001jw}). Furthermore,
a nonlinear Weyl transformation is also easily added to the picture and a unique
Lagrangian appears for this WTDiff nonlinear gauge invariance \cite{Blas:2007pp}. However, as we will
argue, it is not clear whether in this case there are no other possible nonlinear extensions

A consistent
nonlinear extension  of the massive case  may be sought using the
St\"uckelberg or Higgs mechanisms to recover a gauge symmetry at the linear level
\cite{Zinoviev:2006im,Chamseddine:2004dh,ArkaniHamed:2002sp}.
 In both cases, the appearance
of nonlinear terms typically implies the propagation of a new degree of freedom
which makes the theory non-unitary\footnote{A possible solution for this problem
is to impose an additional constraint at the nonlinear level in the spirit of \cite{Hooft:2007bf}.}
\cite{Boulware:1973my}.\\

In this Chapter we will first present some results on the possible nonlinear extensions of the
TDiff theory
and then we will focus on the only consistent possibility that we know about. We will show that
the nonlinear TDiff Lagrangian is completely equivalent to a scalar-tensor theory whereas
the nonlinear WTDiff corresponds to a Lagrangian for unimodular gravity.  We will then
comment on the possible ways in which matter can be coupled to gravity in theories
invariant under TDiff.
The last section of the Chapter is devoted to the first order formalism of WTDiff and
the coupling of the {\em vielbein} to a spin-3/2 field.
This Chapter is partially based
on \cite{Alvarez:2006uu,Blas:2007pp,Blas08}.

%%%%%%%%%%%%%%%%%%%%%%%%%%%%%%%%%%%%%%%%%%%%%%%%%%%%
\section{Non-linear Extensions}
%%%%%%%%%%%%%%%%%%%%%%%%%%%%%%%%%%%%%%%%%%%%%%%%%%%

In this section we will present two different ways of building the nonlinear extension
of the linear Lagrangians of the previous chapters.
We will first say a few words about the techniques that allow to build the interaction terms
constructively and apply a method similar to that suggested by Deser in \cite{Deser:1969wk}
for GR to the WTDiff case.
We will find
that we get an inconsistent\footnote{By inconsistent
we mean that the gauge invariance does not survive at the nonlinear level.}
Lagrangian. Then we
will present the nonlinear extensions of TDiff which we can construct directly from the intuition
gained from the linear theory.

%%%%%%%%%%%%%%%%%%%%%%%%%%%%%%%%%%%%%%%%%%%%%%%%%%%
\subsection{Systematic Extension}\label{systematic}
%%%%%%%%%%%%%%%%%%%%%%%%%%%%%%%%%%%%%%%%%%%%%%%%%%

There are different ways in which the non-linear extensions of the theories of free gravitons
 can be found constructively. The
most direct one is to consider
the energy-momentum tensor of the graviton as a source for its  equations
of motion. This amounts to the
first correction, or three-graviton vertex, for the linear action and for the Diff case it
is not a {\em consistent} way to proceed, as there is
no Lagrangian that gives rise to these equations of motion
\cite{Ogiev:1965,Ortin}. Another way of performing the extension is to first show how the gauge invariance can
be deformed nonlinearly \cite{Ogiev:1965,Wald:1986bj,Boulanger:2000rq} and then build a Lagrangian endowed
with the nonlinear gauge invariance. To find the possible deformations, one benefits from the
nonlinear nature of the closure of the algebra associated
to the gauge invariance, which relates the different orders in a
deformation parameter \cite{Ogiev:1965}.
  For the case of linearized Diff
 symmetry these nonlinear
deformations lead {\em uniquely} to the group of nonlinear diffeomorphisms after some mild assumptions.
The equivalent calculation for TDiff and WTDiff is more cumbersome and  is
currently under research \cite{Blas08} (see also \cite{Pitts:2001jw}).
It is worth noticing that even if the usual techniques for deforming
gauge algebras can be applied (see {\em e.g.} \cite{Henneaux:1997bm}) the fact of dealing
with a {\em  reducible} gauge invariance implies some additional difficulties.\\

An alternative approach for GR which extends easily to the WTDiff case exists \cite{Deser:1969wk,Blas:2007pp}.
This approach is based on the first order (or Palatini's) formulation of gravity \cite{Deser:1969wk}
(see also \cite{Deser:1987uk} for the generalization to a curved background).
The   first order formulation of the second order Lagrangian (\ref{MA}) for the WTDiff case is
built from the action
\be
\label{second}
S^{(1)}_{\WTD}=\frac{1}{\kappa^{n-2}}\int \di^n x\left\{-\hat h^{\m\n}\partial_{[\m}\Gamma_{\phantom{\r}\r]\n}^{\r}+\eta^{\m\n}
\Gamma_{\phantom{\r}\l[\m}^{\r}\Gamma_{\phantom{\l}\r]\n}^{\l}\right\},
\ee
where $\hat h_{\m\n}=h_{\m\n}-h\eta_{\m\n}$ and the metric and the connection
are now considered as independent fields. The equations of motion from the
 variation of $\hat h_{\m\n}$ are the traceless part
of the Fierz-Pauli case, whereas from the variation of $\Gamma_{\phantom{\r}\m\n}^{\r}$ we find a
 constraint for this field which, once solved,
yields (for $n\neq 2$)
\be
\Gamma_{\phantom{\r}\m\n}^{\r}=\frac{1}{2}\eta^{\r\s}\left(\pd_\m \hat h_{\n\s}+\pd_\n \hat h_{\m\s}-\pd_\s \hat h_{\m\n}\right).
\ee
This is just the equation of compatibility of the connection and
the traceless metric at linear order. Substituting this constraint
in the action and after the redefinition $h_{\m\n}\mapsto \sqrt{2}\kappa^{(n-2)/2}h_{\m\n}$,
 we just get the WTDiff Lagrangian for $h_{\m\n}$, (\ref{wtddef}). This is not a trivial result as
the equivalency between the
first and second order formulations without the
use of Lagrange multipliers is not guaranteed {\em a priori} \cite{Iglesias:2007nv,Exirifard:2007da}.
The next step is computing
the energy-momentum tensor of the $h_{\m\n}$ field and couple it to the graviton.
 As it is well known, there is a great amount of ambiguity in the definition of the energy-momentum tensor
 of the gravitational field
 (see {\em e.g.} \cite{Babak:1999dc,Nikishov:2003sq}). Following \cite{Deser:1969wk},
we will use a {\em modified} Rosenfeld's prescription \cite{Ortin}.

Rosenfeld's prescription
consist of substituting the flat space metric $\eta_{\m\n}$ by an auxiliary metric $\g_{\m\n}$
in a way that renders the action invariant under auxiliary non-linear diffeomorphisms. One can
prove that the quantity
$$t_{\m\n}=-\frac{2}{\sqrt{-\gamma}}\frac{\delta S[\g]}{\delta \gamma^{\m\n}}\Big|_{\gamma_{\m\n}=\eta_{\m\n}},$$
is symmetric and conserved on-shell \cite{Babak:1999dc}. Thus, one may identify $t_{\m\n}$
with the energy momentum tensor for the action $S[\eta]$.
To use the previous prescription,
we need to define $\hat h^{\m\n}$ in a curved background
\be
\label{hathg}
\hat h^{\m\n}[\g]\equiv h^{\m\n}-\frac{1}{n}\g^{\m\n}\g_{\a\b}h^{\a\b}
\ee
and
assign a transformation law  under
the auxiliary coordinate transformations to the fields $\hat h_{\m\n}$
and $\Gamma_{\phantom{\r}\m\n}^{\r}$ (this is the strongest assumption of Deser's method \cite{Ortin}).
The general action reads
\ba
\label{Rosenfeld}
{\mathcal S}[\gamma]_{\WTD}=\frac{1}{\kappa^{n-2}}\int \di^n x \left(-|\g|^{a}\hat h^{\m\n}[\g]
\nabla[\gamma]_{[\m} \Gamma_{\phantom{r}\r]\n}^{\r}+
|\gamma|^{b}\gamma^{\m\n}
\Gamma_{\phantom{\r}\l[\m}^{\r}
\Gamma_{\phantom{\r}\r]\n}^{\l}
\right),
\ea
where $a$ and $b$ are arbitrary constants depending on the transformation
rules for the metric and the connection.
The conserved energy-momentum tensor derived from this action differs from the
one of \cite{Deser:1969wk} due to the appearance of $\g_{\m\n}$ in the definition of
$\hat h^{\m\n}$ (\ref{hathg}). However, in the gauge $h=0$,
 $h_{\m\n}=\hat h_{\m\n}$ and the equations of motion for the WTDiff Lagrangian
are the same as the Diff ones. Thus, the quantity
\be
\tilde t_{\m\n}=
-\frac{2}{\sqrt{-\gamma}}\frac{\delta S[\g;\hat h_{\m\n}]_{\D}}{\delta \gamma^{\m\n}}\Big|_{\gamma_{\m\n}=\eta_{\m\n}},
\ee
is also conserved in this gauge. Besides, one
can easily convince oneself that this quantity is conserved
as it corresponds to the energy-momentum tensor associated
with the choice of $\hat h^{\m\n}$ to be
a contravariant tensor density ($a=0$).\\

If we consider $\hat h^{\m\n}$ to be
a contravariant tensor density ($a=0$) and the indices of the connection to behave
like a vector ($b=1/2$), it is easy to see that
the energy-momentum tensor  $\tilde t_{\m\n}$ is
 given by
the usual energy-momentum tensor
of \cite{Deser:1969wk} except for the fact that the tensor $\hat h_{\m\n}$ is now traceless.
 Following \cite{Deser:1969wk},
this energy-momentum tensor can be derived from the term
\be
\label{nlcompl}
{\mathcal S}^{(2)}=-\frac{1}{\kappa^{n-2}}\int \di^n x \hat h^{\m\n}\Gamma_{\phantom{\r}\r[\m}^{\s}
\Gamma_{\phantom{\r}\s]\n}^{\r}.
\ee
as $\hat h^{\m\n}$ is already traceless.
Thus, after the addition of a boundary term, the action at third order simply reads
\be
\label{nllagr}
{\mathcal S}\equiv {\mathcal S}^{(1)}_{\WTD}+{\mathcal S}^{(2)}=
-\frac{1}{2\kappa^{n-2}}\int \di^n x \tilde g^{\m\n} R_{\m\n}\left[\Gamma_{\phantom{r}\a\b}^{\r}\right],
\ee
where we have defined $\tilde g^{\m\n}=\eta^{\m\n}-\sqrt{2}\kappa\hat h^{\m\n}$.
This Lagrangian differs from the Einstein-Hilbert Lagrangian of GR
and is background dependent as $\hat h_{\m\n}$ involves
$\eta_{\m\n}$ in its definition. Besides, the equations of motion coming from the variation
with respect to $g_{\m\n}$ and the connection are not  Einstein's equations but
\be
\label{eomc}
R_{\m\n}[\tilde g]-\frac{1}{n}\eta_{\m\n}\eta^{\a\b}R_{\a\b}[\tilde g]=0,
\ee
where the connection is compatible with the metric associated to the tensor
density $\tilde g^{\m\n}$,
 $$g^{\m\n}\equiv |g|^{-1/2}\tilde g^{\m\n},$$ which satisfies the constraint
\be
\label{cond}
\sqrt{-g}g^{\m\n}\eta_{\m\n}=n.
\ee
We can now wonder about the consistency of this Lagrangian, as the WTDiff gauge invariance
was necessary to go to the $h=0$ gauge and prove the conservation of the tensor $\tilde t_{\m\n}$.
One can show that the action (\ref{nllagr}) is invariant under the
 non-linear diffeomorphisms satisfying
\be
\label{DNLtrans}
\eta_{\m\n}\left(g^{\m\a}\delta^{\n}_{\b}-\frac{1}{2}\delta^{\a}_{\b}g^{\m\n}\right)\nabla_{\a}\xi^{\b}=0,
\ee
which reduces to the transverse condition at the linear level. The algebra of these diffeomorphisms
does not close for a general metric and thus they do not constitute a finite subgroup of Diff. Even if the
algebra may close {\em on-shell}\footnote{The reason why this may happen
is that the transformations
satisfying (\ref{DNLtrans}) are the most general diffeomorphisms that leave the action
(\ref{nllagr}) invariant. This  means that, as their commutator leaves (\ref{nllagr})
invariant, it must correspond to a parameter satisfying (\ref{DNLtrans})
except for a term proportional to the EoM \cite{HenneauxTeit}.}, we expect that the
number of propagating
degrees of freedom will differ from GR. More concretely, as the number of free gauge parameters
is three and they are differentiated in the gauge transformation, we expect that $6$ degrees
of freedom will not be dynamical \cite{Skvortsov:2007kz}.
 As the field $g_{\m\n}$ has $9$ independent components,
we expect the non-linear theory to have $3$ (light) propagating degrees of freedom\footnote{It may happen
that, similarly to what was found for linear WTDiff, a tertiary constraint appears
that kills the extra degree of freedom.}. If this is the case, this
 theory is ruled out phenomenologically. Besides,
the new degree
of freedom that appears may be a ghost, which would mean that the theory is not consistent
at the quantum level.\\

Before finishing this section, it is worth mentioning some of  the assumptions that we made
and which
can be relaxed. First, for the TDiff invariant Lagrangians,
the Bianchi identities are less restrictive than for the
Diff gauge invariance and
it is enough that the source of the EoM is conserved except for a total derivative,
\be
\pd^\m T_{\m\n}=\pd_\n \psi.
\ee
Surprisingly enough, the same is true for the WTDiff case, as far as we consider the coupling
to the traceless part of the tensor.
This opens the possibility for more general energy-momentum tensors than those obtained in any of the
prescriptions of the Diff case. This possible generalization may also be
 helpful to build higher-spin interacting theories \cite{Skvortsov:2007kz}.
 Besides we have
made an assumption on the values of the parameters $a$ and $b$ in (\ref{Rosenfeld}) and
we have used a modified conserved energy-momentum tensor $\tilde t_{\m\n}$.

In the next section we will see that there is a consistent non-linear theory of WTDiff
 equivalent to GR {\em on-shell}. Besides, it is also invariant under a  non-linear
extension of the Weyl symmetry, which casts some doubt in the possibility of finding
it using the method we envisaged. This does not exclude the possibility of
a suitable choice of
variables  at the linear level to perform a consistent non-linear extension in a single step.
We leave the systematic
study of consistent deformations of the TDiff and WTDiff algebras for further research \cite{Blas08}.

%%%%%%%%%%%%%%%%%%%%%%%%%%%%%%%%%%%%%%%%%%%%%%%%%%%
\subsection{Intuitive Extension}
%%%%%%%%%%%%%%%%%%%%%%%%%%%%%%%%%%%%%%%%%%%%%%%%%%

A possible non-linear extension of the linear TDiff is provided by any subgroup of
the non-linear Diff for which an object $f$ which at the linear level reduces to the trace $h$ transforms as
a scalar. That is, given
\be
\label{f}
f\left(\eta_{\m\n},g_{\m\n}\right)=k+\eta^{\m\n}h_{\m\n} + O\left(h_{\m\n}^2\right)
\ee
for $k$ a constant and $h_{\m\n}=g_{\m\n}-\eta_{\m\n}$,
we want to find the subgroup of Diff such that
\be
\delta_\xi f=\xi^\m \partial_\m f,
\ee
for $\delta_\xi g_{\m\n}=2\nabla_{(\m}\xi_{\n)}$. This subgroup, if it exists, will be
background dependent in general. The previous condition can be expressed as
\be
\label{condition}
A^\m _\r \nabla_{\m}\xi^{\r}-\xi^\r \partial_\r f =A^\m _\r \partial_{\m}\xi^{\r}=0,
\ee
where
$$A_{\r}^\m=2\frac{\delta f}{\delta g_{\m\n}}g_{\n\r}.$$
In particular this means that the translations belong always to this subgroup.\\

Let us study the group structure for a generic $f$. From Frobenius theorem applied to the
Diff, the infinitesimal transformations
will be integrable if and only if \cite{Wald:1986bj}
\be
[\xi_1^\m\partial_\m,\xi_2^\n\partial_\n]=\xi_3^\n\partial_\n
\ee
with $\xi_3^\n=\xi_1^\m\partial_\m\xi_2^\n-\xi_2^\m\partial_\m\xi_1^\n$.
The integrability condition that must be satisfied in our case is
\be
A^\m _\r \partial_{\m}\xi_3^{\r}=2A^\m _\r\left(\partial_\m \xi_{[1|}^\a \partial_\a \xi_{|2]}^\r+
\xi_{[1|}^\a\partial_\m  \partial_\a \xi_{|2]}^\r\right)=0,
\ee
for $\xi_1$ and $\xi_2$ satisfying (\ref{condition}). For the term involving second derivatives to cancel,
the only possibility is  $A_{\r}^\m=l(x)S^\r_\m$, with $S_\r^\m$ being a constant matrix, \emph{i.e.}
\be
2\delta f=l(x)g^{\m\n}\delta g_{\m\n}=l(x)g^{-1}\delta g,
\ee
where $g=\det g_{\m\n}$. Thus, $f$ depends just on the determinant of the metric.
The subgroup which preserves these functions is TDiff also at the
non-linear level, {\em i.e.} the subgroup of diffeomorphisms satisfying
\be
\partial_{\m}\xi^\m=0.
\ee
Once integrated, this subgroup gives rise to the diffeomorphisms of Jacobian equal to one, which
are related to unimodular gravity \cite{vanderBij:1982}.

The simplest form of  $f$ is provided by the choice $f=|g|$. As required, this function
satisfies
\be
|g|=1+\eta^{\m\n}h_{\m\n}+O(h_{\m\n}^2),
\ee
which in fact holds for any background. General Lagrangians where $|g|$ is considered as
an independent degree of freedom have been
studied in \cite{vanderBij:1982,Alvarez:2006uu} and (as we will see in section \ref{sectionNLTDiff}) they are
usually equivalent to scalar-tensor theories of
gravity except for an integration constant.

Notice also that the condition $\nabla_\m \xi^\m=0$ is integrable, as its integrability condition
reduces to
\be
\partial_{[\s}\Gamma^{\a}_{\phantom{\a}\r]\a}=0,
\ee
which is automatically satisfied as $\Gamma^{\a}_{\phantom{\a}\r\a}=\partial_{\r}\ln \sqrt{|g|}$.
However, comparing this condition with (\ref{condition}) one realizes that they are inconsistent.
In other words, there is no object $f$ transforming as a scalar under the subgroup
of Diff satisfying  $\nabla_\m \xi^\m=0$.

One can understand the relation between the previous two integrable conditions  from the difference
between the {\em active} and the {\em passive} action of Diff.
The diffeomorphisms act \emph{passively} over (densitized) tensors as
(see {\em e.g.} \cite{AlvarezGaume:1984dr})
\be
\delta^p T(x)=T'(x')-T(x),
\ee
for a Diff: $x\mapsto x'(x)$.
In particular, the integration measure changes under this transformation,
and the integral of a density
is constant for transverse diffeomorphisms (see Appendix \ref{AppendixBig}). Under
 these transformations, the determinant
of the metric transforms infinitesimally as
$$\d^p g= \partial_\m \xi^\m.$$
This means that the transverse subgroup can be understood as the subgroup
of the Diff under which the determinant of a metric transforms as a scalar.

Besides, in every point of the manifold we can also act \emph{actively} with
the diffeomorphism and define the variation
\be
\delta^a T(x)=T'(x)-T(x).
\ee
This is the way in which we usually define symmetries, as
we compare quantities at the same point, \emph{i.e.}  it is a local concept.  Under the previous
{\em active}
transformations, the determinant of the metric changes as
\be
\delta^a g=\nabla^\m \xi_\m,
\ee
which means that the group of symmetries of the determinant is provided by the Diff satisfying
$\nabla_\m \xi^\m=0$.\\

Recall that at the linear level the TDiff gauge invariance
could be enlarged to the Diff or WTDiff groups. At the non-linear level,
the Diff enlargement corresponds to the whole group of the diffeomorphisms
whereas for the WTDiff non-linear transformation we seek a transformation of
the determinant of the form
\be
\delta_{(\phi,\xi)} g=\phi g+\xi^\m\partial_\m g.
\ee
From the previous expression we find that
\be
[\delta_{(\phi_1,\xi_1)},\delta_{(\phi_1,\xi_1)}]=\delta_{(\xi_{[1}\partial\phi_{2]},\xi_3)}.
\ee
If we want the same algebra to hold for the metric field $g_{\m\n}$ then it is clear that the
non-linear Weyl transformation of
the {\em whole} metric must be the usual conformal rescaling, \emph{i.e.}
\be
\delta_{(\phi,\xi)} g_{\m\n}=\phi^{1/n} g_{\m\n}+2\nabla_{(\m}\xi_{\n)}.
\ee
It is interesting to note that once this Weyl invariance
\be
\label{conformal}
 g_{\m\n}\mapsto e^\phi g_{\m\n}
\ee
is added to the TDiff gauge invariance,
we find a unique Lagrangian with just two derivatives of the
 metric\footnote{Notice that this Lagrangian can not be put
in the Einstein frame, as it is invariant under  Weyl transformations.}
\be
\label{WTdiffL}
{\mathcal S}_{\WTD}=-\frac{1}{2\kappa^{n-2}}\int \di^nx \hat g^{\m\n} R_{\m\n}(\hat g_{\m\n})+ S_M(g,\hat g_{\m\n},\psi).
\ee
where $\hat g_{\m\n}=|g|^{-1/n}g_{\m\n}$ and $S_M$ refers to a matter Lagrangian compatible with the WTDiff
invariance.
As we will see in the next section, this Lagrangian yields Einstein's equations of motion in the gauge $|g|=1$
 (even when coupled to matter) except for the origin of the cosmological constant which comes from
 an integration constant \cite{Alvarez:2006uu}.

The reason why we did not
find the previous non-linear extension in the previous section is now
evident: the determinant $g$ is a highly non-linear function
of the field $h^{\m\n}$ and thus the condition $|g|=1$  can not be recovered in
a single step from the variables in the last section (compare it
with the condition (\ref{cond}) which is linear in $h^{\m\n}$).

%%%%%%%%%%%%%%%%%%%%%%%%%%%%%%%%%%%%%%%
\section[Lagrangians and EoM
for Nonlinear TDiff and WTDiff]{Lagrangians and Equations of Motion for Nonlinear TDiff and WTDiff}\label{sectionNLTDiff}
%%%%%%%%%%%%%%%%%%%%%%%%%%%%%%%%%%%%%%%%%%

Non-linear generalizations of $\TD$  invariant theories
in the lines of the previous
subsection have been
discussed in \cite{Buchmuller:1988wx} (see also
\cite{Pitts:2001jw}). The basic idea is to split the metric degrees
of freedom into the determinant $g$, and a new rank-2 object\footnote{If
 we admit non-local splitting of the degrees of freedom, the combination
\be
\check g_{\m\n}\equiv \left[1-\frac{1}{6}\Big(-\nabla_\m \nabla^\m+\frac{1}{6}R\Big)^{-1}R\right]^2g_{\m\n},
\ee
is Weyl invariant and transforms as
a metric    under Diff (cf. \cite{Fradkin:1985am}, p. 319). Besides $R(\check g)=0$.}
 $\hat
g_{\mu\nu}= |g|^{-1/n} g_{\mu\nu}$, whose determinant is fixed
$|\hat g|=1$. Note that $\hat g_{\mu\nu}$ is a tensor density, and
under arbitrary diffeomorphisms (for which $ \delta_{\xi} g_{\m\n}=2
\nabla_{(\m}\xi_{\n)}$) it transforms as
\be \delta_{\xi}\hat
g_{\mu\nu} = 2 \hat g_{\lambda(\mu}\hat\nabla_{\nu)} \xi^{\lambda}
-{2\over n} \hat g_{\mu\nu} \hat\nabla_\lambda \xi^\lambda,
\ee
where $\hat\nabla$ denotes covariant derivative with respect to
$\hat g_{\mu\nu}$. Next, one defines transverse diffeomorphisms as
those which satisfy
\be \hat\nabla_\mu \xi^\mu = \partial_\mu
\xi^\mu =0,
\ee
where in the first equality we have used $|\hat
g|=1$. Under such $\TD$, the new metric transforms as a tensor
$$\delta_{\xi}\hat g_{\mu\nu} = 2 \hat g_{\l(\m}\hat \nabla_{\nu)}
\xi^{\l},$$ while $g$ transforms as a scalar $$\delta_\xi g =
\xi^\lambda
\partial_\lambda g.$$ Moreover \cite{Buchmuller:1988wx}, the only
tensors under $\TD$ which can be constructed from $\hat g_{\mu\nu}$
are the geometric ones, such as $R_{\mu\nu\rho\sigma}[\hat g]$ and
its contractions. It follows that the most general action invariant
under $\TD$ which contains at most two derivatives of the metric
takes the form
\be
S=\int \left(-{\chi^2[g,\psi]\over 2\kappa^{n-2}} R[\hat
g_{\mu\nu}] + L[g,\psi,\hat g_{\mu\nu}]\right) \di^nx.\label{hhh}
\ee
Here, $\chi$ is a scalar made out of the matter fields $\psi$ and
$g$. Thus, the $\TD$ invariant theories can be seen as ``unimodular"
scalar-tensor theories, where $g$ plays the role of an additional
scalar. These are very similar to the standard scalar-tensor
theories, except for the presence of an arbitrary integration
constant in the effective potential. A first restriction
on these Lagrangians is that they must correspond to \emph{healthy}
Lagrangians: if Minkowski space-time is a solution,   at the linear level
 they must reduce to a {\em healthy} form of those discussed in Chapter \ref{chapterLorentz}.

Following
\cite{Buchmuller:1988wx}, we may go to the Einstein frame by
defining $\bar g_{\mu\nu} = \chi^2 \hat g_{\mu\nu}$, and we have
\be
S=-{1\over 2\kappa^{n-2}}\int \sqrt{-\bar g}\ R[\bar g_{\mu\nu}]\ \di^nx + S_M +
\int \Lambda \ \di^n x, \label{here}
\ee
where
\be
S_M =
\int\sqrt{-\bar g}\left[{(n-1)(n-2)\over 2\kappa^{n-2} \chi^2}\ \bar
g^{\mu\nu}\partial_\mu\chi\partial_\nu \chi +\chi^{-n}
L[\chi,\psi,\bar g_{\mu\nu}]-
\chi^{-n}\Lambda\right]\di^nx.\label{there}
\ee
Here, we have first
eliminated $g$ in favor of $\chi$, and we have then implemented the
constraint $\bar g=\chi^{2n}[g,\psi]$ through the Lagrange
multiplier $\Lambda(x)$. Note that the invariance under full
diffeomorphisms which treat $\bar g_{\m\n}$ as a metric and $\chi$
and $\Lambda$ as scalar fields is only broken by the last term in
(\ref{here}). In particular, $S_M$ is $\D$ invariant, and since
$\delta_\xi \Lambda = \xi^{\mu}\partial_\mu\Lambda$, it is
straightforward to show that if the equations of motion for $\psi$,
$\chi$ and $\Lambda$ are satisfied, then
$$
|\bar g|^{1/2} \bar \nabla^\mu T_{\mu\nu} = \partial_\mu \Lambda.
$$
Here, we have introduced $T^{\mu\nu} = -2 |\bar g|^{-1/2} \delta
S_M/\delta \bar g_{\mu\nu}$. On the other hand, the Einstein's
equations which follow from (\ref{here}) imply the conservation of
the source $\bar\nabla^\mu T_{\mu\nu}=0$, and therefore we are led
to
$$
\Lambda=const.
$$
This is the arbitrary integration constant, which will feed into the
equations of motion as an extra term in the potential for $\chi$,
corresponding to the last term in Eq. (\ref{there}). In general,
this will shift the height and position of the minima of the
potential for the scalar fields on which $\chi$ depends. In the
particular case where we have $\chi[g,\psi]=1$ in Eq. (\ref{hhh}),
the effect is just an arbitrary shift in the cosmological constant.

$\D$ invariance is recovered when all terms in $S_M$, given in Eq.
(\ref{there}), except for the last one, are independent of $\chi$.
In that case, $\chi$ is a Lagrange multiplier which sets
$\Lambda=0$, so the freedom to choose the height (or position) of
the minimum of the potential is lost.\\

Likewise, if the action (\ref{hhh}) does not depend on $g$, then the
symmetry is the non-linear $\WTD$ group that we studied in the  last section.
 The situation is exactly the same as in the
$\TD$ case, where now $\chi=\chi[\psi]$. For instance the simple
action
\be S_{\WTD} =-{1\over 2\kappa^{n-2}} \int \di^n x\ R[\hat
g_{\mu\nu}],\label{ul}
\ee
which has $\chi=1$, leads to the
equations of motion
\be
\hat R_{\mu\nu}-\frac{1}{2} \hat R \hat
g_{\mu\nu}= \Lambda \hat g_{\mu\nu},\label{eineq}
\ee
with arbitrary
integration constant $\Lambda$ (note that in this case $\hat
g_{\mu\nu}=\bar g_{\mu\nu}$). This coincides with the standard
Einstein's equations in the gauge $|g|=1$. The same action can be
expressed in terms of the ``original" metric $g_{\mu\nu}$ as
\be
{S}_{\WTD} =-{1\over 2\kappa^{n-2}}\int \di^n x (-g)^{1/n} \left(R[g_{\mu\nu}]
+{(n-1)(n-2)\over 4n^2}\
\partial^\mu\ln g\
\partial_\mu\ln g\right).\label{nlw}
\ee
This is invariant under Weyl transformations (\ref{conformal})
since $\hat g_{\mu\nu}$ is
unaffected by these. Of course, it is also invariant under
transverse diffeomorphisms and provides, therefore, an example of a
consistent non-linear extension of a pure spin-2 Lagrangian,
which is different from GR. It is interesting that
a cosmological constant term is not allowed in the Lagrangian, but as shown before
the cosmological constant is recovered as an integration constant\footnote{A similar action was considered
some time ago in the context of quantum cosmology \cite{Unruh:1988in}.
Besides, this action coupled the scale invariant Standard Model
has been recently considered to describe the evolution of the Universe where all the mass
scales have a common origin \cite{Shaposhnikov:2008xb}.}.

Note that the equations of motion can be derived in two different
ways: directly from (\ref{ul}) under {\em restricted} variations of
$\hat g_{\mu\nu}$ (since by definition $|\hat g|=1$), or from
(\ref{nlw}) under {\em unrestricted} variations of $g_{\mu\nu}$.
Whichever representation is used may be a matter of convenience, but
there seems to be no fundamental difference between the two. In the
latter case, the equations of motion will be completely equivalent
to (\ref{eineq}), although they will only take the same form in the
gauge $|g|=1$.

It is worth mentioning that equations of the form (\ref{eineq}) with
an arbitrary $\Lambda$ can also be derived under {\em unrestricted}
variations of an action which is {\em not} invariant under
(\ref{conformal}). An example is given by\footnote{Related actions
can be found in the case of non-linear Lorentz violating massive
gravity \cite{Grisa:2008um}.}
\be
S=-{1\over 2\kappa^{n-2}} \int
\left[\sqrt{-g} R + f(g)\right]\di^n x,\label{nonsym}
\ee
Here, the
second term breaks $\D$ to $\TD$, and there is no Weyl invariance\footnote{As
we will explain in the Appendix \ref{AppendixQ}, this kind of terms may be induced
quantum mechanically if the usual regularization prescriptions that preserve
the whole Diff group are used.}.
A particular example of these Lagrangians is the standard Lagrangian
of unimodular gravity \cite{Weinberg:1988cp,Henneaux:1989zc}.
However, the equations of motion will give
$$
R_{\mu\nu}-{1\over 2}R g_{\mu\nu}=  \sqrt{-g}\ f'(g)\  g_{\mu\nu},
$$
and from the Bianchi identities it follows that $g$ is an arbitrary
constant (except in the $\D$ invariant case when $f \propto
\sqrt{-g}$ ), a situation identical to (\ref{eineq}). It is unclear
whether the action (\ref{nonsym}) is of any fundamental
significance, since the remaining $\TD$ symmetry does not forbid an
arbitrary function of $g$ in front of $R$, and additional kinetic
terms for $g$. Nevertheless, as we
will see in the next Chapter, Lagrangians similar to (\ref{nonsym})
do arise in the context of certain bigravity theories where the
interaction term between two gravitons breaks $\D \times \D$ to the
diagonal $\D$ times a $\TD$ symmetry \cite{Blas:2007ep}.\\

It should be stressed that it seems to be very
difficult to determine from experiment whether $\D$, $\WTD$ or just
$\TD$ is the relevant invariance of Nature. First, as we have seen
the trace of the equations of motion (except for an integration constant)
 is always recovered
in the WTDiff theory through the Bianchi identity and the
conservation of the energy-momentum tensor. The difference between $\WTD$ and
the rest of $\TD$ theories is just the absence of the extra scalar.
However, this scalar may well have a mass comparable to the cut-off
scale, and in this case it would not be seen at low energies. Also,
at the classical level, the $\WTD$ differs only from $\D$ in that
the cosmological constant is arbitrary. Of course the measurement of
this constant does not reveal too much about its origin. Therefore,
the only ``observable" differences between both theories may be in
the quantum theory
\cite{Alvarez:2006bx,Alvarez:2005iy,Unruh:1988in,Kreuzer:1989ec,Dragon:1988qf,Gabadadze:2005uq}) (see also
the Appendix \ref{AppendixQ}).\\

To conclude, we would like to say a few words about the coupling of matter
to gravity in TDiff invariant Lagrangians. It was shown in \cite{Alvarez:2007nn}
that the relative weight of potential and kinetic energy can be tuned in these models.
Even more, for certain Lagrangians with a GR kinetic term for gravity, consistent models
were found which exhibit non-accelerating solutions even in the presence of  vacuum
energy (see also \cite{Alvarez:2007cp} and the related ideas of \cite{Guendelman:2006af}).

Besides, the action for a particle or
an extended object (like a string) compatible with the WTDiff can be
derived from the substitution
\be
g_{\m\n}\mapsto \hat g_{\m\n}.
\ee
It would be interesting to study whether Einstein's equations (without the
integration constant) are recovered from the consistency of the quantum string
as happens for the Diff case \cite{Polchinski}. Besides,
the extension to the TDiff case deserves further study.

%%%%%%%%%%%%%%%%%%%%%%%%%%%%%%%%%%%%%%%%%%%%
\section{First order formalism of WTDiff}
%%%%%%%%%%%%%%%%%%%%%%%%%%%%%%%%%%%%%%%%%%

We have already seen in section \ref{systematic} that
the first order (or Palatini's) formalism
also applies for the linearized WTDiff Lagrangian without
the need of Lagrange multipliers. One can easily see
that this is also the case for the non-linear extension.
Let us first show it for the metric and the connection. We will
consider the Lagrangian
\be
{\mathcal L}=\hat g^{\m\n} R_{\m\n}[\Gamma_{\phantom{\s}\a\b}^{\s}],
\ee
where $\hat g^{\m\n}=|g|^{1/n}g^{\m\n}$ and $\Gamma_{\phantom{\a}\a\b}^{\s}$ is
an arbitrary connection.  This Lagrangian is invariant
under WTDiff simply imposing that the Weyl transformations do not
change the connection. Varying the action with respect to the connection
one obtains the constraints that make the connection\footnote{This connection
will not transform as a connection under general Diff, but only under TDiff. It is important
to remark that
 the connection is compatible with the object $\hat g^{\m\n}$ for the
covariant derivative $\nabla$. Imposing that the compatibility holds
for other possible covariant derivatives present in TDiff
invariant theories (the $\nabla^w$ to be defined in
(\ref{wcovariant})) does not determine
all the components of the connection in terms of $\hat g^{\m\n}$ \cite{Abbassi:2007bq}.}
compatible with the density $\hat g^{\m\n}$. This means
that once substituted back in the Lagrangian, we obtain the
WTDiff Lagrangian (\ref{WTdiffL}).

If we want to couple the
gravitational field to fermions one must adopt a description
in terms of the {\em vielbein}. The equivalent of the $\hat g_{\m\n}$ field in this case
will be a vielbein $\hat e_{\phantom{a}\m}^a$ with unit determinant. In four dimensions,
\be
\hat e^a_{\phantom{a}\m}=e^{-1/4}e_{\phantom{a}\m}^a,
\ee
where $e=\det e_{\phantom{a}\m}^a$.
Notice that
this condition is compatible with the local $SO(3,1)$ invariance,
and thus the use of $\hat e_{\phantom{a}\m}^a$ just breaks the Diff invariance to TDiff. The
action in four dimensions can be written as
\be
S=-\frac{1}{2\kappa^2}\int \di^4 x  \hat e^{a\m}\hat e^{b\n}R_{\m\n ab}[\omega_\n^{\phantom{\n}ab}],
\ee
where $\omega_\n^{\phantom{\n}ab}$ is an arbitrary spin-connection.
The variation of this action reads
\ba
\delta {\mathcal S^{(2)}}&=&-\frac{1}{2\kappa^2}\int\di^4x \ e^{-1/4}\left(
\hat  R^{a\m}-\frac{1}{4}e^{a\m} \hat R\right)
\delta e_{a\m}\nonumber\\
&&-\frac{1}{16\kappa^2}\int \di^4 x\ \hat\e^{\m\n\l\r}\tilde \e_{abcd}\hat
e^c_{\phantom{c}\l}\hat e^d_{\phantom{d}\r}({\mathcal D}_\m \delta \omega_\n^{\phantom{\n}ab}
-{\mathcal D}_\n \delta \omega_\m^{\phantom{\m}ab})
\ea
where $\tilde \e^{abcd}$ is a totally antisymmetric frame tensor and
\be
\hat \e^{\m\n\l\r}=\hat e_a^{\phantom{a}\m}\hat e_b^{\phantom{a}\n}
\hat e_c^{\phantom{a}\l}\hat e_d^{\phantom{a}\r}\tilde \e^{abcd}.
\ee
Notice that we use the vierbein $\hat e^a_{\phantom{a}\m}$ and its inverse to handle with indexes,
so that
$$
\hat R^{a\m}=\hat e^{a\l}\hat e^{c\m}\hat e^{d\r}R_{\l\r cd}, \quad \hat R=\hat e_{a\m}\hat
R^{a\m}.
$$
Following the standard derivation (see {\em e.g.} \cite{deWit:1985aq}) the equations of motion imply
\be
\omega_{\m ab}=\omega_{\m ab}(\hat e), \quad \hat R^{\m\n}(\hat g)-\frac{1}{4}\hat g^{\m\n}\hat R(\hat g)=0,
\ee
where $\hat g^{\m\n}=\hat e_a^{\phantom{a}\m}\hat e_b^{\phantom{a}\n}\eta^{ab}$ and
$$\omega_{\m ab}(\hat e)=\frac{1}{2}\left[\hat e^{\phantom{a}\n}_a
\left(\pd_\m \hat e_{b\n}-\pd_\n\hat e_{b\m}\right)
-\hat e^{\phantom{a}\n}_b
\left(\pd_\m \hat e_{a\n}-\pd_\n \hat e_{a\m}\right)-\hat e^{\phantom{a}\r}_a
\hat e^{\phantom{a}\s}_b
\left(\pd_\r \hat e_{c\s}-\pd_\s \hat e_{c\r}\right)\hat e_{\phantom{c}\m}^c\right].$$
Besides, we used
$\hat g_{\m\n}$ and $\hat e_a^{\phantom{a}\m}$ to contract indexes. We also find
\be
\hat g_{\m\n}=g^{-1/4}g_{\m\n},
\ee
for $g^{\m\n}=  e_a^{\phantom{a}\m} e_b^{\phantom{a}\n}\eta^{ab}$. As a result, we
find that the first order formalism without the presence
of Lagrange multipliers is well-suited for the WTDiff Lagrangian.\\

Let us finish this Chapter with a brief comment
on supersymmetry. In the previous Chapter we found that there is no minimal supersymmetric action
constructed out of the WTDiff action already at the linear level. For the Diff case, this
minimal supersymmetric action consist of the Diff invariant spin-2 action
together with the Rarita-Schwinger (RS) action, and there is a unique non-linear deformation
that allows to couple the spin-2 and spin-$3/2$ systems and blend the global supersymmetry
transformation with the gauge invariance to reach a local supersymmetric transformation
\cite{Boulanger:2001wq,Deser:1979zb,Deser:1976eh}. The reason why this system is consistent is related to the fact that
once all the Einstein's equations hold, the Bianchi identities related to the
supersymmetric transformation are satisfied  \cite{Deser:1976eh,VanNieuwenhuizen:1981ae}.
If one couples the RS action to the field $\hat e_{\phantom{a}\m}^a$ and use the WRS action,
then, one may hope that the Bianchi identities for the spin-$3/2$ field equations
will imply \emph{all} of the Einstein's equations including the missing trace. In other words, the equations
of motion may imply a vanishing cosmological constant even if the action is not supersymmetric.
 In contrast to what happens
in \cite{Alvarez:2007nn} this result would hold for an action for the spin-$3/2$ field
 invariant
under WTDiff.

Whether the previous naive expectation holds or not is currently under research \cite{Blas08}.

%%%%%%%%%%%%%%%%%%%%%%%%%%%%%%%%%%%%%%%%%%%%%%%%%%%%%%%%%%%%%%%%%%
\chapter{Bigravity: General Aspects and Exact Solutions}\label{chapterbigra}
%%%%%%%%%%%%%%%%%%%%%%%%%%%%%%%%%%%%%%%%%%%%%%%%%%%%%%%%

In the previous Chapter we have studied non-linear extensions of one
of the possibilities to modify the standard theory of gravity at the linear
level. More precisely, we considered theories which are invariant under non-linear
TDiff\footnote{Another way of thinking about this subgroup
is through the introduction of a background volume form as a St\"uckelberg field
that allows for the recovery of the whole Diff group, but reduces to the
TDiff case in the analogous of the unitary gauge \cite{Alvarez:2007nn}.}.
The TDiff gauge invariance
allows for a modification of gravity where a {\em scalar} component of the
metric can be massive and thus it provides a non-linear extension
of the simplest TDiff massive gravity through the introduction of a
fixed background volume \cite{Unruh:1988in,Alvarez:2007nn}.\\

In the next two chapters we will focus on a non-linear extension of
the Lagrangians with {\em massive} spin-2 polarizations.
It is easy to realize that the addition of  scalar or  vector
fields can never render
massive the tensor modes of the graviton unless the
background is not homogeneous.  This is why we will
consider {\em bigravity} ({\em i.e.} theories
with two interacting rank-2 tensors) as the simplest candidate
to provide a {\em mass} to the
tensor modes of the graviton in a covariant way\footnote{A related
possibility that we will not study is
 to  consider one of the metrics as a fixed background \cite{Will}.}.
 In this Chapter
we will study some general issues and  global aspects
of these theories whereas in the next Chapter we will
study perturbations to some exact solutions. This Chapter is
based on \cite{Blas:2005yk,Blas:2005sz,Blas:2007ep,Blas:2007zz}.

%%%%%%%%%%%%%%%%%%%%%%%%%%%%%%%%%%%%%%
\section{Introduction}
%%%%%%%%%%%%%%%%%%%%%%%%%%%%%%%%%%%%%%

Bigravity was first proposed in the seventies in
the context of the strong interactions
as a theory that describes the interaction
of a spin-2 meson with the graviton \cite{Isham:1971gm}. This idea is known
also as \emph{f-g} gravity or \emph{strong} gravity. More
recently, bigravity have been reconsidered
in different contexts.
To list some of them, it is relevant in the presence of extra dimensions with
peculiar compactifications that allow for a mass-gap in the
KK spectrum \cite{Damour:2002ws}; it is also found in braneworlds with certain fine-tuned configurations
\cite{Padilla:2004uj}; two metrics naturally appear in some non-commutative set-ups \cite{Damour:2002ws}.
Bigravity (and its generalization to ``multigravity") is also relevant to the program of
``deconstruction" of gravity \cite{ArkaniHamed:2001ca,Deffayet:2005yn} and
for the area metric gravity
\cite{Punzi:2006nx}.\\

We will
consider \emph{bigravity}
as a simple non-linear model of \emph{massive gravity} that may be useful to understand
whether some of the phenomena found at the linear level (see Chapter \ref{chapterintro})
persist
in the complete theory. An interesting aspect of bigravity (as compared
 to other non-linear infrared modifications of gravity) is
that, as we will discuss, there are exact solutions
which belong to the same category as those of usual GR in the limit
of massless graviton (vanishing coupling). Besides, we will find flat solutions around which the
linear theory does not suffer neither from the vDVZ discontinuity
nor from the strong coupling problem. Finally, it is also interesting
to note that there are accelerated
solutions without the need of introducing dark energy (in a sense
the second metric acts as a sort of dark energy).\\

In dealing
with a space-time with two metrics, it is natural to ask whether
we can make sense of its causal structure. In general, the
light-cones related to the metrics $f$ and $g$ will not agree, and
this may lead to pathologies which may restrict the class of
physically acceptable solutions. We will study the causal structure
of some exact solutions in the last part of this Chapter and
find that the possible pathologies reduce to those which are also present in solutions
of standard GR.

%%%%%%%%%%%%%%%%%%%%%%%%%%%%%%%%%%%%%%%%%%%
\section{Exact Solutions of Bigravity}
%%%%%%%%%%%%%%%%%%%%%%%%%%%%%%%%%%%%%%%%%%%%%

Following \cite{Isham:1971gm}, we consider the action
\begin{equation}
\label{action} S=\int \di^4 x\sqrt{-g}\ \left(\frac{- R_g}{2
\kappa_g} +L_g\right) +\int \di^4 x \sqrt{-f}\ \left(\frac{- R_f}{2
\kappa_f} + L_f\right) + S_{int}[f,g].
\end{equation}
Here $L_f$ and $L_g$ denote generic matter Lagrangians coupled to
the metrics $f$ and $g$ respectively, and subindices $f$ and $g$ on the Ricci scalar
 $R$ indicate which metric we use to
compute it. For the background solutions, we shall restrict attention to the case where there is
only a vacuum energy term in each matter sector $L_f=-\rho_f,
L_g=-\rho_g$, where $\rho_f$ and $\rho_g$ are constant.
The kinetic terms are invariant under independent diffeomorphisms of the metrics $f$ and $g$,
 but the interaction term is invariant under ``diagonal" diffeomorphisms\footnote{In
  principle, we might also include derivative interactions between the two metrics compatible with the diagonal
symmetry, but in general these terms yield a ghost in the vector sector and we
will not consider them here (see {\em e.g.} \cite{Nibbelink:2006sz,Drummond:2001rj} for other
bigravity actions). This fact implies  that the modifications to GR
will happen at a certain length scale, and it seems to indicate that
 derivative couplings may be compulsory
to get modifications of GR closer to MOND theories.
As it is clear from the previous Chapter,
another interesting possibility would be to preserve
the independent unimodular diffeomorphisms in the kinetic terms, in which case the derivative
coupling may be possible.}, under which both metrics transform.

The most general interaction
potential which preserves the ``diagonal" diffeomorphism takes the
form
 \cite{Damour:2002ws}
\be
S_{int}=\zeta\int \di^4 x(-g)^u(-f)^{v} V[\{\t_n\}],\label{generi}
\ee
where $\t_n=\tr[{\mathcal M}^n], \ n:1,...,4$ correspond to the traces
of the first four powers of the matrix ${\mathcal M}^{\m}_{\n}=f^{\m\a}g_{\a\n}$, and $V$ is an
arbitrary function.

There is also some arbitrariness in the way one introduces matter fields, since one has two different metrics
 at hand.
This opens the possibility to have two
   types of matter\footnote{This possibility
   is known as the \emph{weakly coupled worlds} assumption \cite{Damour:2002ws}.},
    one which feels the metric $g$ and the other which feels the metric $f$.
 Those two choices
 correspond to the two matter Lagrangians $L_g$ and $L_f$, of action (\ref{action}), where it is understood
  that the matter fields entering into $L_g$ and $L_f$ are different.
  In fact one can imagine more complicated situations in which matter fields would be coupled
to some composite metric built out of the two metrics $f$ and $g$.  If one wishes to recover
the standard {\em equivalence principle}, one should obviously ask that standard matter only couples
to one metric, and a minimal choice is, {\em e.g.}, that all matter fields appear say in $L_f$
(respectively $L_g$), while $L_g$ (respectively $L_f$),
will be simply given by a cosmological constant.
 With such a choice, matter moves along geodesics of the metric $f$ (respectively $g$), and,
  provided the solutions for the metric $f$ are the same as in standard GR
  (which turns out to be possible as will be seen below), there would be no deviations from
    GR seen in matter motion.
In this case, the other metric can be regarded as some kind of exotic new type of matter
which may violate the {\em equivalence principle}.

Finally, notice that a consequence of the invariance of the action (\ref{action})
under diagonal diffeomorphisms is that the total Hamiltonian will cancel. This may alleviate
the problem of the Boulware-Deser instability in non-linear massive gravity \cite{Boulware:1973my},
but it does not guarantee the absence of ghosts in the spectrum of
the theory (see Chapter \ref{chapterperturbbigrav}).\\

For arbitrary metrics $f$ and $g$, the contribution to the
energy-momentum tensors coming from the interaction term
in (\ref{action}) will be
\ba
\label{emf}
f^{\m\a}T^f_{\a\n}\equiv   {-2\over \sqrt{ -f}}
{\delta S_{int}\over \delta f^{\a\nu}}f^{\m\a}=-2\zeta(g/f)^{u}\left(v V \delta^\m_\n-\sum_{n} n ({\mathcal M}^n)^\m_\n\ V^{(n)}\right),\\
\label{emg}
g^{\m\a}T^g_{\a\n}\equiv {{-2\over \sqrt{ -g}} {\delta S_{int}\over \delta g^{\a\nu}}}g^{\m\a}
=-2\zeta(g/f)^{-v}\left(u V \delta^\m_\n+\sum_{n} n({\mathcal M}^n)^\m_\n\ V^{(n)}\right),
\ea
where we have introduced the notation
$$
V^{(n_1,...,n_l)}\equiv\frac{\pd^l V}{\pd \t_{n_1}\cdots\pd\t_{n_l}},
$$
where $l$ is the number of derivatives.
Moving to the frame where both metrics are diagonal (which can always be done locally),
the matrix ${\mathcal M}=f^{-1}\cdot g$ can be put to the diagonal form with eigenvalues $\l_i$.
Two arbitrary metrics $g_{\m\n}$
and $f_{\m\n}$ which are solutions of the vacuum Einstein's equations, {\em i.e.} such that
\be
\label{getasolu}
g^{\m\a}G_{\a\n}^g/\Lambda_g = f^{\m\a}G_{\a\n}^f/\Lambda_f= \delta^\m_\n,
\ee
will be solutions for bigravity
if all the $\t_n$ are constant and the eigenvalues of the matrix
\be
\sum_n n({\mathcal M}^n)^\m_\n\ V^{(n)},
\ee
entering (\ref{emf}-\ref{emg}) are all equal to each other. Note that for a given \emph{ansatz},
the constancy of the traces (or of the eigenvalues) is a frame independent notion.
The equations of motion will be then satisfied for vacuum solutions $f$ and $g$ with
cosmological constants $\Lambda_f$ and $\Lambda_g$ satisfying
\ba
\label{Lambdaf}
\Lambda_f=-2\kappa_f\zeta(g/f)^{u}\left(v V-\frac{1}{4}\sum_{n} n\t_n\ V^{(n)}\right)+\kappa_f\r_f,\\
\label{Lambdag}
\Lambda_g=-2\kappa_g\zeta(g/f)^{-v}\left(v V+\frac{1}{4}\sum_{n} n\t_n\ V^{(n)}\right)+\kappa_g\r_g.
\ea

%%%%%%%%%%%%%%%%%%%%%%%%%%%%%%%%%%%
\subsection{Type I Solutions}
%%%%%%%%%%%%%%%%%%%%%%%%%%%%%%%%%%%%

Let us  introduce some concrete exact solutions.
The general static spherically symmetric \emph{ansatz} for bigravity can be written as \cite{Isham:1977rj}
\ba
\label{formg} g_{\mu \nu} \di x^\mu \di x^\nu &=&
J \di t^2 - K \di r^2 - r^2 \left(\di \theta^2 + \sin ^2 \theta \; \di\phi^2 \right), \\
f_{\mu \nu} \di x^\mu \di x^\nu &=& C \di t^2 - 2 D \di t\di r - A \di r^2 -
B\left(\di \theta^2 + \sin ^2\theta \; \di\phi^2\right),
\label{formf}
\ea
where the metric coefficients are functions of $r$.
Note that in general it is not possible to write both metrics in
diagonal form in the same coordinate system and that
we have also assumed that the axes for the $SO(3)$ symmetry are shared
by both metrics.

A particularly interesting class of spherically symmetric configurations is
provided by the
 solution\footnote{Recently,
more general non-linear solutions of bigravity which deviate from GR
have been found for certain potentials \cite{Berezhiani:2008nr}.}
\ba
g_{\mu
\nu} \di x^\mu \di x^\nu &=& \left(1-q\right) \di t^2
- (1-q)^{-1} \di r^2 - r^2 (\di \theta^2 + \sin^2 \theta \di \phi^2), \label{spheg}\\
f_{\mu \nu} \di x^\mu \di x^\nu &=& \frac{\g}{\beta} (1-p) \di t^2 -2
D \di t \di r
 - A \di r^2  - \g r^2 (\di\theta^2 + \sin^2 \theta \di\phi^2), \label{sphef}
\ea
where
\ba
A &=&  \frac{\g}{\beta}(1-q)^{-2}\left(p + \beta - q- \beta q \right),\\
\label{Dequation} D^2 &=& \left(\frac{\g}{\beta}\right)^2(1-q)^{-2}(p-q)(p+\beta-1 -\beta q).
\ea
Here $\b$ and $\g$ are arbitrary positive constants and $p$ and $q$ are functions of $r$
to be determined latter.
 Solutions of the form (\ref{spheg}-\ref{sphef}) are called Type I (cf. \cite{Isham:1977rj}).
 Notice also that in the flat limit $p=q=0$, even if the $f$ metric is flat, it
 does not reduce to a Minkowski metric in these coordinates.
As we will see, this breaking of Lorentz invariance will be
crucial for certain properties of the perturbations to these solutions like
the absence of vDVZ discontinuity. Besides, it means that matter cannot be coupled
to the massless combination of the metrics (see next Chapter) as this would imply the violation
of Lorentz invariance in the matter sector.

Remarkably, the non-trivial background (\ref{spheg}-\ref{sphef}) has the property that the
eigenvalues of ${\mathcal M}$ are constant
$$
\l_i=\{\g^{-1},\g^{-1},\g^{-1},\b\g^{-1}\},
$$
which implies
$$
\t_n=\g^{-n}(3+\b^n), \quad \det[{\mathcal M}]=\b\g^{-4}.
$$
Thus, to get a solution of (\ref{action}), it is enough to impose
$$
\sum_{n} n({\mathcal M}^n)^\m_\n\ V^{(n)} \propto \delta^\m_\n,
$$
and that (\ref{getasolu}) holds.

In the frame where ${\mathcal M}$
 is diagonal the previous
combination is a constant diagonal matrix with only two different constant eigenvalues
$$
\left\{\sum_n n\b^n\g^{-n} \ V^{(n)},
\sum_n n \g^{-n}\ V^{(n)}\right\}.
$$
Both eigenvalues will coincide when
\be
\label{typeIcondo}
\sum_n n \g^{-n}(-1+\b^n)\ V^{(n)}=0.
\ee
This tells us that for
any potential there will exist non-trivial solutions with certain $\g$ and $\b$ satisfying
(\ref{typeIcondo}) (note that the values of $V^{(n)}$ depend also on $\b$ and $\g$)
for which,
without assuming any specific form for the functions $p(r)$ and $q(r)$,
\ba \label{EQMMOT}
T^f_{\mu \nu} ={\tilde\Lambda_f \over \kappa_f} f_{\mu\nu}, \;\;\;  T^g_{\mu \nu}
={\tilde\Lambda_g\over \kappa_g} g_{\mu\nu},
\ea
where $\tilde \Lambda_X$ are constant. Thus, (\ref{Lambdaf}-\ref{Lambdag})
translate into
\be
\Lambda_f=\tilde \Lambda_f+\kappa_f \rho_f,\quad
\Lambda_g=\tilde \Lambda_g+\kappa_g \rho_g.
\ee
These are three equations for the parameters
$\Lambda_f$, $\Lambda_g$, $\b$ and $\g$. Therefore,
one of the effective cosmological constants can be chosen arbitrarily.
It has the status of an integration constant which allows
for a {\em see-saw} mechanism that makes one of the metrics
to be flat whereas the other can be highly curved.

 It is clear from the previous discussion and (\ref{getasolu}),
 that the metrics $f$ and $g$ must belong to
the Schwarzschild-(A)dS family. Note that the corresponding
cosmological constants (\ref{Lambdaf}-\ref{Lambdag}) are not
determined solely by the vacuum energies $\rho_f$ and $\rho_g$.
They also contain a contribution from the interaction term in the
Lagrangian. This contribution depends not only on the parameters
$\zeta$ and $u$ (recall that $v=1/2 -u$), but also on the
arbitrary integration constant $\beta$ (recall that
$\g$ is fixed by the condition (\ref{typeIcondo})).

 It is somewhat surprising that the cosmological
constants depend on an integration constant. This situation is
reminiscent of the \emph{unimodular gravity} case that we presented
Chapter \ref{chapternl}. One difference here is that we have
two cosmological constants $\Lambda_f$ and $\Lambda_g$, and we can
only choose the value of one of them at will.\\

The metric (\ref{sphef}) can be put in a more familiar form
defining a new time coordinate $\tilde{t}$ by
\be
\label{ttilde}
\di\tilde{t} = \frac{1}{\sqrt{\beta}}\left \{\di t + \epsilon_D
\frac{\sqrt{(p-q)(p+\beta-1-\beta q)}}{(1-q)(1-p)}\di r \right\},
\ee  where $\epsilon_D = \pm 1 $ is defined by the sign
retained for $D$ from equation (\ref{sphef}), namely by \be D = -
\epsilon_D \frac{\g}{\beta} (1-q)^{-1}\sqrt{(p-q)(p+ \beta - 1 -
\beta q)}. \ee  With such a coordinate change, the line
element (\ref{sphef}) now reads  \ba \label{fdiagonal} f_{\mu
\nu} \di x^\mu \di x^\nu = \g \{(1-p) \di\tilde{t}^2 -
(1-p)^{-1} \di r^2 - r^2 (\di\theta^2 + \sin^2 \theta \di\phi^2)\}.
\ea As is clear from the previous discussion, the potentials $p$
and $q$ will be given by the familiar Schwarzschild-(A)dS forms
\ba
\label{SchwdsI}
p&=& \frac{2 M_f}{r} + \frac{\g \Lambda_f}{3}r^2, \\
\label{SchwdsII}
q &=& \frac{2 M_g}{r} + \frac{\Lambda_g}{3} r^2,
\ea where $M_f$ and
$M_g$ are two additional integration constants with the
interpretation of mass parameters.

It is tempting to conclude that this non-linear ``theory of massive
gravity" is phenomenologically sound, since the vacuum solutions
of GR with a cosmological term are recovered,
without a trace of the vDVZ discontinuity. In this sense, the mass
term does not seem to act as an exponential cut-off at a finite
range\footnote{This argument is not completely correct as
even if we find the same solutions, the interpretation
of the integration constants may differ from that of GR
due to some mass-screening effects \cite{Gabadadze:2007as,Berezhiani:2008nr}.
 To clarify this point, the whole solution representing a star is required.}. Rather, it contributes to the effective cosmological
constant, which tends to bend space-time on a length-scale of the
order of the inverse mass of the graviton (which is of order $m^2
\sim \kappa \zeta$)\footnote{See also the related discussion of
\cite{Gabadadze:2003jq}.}. On the other hand, this
contribution from the interaction term can be compensated for by a
finely-tuned contribution from the vacuum energy of matter fields,
and then we can have an asymptotically flat solution with exactly
the same form as for massless gravity.

It is therefore of some interest to understand the global
structure of the solutions (\ref{sphef}-\ref{spheg}) with
(\ref{SchwdsI}-\ref{SchwdsII}), and we defer this  analysis to the next
section. The study of perturbations and the investigation of
stability of these solutions are left for the next Chapter.\\

Before studying other exact solutions it is worth
mentioning that the solution of the form (\ref{spheg}-\ref{sphef})
was discovered in the context of the potential
\cite{Isham:1971gm}
\begin{equation}
\label{interaction}
 S_{int}=-\frac{\zeta}{4}\int
\di^4 x
(-g)^u(-f)^v(f^{\mu\nu}-g^{\mu\nu})(f^{\sigma\tau}-g^{\sigma\tau})(g_{\mu\sigma}
g_{\nu\tau}-g_{\mu\nu}g_{\sigma\tau}),
\end{equation}
with
$$
u+v = \frac{1}{2}.
$$
This potential is a simple choice that reduces to the Fierz-Pauli combination
in the weak field limit
\cite{Isham:1971gm,Damour:2002ws}.
The metrics (\ref{spheg}-\ref{sphef}) are a solution for $\g=2/3$ and it can be shown that
they are the most general solution for $D(r)\neq0$ \cite{Isham:1977rj} (see also
\cite{Salam:1976as}). This is the origin of the name Type I.
Unfortunately, if $D(r)=0$ the general solution is not known
even for this simple potential
\cite{Aragone:1972fn} (see also the Appendix \ref{AppendixBig}).
Furthermore, as we will see in
the next Chapter, for this particular theory the linearized perturbations around asymptotically bi-flat
Lorentz-breaking solutions of this particular theory show a singular behaviour.

%%%%%%%%%%%%%%%%%%%%%%%%%%%%%%%%%%%%%%%
\subsection{Proportional Metrics and Related Solutions}\label{subsectionPropM}
%%%%%%%%%%%%%%%%%%%%%%%%%%%%%%%%%%%%%%

Another interesting class of solutions is obtained by taking $f$ and $g$ proportional to each other,
but otherwise arbitrary
\be
\label{prop}
f_{\mu\nu}=\gamma(x) g_{\mu\nu}.
\ee
In this case, the matrix ${\mathcal M}$
is proportional to the identity ${\mathcal M}^\m_\n=\g^{-1}\delta^\m_\n$
and the energy-momentum
tensors (\ref{emf}-\ref{emg}) read
\ba
\label{Lambtilde}
\tilde \Lambda_f\delta^\m_\n\equiv \kappa_ff^{\m\a}T^f_{\a\n}=
-2\zeta \kappa_f\g^{-4u}\left(v V-\sum_n n\g^{-n}\  V^{(n)}\right)\delta^\m_\n\nonumber\\
\tilde \Lambda_g\delta^\m_\n\equiv \kappa_gg^{\m\a}T^g_{\a\n}=
-2\zeta \kappa_g\g^{4v}\left(u V+\sum_n n \g^{-n}\ V^{(n)}\right)\delta^\m_\n.
\ea
Thus, for any matter content
this term just adds to the vacuum energy. From Bianchi identities
$\tilde\L_f$ and $\tilde \L_g$ must be constant, and $f$ and $g$ must then be solutions of the
vacuum Einstein's equations. Generically, the expressions for $\tilde\Lambda_{f,g}$ depend on $\gamma$, so that
they imply a constant $\gamma$. In this case, the parameter $\gamma$
is determined through Einstein's equations by noting that (\ref{prop}) implies
\be
\label{Lambdas}
R_g=\gamma R_f.
\ee
Clearly, this class will include solutions in the Schwarzschild-(A)dS family, although non-spherically
symmetric solutions are possible as well. Note also that such solutions can easily be
generalized to multigravity theories by deconstructing 5D metrics with a
warp factor \cite{Deffayet:2003zm}. Maximally symmetric solutions of the form (\ref{prop})
have also been considered in \cite{Damour:2002wu}. As in the Type I case, the
proportional metrics will be of the Schwarzschild-(A)de Sitter family and there
is no sign of vDVZ discontinuity either. For the
potential (\ref{interaction}) one can prove that these are the most general
Type II (\emph{i.e.}
diagonal) solutions when one of the metrics is maximally symmetric (see the Appendix \ref{AppendixBig}).\\

The previous proportional solutions can be slightly generalized in factorized space-times.
The generalization consist simply of considering two metrics which are proportional
but with different proportionality factors for the components of each factorized submanifold.
If one of the metrics is maximally symmetric in the factorized submanifolds (but not in the
whole manifold) we can follow the previous steps to find the conditions to
obtain a solution. Other possible generalizations together with a couple
of methods to generate solutions of bigravity can be found in the section
\ref{methodsbi}.\\

%%%%%%%%%%%%%%%%%%%%%%%%%%%%%%%%%%%%%%%%%%%%%%%%%
\section{Global structure of Bigravity Solutions}
%%%%%%%%%%%%%%%%%%%%%%%%%%%%%%%%%%%%%%%%%%%%%%%%%

In dealing with a space-time with two different metrics, it is
natural to worry about their compatibility in some global
aspects\footnote{Remember that both metrics interact through local
terms that break the symmetry group of the kinetic terms
to the diagonal Diff.}. Even if many concepts of ordinary Lorentzian manifolds
may be (almost trivially) generalized, there are some global
issues that can appear. Concepts such as global hyperbolicity,
closed causal curves (CCC) or
geodesic completeness are related to a {\em single}
metric and not to the underlying
manifold structure, and thus their definition in the case of {\em bigravity}
is done for each of the metrics separately. Requiring that
both metrics are globally hyperbolic with common Cauchy surfaces
or geodesically complete may lead
to some surprises\footnote{There are also other
possible pathologies of bigravity solutions that we will not
treat and whose solution is usually a generalization
of a solution for similar pathologies in GR. For instance,
whenever a metric is not \emph{time orientable} in GR, it is customary to use the
double-covering manifold \cite{HawkingEllis}. When the manifold has two
metrics, it is conceivable that closed curves that change the time
orientation of a single metric exist. In the worst situation we
need a forth-covering manifold whose definition is a trivial generalization
of the double-covering manifold.}. Nevertheless, as we will see, for the known
solutions of bigravity
there are no blatant violations of causality (beyond those of GR). \\

For the sake of simplicity we will
restrict ourselves to solutions with a common $SO(3)$ invariance,
which means that it is enough to focus on radial geodesics in
the diagram $r-t$ (see (\ref{formf}-\ref{formg})). Before further
restricting to the solutions of the form (\ref{spheg}-\ref{sphef})
let us say a few words about the methodology we will follow.

We will first consider the issues of causal compatibility,
maximal extensions and geodesic completeness.  To study them
we will make maximal extensions for both metrics through
 geodesics of each metric
that attain their conformal boundary in a finite proper time.
The causal structure will be illustrated by
means of Carter-Penrose diagrams for one of the metrics where
we will include
information about the causal structure of the companion metric. More
concretely,
 once
the causal structure for the first metric, $g_{\m\n}$, is clarified
and we have found its maximal extension,
we will plot in the
light-cones of $f_{\m\n}$ and study their behaviour.
This will inform us about the way in which the
causal structure of the second metric  fits
in the Carter-Penrose diagram of the first one.

Matter that is coupled to one of the metrics
will follow trajectories inside the future light-cone defined
by that metric. However, at any point there are two light-cones
and one of the sectors will typically propagate outside the
null-cones of the other metric. In other words, there
is faster than light propagation. This may give rise
to a series of very interesting phenomena such as
the possibility of scape from a black hole \cite{Dubovsky:2007zi},
 \v{C}erenkov radiation \cite{Altschul:2007kr}
 or may even be useful for the homogeneity problem
 in cosmology. Besides,  superluminal propagation
is usually associated to the appearance of CCC\footnote{This is not necessarily
true if Lorentz symmetry is broken \cite{Babichev:2007dw,Dubovsky:2005xd}.}. The
causal diagrams that we will draw for bigravity allow to study some of these
phenomena. For instance, we will show that it is
possible to define a global time even in the presence
of superluminal propagation.%\\

The conformal compactification allows to extend the geodesics of the metric $g_{\m\n}$
that reach the boundary in a finite proper time to find a maximal extension of this
metric \cite{HawkingEllis}. If the companion metric is already geodesically
complete, the new region to which the geodesics are extended
is not accessible to it. More specifically, if all the geodesics
of the $f_{\m\n}$ finish within the conformal diagram, the extra region can not
be reached in a finite proper time for the $f_{\m\n}$ geodesics. However, the interaction
between both metrics makes possible the passage from the geodesically complete
initial region to the new region for matter coupled
to the $f_{\m\n}$ metric through the $g_{\m\n}$ metric.
 For this matter, the new region is causally disconnected from
the initial region. Even if this may sound
exotic, it is analogous to the appearance of Cauchy horizons in GR where
 the region beyond the horizon does not depend only on the initial
values of the fields, but has a new dependence on completely arbitrary
boundary conditions\footnote{See also \cite{Racz:1995nh} for related work
on extensibility of matter fields through Killing horizons.}.\\

The global structure of solutions where the metrics
are related by a conformal factor, $f_{\m\n}=\Omega^2(x)g_{\m\n}$ can
also become complicated. In this
case, even if the local structure of the null cones will be the same, there may be global
differences. Remember, for instance, that given a metric with singularities and satisfying
certain plausible physical conditions, a conformal factor exists that sets the singularities
at an infinite distance \cite{HawkingEllis}. However, this is not guaranteed in our
more general set-up if the conformal factor $\Omega(x)$ has some additional singularities. Besides,
depending on the conformal factor the proper time that a causal curve employs  to reach
the boundary may change dramatically. In this case, the
metric $f_{\m\n}$ may be extended beyond the region where $g_{\m\n}$ is already geodesically complete
and the other way around. Beyond this
point the $g_{\m\n}$ metric is not determined by the initial metric in the first region.
The existence of a global common Cauchy surface is not guaranteed even if $f_{\m\n}$
is globally hyperbolic. These are some of the
problems that can appear in general, and we will study them in some detail in the
 examples in the next subsections.
In the trivial case when both metrics are proportional with a constant proportionality factor
 both causal structures coincide.\\

For the rest of this section, we will consider
 solutions of the form (\ref{spheg}-\ref{sphef}).
It is worth mentioning a
 particular type of ``singularity" which arises in
some of these solutions (even in cases where both
metrics are separately smooth). Note that the metric (\ref{sphef})
becomes complex in regions where $D^2<0$. As noted in
\cite{Isham:1977rj}, the coordinate singularity at $D=0$ can be
removed by a change of variables. This is of course true, since
$f$ is in the family of Schwarzschild-(A)dS metrics, which are
everywhere smooth (except perhaps at $r=0$ when $M_f \neq 0$).
 However, it does not seem to be possible to find a
change of variables which would remove the singularity from both
metrics at once, in the vicinity of the point at which $D^2$
changes sign, and which would make both metrics real. The reason
is that there are geodesics of $g$ which invade the regions
$D^2<0$ (with arbitrary slope, in fact). On such geodesics, the
line element with respect to $f$ is generically complex, and since
the line element is a scalar, this fact cannot be changed by a
coordinate transformation.
To avoid a complex metric, we could try matching Type I solutions
with Type II solutions at $D=0$ but this possibility has not yet
been clarified.

Henceforth, we will restrict to real
Type I solutions of the form (\ref{spheg}-\ref{sphef}).
 We shall assume $\beta=1$, which ensures
positivity of $D$ for all choices of the potentials $p$ and $q$,
and therefore seems to be the most natural choice
\cite{Isham:1977rj}. For certain potentials, however, there may be
other special values of $\beta$ for which the metric is everywhere
real. We will say more about it later on. We shall also choose $\g=2/3$
which is a solution for the potential (\ref{interaction}). For definiteness,
we remind that for this interaction term the conditions
that must satisfy the cosmological constants (\ref{Lambdaf}-\ref{Lambdag})
reduce to
\ba
\frac{\Lambda_f}{\kappa_f} &=& \frac
{\zeta }{4} \left( \frac{3}{2}\right)^{4u}\beta^u \left\{3 v +
9\beta(1-v)\right\}
+ \rho_f,\label{lambafI} \\
\label{lambagI} \frac{\Lambda_g}{\kappa_g} &=& \frac
{\zeta}{4}\left( \frac{2}{3}\right)^{4v} \beta^{-v} \left\{ 3 u -
9 \beta (1+u)\right\}+\rho_g.
\ea

%%%%%%%%%%%%%%%%%%%%%%%%%%%%%%%%%%%%%%%%%%%%%%%%%%
\subsection{de Sitter with Minkowski}\label{desittermink}
%%%%%%%%%%%%%%%%%%%%%%%%%%%%%%%%%%%%%%%%%%%%%%%%

\begin{figure}[h]  \centering
\psfrag{r0}[][]{$r=0$}\psfrag{r1}[][]{$r=\infty$}\psfrag{t0}[][]{$t=0$}
\psfrag{U}[][]{$U$}\psfrag{V}[][]{$V$}\psfrag{Vc}[][]{$V=ct.$}\psfrag{r2}[][]{$r=ct.$}
\psfrag{rH}[][]{$r=r_H$}\psfrag{a}[][]{$(a)$}\psfrag{b}[][]{$(b)$}
\includegraphics[width=0.6\textwidth]{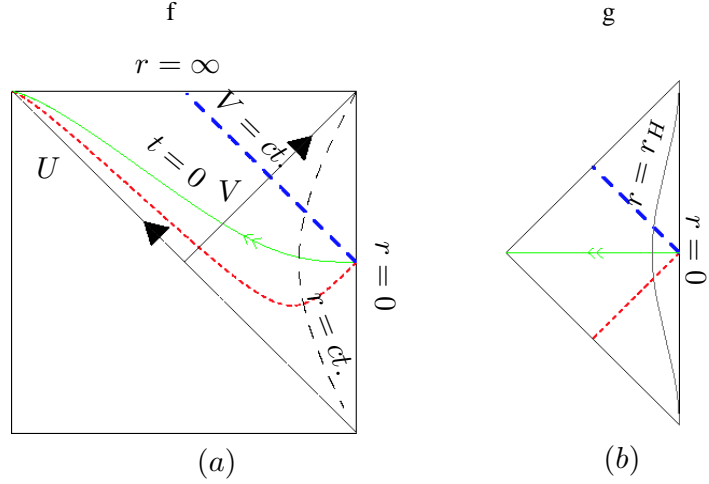}
\caption{\small{Causal diagrams when the $f$ metric is de Sitter
(left diagram) while the $g$ metric is Minkowski (right diagram)
and $\beta=1$. The dashed curly vertical line of the left diagram
represents a sphere of constant radial coordinate $r$. The solid
curly vertical line of the right diagram represents the de Sitter
horizon $r=r_H$ plotted in the Minkowski space-time. We also
plotted three radial geodesics of Minkowksi space-time emanating
from the origin $r=0$ at $t=0$: the thick dashed (blue) curve is a
future-directed radial null ray from the origin (notice it is also
a null geodesic ($V=$ constant) of the  de Sitter space-time), the
thin solid (green) curve with two arrows is a $t=0$ radial
geodesic, the thin dashed (red) curve is a  past-directed null ray
from the origin. The last two curves are radial geodesics of
Minkowski space-time but not of de Sitter space-time. The whole of
the Minkowski space-time is mapped onto the half of the de Sitter
diagram verifying $V>0$. Note that the past directed null
geodesics of Minkowski turn around and start moving towards the
future boundary of de Sitter space. This behaviour, however, does
not lead to closed time-like curves, as discussed in section
\ref{CTLC}}} \label{figure1}
\end{figure}

Let us choose parameters in (\ref{lambafI}-\ref{lambagI}) so that
$\Lambda_g=0$ and $\Lambda_f>0$. Then there is a Type I solution
where $g$ is Minkowski and $f$ is de Sitter. The corresponding
potentials in Eqs. (\ref{spheg}-\ref{sphef}) are given by \be
p=\frac{2\Lambda_f}{9}\ r^2\equiv H^2 r^2, \quad\quad
q=0.\label{potentials}\ee Note that each of the spacetimes,
characterized respectively by the metrics (\ref{spheg}) and
(\ref{sphef}) with the above defined potentials, has a maximal
extension which is geodesically complete (trivial in the case of
Minkowski). However, combining both together will be non-trivial
because the static coordinates $(t,r)$ (where we also include
implicitly the angular part) cover the whole of Minkowski space,
but not the whole of de Sitter.  Hence, the conformal diagram for
the extended de Sitter space accommodates all points for which the
metric $g$ is defined, but the converse is not true. To illustrate
the causal structure, let us represent the light-cones of metric
$g$ in the conformal diagram of $f$. To this end, it is convenient
to use Kruskal-type coordinates, (see {\em e.g.} \cite{HawkingEllis})
\be
\label{UVdS1} U=-\left(\frac{1-Hr}{1+Hr}\right)^{1/2} e^{-H
\tilde{t}},\quad\quad V=\left(\frac{1-Hr}{1+Hr}\right)^{1/2} e^{H
\tilde{t}}.
\ee
Note that this involves $\tilde t$ (and not $t$),
the temporal coordinate in which $f$ is diagonal (see (\ref{ttilde})). Eq.
(\ref{UVdS1}) maps the interior of the de Sitter horizon $Hr<1$
into the quadrant $U<0, V>0$ of the plane $(U,V)$. The future
event horizon for an observer at $r=0$ corresponds to $U=0$,
whereas the past event horizon corresponds to $V=0$ (see Fig. \ref{figure1}).
The quadrant $U>0, V>0$ which lies beyond the future event
horizon, is similarly covered by the change of coordinates \be
\label{UVdS2} U=\left(\frac{Hr-1}{Hr+1}\right)^{1/2} e^{-H
\tilde{t}},\quad\quad V=\left(\frac{Hr-1}{Hr+1}\right)^{1/2} e^{H
\tilde{t}}.\ee The remaining quadrants can be obtained by changing
the sign in the right hand side of Eqs. (\ref{UVdS1}-\ref{UVdS2}).
As usual, we may perform the conformal re-scaling
$$ T=\arctanh\ V+
\arctanh\ U, \quad R=\arctanh\ V - \arctanh\ U,$$ so that the in
the new coordinates the four quadrants lie in a square of finite
size (see Fig. \ref{figure1}). The vertical boundaries correspond to $r=0$,
while the past and future boundaries of the diagram correspond to
$r=+\infty$ (which is a spacelike boundary). Note further that the
coordinate system $(t,r)$ only covers the $V>0$ corner of the
maximally extended de Sitter spacetime but also that it
accomodates positive and negative values of $U$, so that it goes
beyond the future event horizon. Thus, this coordinate system is
similar, as far as the de Sitter metric is concerned, to the
Eddington-Finkelstein coordinates of a black hole. At this point
one might worry about a possible singularity due to the presence
of the horizon. Indeed, as we discussed above, a coordinate
singularity in one of the two metric cannot always be removed by a
coordinate change that renders both metrics non singular. Here the
situation is different, and in the coordinates ($t,r$), both
metrics are smooth and regular everywhere where $t$ and $r$ take
finite values. So the $U=0$ part of the de Sitter horizon in the
$V>0$ corner does not result in a singularity in the bimetric
theory. Things are however more involved for the $V=0$ part of the
horizon, as we will now see.

\begin{figure}[h]  \centering
\psfrag{ip}[][]{$i^+$}\psfrag{im}[][]{$i^-$}\psfrag{imm}[][]{$i^-_{(r<r_H)}$}
\psfrag{imp}[][]{$i^-_{(r>r_H)}$}\psfrag{i0}[][]{$i^0$}
\psfrag{Ip}[][]{$\mathcal{I}^+$}\psfrag{Im}[][]{$\mathcal{I}^-$}
\psfrag{t1}[][]{$t_1$}\psfrag{t2}[][]{$t_2$}
\psfrag{t3}[][]{$t_\epsilon$}
\psfrag{imh}[][]{$i^-_{(r=r_H)}$}
\includegraphics[width=0.6\textwidth]{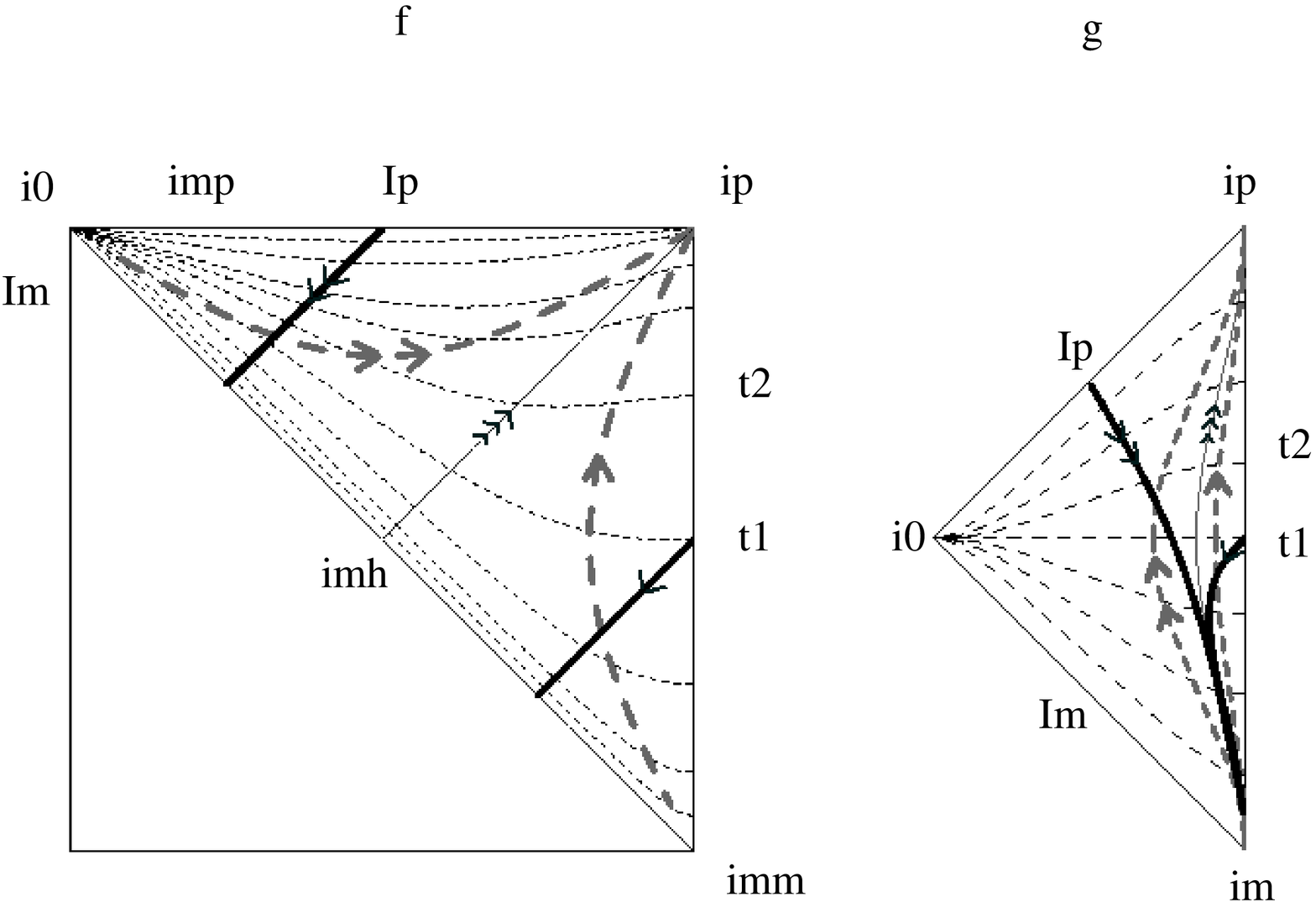}
\caption{\small Causal diagram for de Sitter with Minkowski, for
$\beta=1$. The left diagram is for de Sitter with horizon radius
$r_H$, while the right diagram is for Minkowski. The dashed thin
lines (with no arrows) are $t =$ constant lines. The  dashed thick
line with one (resp. two) arrow is an $r =$ constant curve, with
$r<r_H$ (resp. $r>r_H$). The thin solid line with three arrows
represents the trajectory of an observer sitting at constant
radius $r=r_H$ in Minkowski spacetime. The thick solid lines with
arrows are past directed null geodesics of de Sitter space-time
$U= constant$ curves. The mapping of the infinities (null,
spacelike, timelike) of Minkowski spacetimes ($i^{\pm,0}$, ${\cal
I}^{\pm}$) has been indicated on the de Sitter diagram. One of the
stricking feature of those diagrams, is that the past
time-like infinity of Minkowski is split between the upper left
corner (for $r>r_H$), the lower right corner (for $r<r_H$) and the
diagonal ($r=r_H$) of the de Sitter space-time.}
\label{newfigure1}
\end{figure}

To this end, let us consider the light-cones in the Minkowski
metric. Radial null geodesics are simply given by \be \label{MGeo}
t=\epsilon r +k \ee where $\epsilon=\pm 1$ corresponds to future
and past directed null rays respectively. For $\epsilon=0$ we
obtain the space-like $t=k$ slices. In order to represent such
geodesics in the conformal diagram for metric $f$, let us first
express them in terms of $\tilde t$. For the potentials
(\ref{potentials}), Eq. (\ref{ttilde}) reads \be \label{dtilde}
\di\tilde t = \beta^{-1/2} \di t + {H r \over 1-H^2 r^2}
(\beta-1+H^2 r^2)^{1/2}\ \beta^{-1/2} \di r. \ee For $\beta=1$ this
yields \be \label{nullMdS} \tilde{t}= t - r -
\frac{1}{2H}\ln\left|{1-Hr \over 1+ Hr}\right|. \ee The
integration constant has been chosen so that $\tilde t= t$ at
$r=0$. For $\beta\neq 1$, Eq. (\ref{dtilde}) can also be
integrated, but the expressions are a bit more cumbersome and we
shall omit them in what follows. Note that the change of variables
(\ref{nullMdS}) is discontinuous at the de Sitter horizon. This is
just as well, since the coordinates ($\tilde t, r$) become singular
at $r \equiv r_H=H^{-1}$,  and we need to consider the
Kruskal-type coordinates anyway. Substituting in (\ref{UVdS1}) or
in (\ref{UVdS2}), we have \be U=\Big(\frac{Hr-1}{Hr+1}\Big)
e^{-H(t - r)}, \quad\quad V= e^{H(t - r)}. \label{change}\ee As
noted above,  these expressions are valid both for $U\leq0$ and
$U\geq0$ (with $V>0$), and so they cover both quadrants
(\ref{UVdS1}) and (\ref{UVdS2}) at once. Now, the radial geodesics
are easily given in the $U,V$ chart (as a curve parametrized by
$r$) by substituting (\ref{MGeo}) into (\ref{change}), \be
U=\Big(\frac{Hr-1}{Hr+1}\Big) e^{-Hk} e^{-H(\epsilon - 1)r},
\quad\quad V=  e^{Hk} e^{H(\epsilon - 1)r}. \label{change2}\ee
Future directed null rays of the Minkowski metric $t=r+k$, are
simply straight lines at 45 degrees,
$$
V= e^{Hk} = const.
$$
On the other hand, past directed null geodesics $\epsilon=-1$, as
well as the spacelike geodesics $\epsilon=0$, have a rather
non-trivial behavior which is illustrated in Fig. \ref{figure1}.
For $Hr \ll 1$, the light-cone emanating from $r=t=0$ (\emph{i.e.} $k=0$)
has the same shape as in Minkowski space. However, at $Hr \sim 1$
the past directed light-cone opens up and turns around in the
$U,V$ plane. Beyond this turning point, ``past directed" null rays
of Minkowski start progressing towards the future in the de Sitter
diagram! In particular, at large affine parameter, $Hr \to
\infty$, both space-like and past directed null geodesics of
Minkowski meet at the upper left corner of the conformal diagram,
$U\to +\infty, V\to 0$, which belongs to the future boundary of de
Sitter. In fact, the future timelike infinity $i^+$ of Minkowski
is mapped into the upper right corner of the de Sitter
diagram, the future null infinity ${\cal I}^+$ of Minkowski is
mapped into the future null infinity of de Sitter (which is
spacelike), the spacelike infinity $i^0$ and null past infinity
${\cal I}^-$ of Minkowski are both mapped to the upper left corner
of the de Sitter diagram (see Fig. \ref{newfigure1}). The
situation is more complicated for the past timelike infinity $i^-$
of Minkowski. The latter is split into three pieces: a particle
moving back in time along a $r = constant$ geodesic of Minkowski
space-time would either go to the upper left corner of the
de Sitter diagram if $r > r_H$, to the lower right corner if
$r<r_H$, or to the $U=0, V=0$ central point if $r=r_H$. However, a
given timelike trajectory in Minkowski, stemming from the infinite
past ($t=-\infty, r=r_H$) can emanate in the de Sitter diagram
from any point along the diagonal $V=0$. The latter diagonal is
then representing the whole of the past $r=r_H$ infinity of
Minkowski. This can be better seen, plotting the  null geodesics
of de Sitter into a conformal diagram for Minkowski. Inverting
(\ref{change}), \be t=r + H^{-1} \ln V, \quad\quad
r=\frac{UV+1}{H(1-UV)}, \ee outgoing (or incoming) null curves are
given parametrically in terms of $U$ (or $V$) by taking $V=k$ (or
$U=k$). These are represented in Fig. \ref{newfigure1}. In
particular, one sees that past directed $U=constant$ null lines
can intersect the $V=0$ curve anywhere, while they all asymptote
the $r= r_H$ curve in the Minkowski diagram as $t$ goes to
$-\infty$.

We may then ask whether it is possible to construct a closed
time-like curve by combining signals which propagate in the $f$
metric with those propagating in the $g$ metric. We defer this
discussion to section \ref{CTLC}, where we show that this is not possible
for general Type I solutions.\\

\begin{figure}[h]  \centering
\psfrag{a}[][]{$(a)$}\psfrag{b}[][]{$(b)$} \psfrag{D}[][]{$D(r)\in
\mathbb{C}$}
\includegraphics[width=0.6\textwidth ]{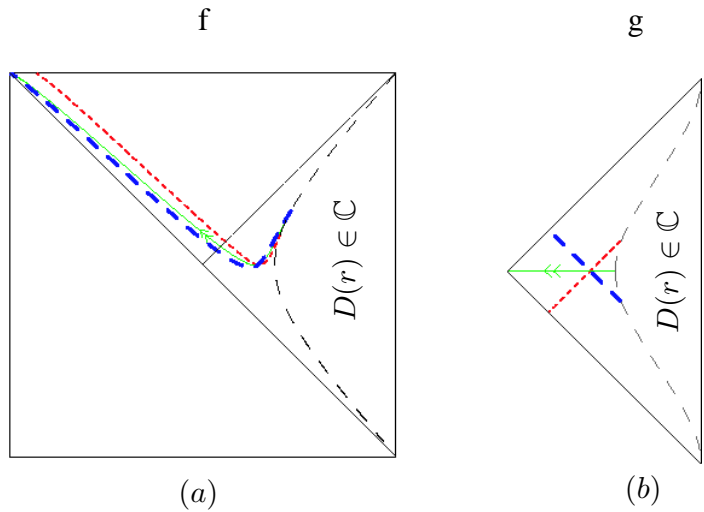}
\caption{\small{ Causal diagrams
when the $f$ metric is de Sitter (left diagram) while the $g$ metric
is Minkowski (right diagram) and $\beta=1/6$. Thick dashed (blue)
curve, thin dashed (red) curve, and thin solid (green) curve with
two arrows,  are respectively null (for the two first) and spacelike
(for the last) radial geodesics of Minkowski space-time. The dashed
curly vertical line in both diagram is an $r=$ constant curve which
is the boundary of the region where one of the metrics becomes
complex.}} \label{figure2}
\end{figure}

A similar analysis can be performed for other values of $\beta$.
For $\beta>1$, $D$ is everywhere real and the causal structure is
quite similar to the one described above. A minor difference is
that the light-cones of Minkowski geodesics are not at 45 degrees
near the origin (as they were in Fig. \ref{figure1}). This can be easily seen
from Eq. (\ref{ttilde}). On the other hand, for $\beta<1$ the
metric becomes complex in the region $H^2r^2 < 1-\beta$ (see Fig.
\ref{figure2}).

\begin{figure}[h]  \centering
\psfrag{a}[][]{$(a)$}\psfrag{b}[][]{$(b)$}
\psfrag{c}[][]{$(c)$}\psfrag{I}[][]{I}
\psfrag{II}[][]{II}\psfrag{III}[][]{III} \psfrag{IV}[][]{IV}
\includegraphics[width=0.7\textwidth]{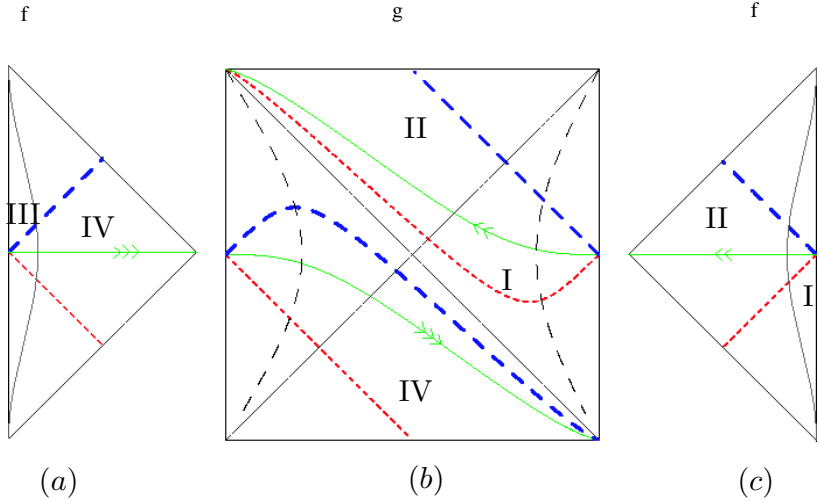}
\caption{\small{Diagram showing the extension proposed in the text
for the de Sitter with Minkowski solution. Notations are the same as in
Fig. \ref{figure1}. By using a second Minkowski space-time, we can
extend the de Sitter diagram of Fig. \ref{figure1}, represented by
region I and II above,  to the lower half, represented by region
III and IV above. The de Sitter space-time is now geodesically
complete, however the whole space-time it is not globally
hyperbolic, when both metric are considered on the same footing.
If we draw a Cauchy surface for all the de Sitter geodesics [such
as a horizontal line cutting across the diagram $(b)$], this
surface will intersect some of the Minkowski geodesics twice,
while it will fail to intersect some others.}} \label{figure3}
\end{figure}

\begin{figure}[h]  \centering
\psfrag{a}[][]{$(a)$}\psfrag{b}[][]{$(b)$} \psfrag{I}[][]{I}
\psfrag{II}[][]{II}\psfrag{III}[][]{III} \psfrag{IV}[][]{IV}
\includegraphics[width=0.6\textwidth ]{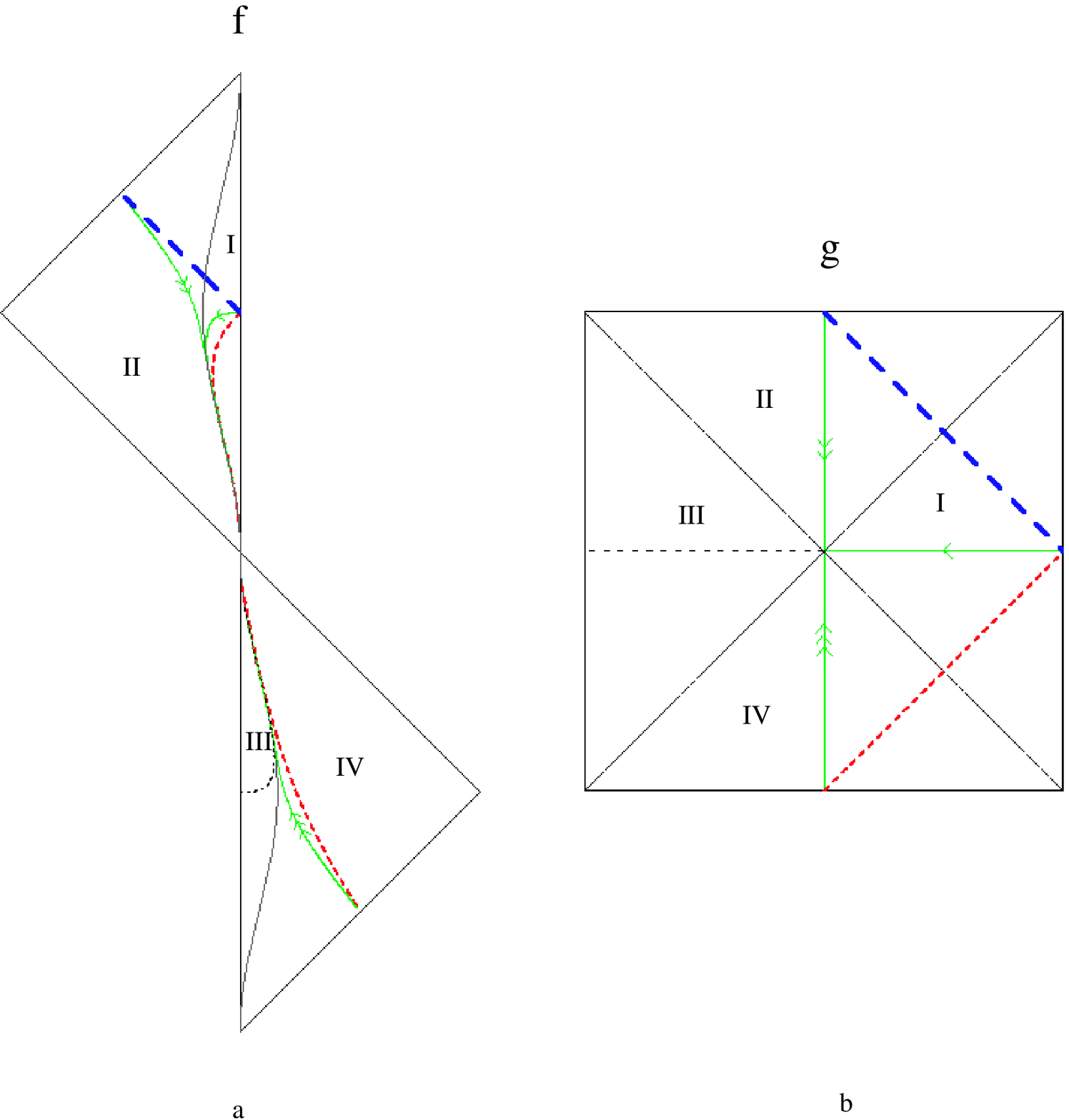}
\caption{\small{Same as Fig. \ref{figure3}, with radial geodesics
of de Sitter plotted instead of those of Minkowski.
 The thick dashed (blue) curve is a future-directed radial null ray from the origin $(r=0, \tilde{t}=0$). The thin solid (green) curve is a $\tilde{t}=0$ radial geodesic of de Sitter.
The thin dashed (red) with one arrow curve a the past-directed
null geodesic from the origin. We also plotted, as thin solid
(green) curves  with two and three arrows, the continuation of the
$\tilde{t}=0$ curve beyond the horizon $r=r_H$.
 When mapped into the Minkowski
diagram, the past directed null geodesics of de Sitter, of region
I, reach the timelike past infinity of the Minkowski space-time at
a finite value of their affine parameter in de Sitter, namely when
they cross the de Sitter horizon $r=r_H$. Nevertheless, we can
``smoothly" continue them in the newly added Minkowski solution
onto which regions III and IV of de Sitter space-time are mapped.
\label{altra}}}
\end{figure}

Let us now consider the issue of global structure. As was stressed
above, the coordinates $(r,t)$ cover the full Minkowski space
corresponding to the metric $g$, but only half of the conformal
diagram for the extended de Sitter metric, corresponding to $V>0$
(see Fig. \ref{figure1} (a)). This portion is by itself globally hyperbolic,
since the $t=k$ surfaces are Cauchy surfaces for all geodesics of
both metrics in this region. However, the region $V>0$ is not
geodesically complete, since the null geodesics $U=const.$ of
de Sitter reach $V=0$ at finite affine parameter. To obtain a
geodesically complete space-time, we can match the solution in the
upper half of the conformal diagram with a solution in the lower
half of the diagram. For this purpose we introduce a {\em second}
Minkowski space, with metric $g'$, which will be covered with
coordinates $r'$ and $t'$. The change of variables (\ref{UVdS1})
and (\ref{UVdS2}) with the substitutions $t\to -t'$, $U\to -U$,
$V\to -V$, maps the full range of the coordinates $(r',t')$ into the
lower half of the de Sitter conformal diagram, below the diagonal
$V=0$. The full diagram, represented in Fig. \ref{figure3} and \ref{altra}, is now
geodesically complete. In doing such an extension,  we mean we are
gluing together one Minkowski spacetime to the other along the
past infinity of the $r=r_H$ sphere of the former to the future
infinity of the $r=r_H$ sphere of the latter. These infinities do
not belong to the Minkowski spacetimes, but to their boundaries,
while they are located in the interior of the de Sitter spacetime.
This provides indeed a perfectly fine geometric maximal extension,
where all geodesics are complete.

We should add, however, that a maximal extension is usually
required to satisfy the equations of motion. The bigravity
equations of motion are certainly satisfied everywhere in regions
I, II, III and IV of Fig. \ref{figure3}, but it is unclear in which sense they
are satisfied along the diagonal $V=0$. The problem is precisely
that we are joining two Minkowski spacetimes [$(a)$ and $(c)$ of
Fig. \ref{figure3}] at a locus which lies at their conformal boundary. It is
conceivable that promoting our maximal extension to a solution of
the equations of motion might necessitate additional input, such
as the inclusion of some source at the time-like infinity of
Minkowski.
Note further, that there is some arbitrariness in
the extensions which are possible, as the already geodesically complete
companion can be extended by any other companion to the
metric that we are extending.
As we have already  commented,
 a similar ambiguity is present in usual General
Relativity when a metric must be continued beyond a Cauchy
horizon.

The  extended diagram, Fig. \ref{figure3}, is not
globally hyperbolic. The $t=k$ surfaces of the region $V>0$ are no
longer Cauchy surfaces for the whole space-time, since they do not
intersect causal geodesics in the lower half of the diagram. A
surface which intersects all causal geodesics should cut through
both regions, $V>0$ as well as $V<0$. One such surface is, for
instance, the horizontal line $U=V$. The problem is that, as can
be seen in Fig. \ref{figure3}, there are causal geodesics which intersect
this surface twice (such as the past directed null rays from
$r=t=0$).  A formal proof that the maximally extended diagram of
Fig. \ref{figure3} is not globally hyperbolic runs as follows. Let
us restrict attention to radial geodesics. A Cauchy surface must
intersect all causal geodesics once and only once. Let us assume
that such a surface $\Sigma$ exists. In particular, $\Sigma$ must
intersect the null geodesic $V=0$ of de Sitter space. By
continuity, it will also intersect the null geodesics $V=const.$,
in the range $-\delta< V < \delta$, where $\delta$ is an
arbitrarily small positive number. Let us now consider the null
geodesic of Minkowski space, parametrized by $r$ in Eq.
(\ref{change2}), and let us choose the constant $k < H^{-1}
\ln \delta$. It is clear that the incoming radial geodesic (with
$\epsilon=-1$) will start at the upper left corner of the de
Sitter diagram (at $r\to \infty$), and work its way down towards
the right boundary of the diagram (at r=0), while $V$ will always
remain in the interval $0<V<\delta$). Hence, the incoming null
geodesic must intersect $\Sigma$ at least once before it reaches
$r=0$. At $r=0$ it bounces and becomes the outgoing null geodesic
$V=e^{Hk}<\delta$, which will intersect $\Sigma$ once more before
it reaches null future infinity. Hence, there are geodesics of
Minkowski which intersect $\Sigma$ twice, which simply means that
this is not a
good Cauchy surface for all geodesics in the extended diagram.
We will have more to say about the tension between global hyperbolicity
and geodesic completeness in section \ref{vs}.\\

Let us compare the present situation to that in usual GR. As
mentioned above, Cauchy horizons are also present in certain
maximally extended solutions of GR, such as Reissner-Nordström or
anti-de Sitter space. Whenever there is such a horizon, the
equations of motion do not suffice to continue the solution past
it, and we need additional input. Usually, analytic continuation
is used, or else some boundary conditions at certain time-like
boundaries of spacetime are introduced. As mentioned above, in the
present context it is not clear whether the equations of motion
are satisfied or not at the Cauchy horizon of the maximally
extended solution, but this is precisely because this horizon
corresponds to a point in the conformal boundary of one of the
metrics. In this sense, the situation is no worse than in GR,
where we have to prescribe data on certain boundaries in order to
determine the maximal extension. Another point to consider is
that, physically, Cauchy horizons tend to be unstable to
perturbations, because of large blueshift effects
expected from the accumulation of perturbations close to the horizon
 \cite{Simpson:1973ua,Chandra:1982}. The same is
expected to happen in the present context. Note, {\em e.g.}, from Fig.
\ref{figure3}, that all future directed null geodesics of
Minkowski in regions III and IV tend to pile up near the Cauchy
horizon at $V=0$, suggesting that there will be a large
backreaction near that surface once we include perturbations.

Another interesting fact of the bi-metric solution is that the
concepts of causal past and future are ``broadened", since signals
can be transmitted by matter coupled to both metrics. For
instance, the observers at $r=0$, with $V>0$ can see signals
emitted by all other observers, and hence they have no future
event horizon. Likewise, observers at $r=0$, with $V<0$, can emit
signals which will eventually reach all other observers, and hence
they have no past event horizon. It is tempting to speculate that
cosmological bi-gravity solutions, if they can be made sense of,
could in principle be relevant to the horizon problem.
%

%%%%%%%%%%%%%%%%%%%%%%%%%%%%%%%%%%%%%%%%%%%%%%%%
\subsection{de Sitter with Schwarzschild}
%%%%%%%%%%%%%%%%%%%%%%%%%%%%%%%%%%%%%%%%%%%%%%%

Let us now replace the Minkowski metric by the Schwarzschild one.
In this case, the potentials of the Type I solution are given by
\ba \label{explipq}
 p= H^2 r^2,\quad\quad q=\frac{2 M}{r}. \\ \ea
\begin{figure}[h]  \centering
\psfrag{ip}[][]{$i^+$}\psfrag{im}[][]{$i^-$}\psfrag{imm}[][]{$i^-_{(r<r_H)}$}
\psfrag{imp}[][]{$i^-_{(r>r_H)}$}\psfrag{imh}[][]{$i^-_{(r=r_H)}$}
\psfrag{iz}[][]{$i^0$}\psfrag{r0}[][]{$r=0$}
\psfrag{Ip}[][]{$\mathcal{I}^+$}\psfrag{Im}[][]{$\mathcal{I}^-$}
\includegraphics[width=0.8\textwidth ]{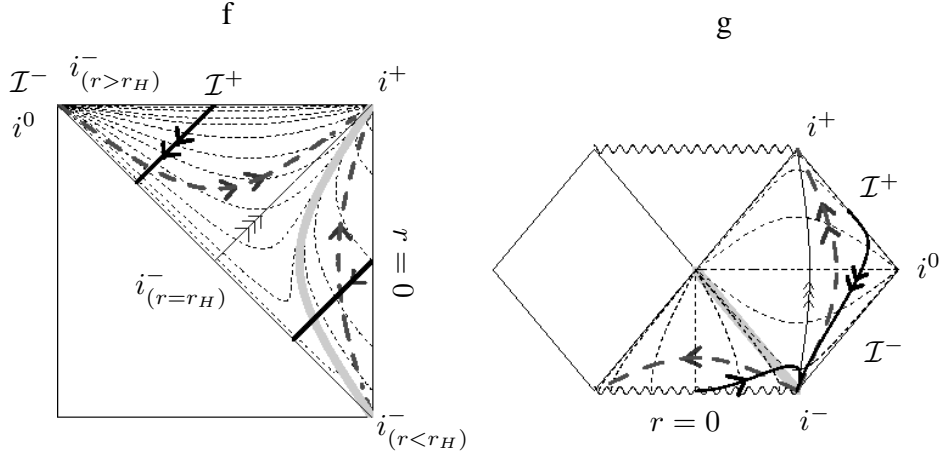}
\caption{\small{Causal diagrams when the $f$ metric is de Sitter (right)
and the $g$ metric is Schwarzschild (left). The notations are the same
as in figure \ref{newfigure1}. The main difference with the case depicted
in this last figure is the presence of the Schwarzschild horizon. The part of
the Schwarzschild horizon shown as a thick gray line on the right diagram above
is mapped to the thick gray line of the left diagram.
The part of the Schwarzschild horizon which is the diagonal of the right
diagram orthogonal to the thick gray line is mapped to the upper right corner of the
de Sitter diagram in analogy to what was found to happen for the de Sitter horizon when the other metric is Minkowski. This shows the possibility to extend the Schwarzschild space-time through another de Sitter spacetime joined to the other by the future infinity of a $r=r_S$ sphere ($r_S$ being the Scharzschild horizon)\label{SCHDSmat}}}
\end{figure}
Both metrics have now horizon singularities whenever $p=1$ and
$q=1$, corresponding respectively to $r=r_H$ and $r=r_S\equiv 2M$.
Those are coordinate singularities from the point of view of each
metric considered separately from the other. However, one might be
concerned by the possibility to remove such singularities from
both metrics at the same time. To study this issue, we first keep
$p$ and $q$ unspecified, and note that the coordinate change
(\ref{ttilde}) reads (with $\beta=1$, which we shall assume in the
following)\footnote{We only discuss here the case
$\epsilon_D=+1$, the other case, which corresponds to a change in
the sign of time, follows similarly}
\be
\label{trrstar}
\di \tilde{t}= \di t - \di r^* + \di\tilde{r}^*,
\ee
$r^*$ and
$\tilde{r}^*$ defining ``tortoise" coordinates associated with
metric $f$ and $g$ respectively by
\ba
\di r^* &=& \frac{\di r}{1-q},
\label{drs}\\
\label{drts}
\di\tilde{r}^* &=& \frac{\di r}{1-p}.
\ea
Thus, introducing the null coordinates $v = t- r^*, u = t+r^*$ for
the metric $g$, and $\tilde{v}= \tilde{t}-\tilde{r}^*, \tilde{u} =
\tilde{t}+ \tilde{r}^*$, for the metric $f$, one has from the
above expression (\ref{trrstar}) \be \di\tilde{v} = \di v.\ee This
means that $v$ is null for both metrics, but also that
$(v,r,\theta,\phi)$ are Eddington-Finkelstein coordinates for both
metrics. In such a coordinates system none of the metric is
singular at the horizons.

\begin{figure}[h]  \centering
\psfrag{a}[][]{$(a)$}\psfrag{b}[][]{$(b)$} \psfrag{I}[][]{I}
\psfrag{II}[][]{II}\psfrag{III}[][]{III} \psfrag{IV}[][]{IV}
\includegraphics[width=0.6\textwidth ]{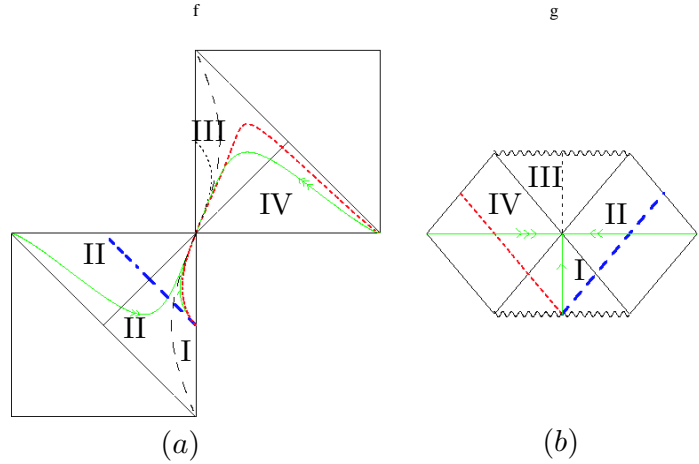}
\caption{\small{Causal diagrams when the $f$ metric is de Sitter (right) and the $g$ metric is Schwarzschild (left) showing the extension proposed in the text for the Schwarzschild space-time. Various radial geodesics of Schwarzschild are mapped onto the de Sitter diagram The dashed vertical curly line in
the de Sitter diagrams indicates the Schwarzschild horizon.
 Note that we can
``send a signal" from region I of the lower de Sitter space to
region IV of the upper de Sitter space by using the left-moving
null geodesic of Schwarzschild (thin dashed (red) line). \label{SCHDSmatched} }}
\end{figure}

Coming back to the explicit expressions for $p$ and $q$
(\ref{explipq}) and substituting those in (\ref{ttilde}) we find
\be \di\tilde{t} = \frac{1}{\sqrt{\beta}}\left\{\di t +
\frac{\sqrt{(H^2r^3-2 M)(H^2 r^3+(\beta-1)r-2\beta
M)}}{(r-2M)(1-H^2r^2)}\di r \right\}, \ee For $\beta=1$, we have
\be \tilde t=t-r^*-\frac{1}{2H}\ln\left|{1-Hr \over
1+Hr}\right|.\ee This matches equation (\ref{trrstar}) where, the
Schwarzschild ``tortoise" coordinate reads
\begin{equation}
r^*=r+2M \ln|1-r/2M|.
\label{tortoise}
\end{equation}
The analog of Eq. (\ref{change}) is now \be
U=\Big(\frac{Hr-1}{Hr+1}\Big) e^{-H(t - r^*)}, \quad\quad V=
e^{H(t - r^*)}, \label{changesch}\ee which, again, is valid both
for $U>0$ and $U<0$ (with $V>0$), covering both quadrants
(\ref{UVdS1}) and (\ref{UVdS2}) of de Sitter,  that is to say the
region covered by the Eddington-Finkelstein coordinates $(v,r,
\theta, \phi)$. The null and spacelike radial geodesics of
Schwarzschild can be written as \be t= \epsilon r^*+k,\label{geo}
\ee  this being obviously valid in the whole region covered by
coordinates $(v,r, \theta, \phi)$.
In the $U,V$ chart these geodesics are given by \be
U=\Big(\frac{Hr-1}{Hr+1}\Big) e^{-Hk} e^{-H(\epsilon - 1)r^*},
\quad\quad V=  e^{Hk} e^{H(\epsilon - 1)r^*}. \label{change22}\ee
Again, we find that the null geodesics $t=r^*$ correspond to
$V=const.$, (or $v = const$) so $V$ is a null coordinate both in
Schwarzschild and in de Sitter. The other radial geodesics, with
$\epsilon=-1,0$ have a more complicated form, which is
qualitatively represented in Fig. \ref{SCHDSmatched}. Note that for this
figure, we have assumed that the Schwarzschild radius $r_S$ is
smaller that the de Sitter horizon radius $r_H$.

As we discussed previously, and is manifest from {Fig. \ref{SCHDSmat}}, half of the de Sitter
diagram (above the diagonal) is mapped onto half of the
Schwarzschild diagram (below the diagonal), corresponding to the
region mapped by the Eddington-Finkelstein coordinates
$(v,r,\theta, \phi)$. Both half-diagrams are geodesically
incomplete, since some geodesics reach the horizons (which
dissects the diagrams in two) at finite affine parameter. These
geodesics can of course be extended by adding new regions of
space-time. If one adds de Sitter and Schwarzschild regions, one
obtains a ``stair-case" diagram with an infinite chain of de Sitter
and Schwarzschild space-times, two adjacent de Sitter (resp.
Schwarzschild) space-times being linked together by a common
Schwarzschild (resp. de Sitter) space-time. Needless to say, there
is also a tension in this case between geodesic completeness and
global hyperbolicity, as we found in the Minkowski-de Sitter case.

As we will discuss, this applies to more general situations where one of the
metrics has a horizon which is not shared by the other one.
 As noted previously, the new metric (new ``step")
which can be added to the stair does not necessarily correspond to
the same solution as the one of the last step of the stair, since one of the two metrics does not determine
uniquely the form of the other. Thus, in general we can construct
``stair-case" diagrams with steps having different forms. Note
further, that in the case considered here, the stairs can always be finished by adding a Minkowski spacetime, linked to a Schwarzschild space-time along a sphere of radius $r_H$ at time-like infinity.

\begin{figure}[h]  \centering
\psfrag{a}[][]{$(a)$}\psfrag{b}[][]{$(b)$}  \psfrag{I}[][]{I}
\psfrag{II}[][]{II}\psfrag{III}[][]{III} \psfrag{IV}[][]{IV}
\includegraphics[width=0.5\textwidth ]{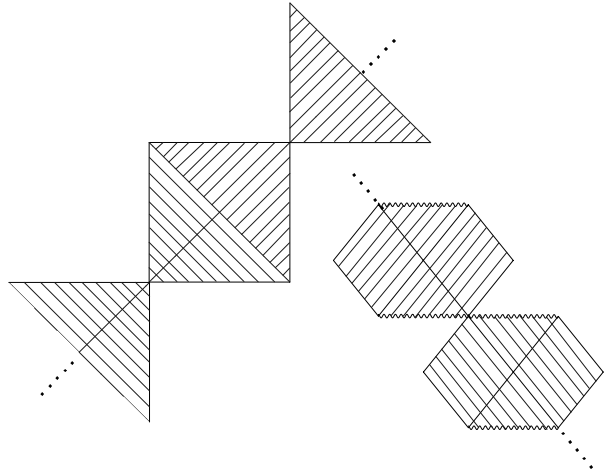}
\caption{\small{This shows a possible maximal extension of the bi-metric space-times, following the procedure given in the text, when one of the metric is de Sitter while the other is Schwarzschild.
 We are led to the ``stair-case" diagram, an
infinite chain of de Sitter spaces linked to each other through a
common Schwarzschild diagram.\label{Stairs}}}
\end{figure}

\begin{figure}[h]  \centering
\psfrag{a}[][]{$(a)$}\psfrag{b}[][]{$(b)$} \psfrag{I}[][]{I}
\psfrag{II}[][]{II}\psfrag{III}[][]{III} \psfrag{IV}[][]{IV}
\includegraphics[width=0.6\textwidth ]{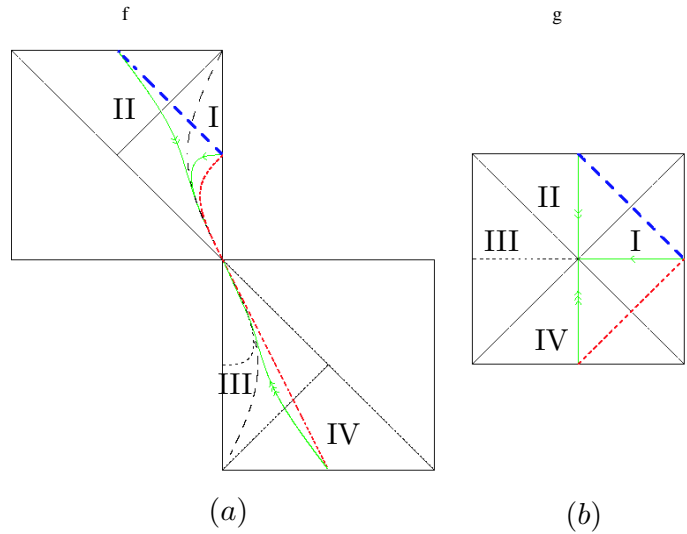}
\caption{\small{Causal diagram when both metric are de Sitter and $\beta=1$.
Notations are the same as in figure \ref{altra}.}\label{dSdS}}
\end{figure}
%%%%%%%%%%%%%%%%%%%%%%%%%%%%%%%%%%%%%%%%%%%
\subsection{de Sitter with de Sitter}
%%%%%%%%%%%%%%%%%%%%%%%%%%%%%%%%%%%%%%%%%%%

When both metrics are de Sitter, the potentials are given by \be
p= H_1^2 r^2  \quad\quad q=H^2_2 r^2. \ee For $\beta=1$, the
analysis proceeds along the same lines as in the previous
subsection, with the only difference that the (de Sitter) tortoise
coordinate is now given by
\begin{equation}
r^* = -{1\over 2H_2} \ln\left|{1-H_2 r \over 1+ H_2 r}\right|.
\label{torto}
\end{equation}
The corresponding causal diagram is represented in Fig. \ref{dSdS}

\begin{figure}[h]  \centering
\psfrag{a}[][]{$(a)$}\psfrag{b}[][]{$(b)$} \psfrag{I}[][]{I}
\psfrag{II}[][]{II}\psfrag{III}[][]{III} \psfrag{IV}[][]{IV}
\includegraphics[width=0.6\textwidth ]{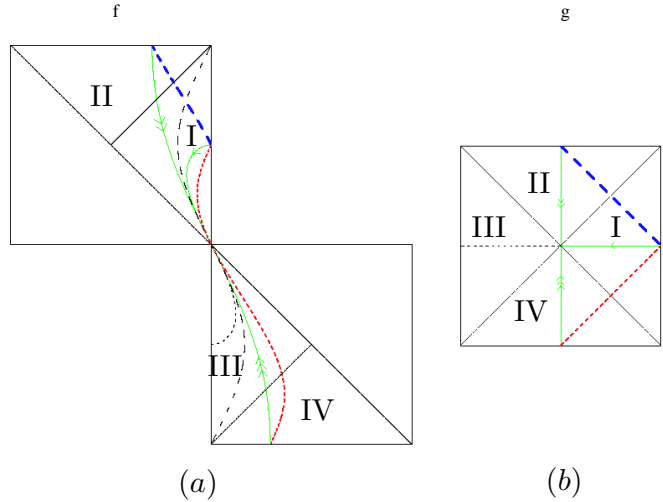}
\caption{\small{Causal diagram when both metric are de Sitter and $\beta=1/4$. Notations are the same as in
Fig. \ref{altra}.}\label{dSdSbeta}}
\end{figure}

Aside from the choice $\beta=1$, the de Sitter with de Sitter solution
allows for another way of having $D^2>0$ for the entire range of
$r$.  Indeed, it is enough to have $H_1^2 \geq \beta H_2^2$ and
$\beta \geq 1$ or $H_1^2 \leq \beta H_2^2$ and $\beta \leq 1$.
Choosing for example $\beta$ given by
\begin{equation}
\beta= {H_1^2\over H_2^2},
\end{equation}
we have \be H_1\tilde t = H_2 t - \frac{1}{2}\ln \left|1-H_1^2r^2
\over 1-H_2^2r^2\right|, \ee or $H_1(\tilde t -\tilde
r^*)+\ln(1+H_1 r) = H_2 (t-r^*)+\ln(1+H_2 r)$. Thus, the Kruskal
coordinates (\ref{UVdS1}-\ref{UVdS2}) for the metric $p$ can be
expressed in terms of coordinates $t$ and $r$ as \be
U=\left(\frac{H_1r-1}{H_2r+1}\right) e^{-H_2 (t-r^*)},\quad\quad
V= \left({H_2 r + 1 \over H_1 r +1}\right) e^{+H_2 (t-r^*)},\ee
where $r^*$ is given by (\ref{torto}). The corresponding diagram
is given in Fig. \ref{dSdSbeta}.

%%%%%%%%%%%%%%%%%%%%%%%%%%%%%%%%%%%%%%%%%%%
\subsection{Closed time-like curves?}
%%%%%%%%%%%%%%%%%%%%%%%%%%%%%%%%%%%%%%%%%%
\label{CTLC}

An interesting question regarding the bigravity
solutions is whether we can construct closed time-like curves (CTC) or
closed causal curves (CCC) by
patching together future directed geodesics corresponding to both metrics. The existence
of these curves is seen as a serious pathology of a solution and
they are forbidden by the
chronology protection conjecture which basically states that
quantum effects and vacuum polarization effects
 prevent the formation of CCC, as this curves lead to instabilities due
 to the piling of modes \cite{Babichev:2007dw}.\\

For $\beta=1$ it is easy to show that CTC cannot be constructed by
using the ``tortoise" coordinates $r^*$ and $\tilde{r}^*$ that we
defined in equations (\ref{drs}) and (\ref{drts}), as well as the
null (for both metric) coordinate $v$ (in all this subsection, we
keep the functions $p$ and $q$ unspecified). The radial null and
time-like geodesics of both metrics are given by
$$
t=\epsilon r^* + k,\quad\quad \tilde t= \tilde\epsilon \tilde
r^*+\tilde k,
$$
(Here $\epsilon=\pm 1, 0$ for outgoing and incoming null rays, or
for spacelike geodesics, respectively, and similarly for
$\tilde\epsilon$).
Thus,  any future directed causal curve with respect to $f$ or $g$
has the property that $\di v \geq 0$, and $\di v$ vanishes only
along the outgoing null radial geodesic. Once $v$ increases, even
if it is by just a little bit, it is impossible to go back to the
original value by following a future directed time-like curve,
which means that such curve cannot be closed.

Here, we disregard the possibility of making global
identifications in the coordinate $v$, which might allow for the
construction of a closed loop. Of course, even in flat space with
a single metric, closed time-like curves could be constructed by
global identifications, and in what follows we shall ignore this
somewhat artificial setup. We shall only be concerned with the
possibility of locally constructing closed time-like curves within
a given coordinate patch of space-time, without identifications.

To analyse the general case $\beta\neq 1$ it is convenient to
separately consider the following regions of space-time:

{\em a:} For $(1-p)<0$, and $(1-q)<0$ the condition $\di r=0$ defines
a space-like surface for both metrics $f$ and $g$. This means that
$r$ can only change monotonically along time-like curves of both
metrics, making it impossible to close them in this region.

{\em b:} For $(1-p)<0$ and
$(1-q)>0$, the condition $\di t=0$ defines a space-like surface for
the metric $g$. Also, from (\ref{ttilde}) with $\di t=0$, we have
\begin{equation}
\left|{\di\tilde t \over \di\tilde r^*}\right|^2 =1+{1\over \beta}
\left({1-p \over 1-q}\right)^2-{\beta+1\over\beta}\left({1-p \over
1-q}\right)
> 1.
\label{puri}
\end{equation}
Since $\tilde t$ is space-like in metric $f$ this means that the
surface $\di t=0$ [which is also defined by Eq. (\ref{puri})] is
space-like in metric $f$ too. Hence, $t$ changes monotonically
along time-like curves of both $f$ and $g$, and as a consequence
such curves cannot be closed.

{\em c:} If $(1-p)>0$ and $(1-q)<0$, then the surface $\di\tilde
t=0$ is space-like for $f$. From (\ref{ttilde}) with $\di\tilde
t=0$, we have
\begin{equation}
\left|{\di t \over \di r^*}\right|^2 =1+\beta \left({1-q \over
1-p}\right)^2-(\beta+1)\left({1-q \over 1-p}\right)
> 1.
\label{pili}
\end{equation}
Since $t$ is space-like in metric $g$, Eq. (\ref{pili}) means that
the surface $\di\tilde t=0$ is space-like in metric $g$ too, and
$\tilde t$ must be monotonic on time-like curves, which therefore
cannot close.

{\em d:} Finally, if $(1-p)>0$ and $(1-q)>0$, then  we must
distinguish two cases. For $p \geq q$, it is easy to see that
$A>0$ in Eq. (\ref{sphef}), and therefore $\di t=0$ is space-like for
both metrics $f$ and $g$. Hence, $t$ is monotonic for time-like
curves of both metrics. On the other hand, for $p\leq q$, Eq.
(\ref{pili}) for $\di\tilde t=0$ leads to
\begin{equation}
\left|{\di t \over \di r^*}\right|^2 < 1.
\end{equation}
Since now $t$ is time-like in metric $g$, this means that $\di\tilde
t=0$ is a space-like surface for this metric. Of course $\di\tilde
t=0$ is also space-like for $f$, and so $\tilde t$ is monotonic
along causal curves for both metrics.

This completes the proof for the individual regions listed above.
It is remarkable that in spite of the strong differences in the
light-cone structure of both metrics, it is not possible to draw
closed time-like curves in any of the regions. The reason is that
the future light-cone for one of the metrics never contains a part
of the past light-cone for the other metric. Thus, we can always
find a coordinate which labels hypersurfaces which are space-like
for both metrics. This coordinate must grow monotonically along
time-like curves.

By continuity, at the boundaries in between the regions, the
future light-cone of one of the metrics can at most touch the past
light-cone of the other metric, sharing perhaps a common null
direction for both metrics. Even if this were the case, a future
directed time-like geodesic with respect to one of the metrics can
never get to the inside of the past light cone with respect to the
other metric, and closed time-like curves cannot be constructed
even if we cross the boundaries between the individual
regions\footnote{In the examples we have examined, the situation where
the future light-cone of one of the metrics marginally touches the
past light-cone of the other metric at the boundary between
regions does not arise. If it did, then there might be closed
future-directed {\em null} curves at such boundary. Note, however,
that since the boundary is at $r=const.$, this situation can only
happen when both metrics have a common event horizon at the same
value of $r$. The possibility of having closed null curves on
these boundaries may require a case by case analysis, and is left
for further research.}.

%%%%%%%%%%%%%%%%%%%%%%%%%%%%%%%%%%%%%%%%%%%%%%%%%%%%%%%%%%%%%
\subsection{Global Hyperbolicity vs. Geodesic Completeness}\label{vs}
%%%%%%%%%%%%%%%%%%%%%%%%%%%%%%%%%%%%%%%%%%%%%%%%%%%%%%%%%%%%%%

In section \ref{desittermink}, we showed that global hyperbolicity may be lost
when a solution of bigravity is maximally extended to obtain a geodesically
complete metric (not
necessarily a solution of the equations of motion).
\begin{SCfigure}[][h]  \centering
\psfrag{gn}[][]{$\g_n$}\psfrag{s}[][]{$\Sigma$}\psfrag{p}[][]{$p$}
\includegraphics[width=0.5\textwidth]{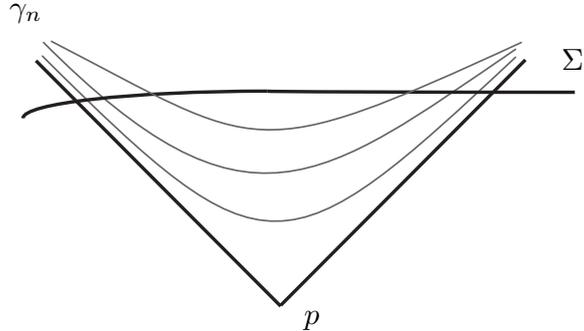}
\caption{\small This figure gives a general idea of the settings in this section. $\Sigma$ is
a Cauchy surface for the metric $g$ for which the lightcone from $p$ is drawn. $\{\g_n\}$ is a
series of spacelike curves for $g$ which converge to a curve in the lightcone
$T^{+g}_p$ and to a timelike curve
for $f$.}
\label{Cauch}
\end{SCfigure}

The main idea of the proof can be easily generalized to other situations\footnote{We
will use the notation and conventions of \cite{HawkingEllis}. A subindex
$f$ or $g$ will
indicate that the concept refers to the metric $f$ or $g$ respectively.}
(see Fig. \ref{Cauch} to get an intuitive idea).
Let us consider a time orientable manifold ${\mathcal M}$
endowed with two globally hyperbolic metrics $f$ and $g$.
Let us suppose that there
exists a point $p$ in the boundary of the manifold ($p\in \overline{{\mathcal M}}$)
  through which the
manifold can be extended for the metric $g$
through the past (future). Any Cauchy
surface $\Sigma$ for the metric $g$ will have to intersect the causal future or causal
past of
$p$, $J_g(p)$. If for any such a surface there is a  non-causal curve for $g$ which intersects $\Sigma$
more than once and which is timelike for $f$, $\Sigma$ will not be a
Cauchy surface for $f$.\\

Let us see with some examples that the existence of this curve $\g$ for
any Cauchy surface $\Sigma$ is a generic feature when one extends the non-geodesically complete
manifold through a horizon which is not shared by both metrics
 or when both metrics share a horizon but it is
 of different type for each of them.

First, take
the future null cone for the metric $g$ at a point $p$ of the boundary of a manifold ${\mathcal M}$,
\emph{i.e.}, $p\in \overline {\mathcal M}$.
 If $\overline {\mathcal M}$ is b-complete\footnote{A manifold $\M$ endowed with
 a metric $g$ is {\em b-complete} if there is an endpoint for every continuous curve
 of finite length as measured by a generalized affine parameter \cite{HawkingEllis}.},
the light rays in the null cone
can be approached by
both connected timelike and connected spacelike curves in all the disconnected parts in which $\overline {\mathcal M}$
is divided by the cone. When the manifold is maximally extended for $g$ through
the past at $p$ the future lightcone $T^{+g}_p$ can be approached
by spacelike curves $\{\g_n\}\in \M$ (see Fig \ref{Cauch}). This means that they must converge to a curve $\g_g$ in
 $\overline{{\mathcal M}}_g$ and similarly to a curve $\g_f$ in $\overline{{\mathcal M}}_f$\footnote{The
  map from one of this limit curves to the other one is not necessarily continuous
 as the topology of $\overline{{\mathcal M}}$ depends on the metric which is used to make the conformal
 compactification.}. For the $g$ metric,
 this curve is composed of two future directed null curves stemming from $p$, and thus every Cauchy surface
 $\Sigma$
 will have to intersect both curves in $J_g^+(p)$ or $J_g^-(p)$ or at $p$. Let
 us suppose that it intersects $J_g^+(p)$.
 As the surface $\Sigma$ must be spacelike for both $f$ and $g$, there exists $m\in\mathbb{N}$ such that
 it will also intersect twice
 the curves $\g_n$ for $n\geq m$. The curve $\g_f\cap {\mathcal M}$
 will be null as for the $g$ metric. If it is timelike for the $f$ metric so will be the curves $\g_n$
 for $n\geq q$ for a certain $q\in\mathbb{N}$. Now consider a curve $\g\in \M$ in $\{\g_n\}$ for $n\geq \mathrm{max}(q,m)$. This will be
 a timelike curve for $f$ which intersects twice $\Sigma$, which will not be an appropriate Cauchy surface.

In more abstract terms, the curve $\g$ can be characterized as follows.
Let us consider a family $\l_p$ of future (past) directed non-spacelike
curves for the $g$ metric stemming from $p\in \overline\M$.
Given a non-causal curve for $g$ in the future domain
of dependence of $\l_p$, $\gamma\in \mathrm{int} \left(D_g^+(\l_p,{\mathcal
M})\right)$, such that
$\g$ is non-compact and without boundary in the open set
$\mathrm{int}\left(D_g^+(\l_p,{\mathcal M})\cap D^-_g(\Sigma)\right)$ but it is compact in
$D_g^+(\l_p,{\mathcal M})\cap D^-_g(\Sigma)$, if
$\g \cap D^-_g(\Sigma)$
is timelike for the companion $f$ metric, this will be such a curve.
To see it, it is enough to realize that
being  timelike for $f$ which is globally hyperbolic, $\g$ can not be a self intersecting curve.
Thus, being compact and not-self intersecting, $\g$ will have two boundary points $q_1$ and
$q_2$ in $D_g^+(\l_p,{\mathcal M})\cap D^-_g(\Sigma)$ (which may coincide).
 As $\g\cap\dot{D}^+(\l_p,{\mathcal M})={\O}$ and $\g$
 is non compact and without boundary in
$\mathrm{int}\left(D_g^+(\l_p,{\mathcal M})\cap D^-_g(\Sigma)\right)$, these
points can only be in $\Sigma$. Thus, the curve intersects the Cauchy surface at least
twice.\\

It is not hard to identify other pathological situations where
 global hyperbolicity is lost once bigravity solutions are extended (see {\em e.g.} \cite{Blas:2007zz}).
They refer to particular situations and we shall not
elaborate on them.

%%%%%%%%%%%%%%%%%%%%%%%%%%%%%%%%%%%%%%%%%%%%%%%%%%%%%%%%%%%%%%%%%%
\chapter{Perturbations around Bigravity Solutions}\label{chapterperturbbigrav}
%%%%%%%%%%%%%%%%%%%%%%%%%%%%%%%%%%%%%%%%%%%%%%%%%%%%%%%%%%%%%%

In the previous Chapter we have considered a non-linear extension
of massive gravity consisting of two interacting metrics that
at the linear level reduce to certain models of massive gravity.
Here we will study the linear regime of perturbations to
 some of the solutions more
closely. We will be interested in two cases. First, there are
some Type I solutions that reduce to two diagonal flat metrics
which are not proportional to each other. This \emph{bi-flat}
solution is very interesting as Lorentz invariance is broken
in the vacuum. This will give rise to mass terms
which do not suffer from neither vDVZ discontinuity
and strong coupling
nor ghost states \cite{Rubakov:2008nh,Rubakov:2004eb}. As we will see,
the dispersion relations are also modified in this set-up (there
are two ``speeds of light"). This solution is also interesting
because  it corresponds to the field far from the sources
in a wider class of spherically symmetric exact solutions of the
Schwarzschild form.

Besides, even when both metrics are proportional,
 the mass term of the perturbations for a generic
 potential $V[\{\t_n\}]$  is not FP.
For Minkowski spacetime this means that only the case where
the FP condition is satisfied can be considered as a
stable vacuum of the theory. For other mass terms,
 a Lorentz breaking cut-off is necessary to
 regularize the decay rate \cite{Cline:2003gs}. As the cut-off must be
of the order of the mass scale, the theory is effectively equivalent
to GR within its range of validity. For non-trivial
backgrounds
the appearance of a curvature scale suggests the possibility of a
softer cut-off which would allow more general mass
terms. We will study this possibility in the second
part of this Chapter and find that this possibility does not happen
for bi-de Sitter solutions. Finally, we will study the case
of two de Sitter solutions with a common $SO(3)$ invariance.
This Chapter is based on \cite{Blas:2007ep} (see also
\cite{Blas:2007zz,Blas:2005sz}). A potentially interesting
possibility which we leave for future research is a background with
a black hole for one of the metrics \cite{Blas08}. Black holes
are not yet well understood in the theories of massive gravity
and bigravity provides a simple scenario to study some of their features
(see also \cite{Dubovsky:2007zi}  for the ghost condensate case
and  \cite{Jacobson:2008yc} for some problems of black holes
when Lorentz invariance is broken).
 Besides, it is well known that in GR stationary black holes can not carry
massive tensor field (\emph{no hair} theorem \cite{Bekenstein:1971hc}).
It would be interesting to study whether it can support
a non-covariant massive tensor hair.

%%%%%%%%%%%%%%%%%%%%%%%%%%%%%%%%%%%%%%%%%%%%%%%%%%%%%%%%%%%
\section{Perturbations around Lorentz-breaking bi-flat metrics}
%%%%%%%%%%%%%%%%%%%%%%%%%%%%%%%%%%%%%%%%%%%%%%%%%%%%%%%%%%%

In a theory with two metrics with Einstein-Hilbert kinetic terms and
no interaction, there are 4+4 ADM Lagrange multipliers\footnote{For the ADM
analysis of massive gravity see \cite{Boulware:1973my,Deffayet:2005ys,Gabadadze:2004iv}.}.
 When we add
a non-derivative interaction which preserves diagonal
diffeomorphisms, only 4 combinations of these may in principle appear
non-linearly in the action \cite{Damour:2002ws}. For these, their
equation of motion relates them to the other variables, but they do
not lead to further constraints. Thus, we have a minimum of 4 and a
maximum of 8 Lagrange multipliers for 20 metric components. Hence,
we generically expect a maximum of $(10-4)+(10-8)=6+2=8$ degrees of
freedom and a minimum of $(10-8)\times 2=2+2=4$. In a
Lorentz-invariant context, the first possibility corresponds to a
massless and a massive graviton, whereas the second would correspond
to two massless gravitons. In the Lorentz breaking
context, it is possible to have a massive graviton with just two physical polarizations
\cite{Dubovsky:2004ud,Gabadadze:2004iv}.\\

Let us consider a general potential $V[\{\t_n\}]$ as in
(\ref{generi}). As we showed in the previous Chapter, the vacuum
energies $\r_f$ and $\r_g$ can be tuned so that the previous
 potential  has asymptotically bi-flat solutions.
At large distances from the origin, these take the form \ba
\label{BIMETMIN} g_{\mu\nu}=\eta_{\mu\nu}, \quad f_{\mu\nu}=\gamma
\tilde \eta_{\mu\nu}, \ea where
 \be
\tilde{\eta}_{\mu\nu}=\eta_{\mu\nu}-\frac{\beta-1}{\beta}
\delta_\mu^0\delta_\nu^0, \ee and $\eta_{\mu \nu} =
\mathrm{diag}(1,-1,-1,-1)$. The parameters $\gamma$ and $\beta$ are
related by Eq. (\ref{typeIcondo}). For $\beta\neq 1$, we cannot simultaneously
write both metrics in the canonical form $\eta_{\mu \nu}$, and
Lorentz invariance breaks down to spatial rotations\footnote{For
$\beta =1$, we have proportional flat metrics the perturbations of
which can be obtained from the flat space-time limit of the
calculations done in the next section.}. It will be
convenient to introduce the general perturbation in the form \ba
f^{\mu\nu}&=&\gamma^{-1}\big(\tilde{\eta}^{\mu\nu}+h_{f}^{\phantom{f}\mu\nu}\big),\\
g_{\mu\nu}&=&\eta_{\mu\nu}+h^g_{\phantom{f}\mu\nu}, \ea where
$\tilde \eta^{\m\n}$ is the inverse of $\tilde \eta_{\m\n}$. The
perturbation to the metric $f$ has been defined with the upper
indices, just because this simplifies the manipulations which yield
the action quadratic in the perturbations shown below. For the
remainder of this section, all space-time indices will be raised and
lowered with the canonical Minkowski metric $\eta_{\mu \nu}$. The
interaction Lagrangian quadratic in perturbations then reads
\ba
\tilde L_{int}&\equiv& L_{int}-\sqrt{-g}\r_g-\sqrt{-f}\r_f\nonumber=\\
&&-\frac{M^4}{8}\Big\{
n_2(h^g_{\phantom{f}ij}+h_f^{\phantom{f}ij})(h^g_{\phantom{f}ij}+h_f^{\phantom{f}ij})
+n_0
(h^g_{\phantom{f}00}+\beta^{-1}h_f^{\phantom{f}00})(h^g_{\phantom{f}00}+\beta^{-1}
h_f^{\phantom{f}00})\nonumber\\
&&
-2n_4(h^g_{\phantom{f}00}+\beta^{-1}h_f^{\phantom{f}00})(h^g_{\phantom{f}ii}
+h_f^{\phantom{f}ii})+n_3
(h^g_{\phantom{f}ii}+h_f^{\phantom{f}ii})^2\Big\},
\label{lame}
\ea
where, after imposing (\ref{typeIcondo}),
\ba
M^4&=&4\zeta\left(\frac{\g^4}{\b}\right)^v, \quad
n_0=3n_3-2n_4-n_2+\g\frac{\pd}{\pd \g}\left(\sum_n n \g^{-n}(-1+\b^n)V_0^{(n)}\right),\nonumber\\
n_2&=&-\sum n^2\g^{-n}V_0^{(n)},\quad n_3= u v V_0+\sum_n n[v-u]
\g^{-n}V^{(n)}_0
-\sum_{m,n} n m \g^{-(n+m)}V_0^{(n,m)},\nonumber\\
n_4&=&n_0+\b \frac{\pd}{\pd \b}\left(\sum_n n
\g^{-n}(-1+\b^n)V_0^{(n)}\right).\label{coefftypeI}
\ea
For the sake
of simplicity, we will restrict to potentials $V[\{\t_n\}]$ for
which  Eq. (\ref{typeIcondo}) is independent\footnote{The case
where (\ref{typeIcondo}) is satisfied independently of $\b$ and $\g$
leads to the condition \be \label{noNcorr} 3n_3-3n_0-n_4=0, \quad
n_4=n_0, \ee which, as we shall see, corresponds to the case of no
corrections to the Newton's law. An example of an interaction where
these conditions are satisfied is a potential which is only a
function of the ratio of determinants of $f$ and $g$; that is
$V\left[\{\tau_n\}\right] = V[f/g]$. In this particular case, there
is an enhanced symmetry under independent ``non-diagonal" transverse
diffeomorphisms, which do not change the value of the determinants
of the respective metrics.} of $\beta$, and determines $\gamma$.
From equation (\ref{coefftypeI}), this implies $n_0=n_4$.
 In particular, this class includes
the interaction (\ref{interaction}), which, as we shall see, leads
to a rather pathological behaviour for the perturbations. On the
other hand, it is general enough to be representative of generic
choices of potentials.

In the works \cite{Rubakov:2004eb,Dubovsky:2004sg} the case of a single
graviton with a Lorentz violating mass term has been discussed. For
comparison with those references, it will be useful to introduce
$$m_0^2=-c n_0, \ m_1^2=0, \ m_2^2=c n_2, \ m_3^2=-c n_3, \ m_4^2=-c n_4,$$
where $c>0$ is an irrelevant constant which has the dimensions of
mass squared.

Note that the components $h^g_{\phantom{f}0i}$ and
$h_f^{\phantom{f}0i}$ are absent from (\ref{lame}). As noted in
\cite{Berezhiani:2007zf} the absence of such terms is a consequence
of invariance under diagonal diffeomorphisms in this background (see below). In
the case of a single graviton (with a Fierz-Pauli kinetic term), the
absence of $h_{0i}$ in the mass term leads to a very interesting
behaviour \cite{Dubovsky:2004sg,Dubovsky:2004ud,Dubovsky:2005dw},
where the two polarizations of the massless graviton acquire mass,
while all the other modes
 do not propagate\footnote{It should be stressed that
the absence of $0i$ components is a peculiarity of the background
considered. By suitable adjustment of the vacuum energies, the
theory we are considering also admits the Lorentz preserving vacuum
of type II, where $f_{\mu\nu}=g_{\mu\nu}=\eta_{\mu\nu}$. In that
case, the interaction term leads to the Fierz-Pauli mass term for a
combination of the two gravitons. This mass term does contain the
$0i$ components.}.

Let us now investigate whether a similar phenomenon occurs in our
model. The situation is not directly reducible to that of a single
graviton, since the equations of motion are not diagonal. Also, the
kinetic term breaks the Lorentz invariance. It is convenient to
decompose the perturbations into irreducible representations of the
spatial rotations, \ba \label{decompAph}
 h^X_{\phantom{f}00}&=&2 A^X\nonumber,\\
 h^X_{\phantom{f}0i}&=& B^X_{,i}+V^X_i\nonumber,\\
 h^X_{\phantom{f}ij}&=&2\psi^X\delta_{ij}-2E^X_{,ij}-2F^X_{(i,j)}-t_{ij}^X,\label{irrep}
\ea where $t^X_{\phantom{X}ii}= t^X_{\phantom{X}ij,i}=
V^X_{\phantom{X}i,i}= F^X_{\phantom{X}i,i}=0$ for $X=f,g$, and all
space-time indices are raised and lowered with
the metric $\eta_{\mu\nu}$.\\

To second order in the perturbations, the kinetic terms in
(\ref{action}) can be written in terms of these scalar, vector and
tensor variables as: \ba
L_K&=&\frac{1}{2\kappa_g}\Big\{-\frac{1}{4}t^g_{ij}\Box t^g_{ij}
-\frac{1}{2}\left(V_i^g+\dot F_i^g\right)\Delta
\left(V^g_i+\dot F^g_i\right)+4\Delta \psi^g \left(A^g-\dot B^g-\ddot E^g\right)\nonumber \\
&&-2\psi^g \Delta \psi^g-6 (\dot\psi^g)^2\Big\} +\frac{1}{2\tilde
\kappa_f}\Big\{-\frac{1}{4}t^f_{ij}\tilde \Box t^f_{ij}
-\frac{\beta^{-1}}{2}\left(V^f_i+\b\dot F^f_i\right)\Delta
\left(V^f_i+\b\dot F^f_i\right)\nonumber \\
&&\hspace{1cm}+ 4\beta^{-1} \Delta \psi^f \left(A^f-\b\dot
B^f-\b^2\ddot E^f\right) -2 \psi^f \Delta \psi^f-6 \beta(\dot
\psi^f)^2\Big\},
\ea
where $\tilde \Box=\tilde \eta^{\m\n}\partial_\m
\partial_\n$,
 $\tilde \kappa_f=\gamma^{-1} \beta^{1/2}\kappa_f$ and dot means a derivative with respect to time.
At the linear level, the transformations generated by independent
diffeomorphisms $\delta x^\mu=\xi^\mu_X$ in each one of the metrics
can be expressed as
\ba \delta
h^g_{\phantom{g}\m\n}=2\partial_{(\m}\xi^g_{\n)}, \quad \delta
h^f_{\phantom{g}\m\n}=2\eta_{\b(\m|}\tilde \eta^{\a\b}\partial_\a
\xi^f_{|\n)}. \ea Note that the kinetic term is written in terms of
the following quantities:
\ba
t_{ij}^g, \ V_i^g+\dot F_i^g, \ \psi,\ A^g-\dot B^g-\ddot E^g, \nonumber\\
t_{ij}^f, \ V_i^f+\b\dot F_i^f, \ \psi,\ A^f-\b\dot B^f-\b^2\ddot
E^f, \ea
which are invariant under both gauge transformations. On the
other hand, the full action (including the mass terms), is invariant
only under the diagonal gauge invariance \be \xi_\m^g=\xi^f_\m. \ee No
second order scalar combination of $h^X_{\phantom{X}0i}$ is
invariant under this gauge invariance, which implies that those terms
are always absent (cf. (\ref{lame})). We may now analyze the
propagating degrees of freedom.

% %%%%%%%%%%%%%%%%%%%%%%%%%%%%%%%%%%%%%%%%%%%%%%%%%%%%%%%%
%%%%%%%%%%%%%%%%%%%%%%%%%%%%%%%%%%%%%%%%%%%%%%%%
\subsection{Tensor Modes}
%%%%%%%%%%%%%%%%%%%%%%%%%%%%%%%%%%%%%%%%%%%%%%

The linearized Lagrangian for the tensor and vector modes can be
expressed as
\ba \label{tensandvec}
 L_{t,v}&=&\frac{1}{2\kappa_g}\Big\{-\frac{1}{4}t^g_{ij}\Box t^g_{ij}
-\frac{1}{2}\left(V_i^g+\dot F_i^g\right)\Delta
\left(V^g_i+\dot F^g_i\right)\Big\}\nonumber\\
&&+\frac{1}{2\kf}\Big\{-\frac{1}{4}t^f_{ij}\tilde \Box t^f_{ij}
-\frac{\beta^{-1}}{2}\left(V^f_i+\b\dot F^f_i\right)\Delta
\left(V^f_i+\b\dot F^f_i\right)\Big\}\nonumber\\
&&-\frac{M^4}{8}\Big\{
n_2(t_{ij}^g+t_{ij}^f)^2-2n_2(F_i^g+F_i^f)\Delta(F_i^g+F_i^f)\Big\},
\ea
where $\tilde\kappa_f=\gamma^{-1} \beta^{1/2}\kappa_f$. The
corresponding equations of motion in Fourier space read
\ba
\label{DIS1}
\omega^2 t_{ij}^g&=&{\bf k}^2t_{ij}^g+ \kappa_g
M^4n_2(t_{ij}^g+t_{ij}^f),\\
\label{DIS2}
\beta\omega^2 t_{ij}^f&=&{\bf k}^2t_{ij}^f+ \kf M^4 n_2(t_{ij}^g+t_{ij}^f),
\ea
from which we obtain the dispersion relations \be \label{dispete}
\omega^2_\pm=\frac{1}{2\beta}\Big((\beta+1){\bf  k}^2+ \kappa_0
M^4\pm\sqrt{ ((\beta+1){\bf k}^2+ \kappa_0 M^4)^2-4\b {\bf
k}^2(\kappa_1 M^4+{\bf k}^2)}\Big),
\ee
where  $\kappa_0=n_2
(\b\kappa_g+\kf)$ and $\kappa_1= n_2  (\kappa_g+\kf)$.

At high energies, we have
\be \omega^2_+\approx{\bf k}^2,\quad
\omega^2_-\approx{\beta}^{-1 }{\bf k}^2.
\ee
In this limit, each one
of the two gravitons propagates in its own metric
 (with the corresponding ``speed of light"\footnote{Superluminal propagation has
 previously been considered in several contexts (see {\em e.g.} \cite{Babichev:2006vx,Babichev:2007dw} for a
 recent discussion). Clearly, such propagation cannot by
 itself be considered pathological. Indeed, in the present case
 we always have superluminal propagation from the point of view of one of the metrics,
 whereas there is not any superluminal propagation from the point of view of the other
 metric. Nevertheless, as we have seen in the
 previous Chapter, the global structure of non-linear bi-gravity solutions is complicated in
 general, and its interpretation is far from trivial. Even more, instantaneous
 interaction is also present in certain theories of Lorentz breaking massive
 gravity \cite{Bebronne:2008tr}.})
along null directions $k^{\mu}=(\omega,{\bf k})$ satisfying
$$g^{\m\n}_X k_\m k_\n \approx 0.$$
The low energy expansion of (\ref{dispete}) is given by \ba
\omega^2_-&=&\frac{\kappa_1}{\kappa_0} {\bf k}^2 +O({\bf k}^4),\\
\label{Ommexp} \omega^2_+&=&\frac{\kappa_0 M^4}{\beta}+
\left({\tilde \kappa_f+\beta^2\kappa_g \over \beta\tilde\kappa_f+
\beta^2\kappa_g}\right){\bf k}^2+ O({\bf k}^4).\label{masgru} \ea
The first dispersion relation corresponds to two massless
polarizations
 which propagate at the ``intermediate" speed
$$
c_s^2 = {\omega_-^2\over {\bf k}^2}= {\kappa_1 \over \kappa_0} =
{\kappa_g +\tilde\kappa_f \over \beta \kappa_g +\tilde \kappa_f}.
$$
Note that for $\beta>1$ we have $\beta^{-1}<c_s^2<1$, while for
$\beta<1$ we have $1<c_s^2 <\beta^{-1}$. The second dispersion
relation, Eq. (\ref{masgru}), corresponds to two massive
polarizations. It is easy to check that the graviton polarizations
are stable and tachyon free as long as $\kappa_0>0$, in the whole
range of momenta ${\bf k}$. The second dispersion relation
(\ref{masgru}) corresponds to the massive graviton.

%%%%%%%%%%%%%%%%%%%%%%%%%%%%%%%%%
\subsection{Vector Modes}
%%%%%%%%%%%%%%%%%%%%%%%%%%%%%%%%

From the Lagrangian (\ref{tensandvec}), we find that $V_i^g$ and
$V_i^f$ do not appear in the interaction term. Varying with respect
to the vector fields we have,
\ba \label{vector}
\Delta(V_i^g+\dot{F}_i^g)&=& 0, \label{tor1}\\
\Delta \left(\dot{V}^g_i+\ddot{F}^g_i\right)  &=& -M^4n_2\kappa_g \Delta \left(F^g_{i} +F^f_{i} \right),\label{tor2} \\
\Delta(V_i^f+\beta \dot{F}_i^f) &=& 0,\label{tor3}\\
\Delta \left(\dot{V}^f_i+\beta\ddot{F}^f_i\right)  &=& - M^4 n_2\kf
\Delta \left(F^g_{i} +F^f_{i} \right).
\ea
We can always use the
diagonal diffeomorphism invariance to work in the gauge where
$V_i^g=0$. It then follows from (\ref{tor1}) that $F_i^g=F_i(\vec
x)+f_i^g(t)$, where $F_i$ are arbitrary functions of position and
$f_i$ are arbitrary functions of time. The latter are in fact
irrelevant, because $F_i^X$ enters the metric only through spatial
derivatives. Formally, we may describe this as a gauge invariance
$F_i^X\mapsto F_i^X + f_i^X(t)$, which we can use in order to write,
without loss of generality,
$$F_i^g=F_i(\vec x).$$ It then follows from
(\ref{tor2}) that $$F_i^f= - F_i(\vec x),$$ where again we eliminate
the additive time dependent part. Finally, from (\ref{tor3}) we
obtain $$V_i^f=\tilde f_i(t),$$ where $\tilde f_i$ are new arbitrary
functions of time. This is not a desirable situation, since it means
that the initial conditions do not determine the future evolution of
$V_i^f$. Technically, the absence of the fields $V_i^g$ and $V_i^f$
in the mass term leads to an enhanced gauge invariance in the {\em
linearized} Lagrangian. Indeed, we can consider independent gauge
transformations for each of the metrics
\be h_{\m\n}\mapsto
h_{\m\n}+2\partial_{(\m}\xi^h_{\n)},\quad l_{\m\n}\mapsto
l_{\m\n}+2\partial_{(\m}\xi^l_{\n)},
\ee
of the form
$\xi_i^X=\xi_i^X(t)$. As we have discussed, these do not affect the
$F_i^X$, but can be used to give both of the $V_i^X$ an arbitrary
time dependence.

%%%%%%%%%%%%%%%%%%%%%%%%%%%%%%%%%%%%%%%%%%%%%%%%
\subsection{Scalar Modes}
%%%%%%%%%%%%%%%%%%%%%%%%%%%%%%%%%%%%%%%%%%%%%%

The Lagrangian for the scalar modes can be expressed as
\ba
L_s&=&\frac{1}{\kappa_g}\Big\{2\Delta \psi^g \left(A^g-\dot
B^g-\ddot E^g\right)-\psi^g \Delta \psi^g-3 (\dot\psi^g)^2\Big\}\nonumber \\
&&+\frac{1}{\kf}\Big\{
 2\beta^{-1} \Delta \psi^f \left(A^f-\b\dot B^f-\b^2\ddot E^f\right)
 -\psi^f \Delta \psi^f-3 \beta(\dot
\psi^f)^2\Big\}\nonumber\\
&&-\frac{M^4}{2}\Big\{
n_2\{3(\psi^g+\psi^f)^2+(\Delta(E^g+E^f))^2-2(\psi^g+\psi^f)\Delta (E^g+E^f)\}
\nonumber\\
&&\hspace{1cm}+n_0
\{(A^g+\beta^{-1}A^f)\left(A^g+\beta^{-1}A^f-2[3(\psi^g+\psi^f)-\Delta(E^g+E^f)]\right)\}\nonumber\\
&&\hspace{1.3cm}
+n_3
\{3(\psi^g+\psi^f)-\Delta(E^g+E^f)\}^2\Big\}.\nonumber \ea Let us
first study the non-homogeneous modes. The mass terms do not depend
on $B^g$ nor on $B^f$, so those fields are Lagrange multipliers,
just as in Einstein's gravity. Variation with respect to these
fields yields \be \Delta \dot \psi^g=\Delta \dot \psi^f=0. \ee

The variation with respect to $A^g$ and $A^f$ yields the constraints
\ba
A^g&=&-\b^{-1} A^f+3(\psi^g+\psi^f)-\Delta(E^g+E^f)+\frac{2}{M^4 n_0\kappa_g}\Delta \psi^g, \nonumber\\
\label{Af} \psi^g&=& \frac{\kappa_g}{\kf} \psi^f+f(t). \ea Once we
substitute the first of these constraints in the Lagrangian, the
quadratic term in $E^h$ and $E^l$ takes the form \be
(n_2-n_0+n_3)(E^h+E^l)^2. \ee We can now distinguish two different
cases, neither of them with propagating scalar degrees of freedom.
First, if the coefficient $n_2-n_0+n_3$ does not cancel, the
equations of motion for $E^h$ and $E^l$ result in a new constraint
which determines these fields, and upon substitution into the
Lagrangian we are left without any scalar degrees of freedom. If the
coefficient cancels, as happens for the potential
(\ref{interaction}),
 $E^g$ and $E^f$ are Lagrange multipliers appearing
in the gauge invariant combination $E^h+E^l$. After using
(\ref{Af}), the variation with respect to $E^h$ yields \be \Delta
\psi^g=\Delta \psi^f=0. \ee The Lagrangian cancels after
substitution of these constraints, and there are no propagating
degrees of freedom. Note that in this last case the combination
$E^h+E^l$, is not determined by the equations of motion. Again, this
is not a desirable feature, since it means that the value of this
combination, which is gauge invariant under the diagonal
diffeomorphisms, is not predicted by the linear theory.
Nevertheless, we expect that higher order terms in the expansion
will determine $E^h+E^l$, since there is no symmetry in the
non-linear Lagrangian under which this quantity can be ``gauged" to
arbitrary spacetime dependence (see section \ref{thirdorder}).

Concerning the homogeneous modes, after using the constraints we are
left with two modes $\psi^{f}$ and $\psi^g$ which have a negative
definite kinetic term. Nevertheless, the dispersion relations for
the degrees of freedom which diagonalize the equations of motion are
$\omega^2=0$ and $\omega^2= M^4 n_2(\tilde \kappa_f +\kappa_g)>0$,
so there is no classical instability associated to these modes.

%%%%%%%%%%%%%%%%%%%%%%%%%%%%%%%%%%%%%%%%%%%%%%%
\subsection{A comment on third order perturbations}\label{thirdorder}
%%%%%%%%%%%%%%%%%%%%%%%%%%%%%%%%%%%%%%%%%%%%%%%

As we have seen in the previous section there are some interaction
terms of bigravity that have ill-defined perturbation theory at second
order. In particular, when the condition
$$(n_2-n_0+n_3)=0$$
is satisfied, the gauge invariant combination $E^f+E^g$ is not determined
by the equations of motion from the boundary conditions.
 The absence of a non-linear gauge invariance that accounts
for this behaviour makes one
expect that the next order in perturbation theory will determine this combination
from the initial conditions.

Third order perturbation theory is a thorny issue in GR (see {\em e.g.} \cite{D'Amico:2007iw}
and references therein). Contrary to what happens at second order, at third
order the tensor, vector and scalar perturbations mix, which makes the
general formalism very involved. For massive gravity the previous problem
is alleviated by the {\em strong coupling}. In fact, as the scalar perturbations
have a strong coupling energy scale smaller than that of the other perturbations,
at this scale the only strongly interacting field will be the scalar. This
allows to consistently study the third order perturbations in certain models
such as DGP in a certain regime \cite{Nicolis:2004qq}. Unfortunately, we
are not so lucky in the bigravity case. As it is clear from the previous
section the combination $E^g+E^f$ is not strongly coupled, but directly absent
at the linear level. Thus, if we want to push the theory till the scale
where this mode is dynamical, we need to take into account all the plethora of
vector, scalar and tensor modes (which, furthermore, are coupled at third order).
We studied other possibilities, such as a the imposition of a hierarchy in the perturbations
$E^2\sim \e^2$, where $\e$ is the scale of the rest of the perturbations, but
we could not find a consistent scheme with a simple perturbation
theory at third order (we will, however, present a heuristic argument on the
behaviour of third order perturbations in the next subsection).

From the previous arguments, it seems clear that it is more convenient
to work with Lagrangians where $(n_2-n_0+n_3)\neq0$.  We will assume this condition unless
otherwise stated.

%%%%%%%%%%%%%%%%%%%%%%%%%%%%%%%%%%%%%%%%%%%%%%%%%%%%%%%
\subsection{Coupling to Matter and vDVZ discontinuity}
%%%%%%%%%%%%%%%%%%%%%%%%%%%%%%%%%%%%%%%%%%%%%%%%%%%%%%%

The explicit and non-singular exact solutions of bigravity
which we reviewed in Chapter \ref{chapterbigra} are also solutions of GR\footnote{Recently,
solutions which deviate from GR have been found in \cite{Berezhiani:2008nr}.}. This
immediately suggests that the vDVZ discontinuity may be absent
altogether in this theory at the non-linear level. Also, from the
analysis of perturbations done in the previous section around the
Lorentz breaking background, it is clear that the situation here is
very different from that of ordinary massive gravity. The massive
spin-2 graviton has only two physical polarizations (as opposed to
the five polarizations of the ordinary FP massive graviton), and
there are no propagating vector or scalar modes.

Let us consider the coupling of the linearized theory to conserved
sources. To this end, we introduce the couplings \be
S_{matt}=\frac{1}{4} \int \di^4 x \left(\lambda_g
h^g_{\phantom{g}\m\n} T_g^{\m\n}+\lambda_f h^f_{\phantom{g}\m\n}
T_f^{\m\n}\right), \ee where $T_g^{\m\n}$ and $T^f_{\m\n}$ are
conserved, \emph{i.e.} $\partial_{\m}T_g^{\m\n}=0$ and $\eta_{\r\m}\tilde
\eta^{\r\a}\partial_\a T_f^{\m\n}=0$. In terms of the decomposition
(\ref{irrep}), we have \ba S_{matt}&=&\frac{\lambda_g}{4}\int \di^4
x \left(-t_{ij}^g T_g^{ij}+2 T_g^{0i}(V^g_i+\dot F^g_i)+2
T_g^{00}\Phi^g+2T_g^{ii}\psi^g\right)
\nonumber\\
&+&\frac{\lambda_f}{4}\int \di^4 x \left(-t_{ij}^f T_f^{ij}+2
T_f^{0i}(V^f_i+\b\dot F^f_i)+2
T_f^{00}\Phi^f+2T_f^{ii}\psi^f\right). \ea where we have introduced
the gauge invariant combinations
$$\Phi^g\equiv A^g-\dot B^g-\ddot E^g, \quad \quad \Phi^f\equiv
A^f-\b\dot B^f-\b^2\ddot E^f.$$ Inverting the equations of motion
for the tensor modes in the presence of the source $T^{ij}$, we find
\be t_{ij}^g=\frac{\l_g ({\bf k}^2-\b \omega^2+\kf
M^4n_2)T^g_{ij}-\l_f \kappa_g M^4 n_2 T_{ij}^f}{\omega^2\{\b
\omega^2- (\kf+\b\kappa_g)M^4n_2\}+{\bf k}^2\{
(\kf+\kappa_g)M^4n_2-(\b+1) \omega^2\}+{\bf k}^4}, \ee and an
analogous expression for $t_{ij}^f$: \be t_{ij}^f=\frac{\l_f ({\bf
k}^2- \omega^2+\kappa_g M^4n_2)T^f_{ij}-\l_g \tilde\kappa_f M^4 n_2
T_{ij}^g}{\omega^2\{\b \omega^2- (\kf+\b\kappa_g)M^4n_2\}+{\bf
k}^2\{ (\kf+\kappa_g)M^4n_2-(\b+1) \omega^2\}+{\bf k}^4}. \ee In the
limit $M^4\rightarrow 0$ this reduces to the standard expression for
linearized GR.

For the vector modes, the equations of motion read \ba
\Delta(V_i^g+\dot F^g_i)&=&\lambda_g \kappa_g T^{0i}_g  \nonumber \\
\Delta \left(\dot{V}^g_i+\ddot{F}^g_i\right)  &=& -M^4n_2\kappa_g
\Delta\left(F^g_{i} +F^f_{i} \right)+\lambda_g \kappa_g\dot
T^{0i}_g \\
\Delta(V_i^f+\b\dot F^f_i)&=&\lambda_f \b\kf T^{0i}_f  \nonumber \\
\Delta \left(\dot{V}^f_i+\b\ddot{F}^f_i\right)  &=& -\kf
\b^{-1}M^4n_2 \Delta\left(F^g_{i} +F^f_{i} \right) +\lambda_f
\kf\b\dot T^{0i}_f. \ea It follows immediately that
$\Delta(F_i^g+F_i^f)=0$, and therefore the term proportional to
$M^4$ vanishes. This means that there is no difference with the GR
results for each one of the metrics.

For the scalar part, we may start with variation with respect to
$B_i^X$, which yields the constraints \be \dot C_X = 0 \label{consu}
\ee where
$$
C_g\equiv 4\Delta \psi^g +\lambda_g\kappa_g T_g^{00}, \quad\quad
C_f\equiv 4\Delta \psi^f +\lambda_f\tilde\kappa_f\beta T_f^{00}.
$$
Variation with respect to $A^X$ gives \be C_f = C_g \ee and \be
C_+\equiv C_f + C_g = 2 M^4(\tilde\kappa_f + \kappa_g )(A_+ -3
\psi_+ + \Delta E_+) n_0,\label{nnn} \ee where $A_+= A_g +
\beta^{-1} A_f$, $\psi_+ = \psi_f + \psi_g$, and $E_+ = E_f + E_g$.
Variation with respect to $\Delta E^X$ yields, with the help of
(\ref{consu}), \be n_0 A_+ = (n_2 + 3n_3) \psi_+ - (n_2 + n_3)
\Delta E_+. \label{cep} \ee Substituting into (\ref{nnn}), we have
$$
C_+ = 2 M^4(\tilde\kappa_f + \kappa_g )[(n_2+3n_3-3n_0)\psi_+ -
(n_2+n_3-n_0) \Delta E_+]
$$
and using (\ref{consu}), we have \be 4(n_2 - n_0 + n_ 3) \Delta^2
\dot E_+ = - (n_2+3 n_3 -3 n_0) (\lambda_f\tilde\kappa_f \beta\dot
T_f^{00}+\lambda_g\kappa_g \dot T_g^{00}). \label{consu2} \ee For
$(n_2-n_0+n_3)\neq 0$, this determines $\dot E_+$ in terms of the
sources. The solution will depend on an arbitrary time independent
mode $E_0(x)$.

For the singular case $(n_2-n_0+n_3)=0$, Eq. (\ref{consu2}) do not
determine $E_+$ at all. Instead, it imposes some non-trivial
equations to be satisfied by the sources, \be
\lambda_f\tilde\kappa_f \beta\dot T_f^{00}=-\lambda_g\kappa_g \dot
T_g^{00} \label{urest} \ee which seem hard to motivate. Thus,
coupling to the sources seems rather inconsistent in this case,
unless $(n_2+3 n_3-3n_0)=0$ as well. But this would imply $n_2=0$,
in which case the tensor modes are massless. As we have already
stated, this problem is likely to disappear at the third order
in perturbation. Concerning the exact non-linear
solutions, they do not require any condition on the matter content
but for the studied case of constant energy they satisfy (\ref{urest}).

In the generic case, the solution for the $\psi$ potentials is of
the form
\ba
\Delta \psi^g &=& -{\kappa_g \lambda_g \over 4} T^{00}_g + {1\over 8} C_+ (\vec x),\nonumber\\
\Delta \psi^f &=& -{\tilde\kappa_f \lambda_f \beta \over 4} T^{00}_f +
{1\over 8} C_+ (\vec x).
\label{mosca}
\ea
where $C_+(\vec x)$ is
entirely determined by initial conditions.

Finally, variation with respect to $\psi_f$ and $\psi_g$ leads
[after use of (\ref{mosca})] to the following equations for the
gauge invariant potentials:
\ba
\Delta
\Phi^g&=&-\frac{\kappa_g\lambda_g}{4}\left(
T^{00}_g+T^{ii}_g-{3\over \Delta} \ddot T^{00}_g\right)+ {1\over 8}
C_+
+\kappa_g M^4 n_2 \Delta E_+ ,\\
\beta^{-1}\Delta \Phi^f&=&-\frac{\tilde\kappa_f\lambda_f}{4}\left(
\beta T^{00}_f+T^{ii}_f-{3\over \Delta} \beta^2 \ddot
T^{00}_f\right) +{1\over 8} C_+ +\tilde \kappa_f M^4 n_2 \Delta
E_+,
\ea
where
\ba
\Delta E_+= - {1\over n_2 +n_3 -n_0}&&
\Big[ {1\over 2 M^4 (\tilde\kappa_f + \kappa_g)} C_+\nonumber\\
&&+{n_2 + 3 n_3 -3 n_0 \over 4 \Delta} \left( \kappa_g\lambda_g
T^{00}_g + \tilde\kappa_f\lambda_f \beta T^{00}_f - C_+\right) \Big].
\label{ep}
\ea
In general, the solution depends on an arbitrary ``initial"
function $C_+(\vec x)$. This corresponds to a mode with dispersion
relation $\omega^2=0$ in the linear theory. It was argued in
\cite{Dubovsky:2004sg} that in such cases, from higher order terms
the expected dispersion relation will be of the form $\omega^2\sim
p^4$, and in this sense $C_+$ corresponds to a slowly varying
``ghost condensate" \cite{ArkaniHamed:2003uy}. In what follows, we
shall take the initial condition $C_+(\vec x)=0$.

For $n_2-n_0+n_3 \neq 0$, the solution is of the form
\be \Delta
\psi^g = -{\kappa_g \lambda_g \over 4} T^{00}_g,\quad \Delta \psi^f
= -{\tilde\kappa_f \lambda_f \beta \over 4} T^{00}_f, \ee and
\ba
\Delta \Phi^g=&-&\frac{\kappa_g\lambda_g}{4}\left(
T^{00}_g+T^{ii}_g-{3\over \Delta} \ddot T^{00}_g\right)\nonumber\\
&&\hspace{1.5cm}
-\left({\kappa_g M^4 n_2\over 4\Delta}\right) {n_2 + 3 n_3 -3 n_0
\over n_2 +n_3 -n_0} \left( \kappa_g\lambda_g T^{00}_g +
\tilde\kappa_f\lambda_f \beta T^{00}_f \right)
,\nonumber\\
\Delta \Phi^f=&-&\frac{\tilde\kappa_f\lambda_f\beta}{4}\left( \beta
T^{00}_f+T^{ii}_f-{3\over \Delta} \beta^2 \ddot T^{00}_f\right)
\nonumber\\
&&\hspace{.9cm}-\left({\tilde\kappa_f\beta M^4 n_2\over 4\Delta}\right) {n_2 + 3
n_3 -3 n_0 \over n_2 +n_3 -n_0} \left( \kappa_g\lambda_g T^{00}_g +
\tilde\kappa_f\lambda_f \beta T^{00}_f \right).
\label{mosco}
\ea
 Hence, there is a well behaved massless limit,
with corrections of order $M^4\Delta^{-2}$ to the gauge invariant
potentials $\Phi$ and $\psi$. This means, in particular, that there
is no vDVZ discontinuity. This is quite analogous to the ``half
massive gravity" model discussed in
\cite{Gabadadze:2004iv} (see also \cite{Dubovsky:2004ud}). The
additional terms lead to corrections to the Newtonian potential. The
sign of this correction can be positive or negative, depending on
the values of the numerical coefficients $n_i$. For isolated
sources, such corrections scale like the square of the graviton mass
$m^2 \sim \kappa M^4$ times the ``Schwarzschild" radius $r_s$
corresponding to the given source, and grow linearly with the
distance $r$. Parametrically, the potential takes the form
$$\Phi \sim \phi_N + m^2 r_s r,$$
where $\phi_N$ is the standard Newtonian potential. Linear theory
breaks down at large distances, when the second term is of order
unity. It would be interesting to try and match this solution to a
non-perturbative exact solution which is well behaved at infinity.

As we stated before, the case of no correction to the Newton's law
corresponds to the case where (\ref{typeIcondo}) is independent of
$\b$ or $\g$ (cf.
(\ref{noNcorr})). One possibility for this is a potential which depends
only on the determinants $g$ and $f$. From the arguments in Chapter \ref{chapternl},
it is easy to show that this
kind of interaction leads also to two independent massless metrics. Indeed,
notice that the gauge group is Diff$\times$TDiff.
\\

Finally, we note that the simple interaction term
(\ref{interaction}) first considered in \cite{Isham:1971gm,Isham:1977rj}
happens to land on the special case $$n_2-n_0+n_3=0,$$ where the
above expressions for the gauge invariant potentials are singular.
The origin of the singularity is the following. After substitution
of the constraints (\ref{cep}), the linearized action no longer
depends on $\Delta E_+$. In particular, the absence of this variable
results in the unwanted restriction (\ref{urest}) on the
sources\footnote{This accidental symmetry is similar to that which
exists in ordinary massive gravity where the linear action
has 5 PPoF whereas a new ghost-like PDoF appears at the
non-linear level \cite{Deffayet:2005ys,Boulware:1973my}.
 However, in that case the accidental symmetry corresponds
to a symmetry of the massless theory and no further constraints are
needed in the sources.}. Nevertheless, beyond the linear order, the
action will depend on $\Delta E_+$, and hence the ``restriction"
will no longer exist. Rather, a nonlinear equation will determine
the value of $\Delta E_+$. Can we nevertheless try to find classical
solutions in a perturbative expansion? The above considerations
suggest an expansion scheme for the singular case $n_2-n_0+n_3= 0$,
where $E_+$ is treated as a much bigger quantity than the rest of
the linearized fields\footnote{Some of the linearized fields will be
of the order of $E$ as is clear from (\ref{cep}).} (such as $\psi$).
Heuristically, the size of $\Delta E_+$ can be estimated as follows.
Instead of perturbing the flat solution Eq. (\ref{BIMETMIN}), we may
consider the quadratic action for perturbations around a solution
which differs from the original by $O(h)$. The expansion around this
new solution will have\footnote{All the coefficients will have
corrections of order $O(h)$. However, for the rest of coefficients
one expects that they will yield second order small corrections. }
 $$n_2-n_0+n_3=O(h).$$
From (\ref{mosca}), we have
$$
\Delta\psi \sim \kappa T \equiv\Delta \phi_N,
$$
where $\phi_N$ stands for the potential corresponding to the given
source in Newton's theory. From (\ref{mosco}), $\Delta\Phi \sim
O(\kappa T) + O(m^2 \Delta E)$, where $m^2\sim \kappa M^4$ denotes
the graviton mass squared. From (\ref{ep}), we have  $\Delta E
\times O(h) \sim O( \kappa T/\Delta )$. This suggests the hierarchy
$$
\Delta E \gg \psi, \quad \Phi\sim \max (\psi, m^2 E).
$$
Taking
 $n_2-n_0+n_3 \sim \max (\Phi,\Delta E)\sim \max (\psi, m^2 E,\Delta E)\sim \Delta E\left(
1+m^2/\Delta\right)$, this leads to the estimate
\be
\label{estimate}
(\Delta E)^2 \sim {\psi \over 1+ m^2/\Delta}.
\ee
For distances shorter than the inverse graviton mass, we have $
\Delta E \sim \phi_N^{1/2}, $ and hence we may expect
$$
\Phi \sim \phi_N + (m^2/\Delta) \phi_N^{1/2}.\quad\quad (\Delta \gg
m^2)
$$
At distances which are large compared with the inverse graviton
mass, the estimate (\ref{estimate}) yields $\Delta E \sim (\Delta \phi_N /m^2)^{1/2}$, and we expect
$$
\Phi \sim (m^2/\Delta)^{1/2} \phi_N^{1/2}.\quad\quad (\Delta \ll
m^2)
$$
These very crude arguments seem to indicate that, also in this
special case, there is no vDVZ discontinuity. However, for finite
$m$, there are significant modifications to the value of the ``gauge
invariant" potential $\Phi$ which determines the motion of slowly
moving particles. For isolated sources, such modifications scale
like $r_s^{1/2}$, where $r_s$ is the ``Schwarzschild" radius
corresponding to the given source. They grow with the distance as
$r^{3/2}$ below the graviton Compton wavelength $m^{-1}$, and as
$r^{1/2}$ for larger distances. The potential $\Phi$ becomes of
order one for $r\gtrsim m^{-2} r_s^{-1}$, beyond which we enter a
non-perturbative regime. It would be interesting to confirm this
heuristic analysis in a numerical study of a spherically symmetric
solution with sources. This is left for further research.\\

Perturbations around Lorentz-breaking bi-flat
solutions lead to gravitons with Lorentz-breaking mass terms. Because of the invariance under
diagonal diffeomorphisms, mass terms with components $h_{0i}$ are absent from the
second order Lagrangian\cite{Berezhiani:2007zf}. This, in turn, leads to a well behaved theory of linearized perturbations
\cite{Berezhiani:2007zf,Gabadadze:2004iv},
which is not afflicted by the vDVZ discontinuity. It is somewhat puzzling that in the linear theory,
there are corrections to the Newtonian potential which are proportional to the square of the graviton mass
and which grow linearly with the distance to the origin. On the other hand, as mentioned above,
these theories admit the Schwarzschild metric as an exact solution for the same values of the parameters.
Thus, the linearized solutions for a static spherically symmetric sources do not coincide with the
linearization of the known vacuum solutions\footnote{It has
recently been argued in \cite{Berezhiani:2008nr} that the linear theory
is not appropriate to describe  bigravity at large distances. In this work they
also propose an exact solution relating the interior of a star (where perturbation theory is valid)
to an exterior solution which presents modifications to GR. See also \cite{Damour:2002gp}.}. This
 seems to indicate that this theory has a linearization
instability such as the one which is found in other
contexts \cite{Moncrief:1976un,Kastor:1991ir,Higuchi:1991tp}, some of which
are related to massive gravity and may have important phenomenological
consequences \cite{Deffayet:2006wp}. Another possibility is that
there may be other exact solutions which coincide with the linearized approximation at large distances, and
those may be the relevant ones which can be matched to spherically symmetric matter sources near the origin.
This issue clearly deserves further investigation.

%%%%%%%%%%%%%%%%%%%%%%%%%%%%%%%%%%%%%%%%%%%%%%%%%%%%%%%%%%
\section{Perturbation theory of Proportional  de Sitter Metrics}
%%%%%%%%%%%%%%%%%%%%%%%%%%%%%%%%%%%%%%%%%%%%%%%%%%%%%%%%%%

As stated in the previous Chapter, another interesting class of solutions of
bigravity can be constructed from two proportional metrics with a
constant proportionality factor. Let us define our perturbations as
\ba
g_{\mu\nu}&=&\Omega_{\mu\nu}+h^g_{\mu\nu},\\
f^{\mu\nu}&=&\gamma^{-1}(\Omega^{\mu\nu}+h_f^{\mu\nu}). \ea
All indices will be handled with the $\Omega_{\mu\nu}$ metric. \\

We first focus on the interaction term for a general potential
(\ref{generi}). Using (\ref{Lambtilde}) we can write
\ba
\label{lgenerbi}
 \tilde L_{int}&=&\zeta (-g)^u(-f)^v V[\{\t_n\}] +
\sqrt{-g} \frac{\tilde \Lambda_g}{\kappa_g}+ \sqrt{-f}
\frac{\tilde \Lambda_f}{\kappa_f}\nonumber\\
&&=-\frac{1}{8\kappa_+}\sqrt{-\Omega}
\left\{m^2_t(h_g^{\m\n}+h_f^{\m\n})(h^g_{\m\n}+h^f_{\m\n})-m^2_s(h^g+h_f)^2\right\},
\ea
where indices are manipulated with the metric $\Omega_{\m\n}$,
{\em e.g.} $h^g=\Omega^{\m\n}h^g_{\m\n}$,
 and
\ba
\label{def_ms}
 m^2_s&=&4\kappa_+\zeta \g^{4v}\left(-uvV_0 +(u-v)\sum_n n\g^{-n}
V^{(n)}+
\sum_{n,m}  nm\g^{-(n+m)}V^{(n,m)}\right),\nonumber\\
m^2_t&=&-4\kappa_+\zeta \g^{4v}\sum_n n^2 \g^{-n}V^{(n)}.
\ea%
We have also introduced an effective Newtons's constant $\kappa_+$
for later convenience.

Note that the massive graviton corresponds to
$h^+_{\mu\nu}=(h_g+h_f)_{\mu\nu}$. This is to be expected, as for
$h^g_{\mu\nu}=-h^f_{\mu\nu}$ the metrics are still proportional and
therefore the perturbations are standard massless gravitons of GR in
vacuum. Also, in the present set-up, $h^+_{\mu\nu}$ are the
quantities invariant under the diagonal diffeomorphisms. Notice also
that the mass term does not have in general a Pauli-Fierz form, \be
m^2( h_+^2- h_+^{\m\n} h^+_{\m\n}). \ee This particular form can
only be achieved by properly tuning the parameters. This is in
contrast with other ways of getting massive gravitons, such as
dimensional reduction, where the original symmetry group is much
larger. Here, the degrees of freedom of the original theory are $8$
which can be split into a massless graviton with $2$ polarizations
and a massive graviton with $6$ polarizations\footnote{The number
of degrees of freedom coincides with that of higher derivative
gravity \cite{Stelle:1977ry}.}. The expression of the massless
graviton as a linear combination of the metric perturbations will be
given below.

From Eq. (\ref{lgenerbi}) we note that whenever $m_t=0$ there is an
enhancement of the gauge invariance, which now admits all
transformations which leave the traces $h_g$ and $h_f$ invariant\footnote{This
 happens in the case when the derivative of Eq.
(\ref{typeIcondo}) with respect to $\beta$ vanishes at $\beta=1$.
For the case (\ref{interaction}) this amounts to $\g=2/3$}. This
corresponds to the transverse subgroup of the diffeomorphisms, which
we considered in the first part of the Thesis.  In this
special case the gauge invariance is enough to have just two massless
gravitons propagating\footnote{At first sight, this seems to
contradict the results of Ref. \cite{Boulanger:2000rq}, where it is
shown that we cannot have two massless interacting gravitons.
However, the starting point in \cite{Boulanger:2000rq} is a free
Lagrangian invariant under linearized diffeomorphisms. As we
showed in the first part of this Thesis (Chapter \ref{chapterLorentz}), there are Lagrangians invariant under
transverse diffeomorphisms which propagate just massless spin-two
particles. An extension of the analysis of \cite{Boulanger:2000rq}
to the transverse subgroup is currently under investigation \cite{Blas08}.}.
\\

Let us now consider the case of generic $m_s$ and $m_t$. For
simplicity we will concentrate on perturbations around de Sitter
solutions which will be foliated by spatially flat sections, \be
\label{deSitter} \Omega_{\mu\nu}\di x^\mu \di x^\nu=a(\eta)^2(\di
\eta^2 -\delta_{ij} \di x^i\di x^2), \ee where
$a(\eta)=-(H\eta)^{-1}$, $H^2=\Lambda_g/3$ being a constant and
$\eta\in (-\infty, 0)$. The kinetic term in (\ref{action}) will be
given by (cf. (\ref{lgenerbi}))
\ba
L_K\equiv-\frac{1}{2\kappa_g}
\sqrt{-g}\ (R_g+2\Lambda_g) -\frac{1}{2\kappa_f} \sqrt{-f}\
(R_f+2\Lambda_f), \ea with $\L_f=\g^{-1}\L_g$. To second order in
perturbations we can rewrite the kinetic term in terms of a massive
and a massless field,
\ba \label{massivemasslesss}
L_K=-\frac{1}{2\kappa_+} \sqrt{-g_+}\ (R_{g_+}+2\Lambda_g)
-\frac{1}{2\kappa_-} \sqrt{-g_-}\ (R_{g_-}+2\Lambda_g)+o(h^3),
\ea
where $\kappa_-=\frac{\kappa_g}{1+\kappa}$,
 $\kappa_+=\kappa_g\kappa^{-1}(1+\kappa)$, with $\kappa=\g \kappa_g \kappa_f^{-1}$,
  $g_-{}_{\mu\nu}=\Omega_{\mu\nu}+h^-_{\mu\nu}$ and $g_+{}_{\mu\nu}=\Omega_{\mu\nu}
+ h^+_{\mu\nu}$. Besides, we have introduced the  massive and
massless combinations
\be \label{massless}
 h^+_{\m\n}=h^g_{\m\n}+h^f_{\m\n},\quad h^-_{\m\n}=(1+\kappa)^{-1}\left(h^g_{\m\n}-\kappa h^f_{\m\n}\right).
\ee
The dynamics of the massless part is well known. One easily
finds that only the tensor modes are dynamical. For the generic
massive theory in de Sitter space, studying the longitudinal mode of
the massive representation we would argue that the only ghost-free
possibility is the Fierz-Pauli mass term, $m^2_t=-m^2_s$
\cite{Fierz:1939ix,ArkaniHamed:2002sp}. However, in general, this
mode decouples only at high energies (larger than a combination of
the rest of relevant mass scales). For intermediate energy scales,
the longitudinal mode is coupled to another scalar mode which can
modify this picture \cite{Dubovsky:2004sg, Creminelli:2005qk}. Also,
the curvature scale $H$ could play a role in making these
intermediate scales phenomenologically relevant\footnote{Recently
a consistent  model of Lorenz invariant
massive gravity with a mass term different from the FP mass term has been discovered
in certain local brane models with two extra dimensions \cite{deRham:2007xp}.
In this case there is a momentum dependence in the mass parameters.}. We will study this
possibility directly in the unitary gauge\footnote{Notice that
the St\"uckelberg formalism is more useful to determine the {\em strong
interacting scale} and the {\em cut-off} of the theory \cite{ArkaniHamed:2002sp}. Nevertheless, as
we are interested in the validity of the linear theory, it is enough to
work in the {\em unitary}
gauge.}.

Let us first split the degrees of freedom of the massive combination
into scalar, vector and tensor modes, \ba
h^+_{00}&=&2a(\eta)^2 A,\nonumber\\
h^+_{0i}&=&a(\eta)^2(B_{,i}+V_i),\nonumber \\
h^+_{ij}&=&a(\eta)^2(2\psi \delta_{ij}-2 E_{,ij}-2
F_{(i,j)}-t_{ij}), \label{decomp} \ea where $\psi$, $B$, and $E$ are
the scalar modes, $F_i$ and $V_i$ are vector modes, and $t_{ij}$ is
a tensor mode. The vector modes are divergenceless and the tensor
modes are transverse and traceless.

The expansion of the kinetic term in this foliation can be extracted
from the usual expansion in de Sitter space (see {\em e.g.}
\cite{Mukhanov:1990me}, notice however the difference of
convention). One finds
\ba
\label{kinetic}
&&-\frac{1}{2\kappa_+} \int
\di^4 x\sqrt{-g_+} (R_++2\Lambda_g)=
-\frac{1}{2\kappa_+}\int \di^4 x a^2(\eta)\Big\{\frac{1}{4}t_{ij}\Box t_{ij} \\
&&+\frac{1}{2}(V_i+F'_i)\Delta (V_i+F'_i)+6(\psi'+\H A)^2-2 \Delta \psi(2A-\psi)-4\Delta(B+E')(\psi'+\H A)
\Big\},\nonumber
\ea
where $\H=a(\eta)'/a(\eta)=a(\eta)H$ and the prime
refers to derivative with respect to the conformal time $\eta$. We
have also introduced the d'Alembertian $\Box=\eta^{\m\n}\pd_\m
\pd_\n$ and the Laplacian $\triangle=\pd_i \pd_i$.
 The interaction term (\ref{lgenerbi}) reads
\ba \label{linter}
\tilde L_{int}&=& \frac{1}{2\kappa_+}a(\eta)^4\Big\{m^2_s(A+\Delta E-3\psi)^2\\
&&-\frac{1}{4}m^2_t\Big(t_{ij}t_{ij}-2(V_i V_i+F_i \Delta F_i)
 +4 (A^2+\frac{B \Delta B}{2}+(\Delta E)^2+3\psi^2-2\psi \Delta E)\Big)\Big\}.\nonumber
\ea
We can now analyse the different components in turn.

%%%%%%%%%%%%%%%%%%%%%%%%%%%%%%%%%%%%%%%%%%%%%%%%%%%%%%%%%%%%%%%%%%
 \subsection{Tensor and Vector Modes}
%%%%%%%%%%%%%%%%%%%%%%%%%%%%%%%%%%%%%%%%%%%%%%%%%%%%%%%

The action for the massive tensor modes is simply
\be
{}^{(t)}\delta S_2=-\frac{1}{8\kappa_+}\int \di x^4
a^2(\eta)\Big(t_{ij}\Box t_{ij}+a(\eta)^2m^2_t  t_{ij}t_{ij}\Big).
\ee From this equation we can read the mass of the graviton which
will be given by $ m^2_t,$ and the tachyon-free condition will
simply read
$$m^2_t\geq 0.$$
Regarding the vector modes, their action is
\be {}^{(v)}\delta
S_2=-\frac{1}{4\kappa_+}\int \di x^4  a^2(\eta)\Big((V_i+ F'_i)\Delta
(V_i+ F'_i)-a^2(\eta)m^2_t (V_i V_i+F_i\Delta F_i)\Big).
\ee
The
field $V_m$ enters the action without time derivatives, and thus its
variation yields the constraint, \be \triangle (V_i+ F'_i)=
a(\eta)^2 m^2_t V_i\equiv m^2(\eta) V_i.
\ee
Taking this constraint
into account, the action for the vector modes up to second order
can be written as
\be
{}^{(v)}\delta S_2= \frac{1}{4\kappa_+} \int
\di^4 x a^2(\eta)m^2(\eta)\Big( F'_i \frac{\Delta}{\Delta-m(\eta)^2}
F'_i +F_i\Delta F_i\Big).
\ee
This Lagrangian has the usual signs,
and thus no ghost or tachyons appear in the theory for $m^2_t\geq
0$. More concretely, we can canonically normalize the previous field
equation with the field redefinition \be
F_i^c=m(\eta)\sqrt{\frac{\Delta}{\kappa(\Delta-m(\eta)^2)}}F_i. \ee
We conclude that the only constraint we get form the analysis of the
vector and tensor modes is $m^2_t\geq0$.

%%%%%%%%%%%%%%%%%%%%%%%%%%%%%%%%%%%%%%%%%%%%%%%%%%%%
\subsection{Scalar Modes}
%%%%%%%%%%%%%%%%%%%%%%%%%%%%%%%%%%%%%%%%%%%%%%%%%%%%%%%
From (\ref{kinetic}) and (\ref{linter}), the second order Lagrangian
for the scalar part reads
\ba&& {}^{(s)}\delta S_2=
\frac{1}{2\kappa_+}\Big[\int \di^4 x a^2(\eta)\{ -6( \psi'+
{\mathcal{H}}A)^2
+2 \Delta \psi(2A-\psi)+4\Delta(B+E')( \psi'+{\mathcal{H}}A)\}\nonumber\\
&&+\int\di^4x a^4(\eta)\Big(m^2_s(A+\Delta E-3\psi)^2-m^2_t\{3 \psi^2
+(\Delta E)^2 -2 \psi \Delta E+ \frac{B\Delta B}{2}+A^2\}\Big)\Big].\nonumber\\
\ea $B$ is non-dynamical, and for $m^2_t\neq 0$ it is determined in
terms of the other fields. For $m^2_t=m^2_s$, $A$ appears only
linearly in the mass term. For the flat case $H=0$ and $a(\eta)=1$,
this makes $A$ a Lagrange multiplier and thus its variation gives
rise to a constraint between the fields $E$ and $\psi$, leaving just
one scalar propagating degree of freedom. In the de Sitter case, the
result is the same, although this is not so obvious from the
previous expression
for the action until one substitutes the constraints. \\

The variation with respect to $A$ and $B$  yields the constraints
\ba
B&=&\frac{4(\psi'+{\mathcal{H}}A)}{ a(\eta)^2 m^2_t},\\
A&=&\frac{ -2a(\eta)^2 m^2_t ({\mathcal{H}}(\phi'-3\psi')
+\Delta\psi)- a(\eta)^4 m^2_s  m^2_t (\phi-3 \psi)
-8\Delta{\mathcal{H}} \psi'}{  m^2_t( m^2_s- m^2_t)a(\eta)^4+8\Delta
\H^2-6  m^2_t a(\eta)^2 \H^2}, \ea where $\phi=\Delta E$. Let us
first consider the kinetic part of the action, which after insertion
of the constraints reads
\ba
\label{kinetwofields}
K=
\frac{a(\eta)^2}{2\kappa_+}(M_1(\eta) \phi \psi' +M_2(\eta)
\psi'^2+M_3(\eta) \psi'\phi'+M_4(\eta) \phi'^2),
\ea
where we have
performed a partial integration to eliminate the term $\phi'\psi$.
The functions $M_i(\eta)$ are given by \ba M_1(\eta)&=&\frac{
8\Delta (m^2_t-2  m^2_s) a(\eta)^2 \H}
{ m^2_t( m^2_s- m^2_t)a(\eta)^4+8\Delta \H^2-6  m^2_t a(\eta)^2 \H^2},\\
M_2(\eta)&=&\frac{2 ( m^2_s-m^2_t)a(\eta)^2(4\Delta -3m^2_t
a(\eta)^2)}
{ m^2_t( m^2_s- m^2_t)a(\eta)^4+8\Delta \H^2-6  m^2_t a(\eta)^2 \H^2},\\
M_3(\eta)&=&\frac{4 m^2_t(m^2_s-m^2_t) a(\eta)^4}
{ m^2_t( m^2_s- m^2_t)a(\eta)^4+8\Delta \H^2-6  m^2_t a(\eta)^2 \H^2},\\
M_4(\eta)&=&\frac{-4 m^2_t a(\eta)^2 \H^2}{ m^2_t( m^2_s-
m^2_t)a(\eta)^4 +8\Delta \H^2-6  m^2_t a(\eta)^2 \H^2}. \ea A
difference between the flat and the de Sitter backgrounds is that
the coefficients $M_1(\eta)$ and $M_4(\eta)$ cancel in the former
case, and this automatically yields a kinetic term with a negative
eigenvalue unless $M_3(\eta)=0$ which happens for the Fierz-Pauli
combination $m_s^2=m_t^2$. The situation in de Sitter is slightly
more complicated.
\\

Let us now show that the previous kinetic term gives a positive
contribution to the Hamiltonian in the range of parameters
\ba
&&m^2_t \geq 0, \quad 0\leq m^2_s -m^2_t \leq 6 H^2
\label{noghostI}.
\ea
Indeed, the kinetic term can be written as
\be \label{Eulerkin}
K=\frac{a(\eta)^2}{2\kappa_+}\left(M_1(\eta)\phi\psi'+\left(M_2(\eta)
-\frac{M_3^2(\eta)}{4M_4(\eta)}\right)\psi'^2+M_4(\eta)\left(
\phi'+\frac{M_3(\eta)}{2M_4(\eta)}\psi'\right)^2\right).
\ee
In the
range (\ref{noghostI}), $M_4(\eta)$ and $4
M_4(\eta)M_2(\eta)-M_3^2(\eta)$ are positive. By Euler's theorem,
the corresponding Hamiltonian \ba \label{hamiltonian} {\mathrm
H}_K\equiv \Pi_\phi  \phi'+\Pi_\psi  \psi'-K, \ea is numerically
equal to the two last terms in the Lagrangian, which are quadratic
in generalized velocities, and hence it is positive definite. The
second condition in (\ref{noghostI}) for a positive kinetic term
reduces to the usual $m^2_s=m^2_t$ for the Minkowski limit $H=0$.
For  $H>0$ the endpoints of the interval are of different nature:
the condition $m^2_s-m^2_t\geq 0$ is a necessary condition for the
positivity of $M_2-M_3^2/M_4$ at any value of the momentum, whereas
the upper bound on the range of $m_s^2-m_t^2$ can be somewhat
relaxed depending on the value of the momentum. Indeed, what we need
is that \be m_s^2-m^2_t \leq 6 H^2 \left[ 1-{4\Delta\over
3a^2m^2_t}\right], \ee
so the condition is considerably relaxed at wavelengths shorter than the inverse graviton mass.\\

Once we have established the positivity of part of the Hamiltonian,
let us see what happens to rest of it, namely to the potential part.
This part will be given by \be V\equiv K-L
%=-(a(\eta)^2M_1(\eta))'\frac{M_3(\eta)}
%{8\kappa_+M_4(\eta)}\psi^2
=\frac{a(\eta)^2}{2\kappa_+}(M_5(\eta)\phi^2
+M_6(\eta)\phi\psi+M_7(\eta)\psi^2), \ee
%$$\Phi=\phi+\frac{M_3}{2M_4}\psi,$$
where the coefficients are rather cumbersome and we omit them.
Before proceeding, it should be noted that the Hamiltonian we are
considering is time dependent, and hence not conserved. Its
positivity and boundedness is a useful criterion only as long as we
consider time-scales shorter than the expansion time, or energies
larger than $H$. This is what we may call the adiabatic limit.
Hence, let us assume that $m_s,m_t \gg H$, even if their difference
is much smaller $m_s^2 - m_t^2 \lesssim H^2$, so that we can satisfy
the positivity of the kinetic term as discussed above. We have
checked that within this adiabatic limit, the potential $V$ grows
negative and unbounded below for $-\Delta/a^2 \gg m^2$.
Instabilities at high momenta have been previously studied in
\cite{Dubovsky:2005xd}, and they are just as bad as ghost
instabilities. Unlike the case of tachyons, the phase space for
instability is infinite and this yields infinite decay rates.

If the masses $m_t$ and $m_s$ are small, of order of the expansion
rate $H$, then we are outside of the adiabatic limit, and the
Hamiltonian above is not a very useful indicator of stability.
Instead, we should use a conserved charge associated to the
time-like Killing vector for length scales smaller than the horizon
\cite{Abbott:1981ff}. Due to the existence of the cosmological
scale, it is in principle possible (although by no means clear) that
there may be some range \be \label{range} m^2
\left(\frac{H}{m}\right)^{\a}\gtrsim -\Delta/a^2\gtrsim H^2 \gtrsim
m^2, \ee (with $\alpha>2$), where this conserved charge is positive
definite. The effective theory would then be well defined for
momenta larger than $H$ (corresponding to modes within the horizon),
provided that the theory is cut-off at the energy scale
$m(H/m)^{\alpha/2}$. We leave the study of this conserved charge for
further research. We note, however, that we need a theory which is
applicable to wavelengths {\em much} smaller than the horizon
$-\Delta/a^2\gg H^2$, where the adiabatic approximation should again
be valid. We have checked that for $-\Delta/a^2\gg H^2 \gtrsim m^2$,
the potential $V$ grows negative and unbounded below, so the
possibility of a range of the form (\ref{range}) where the conserved
charge is positive does not look particularly promising.
\\

Finally, for the case $m^2_s=m^2_t$ the analysis of the degrees of
freedom has already been performed in another foliation in
\cite{Deser:2001wx} (see also \cite{Bengtsson:1994vn} and \cite{Gabadadze:2008ha} for
related recent work). In our
analysis for this case we find
 $M_2(\eta)=M_3(\eta)=0$ and thus $\psi$ is not a
propagating field. After varying the action with respect to $\psi$
we obtain a constraint which after substitution yields the
Lagrangian
\ba
{\mathcal L}=\Upsilon\Big( \phi'^2 +\frac{9m^4_t a(\eta)^4\m_H-21
m^2_t a(\eta)^2\mu_H\Delta+4(\m_H-6m^2_t
a(\eta)^2)\Delta^2+2\Delta^3}{(9m^2_t a(\eta)^2\m_H- 6\m_H\Delta
-2\Delta^2)}\ \phi^2\Big),\nonumber \ea where $\mu_H=2\H^2-m^2_t
a(\eta)^2$ and
$$
\Upsilon=\frac{3a(\eta)^4m^2_t \mu_H}{\kappa_+ (9m^2_t
a(\eta)^2\m_H- 6\m_H\Delta -2\Delta^2)}.
$$
This Lagrangian will be ghost-free and tachyon-free for $\m_H\leq0$.
This reduces to the well known condition $m^2\geq 2H^2$
\cite{Higuchi:1986py}.

%
%%%%%%%%%%%%%%%%%%%%%%%%%%%%%%%%%%%%%%%%%%%%%%%%%%%%%%%%%%%%%%%%%%
\subsection{Offloading the Cosmological Constant}
%%%%%%%%%%%%%%%%%%%%%%%%%%%%%%%%%%%%%%%%%%%%%%%%%%%%%%%%

In Chapters \ref{chapterbigra} and \ref{chapterperturbbigrav}, we
have considered a couple of
interacting metrics and found that there are
cosmological solutions where the cosmological constant
is not only determined by the vacuum energy (cf. (\ref{Lambdaf}-\ref{Lambdag})). For the Type I metrics,
we saw that the solution includes an integration constant that can be chosen so that
one of metrics does not feel the vacuum energy whereas the other one is highly curved.
This {\em see-saw} mechanism is
reminiscent
of unimodular gravity (see Chapter \ref{chapternl}).

Besides,
we found proportional solutions
for which
\be
\label{propor}
\Lambda_g=\gamma \Lambda_f,
\ee
where $\gamma$ is  the proportionality factor and the
cosmological constants are functions of the parameters of the
theory (in particular of $\gamma$) (cf. section \ref{subsectionPropM}).
The previous equation {\em fixes} the relative curvature of both metrics and
 we may hope that the fact
of dealing with two different scales $\zeta$ (related to the
mass of the massive graviton) and $\rho$ (the vacuum energy) can lead
 to a {\em see-saw} mechanism (this time dynamical)
  yielding $\gamma \ll 1$ or $\gamma \gg 1$
 for ``natural" values of the potential. If this were the case,
the way in which the system would react to the presence of a vacuum
energy would be by producing a couple of solutions, one of which
with a very
small cosmological constant. In other words, the mechanism would
achieve the
\emph{off-loading} of one of the cosmological constants towards the other
metric. It is
easy to understand that this possibility is not present in our models (except
in  very finely tuned situations)
for Lagrangians which are ghost and tachyon free. To see it, just notice that
the condition that makes the theory free from rapid instabilities is
$$m_s=m_t,$$
where $m_s$ and $m_t$ are defined in (\ref{lgenerbi}) and (\ref{def_ms}).
This condition fixes $\g$, and as it does not involve
neither the vacuum energy, nor the mass of the graviton, $\g$
will be of the same order as the parameters in the interaction term.
This hinders the possibility of a see-saw mechanism.

It is important to notice that, $\g$ is also
determined by the condition (\ref{propor}), which means that in general
the proportional {\em solutions} suffer from instabilities, as they are not
of the FP form.
It is always possible to build a  finely tuned interaction term with a
healthy solution with small cosmological constant for one of the metrics
(see (\ref{Lambdaf}-\ref{Lambdag})), but
this is not very different from the addition of an arbitrary
cosmological constant to the original Lagrangian.

Besides the previous argument, we studied the behavior of the
factor $\g$ for specific interaction terms, like those
appearing in \cite{Isham:1971gm} (and a slight generalization) or
those inspired in brane
interactions or \emph{FP augmented} of \cite{Damour:2002ws}. As expected,  we did not
find the desired {\em off-loading} for a stable solution
in any of these cases.

%%%%%%%%%%%%%%%%%%%%%%%%%%%%%%%%%%%%%%%%%%%%%%%%%%%%%%%%%
\section{Non Covariant Mass Term in de Sitter Space}
%%%%%%%%%%%%%%%%%%%%%%%%%%%%%%%%%%%%%%%%%%%%%%%%%%%%%%%%

Another possible mass term for the gravitons  which
 differs from the usual FP term and  may be
still well defined is provided by Lorentz-breaking mass terms
\cite{Rubakov:2004eb,Dubovsky:2004sg,Rubakov:2008nh}. In bigravity solutions,
these non-covariant mass terms can appear when one of the
metrics is de Sitter whereas the
companion background metric around which we perform the perturbations breaks the de Sitter invariance
of the first metric.

A simple possibility  would
be given by the Type I solutions (\ref{formg}-\ref{formf}) with $p=q=H^2 r^2$.
Here we are going to perform a general analysis of the mass terms which
still preserve a $SO(3)$ symmetry without considering a particular solution.
 There are two different phases in the parameter
space for the masses
which are free of ghosts and gradient instabilities. First, we will find that the
possibilities which satisfy these conditions for Minkowski space-time (see
\cite{Rubakov:2004eb,Dubovsky:2004sg}) are also fine in de Sitter. Besides,
for the non-covariant mass term in de Sitter (and contrary to what
we found in the previous section for the covariant case) we will find that the curvature
scale allows to find regions in the space of masses which are well defined
as an EFT till a scale which goes to zero as $H\rightarrow0$.\\

Let us consider the most general minimal mass term for a graviton propagating
in a de Sitter background which breaks the covariance to rotational invariance\footnote{The covariant limit
is recovered in the case $m_1^2=m_2^2=m_t^2$, $m_3^2=m_4^2=m_s^2$,
$m_0^2=m_s^2-m_t^2$.},
\be
\tilde L_{int}= \frac{1}{8\kappa_+}a(\eta)^4\{m^2_0 h_{00}
h^{00}-2m_1^2h_{0i}h^{0i} -m_2^2 h_{ij}h^{ij}+m_3^2
h_{ii}h^{jj}-2m_4^2 h_{00}h^{ii}\}
\ee
where we are considering
a flat foliation where the metric is given by (\ref{deSitter}), and
the indexes are risen with the metric $\Omega_{\mu\nu}$. In
terms of the decomposition into scalar, vector and tensor modes
of (\ref{decomp}) ($h^+_{\m\n}\equiv h_{\m\n}$), the previous expression can be written as
\ba
\label{genermassdS}
\tilde L_{int}= \frac{1}{2\kappa_+}a(\eta)^4&\Big\{& m_0^2 A^2+\frac{1}{2}m_1^2( V_i V_i- B \Delta B)
+ m_3^2(\Delta E-3\psi)^2 +2m_4^2 A(\Delta E -3\psi)\nonumber\\
&-&m_2^2\left(\frac{1}{4}t_{ij}t_{ij}-\frac{1}{2} F_i\Delta F_{i}+3\psi^2
-2\psi \Delta E+\Delta E\Delta E\right)\Big\}.
\ea
Concerning the kinetic term, its form is shown in (\ref{kinetic}).

%%%%%%%%%%%%%%%%%%%%%%%%%%%%%%%%%%%%%%%%%%%%%%%%%%
\subsection{Tensor and Vector modes}
%%%%%%%%%%%%%%%%%%%%%%%%%%%%%%%%%%%%%%%%%%%%%%%%%

The analysis of these modes proceeds in the same way as in the
covariant case (see also \cite{Rubakov:2004eb}). For the tensor
modes we find that their action is given by
\be
{}^{(t)}\delta S_2=-\frac{1}{8\kappa_+}\int \di x^4
a^2(\eta)\Big(t_{ij}\Box t_{ij}+a(\eta)^2m^2_2
t_{ij}t_{ij}\Big),
\ee
which imposes the condition $$m_2^2\geq 0.$$
Regarding the vector modes, their action can be written as
\be
{}^{(v)}\delta S_2=-\frac{1}{4\kappa_+}\int \di x^4
a^2(\eta)\Big((V_i+F'_i)\Delta (V_i+F'_i)-a^2(\eta)(m_1^2 V_i
V_i+m_2^2F_i\Delta F_i)\Big).
\ee
The field $V_m$ enters the action without
time derivatives, and thus it yields a constraint, \be \triangle
(V_i+F'_i)=a(\eta)^2 m_1^2V_i\equiv m(\eta)^2V_i
\ee
Substituting this constraint back in the action, we can write
\be
{}^{(v)}\delta S_2=
2\int \di^4 x a^4(\eta)\Big(m_1^2 F'_i \frac{\Delta}{\Delta -m(\eta)^2}
F'_i +m_2^2F_i\Delta F_i\Big).
\ee
This Lagrangian is free of ghosts and tachyons if $m_1^2\geq 0$ and $m_2^2\geq0$.

%%%%%%%%%%%%%%%%%%%%%%%%%%%%%%%%%%%%%%%%%%%%%%%%%%%%%%%%%
\subsection{Scalar modes}
%%%%%%%%%%%%%%%%%%%%%%%%%%%%%%%%%%%%%%%%%%%%%%%%%%%%%%%%%%%
From (\ref{kinetic}) and (\ref{genermassdS}),
the action for the massive scalar degrees of freedom is given by
\ba
\label{generalscalar} {}^{(s)}\delta S_2=\frac{1}{2\kappa+}\Big[\int&
\di^4& x a^2(\eta)\{ -6(\psi'+ {\mathcal{H}}A)^2
+2 \Delta \psi(2A-\psi)+4\Delta(B+E')(\psi'+{\mathcal{H}}A)\}\nonumber\\
+\int\di^4x a^4(\eta)&\Big(&m_0^2 A^2-\frac{m_1^2}{2}  B \Delta B
-m_2^2[(\Delta E)^2+3 \psi^2-2\psi \Delta E]\nonumber\\
&&+ m_3^2(\Delta E-3\psi)^2 +2m_4^2 A(\Delta E -3\psi)\Big)\Big].
\ea

Following \cite{Rubakov:2004eb},
 let us first consider the case $m_0=0$. In the flat case, $m_0=0$ implies that $A$ appears
 linearly in the Lagrangian and its EoM impose a condition between $E$ and $\psi$
 which means that there will be just one PDoF in the scalar sector. Even
 if $A$ is no longer a Lagrange multiplier for $\H\neq0$, we will see that there is also one
  PDoF in the scalar sector.
 Notice that the condition $m_0=0$ is the condition which makes
 the FP case $m_t^2=m_s^2$ special in the Lorentz preserving case,
 but that other similar ghost-free possibilities exist once the
 Lorentz symmetry is broken. In particular,
 the choice $m_1=0$ corresponds to the case where $B$ is a Lagrange multiplier
 and in such a case there is only one scalar PDoF which can be well behaved \cite{Dubovsky:2004sg}.
 We will study this possibility later.\\

For the de Sitter case, $m_0=0$ means that the kinetic term of the
propagating fields (\ref{kinetwofields}) has the values
\be
M_1(\eta)=\frac{4(m_1^2-2m_4^2)\Delta a(\eta)^2}{(4\Delta-3m_1^2
a(\eta)^2)\H}, \quad M_4=\frac{2 m_1^2 a(\eta)^2}{-4\Delta+3m_1^2
a(\eta)^2},
\ee
with all the other terms vanishing. Thus the kinetic term is written as
\be
K=
\frac{a(\eta)^2}{2\kappa_+}(M_1(\eta) \phi \psi' +M_4(\eta) \phi'^2),
\ee
and $\psi$
appears only  linearly in the kinetic term, leaving $\phi$ (recall
that $\phi\equiv \Delta E$) as the only PDoF. Once the equation of
motion for $\psi$ is substituted in the Lagrangian and after partial
integration one finds that the kinetic term reads \be
\label{kmassm0}
K=\frac{a(\eta)^2}{\kappa_+}\left(\frac{4m_4^2(m_4^2-m_1^2)\Delta+3m_1^2(m_4^4
a(\eta)^2+ 2\m^2\H^2)}{m_1^2(2\Delta-3m_4^2 a(\eta)^2)^2-
6\m^2(4\Delta-3m_1^2 a(\eta)^2)\H^2}\right)\phi'^2 \ee where
$\m^2=-m_2^2+3(m_3^2-m_4^2)$. Notice that the denominator is always
positive for $\m^2\geq 0$, and that once this condition is imposed
the numerator is positive provided that $m_1^2\geq m_4^2$. The first
condition is related to the term which multiply the parameter $\H$,
and thus is not present in the Minkowski case\footnote{One
 can argue that for scales inside the de Sitter horizon this
condition is not necessary, but we will not make these considerations
here.} \cite{Rubakov:2004eb}.
Also notice that for $\m=0$ there is no contribution from $\H$.\\

The analysis of the mass
term is more involved. We can write it as
\be
V=-\frac{a(\eta)^4(3b^2+9 m_1 a(\eta)^2c\Delta+6d \Delta^2+4m_1^2e\Delta^3+f
\Delta^4)}{\kappa q(\Delta)^2}\phi^2,
\ee
where
\ba
b&=&3 m_1^2 m_2 a(\eta)^2(m_4^4 a(\eta)^2+2\m^2 \H),\nonumber\\
c&=& -8m_1^2m_2^2m_4^6a(\eta)^4
+m_1^2m_4^8 a(\eta)^4-4m_4^2 \m^2[4m_2^2 m_4^4+m_1^2(4m_2^2-m_4^2)]a(\eta)^2\H^2\nonumber\\
&&\hspace{2cm}+
4(m_1^2-8m_2^2)\mu^4 \H^4, \nonumber\\
d&=&m_1^4 m_4^4(13 m_2^2-3m_3^2-2m_4^2)a(\eta)^4+8(3m_1^2-4m_ 2^2)\m^4 \H^4\nonumber\\
&&\hspace{2cm}
+2 m_1^2 \m^2[16m_2^2m_4^2-6m_ 4^4+
m_1^2(5m_ 2-3m_3^2-2m_ 4^2)],\nonumber\\
e&=&m_1^2m_ 4^2(-10m_2^2+6m_3^2+m_4^2)a(\eta)^2+2(5m_1^2-10m_2^2+6m_3^2-4m_ 4^2)\m\H^2,\nonumber\\
f&=&8m_1^4(m_2^2-m_3^2).\nonumber
\ea
The Lagrangian will be free of gradient instabilities provided that $m_2^2\geq m_3^2$ and has no unstable
modes at intermediate scales. Indeed,  even in the presence of unstable modes
at intermediate scales, the model can be phenomenologically
acceptable if they are set beyond the  horizon \cite{Creminelli:2006xe}.

To study the behaviour at
intermediate momentum we can try to localize the zeros of the numerator
to see when it changes sign. Unfortunately,
the numerator is of forth degree in $\Delta$, and the general solution of the zeros
is not known. Instead,
as we know that both at high and at low momentum the numerator is positive, it is enough to
prove that the minima of the polynomial in the regime $-\infty\leq \Delta\leq 0$
are above zero to ensure the positivity of the potential at any scale.
The minima of the numerator will be located at momenta satisfying
\be
\frac{9}{4}m_1 a(\eta)^2 c/f+3 d/f \Delta+3 m_1^2 e/f\Delta^2+\Delta^3=0.
\ee
The exact solutions of this polynomial can be easily found and imposing that,
when they exist, they are either at $\Delta > 0$ or such that the numerator
evaluated at them is positive we find all the tachyon free possibilities.
As an example one can consider the case $\m=0$. In this case the
Lagrangian is simply
\ba
\mathcal L=\frac{a(\eta)^4}{\kappa_+(2\Delta-3m_4^2 a(\eta)^2)^2}&\Big(&
[4m_4^2(m_4^2-m_1^2)\Delta+3m_1^2 m_4^4 a(\eta)^2]\phi'^2\nonumber\\
&&\hspace{-5cm}+
[2(3m_4^2-2m_3^2)\Delta^2+m_4^2(12m_3^2-13m_4^2)\Delta a(\eta)^2+9m_4^4(m_4^2-m_3^2)a(\eta)^4]\phi^2\Big),
\ea
and it is enough
to impose $m_3^2\geq \frac{3}{2} m_4^2$ to find a perfectly well defined Lagrangian.\\

Another interesting possibility consist of imposing $m_1=0$. As we see from (\ref{generalscalar}),
this condition transforms $B$ into a Lagrange multiplier which fixes $A$ as a function
of $\psi$ (see also \cite{Dubovsky:2004sg}). Again, there is only one scalar field left
 whose Lagrangian is
\be
{\mathcal L}=\frac{a(\eta)^4}{2\kappa_+(m_2^2-m_3^2)\H^2}\left([m_0^2(m_2^2-m_3^2)+m_4^4]\psi'^2
+2m_2^2\m^2 \H^2 \psi^2\right).
\ee
Notice that there are no spatial derivatives and that  for $m_2^2-m_3^2\geq 0$ and $\m^2\leq 0$
the previous Lagrangian is free of instabilities. The case $m_2=m_3$ implies that no scalar degree
of freedom propagates. \\

Finally, in the general case ($m_i\neq0$) we recover the second
propagating field. The parameters in the kinetic term (\ref{kinetwofields}) are now
\ba
M_1(\eta)&=&\frac{
8\Delta (m^2_1-2  m^2_4) a(\eta)^2 \H}
{ m^2_1 m_0^2 a(\eta)^4+8\Delta \H^2-6  m^2_1 a(\eta)^2 \H^2},\\
M_2(\eta)&=&\frac{2 m_0^2a(\eta)^2(4\Delta -3m^2_1
a(\eta)^2)}
{ m^2_1 m_0^2 a(\eta)^4+8\Delta \H^2-6  m^2_1 a(\eta)^2 \H^2},\\
M_3(\eta)&=&\frac{4 m^2_1 m_0^2 a(\eta)^4}
{ m^2_1 m_0^2 a(\eta)^4+8\Delta \H^2-6  m^2_1 a(\eta)^2 \H^2},\\
M_4(\eta)&=&\frac{-4 m^2_1 a(\eta)^2 \H^2}{ m^2_1
m_0^2 a(\eta)^4+8\Delta \H^2-6  m^2_1 a(\eta)^2 \H^2}.
\ea
The kinetic term gives a positive contribution to the Hamiltonian
in the range of parameters
\be
0\leq m_0^2 \leq 6 H^2, \quad m_1^2\geq 0,
\ee
for which $M_4(\eta)$ and $4M_4(\eta)M_2(\eta)-M_3^2(\eta)$ are positive (see
(\ref{Eulerkin})). The same comments that we made in the previous section about the
kinetic term of the scalar part
apply here with the substitution of $m_t$ by $m_1$ and $m_s^2-m_t^2$ by
$m_0^2$.\\

Finally, once the kinetic term has been shown to be positive definite, we can
look for  potential terms free of high-energy instabilities. One can show that
at very large momenta there is always a gradient instability which makes
the theory ill-defined. However, and contrary to what happens in the
covariant case (see before) or in the flat case (see \cite{Dubovsky:2004sg}),
one can make use of the curvature scale $H^2$ to
find regions in the parameter space where the theory is unitary. In particular,
at energies $\Delta$ inside the horizon and such that
\be
-\Delta \ll m_2^2\left(\frac{H}{m_1}\right)^2,
\ee
there exists a hierarchy of parameters where the Hamiltonian is positive definite.
More concretely, if we choose $m_2\sim m_3\sim m_4\sim M$, where $M$ is a mass scale, and
\be
-\Delta \gg M \gg H^2\sim m_0^2\gg m_1^2,
\ee
the potential reduces to
\be
V=-a(\eta)^2\left[(m_2^2-m_3^2)\phi^2+2(3m_3^2-m_2^2)\phi \psi+3(m_2^2-3(m_3^2-m_4^2))\psi^2\right],
\ee
which is negative for
$$m_2^2\geq m_3^2, \quad 2m_2^4-9m_3^2 m_4^2(-6m_3^2+9m_4^2)\geq 0.$$
However, whenever $-\Delta \gg m_1^2$, as happens in the
case under study, only $M_1(\eta)$ and $M_2(\eta)$ in (\ref{kinetwofields})
do not cancel and  the final Lagrangian for the scalar sector have only
one PDoF. Furthermore, the kinetic energy of this scalar is much larger than
its mass, and thus its Lagrangian is simply
\be
L=\frac{a(\eta)^2m_4^4}{2\kappa_+H^2(m_2^2-m_3^2)}\psi'^2.
\ee
The existence of other theories with a Lorentz breaking
cut-off depending on $H$ and free from
ghosts and tachyons is currently under research.

%
%
%%%%%%%%%%%%%%%%%%%%%%%%%%%%%%%%%%%%%%%%%%
\part{Conclusions and Appendixes}
%%%%%%%%%%%%%%%%%%%%%%%%%%%%%%%%%%%%%%%%%%%

%%%%%%%%%%%%%%%%%%%%%%%%%%%%%%%%%%%%%%%%%%
\chapter{Conclusions and Outlook}\label{chapterconclu}
%%%%%%%%%%%%%%%%%%%%%%%%%%%%%%%%%%%%%%%%%%%

In this dissertation we have studied certain modifications of GR motivated
by the possibility of finding a consistent theory which may alleviate the problem
of the {\em cosmological constant} or may suggest new avenues to its resolution (see the introduction).

We first focused on the analysis of the local second order Lagrangians which
are ghost and tachyon free and that include spin-2 particles in their spectrum.
It was shown in Chapter \ref{chapterLorentz} that for the massless case those
Lagrangians must be invariant under a subgroup of the whole Diff group. More concretely,
the analysis of the vector components of the rank-2 object $h_{\m\n}$
shows that the Lagrangian must be invariant under the subgroup
of the Diff satisfying
\be
\label{concl_cond}
\pd_\m \xi^\m=0,
\ee
otherwise the spectrum of the sector will include ghosts. We dubbed this
subgroup TDiff. If TDiff is violated and the ghosts are not
coupled to conserved matter at the linear level
the linear theory may still be unitary. Nevertheless, the linear
theory is not enough to describe gravity and one expects
that the non-linear interactions will include coupling of these
modes both to matter ant to the other PDoF of the graviton itself.
This would render the theory non-unitary at the non-linear level
and thus we required the invariance under TDiff at the linear level
to get a meaningful theory.

The spectrum of perturbations
of the TDiff invariant theories
consists of a spin-2 particle and a scalar field. The spin-2 component is always
well-behaved, whereas the Lagrangian must satisfy certain condition
for the scalar part to be fine (cf. (\ref{chLscalar})). The linear theory
is completely equivalent to a {\em scalar-tensor} theory
except for the appearance  of an integration constant. A mass term for the
scalar component exists which preserves the TDiff invariance, and for
a heavy scalar field the
phenomenology of the theory coincides with that of linearized GR for energy
scales below the mass scale\footnote{Indeed, the
mass term is not protected by any symmetry, and we expect it
to receive radiative corrections that set its scale to the cut-off
scale of the theory.}.

The scalar field disappears when the TDiff symmetry is enhanced in one of two possible
ways. The standard choice is to consider the full group of Diff (\emph{i.e.}
lift the condition (\ref{concl_cond})). We showed
that there is yet another possibility
(which we called WTDiff) where an additional Weyl symmetry is imposed
and the condition (\ref{concl_cond}) still holds. In this last case,  the
action depends only on the traceless part of the field
$\hat h_{\m\n}\equiv h_{\m\n}-\frac{1}{n}h_{\m\n}$. Even if both actions are
not equivalent\footnote{We consider two actions to be equivalent if they are related
by a field redefinition or by the addition of a gauge fixing term.},
 they yield the same equations of motion except for an extra
integration constant in the WTDiff case. This integration constant is  related to the
cosmological constant (we elaborated more on this in Chapter \ref{chapternl}).

It is interesting to note that the similarities between both types of theories do not extend to the case
where the spin-2 components are massive.
Once a Lorentz preserving mass term is added to the action, the only
ghost and tachyon free Lagrangian has the Diff invariant kinetic term and the Fierz-Pauli (FP)
mass term. There is no equivalent
construction with a WTDiff invariant kinetic term.
The root of the difference between the massless and the massive cases is that
 the gauge invariance present in the WTDiff massless case requires the
 imposition of a {\em tertiary} constraint which kills the extra scalar which
 one would expect from a naive counting of the PDoF. Once a generic
mass term is considered, the Lagrangian is no longer gauge invariant and for
a WTDiff invariant kinetic term it is not possible to kill the ghost-like scalar
degree of freedom. In this sense, WTDiff is a more rigid theory than the
standard linearized GR (see also the comments on supersymmetric extensions).\\

The previous analysis can be extended to other higher spin theories. Namely, we can
look for Lorentz invariant Lagrangians
of higher spin fields that yield the same equations of motion as the standard
gauge invariant Lagrangians
once the appropriate initial conditions are imposed. This is precisely
 what happens
when one adds covariant gauge-fixing terms to a gauge invariant Lagrangian (see
{\em e.g.} \cite{Itzykson:1980rh}). For the {\em bosonic} field theories, this
extension can be  performed and it amounts again to replacing
the higher spin field by its traceless part
in the gauge invariant Lagrangians that were proposed in \cite{Fronsdal:1978rb}.
 Both Lagrangians, which are not
equivalent,  yield the same EoM
except for an integration constant \cite{Skvortsov:2007kz}.

For the fermionic field of spin-$3/2$, we have shown in Chapter \ref{chapterRS} that
something similar happens for the $\g$-traceless part of this field. First,
we have shown that there are two possible groups of gauge invariance for the generic Lagrangians
which include spin-$3/2$ particles
in their spectrum. The presence of the gauge invariance is important as it allows to kill some
of the
potentially ghost-like spin-$1/2$ excitation. The first of these possibilities
corresponds to the usual Rarita-Schwinger (RS) Lagrangian which
is known to propagate just the spin-$3/2$ polarizations and to be unitary
once coupled to a conserved source. Besides, the gauge invariance can be
of a Weyl type ($S$-symmetry), $\delta \psi_\m=\g_\m \phi$,
if one works directly
with the $\g$-traceless combination (for $n=4$),
 $$\hat \psi_\m\equiv\psi_\m-\frac{1}{4}\g_\m\g^\a\psi_\a.$$
 The Lagrangian endowed with  this gauge invariance, which we  called WRS
 Lagrangian, yields the same propagator
 as the RS one once coupled to a conserved source\footnote{Again, and
 as happens once the gauge is fixed covariantly \cite{Das:1976ct},
 there is an extra degree of freedom in the WRS case which is decoupled
 from the sources and
 can be consistently set to zero.}.
  Thus, we found a Lagrangian which yields
 the same predictions as the standard RS Lagrangian.

A key difference between both Lagrangians is that their groups of gauge invariance
are different. We have
elaborated a bit on the possibility that this might
alleviate the problem of the consistent
coupling of the spin-$3/2$ field to the electromagnetic field, as the algebraic constraints
that appear once the RS Lagrangian minimally coupled to electromagnetism, are not present
for the  WRS Lagrangian. Nevertheless, the low spin component of the field
$\hat \psi_\m$ that was decoupled in the case of interaction with external sources
  is turned on by this interaction, and this may spoil the
unitarity of the theory.

For the massive spin-$3/2$ field, the results are analogous to those of the spin-2 Lagrangians.
One can show that the {\em only} possibility which just propagates massive spin-$3/2$ is
the massive RS Lagrangian. Besides,
one can consider mass terms that
render some of the spin-$1/2$ polarization massive, leaving the
 spin-$3/2$ components untouched.

Independently of the previous results, it is interesting to study the general
Lagrangian for spin-$3/2$ as a possible partner of the WTDiff Lagrangian to build
a supersymmetric Lagrangian. However, as we proved in the last section of
Chapter \ref{chapterRS}, the WTDiff Lagrangian {\em does not} admit a minimal
supersymmetric extension. A simple argument for this fact is
 that the number of {\em off-shell} and {\em on-shell} degrees of freedom
of the WTDiff case only coincide with those of the RS action, which is already
the supersymmetric counterpart of the Diff invariant action.

A general conclusion of the previous analysis is that, due to the more involved
canonical structure of the theory, it is difficult to deform the WTDiff Lagrangian
consistently. The two examples
that we studied showed that neither the addition of a mass term for the spin-2
polarizations nor of a minimal superpartner
are possible.\\

The previous conclusions apply for the linearized theories. The non-linear
extension of the spin-2 Lagrangians was considered in the second part of the dissertation.
For the TDiff invariant Lagrangians, a systematic derivation of the non-linear
extension is currently absent. In Chapter \ref{chapternl} we found that, for the WTDiff
Lagrangian, a  non-linear extension along the lines suggested by Deser in \cite{Deser:1969wk}
for the Diff case seems to be problematic. In particular, even if the method
can be applied, the non-linear theory that is found
 differs form GR and seems to include a scalar field in
its spectrum, though an explicit calculation has not yet been performed.
Besides, it depends {\em explicitly} on the background Minkowski metric.

The linear {\em reducible} gauge invariance related to TDiff group can be
deformed non-linearly
to the subgroup of non-linear Diff transformations satisfying precisely the condition
(\ref{concl_cond}). Under this subgroup, the determinant of the metric
transforms as a scalar field, which implies that non-linear invariant
 Lagrangians  can be constructed out of the geometrical tensors
 for the metric and arbitrary functions of the determinant.
We proved, following previous results, that these theories are in general equivalent
to {\em scalar-tensor} theories except for the presence
of an integration constant that plays the role of a
cosmological constant.
The mass term compatible with
the TDiff gauge invariance also admits a non-linear extension.
As we said, this term provides a mass for the scalar component
and from a {\em naturalness
criterion}, this mass should be of the order of the cut-off
of the theory.  This implies that the low-energy PDoF of non-linear TDiff coincide
with those of GR.

 Concerning the WTDiff linear Lagrangian, it admits
a {\em unique} non-linear extension  which is also invariant
under non-linear Weyl transformations. We proved
that this Lagrangian yields Einstein's equations in the gauge
$|g|=1$ except for
an integration constant. This property is also shared
by a plethora of TDiff invariant Lagrangians where a term depending
on the determinant of the metric is added to the GR kinetic term. These
additional TDiff
invariant Lagrangians are expected to receive radiative
corrections which may make the scalar component dynamical. However, those
corrections also affect the mass term, which makes one expect
this mass to be at the
cut-off scale of the theory.  We conclude that, if we consider
these effects, the low-energy PDoF of GR, TDiff and WTDiff
theories are generically the same.

In the last part of Chapter \ref{chapternl}, we  studied the
\emph{first order} formulation of the WTDiff invariant Lagrangian.
We  proved that writing
the Lagrangian in terms of the {\em vielbein} and the {\em spin-connection}
is classically equivalent to the WTDiff Lagrangian written
in terms of the metric without the need of Lagrange multipliers. This allows us to couple the WTDiff
invariant Lagrangian to fermionic matter, and in particular to
look for a consistent minimal coupling with a spin-$3/2$ field (which
we know that will not be supersymmetric, as at the linear
level we showed that there is not a minimal supersymmetric action for both fields).
Even if supersymmetry is lacking, one may hope
 that due to the conditions
on the EoM imposed by the gauge invariance of the RS Lagrangian (cf. \cite{VanNieuwenhuizen:1981ae}),
 the
 integration (cosmological) constant will be set to zero. \\

We have devoted the rest of the Thesis to study the concrete non-linear model
of massive gravity provided by {\em bigravity}. We have
focused on the study of the systems with two metrics with
independent Einstein-Hilbert kinetic actions and coupled
through a non-derivative term preserving a ``diagonal"
group of diffeomorphisms. Our aim was to extract some
conclusions about the behaviour of non-linear massive gravity
from this simple set-up.

We first studied some {\em exact} solutions of the non-linear equations.
 For a given pair of metrics which are
 solutions of
the vacuum Einstein's equations with corresponding cosmological constants,
we have derived the
conditions that the interaction term must satisfy for this pair to be a
 solution of the bigravity theory.

Being exact solutions of GR, these solutions
are  important as they constitute a simple candidate
to understand the
way in which non-linearities may cure the vDVZ discontinuity.
We identified a particularly interesting
family of solutions which are static and spherically symmetric
with respect to a common $SO(3)$ group. Interestingly enough,
these solution depend on some integration constants that once fixed
by a condition depending on the potential (cf. (\ref{typeIcondo}))
make them solution for {\em any} potential. In other words, every
potential admits solutions in this family.

Another interesting point about these solutions
is that they can correspond to metrics
with different {\em global structure}. In Chapter \ref{chapterbigra},
we developed a method to visualize the global structure
of the bigravity system by studying
the behaviour of the lightcone of one of the
metrics in the conformal diagram of the companion
metric of the solution. This allowed us to see
how does the conformal structure of the first metric map
into the conformal diagram of the other metric.

A particularly interesting possibility that
occurs in some of the solutions is the presence
of a horizon for just {\em one} of the metrics.
When the companion metric
is already  geodesically complete, this
rises  questions about the meaning of the
maximal extension of the incomplete metric.
By plotting the
null-cones of the geodesically complete metric in the Carter-Penrose
diagram of the incomplete one, we provided a precise
map of the causal structure of the geodesically complete metric
as seen by the incomplete one. We showed in some detail how the geodesics
of the first metric end within the incomplete
patch.  This means that once the geodesically incomplete metric is
maximally extended, the new region of space-time
is causally disconnected from the original space-time patch
for the geodesically complete metric. To get the full {\em extended} bimetric
solution,
 we proposed to choose a
new solution of bigravity in the extended region in a way that the system
preserves causality. There is much freedom in this
possible extension, and this freedom is similar to
the standard situation of GR for solutions
 with a Cauchy horizon.

Given the existence of two different causal structures, we investigated whether
it is
possible that  closed time-like curves (CTC) exist even
if both metrics are globally hyperbolic.  To build these curves,
we need to propagate signals using both metrics.
We showed that
for the solutions considered in Chapter \ref{chapterbigra},
CTC are absent even if the global
notion of time is not trivial. Indeed, it may happen that a certain
Cauchy surface is so only for one of the metrics, even
if there are other common Cauchy surfaces.
We also found an apparent generic
tension between geodesic completeness and global hyperbolicity
in the presence of horizons which are not shared by both metrics.

As a conclusion of our studies on global structure
we can say that the possible pathologies that we identified
in the class of solutions of {\em bigravity} which we considered
are not worse than
those found for certain solutions of GR such as anti-de Sitter
 or Reissner-Nordström.\\

Finally, in the last Chapter
we studied how the presence of a second dynamical
metric gives rise to {\em mass terms}
 for a certain combination of the gravitons. We have first focused on
flat solutions which break the
Lorentz invariance to a common $SO(3)$.
The analysis of perturbations around
this background reveals that the only PDoF
are the tensor components of the metrics, and they satisfy  Lorentz-breaking
dispersion relations with a mass term.
This fact implies corrections to the Newtonian potential
between two sources which are
proportional to the square of the graviton mass and which grow
linearly with the distance to the origin.
We have shown that the system is not strongly coupled in
general and that the massless limit, as expected from the absence
of strong coupling,
 is well defined. We see that the breaking of the Lorentz invariance
by the background allows to avoid the vDVZ discontinuity
and the strong coupling of the scalar mode\footnote{For
certain Lagrangians the linearized perturbation theory is not
well defined and one is forced to go to the next order in
perturbation theory with more than just one {\em strongly coupled} mode,
which complicates the analysis.}.
However, there seems to be a tension between the
perturbative solution and the exact solution. Indeed,
an exact solution which asymptotes to the bi-flat
solution is known but the interacting term
does not give rise to a Yukawa type potential, but to
a contribution to the vacuum energy. This seems to
indicate the presence of a linearization instability
or the existence of other exact solutions which coincide
with the linearized approximation at large distances.

We also analyzed in detail the perturbations around bi-de Sitter vacua.
For generic solutions we found that the spectrum
consist of a massless and a massive graviton. For proportional metrics,
the theory is covariant but the mass term is not in general of the Fierz-Pauli
form. For flat space this means the loss of unitarity at energy scales of the
order of the mass scale. For the de Sitter case, one may think that
the presence of a new energy scale (associated to the curvature scale) could
help to increase the cut-off scale and to find consistent field theories with a
cut-off scale larger than the mass of the tensor modes\footnote{Something
 similar happens for the {\em strong coupling} scale
of the FP Lagrangian.}. Even if we found
that the presence of curvature allows for a healthy kinetic term,
we showed that, in the adiabatic limit,  {\em gradient} instabilities set in
at the scale of the mass of the tensor modes, which makes the
theory non-unitary at this scale.

From the fact that the only Lorentz invariant mass term which is consistent
in the bi-de Sitter case is the FP mass term, we argued that
a dynamical {\em see-saw} mechanism,
where the vacuum energy of one of the metrics weights very little, is not possible
for natural values of the parameters in the solution.

The previous reasonings may be successful once one admits non-covariant (or
Lorentz-breaking)
mass terms for gravitons propagating in de Sitter space. The study of
this kind of Lagrangians reveals that in the presence of curvature
there are new regions in the parameter space which allows for a EFT
description with a cut-off scale
which tends to the mass scale as the curvature goes to zero.\\

We would also like to comment a bit on the contents of the appendices.
Even if they are based on original material, we have decided to defer the
discussion of this work to the appendix due to its preliminary form
or because it corresponds to the study of very concrete models which do
not add much to the main results of the dissertation. In Appendix \ref{AppendixQ},
we study some issues of the quantization of TDiff invariant theories.
We first consider some aspects of the {\em semiclassical} approximation.
In this approximation, one expects the
appearance of differences between Diff and WTDiff theories
 because the gauge invariance of the regularization process
determines the possible counterterms
that may be needed to make the theory renormalizable. A regularization
scheme preserving the Weyl and Diff invariance is not known. We
propose a {\em generalized} Pauli-Villars regularization scheme which can be
used to preserve the WTDiff, Diff or just the TDiff invariance of the theory. The structure
of the counterterms may differ in those three cases, which may imply that
classically equivalent Lagrangians differ at the semiclassical level.
A particular example of the possible differences due to the
regularization scheme is provided by the Weyl anomaly. We  argue that
the Weyl anomaly can appear in the Diff sector if
the regularization process is consistent with the WTDiff invariance.
In other words, the anomaly can be traded
from the Weyl symmetry to the Diff symmetry group (breaking it to TDiff).
The counterterms associated to this regularization will break the WDiff
symmetry to WTDiff but we will argue that the
semiclassical EoM are equivalent in both cases.

We
show a particular example provided by the conformal anomaly in $1+1$ dimensions.
 On the other hand, we could consider regularization schemes that break the symmetry
 of the classical action ({\em e.g.} the Diff preserving scheme for the WTDiff
 invariant action). The breaking of the symmetry by this process will generate
 a small scale in the problem (maybe related to the cosmological constant),
 but the consistency of the model is not clear in this case.

If we want to go beyond the semiclassical approximation and consider
a quantum theory of gravity we first have to worry about the unitarity of the theory.
The first thing we  study is the existence of a nilpotent BRST transformation
in the WTDiff case\footnote{Remind that for most gauge theories, the existence of this
transformation is essential to prove the {\em unitarity} of the theory.}.
 The {\em reducible} nature of the
TDiff transformation, makes the BRST transformation more involved than in the Diff
case and more fields besides the usual Fadeev-Popov ghosts are required
to get a nilpotent transformation. These new fields are the ghost-for-ghost fields,
which are required to  find a covariant gauge-fixed action. It is remarkable
that the study of the BRST transformation can be phrased in terms of forms,
which makes the analysis quite straightforward. We  present a minimal
set of ghost-for-ghost fields together with their BRST transformations
and Grassmanian character. This is a first step towards the covariant
quantization of the WTDiff theory.

We  end this Appendix with some comments on the Euclidean Quantum Gravity formulation
of the WTDiff theory. We  show that, even if the action is Weyl invariant, it is not
bounded from below as there is a mode (a Diff which is not TDiff) which plays the same
role as the conformal mode in the Diff invariant case. This means that the WTDiff
action has no better convergence behaviour than the Diff invariant action.\\

Appendix \ref{AppendixBig} is devoted to the study of further aspects of
{\em classical unimodular gravity} and {\em bigravity}.
Some well-known facts about Diff invariant theories may change
once one restricts the analysis to the TDiff subgroup. In
the first part of Chapter \ref{chapternl} we study some of them. We  show
that the
condition for a metric $g_{\m\n}$ to be related to the Minkowski metric
by a gauge transformation in the WTDiff theory is that the Riemann tensor
associated to the combination $\hat g_{\m\n}=g^{-1/n}g_{\m\n}$ cancels.

Furthermore, the restriction to the TDiff invariant subgroup allows
more freedom to define covariant derivatives, as
the object $\Gamma^\r_{{\phantom r}\a\r}$ transforms
as a vector under TDiff. We extended the usual formalism
of integration of forms on manifolds to the TDiff invariant
case, including Stokes'  theorem.

Concerning {\em bigravity}, we  show that for a certain simple potential of bigravity,
the solutions
consisting of two proportional metrics is the most general diagonal static and
spherically symmetric solution when one of the metrics is maximally symmetric.
This result is a first step in the search of  more general solutions, but the general
static spherically symmetric solution of {\em bigravity} is still unknown even for
simple potentials. The knowledge of the general solution would be very important to understand
why the linear treatment does not agree with the non-linear solution for certain
cases.
We end the Appendix \ref{AppendixBig} with some comments on possible methods
to find solutions of bigravity from solutions of ordinary GR.

%%%%%%%%%%%%%%%%%%%%%%%%%%%%%%%%%%%%%%%%%%%%%%%%%%%%%%%%%%%%%
\section{Outlook}
%%%%%%%%%%%%%%%%%%%%%%%%%%%%%%%%%%%%%%%%%%%%%%%%%%%%%%%%%

Throughout the text we have discussed some possible ways in which our analysis can be
extended. In this section we want to sketch some of them and present
related ideas left for future research.

In the linear analysis of Chapter \ref{chapterLorentz}, we described
the spin-2 field by means of a symmetric rank-2 field $h_{\m\n}$.
An interesting extension would be to study the
 ghost and tachyon free possibilities for linear Lagrangians
 in the metric-affine theories of gravity\footnote{A general analysis for Diff and Local Lorentz
invariant theories was performed
in \cite{Kuhfuss:1986rb} (see also \cite{Sezgin:1981xs,Nibbelink:2006sz} for related work).} (where the
  {\em vielbein} and the {\em connection} are
  considered as independent fields) \cite{Hehl:1994ue}.

Other possible extensions include the addition of terms with
higher derivatives or the breaking of the global Lorentz invariance.
 A model where the four dimensional Lorentz invariance
is consistently broken due to bulk effects was presented in \cite{Dvali:2007ks}.
 One expects that the
massive modes of the KK spectrum in this case will have a Lorentz violating mass term,
which may result in a model of massive gravity lacking the strong coupling problem.
Besides, the Pauli-Fierz structure of the mass term can
also be generalized if one allows for a momentum dependence in the mass parameters \cite{deRham:2007xp}.
The search of other scenarios showing this behaviour is currently under research \cite{Blas08}.

Another source of
consistency problems of the coupling of higher spin states appears in the
 study of the properties of the $S$-matrix \cite{Weinberg:1980kq,Porrati:2008rm}. A first analysis seems
 to indicate that also in the TDiff invariant case, the existence
 of a conserved source implies the absence of massless particles
 of spin-2 \cite{Blas08}. However, as the energy-momentum tensor can be conserved up to
a derivative,  a non-vanishing energy is allowed \cite{Blas08}.

Concerning the theories with spin-$3/2$ fields, we have
outlined a couple of  lines of future research in Chapter \ref{chapterRS}.
First, it would be nice to study the (lack of) unitarity of the theory where
the  WRS Lagrangian is minimally coupled to
a $U(1)$ field.
Besides, we have not studied in detail the coupling of the Rarita-Schwinger
field to the WTDiff field. One may hope that, as happens in GR (cf. \cite{VanNieuwenhuizen:1981ae}),
the consistency of the coupling implies the cancelation
of the cosmological constant, even in the absence of supersymmetry.

There are many open directions related to the
non-linear extensions presented in Chapter \ref{chapternl}.
It would be very interesting to study the
possible non-linear deformations of the TDiff algebra in a more systematic way.
The most powerful formalism for the deformation of gauge algebras
is provided by their cohomological structure \cite{Henneaux:1997bm} (see
also \cite{Ogiev:1965} for earlier related work) and
the application of this formalism to the TDiff case is in progress \cite{Blas08}.
The presence of a relation between the gauge parameters of the theory
imposes some technical difficulties in comparison with the
{\em irreducible}\footnote{On the other hand, we have seen that the TDiff group can be augmented
to the Diff group by the addition of the trace of the field $h_{\m\n}$.
This field plays the role of a St\"uckelberg field and turns the
{\em reducible} symmetry into an {\em irreducible} one, whose
quantization is much simpler. One may wonder whether a similar
possibility exists for other reducible gauge theories.}
case but the general formalism still applies \cite{Henneaux:1999ma,Henneaux:1997ha}.

As we emphasized throughout the Thesis, the structure
of the constraints of the WTDiff invariant theory differs from that of GR.
The canonical formulation of the WTDiff invariant theories, together
with the interpretation of the different constraints has not been clarified yet.
Besides,  the extension of the Lovelock analysis to the
 TDiff or WTDiff invariant theories is still an open issue.

Another of the results that we underscored in this Thesis is the
{\em classical} equivalence of the WTDiff and the Diff invariant
non-linear Lagrangians. In Appendix \ref{AppendixQ} we argue that
the regularization of the energy-momentum tensor at one-loop in matter
fields can be consistent with the WTDiff invariance. Even if the structure
of the counterterms will be different from that of the
regularization that preserves the Diff invariance, we claim that both
 possibilities are physically equivalent for WDiff invariant classical
 theories.
 It would be desirable to find a local
 counterterm that mediates between both possibilities\footnote{The
counterterms account for terms with higher derivatives.  The equivalence of the
EoM coming from the Diff or the WTDiff invariant theories is not clear in this
case.}. Besides,
the regularization procedure may break the symmetry of the theory, which
may be useful to generate a small cosmological constant.

Concerning the structure of perturbative quantum gravity,
from the results of Chapter \ref{chapterLorentz}, we see that
even if the {\em on-shell} propagators of the graviton are the same
for the WTDiff and Diff invariant theories,
 the {\em off-shell} propagators do not coincide in any gauge.
This means that even if the interaction terms of both theories are related,
it is far from clear that the loop computations coincide\footnote{
Furthermore, as
emphasized in \cite{Farajollahi:2008ba,Unruh:1988in},
 the presence of a preferred form may have some consequences in other
 formulations of quantum gravity (see also \cite{Rovelli:89}).}.

There are also many interesting open problems for {\em bigravity} theories.
First, the most general static and spherically symmetric solution is not known
even for the simplest potentials. The knowledge of this solution is very important
as it might help to understand the way in which the linearized solutions are matched
to the non-linear ones. For other theories of non-linear massive gravity,
the exact static and spherically symmetric solutions is not known either.
The simplicity of the {\em bigravity} Lagrangian makes it a good starting
point to try to understand some general features of this solution in
non-linear massive gravity.

Concerning the perturbation theory, there are some exact solutions
of {\em bigravity} whose perturbation theory may yield
interesting results. First, if one (or both) of the
metrics of the solution has  a horizon, one expects
the theory of perturbations to be very different than in GR.
In particular, there is no reason to expect the {\em no hair} theorems
to be still valid.
Some work in this direction has already been done for the ghost condensate,
and many differences with respect to the GR case have been found
\cite{Dubovsky:2007zi}. The bottom line of these studies is that
black holes physics is very different in modified theories of GR\footnote{A
 first intriguing fact is that
there may be some modes that can  exit the horizon.}. A related question
is the possible existence of Lorentz breaking hair for black holes. Nevertheless,
the presence of black holes in Lorentz violating theories seems to be problematic \cite{Jacobson:2008yc},
and this issue deserves further clarification.

Various questions arise, should one wish
to consider bigravity theories as realistic. Among those, the fact that bigravity theories
may suffer from  instabilities
coming from the propagation of ghost modes at the non-linear level \cite{Boulware:1973my,Creminelli:2005qk}
(see however \cite{Gabadadze:2004iv,Damour:2002ws,Damour:2002wu}).

Finally,  we proposed a mechanism that
may offload the cosmological constant for one of the metrics of {\em bigravity} dynamically.
This mechanism does not work for the interaction terms and solutions
that we studied, but yet it is not clear that other {\em bigravity} scenarios
(as for non-proportional accelerating solutions) may enforce it.

%%%%%%%%%%%%%%%%%%%%%%%%%%%%%%%%%%%%%%%%%%%%%%%%%%%%%%%%
\appendix
 %
%%%%%%%%%%%%%%%%%%%%%%%%%%%%%%%%%%%%%%%%%%%%%%%%%%%%%%%%%%%%%%%%%%
\chapter{Remarks on Quantization of WTDiff Theories}\label{AppendixQ}
%%%%%%%%%%%%%%%%%%%%%%%%%%%%%%%%%%%%%%%%%%%%%%%%%%%%%%%%

One of the points stressed throughout this dissertation has been the existence of different
Lagrangians whose equations of motion are equivalent to
Einstein's equations  (except for an integration  constant).
Out of them, there are two which are {\em fixed} by gauge invariance, namely the Diff case of GR and
the WTDiff case whereas the rest consist of adding a function of the determinant
to the Einstein-Hilbert Lagrangian.
The addition of matter does not change this behaviour, which means that all of
 these theories
are classically equivalent\footnote{It is important to remark that the structure
of the constraints is different  for the TDiff, Diff and WTDiff cases.}.

Even if one cannot construct a renormalizable quantum
theory from the GR Lagrangian,  one can pursue its quantization
as an EFT \cite{tHooft:1974bx,Burgess:2003jk,Donoghue:1995cz}
(see also \cite{Holstein:2006bh}).
This programme yields some testable predictions and is valid up to a certain energy scale
beyond which one expects the appearance of new Physics to cure the infinities of quantum GR.

The first step in this programme is to work out the so called ``semiclassical" regime
in which the gravitational field is considered as a background where other quantum fields
propagate \cite{BirrellDavies}. In the first part of this Appendix,
we will sketch how the analysis may be modified in the
WTDiff and TDiff theories.

Once the gravitational field is considered as a quantum dynamical field,
in some situations we can consider it as a quantum perturbation propagating in
a fixed background. The presence of low spin components appearing with the
wrong sign in the {\em
off-shell} propagator, makes one worry about the unitarity of the theory. For
gauge theories, a useful way of proving the unitarity of the theory
is with the help of the  BRST invariance of the gauge fixed action, and we will
embark upon the search of a possible BRST transformation for the
{\em reducible} gauge theories appearing in the TDiff and WTDiff theories.

A different approach to quantum gravity which allows to study non-perturbative phenomena
is the path integral formulation, or Euclidean Quantum Gravity \cite{Hawking:1980gf}. We will show
that for the WTDiff invariant theory, the convergence of the path integral does not
seem to be better than for the (ill-defined) Diff case.

Finally, notice that string theory can also be considered in the WTDiff case
by simply substituting
the background metric $g_{\m\n}$ by the combination $\hat g_{\m\n}$. Following
\cite{Polchinski},
one finds that for the cancelation of the $\b$-function,
$$R_{\m\n}[\hat g_{\a\b}]=0.$$
These are Einstein's equations for $g_{\m\n}$ in the gauge $|g|=1$. Thus, as far
as WTDiff world volume gauge invariance is preserved we find the same result at first
order in $\a'$ as for the Diff case. This does not guarantee that higher
order corrections are the same in both cases.\\

This Chapter is based on unpublished results which have been presented in some
conferences or talks. They constitute a first step towards the
quantization of TDiff and WTDiff theories, but a lot of work is still needed
(for recent work see also \cite{Alvarez:2008zw,Fiol:2008vk}).

%%%%%%%%%%%%%%%%%%%%%%%%%%%%%%%%%%%%%%%%%%%%%%%%%%%%%%%%%%%%%%%%%
\section{Semiclassical Approximation}\label{anomalies}
%%%%%%%%%%%%%%%%%%%%%%%%%%%%%%%%%%%%%%%%%%%%%%%%%%%%%%%

The standard formalism of quantum field theory in curved
space-times can be easily extended to TDiff and
WTDiff invariant theories. Once  the coupling of
matter to gravity is introduced (as we did in
Chapter \ref{chapternl}), the
quantization techniques described in \cite{BirrellDavies}
can be applied.

Recall also that we found the  same {\em on-shell}
propagators and interaction vertices for Diff and WTDiff theories
in a certain gauge.
This implies that both theories yield equivalent predictions
at {\em tree-level}.  In curved space-time, the renormalization of the
theory at one-loop in matter fields (which is the regime we are interested in)
implies the inclusion of geometrical higher order counterterms whose
structure is dictated by the gauge invariance preserved by the
regularization process \cite{BirrellDavies}. No regularization scheme
that preserves both the Weyl and the Diff invariance is known, which means
that the structure of the counterterms will be different for
the schemes that preserve the Diff or the WTDiff invariance. This fact may imply
the discrepancy in the physical predictions of Diff and WTDiff invariant
theories at one-loop in matter fields.

We will present here a regularization scheme depending on some parameters
that can be chosen to preserve
the TDiff, Diff or WTDiff and leave the study of the general counterterms preserving
the TDiff or WTDiff
and their physical predictions for further research (see also below) \cite{Blas08}.\\

For definiteness, let us consider a scalar field coupled to gravity in a WTDiff invariant
theory. The UV divergences of the two-point function will be equivalent to those of the
Diff invariant theory in the gauge $|g|=1$. To cure these divergences, we will
use a modified Pauli-Villars (PV) regularization scheme\footnote{For the application
of  PV regularization
in a Diff invariant way see
\cite{Bernard:1977pq,Vilenkin:1978wc} (see also \cite{Asorey:2003uf}).}.
Recall that this regularization method
resorts to the introduction of massive fields, $\phi_i$, with a Lagrangian
which cancels the UV divergences of the rest of fields. Setting
the mass of these fields beyond the cut-off of the effective field theory at hand,
the theory gives sensible predictions.

The difference between the Diff and WTDiff invariant theories can be
traced to the
 absence of a mass term compatible with
the whole WDiff symmetry. This means that the PV regularization scheme breaks the
WDiff
symmetry (which is the basis of the conformal anomaly). It is customary
to choose a mass term for the regulator field compatible with the Diff invariance,
\be
\label{massterm}
L_m=m^2\int \di^n x \sqrt{-g}\phi_i^2.
\ee
The addition of this mass term to  {\em any}
kinetic term\footnote{By this we mean the Diff, WTDiff or WDiff invariant kinetic terms.} yields a Lagrangian which
is not  invariant under the Weyl transformation\footnote{A similar regularization does not exist
for any field in any dimension. See {\em e.g.} \cite{AlvarezGaume:1983ig} for some
comments on mass terms for chiral fermions.},
\be
\label{generalWtr}
g_{\m\n}\mapsto e^{2\s}g_{\m\n}, \quad \phi \mapsto f(\s,\phi),
\ee
for any function $f(\s,\phi)$. In particular, this means that the
 trace of the energy-momentum of the regularized action
 will be different from zero in general.\\

To adapt the previous prescription to preserve WTDiff, Diff or just TDiff,  it is enough
to modify the mass term to\footnote{Notice that
at high enough energies, much larger than the scale of the variation of the determinant,
$|g|\rightarrow 1$ and the mass term is independent of $\r$ and
we expect it to be equivalent  to a standard mass term.}
\be
\label{massterm_a}
L_m=m^2\int \di^n x |g|^{\r/2} \phi^2.
\ee
For arbitrary $\r$, this term is just compatible with the TDiff subgroup whereas
for $\r=\frac{n-2}{n}$ this mass term is compatible with the Weyl invariance
of conformally coupled scalar fields. Even more, for $\r=0$, if the action of the field
$\phi$ depends just on $\hat g_{\m\n}$, the regularized action is invariant under
the transformation (\ref{generalWtr}) for $f=\phi$.
Finally, for $\r=1$, we recover the mass term (\ref{massterm}).\\

The previous regularization procedure makes one expect the violation
of the Ward identities related to the WTDiff or Diff symmetries at the quantum level.
As an example, if all the fields
 are {\em conformally coupled} (including the PV fields, except for the mass term),
we expect  the expectation value of the energy-momentum tensor to
behave as\footnote{Other regularization methods, such as
point-splitting yield similar violations of the Ward identities \cite{BirrellDavies}
(see also \cite{Guadagnini:1988ck}).
Besides, the previous expectation values do not satisfy all of the Wald's axioms.
There is no problem with this, as in TDiff invariant theories the
energy-momentum tensor is not necessarily conserved. }
\be
\label{Ward}
 \nabla^\m \langle T_{\m\n}\rangle\sim p[(1-\r)]\hbar \pd_\n A ,
  \quad   g^{\m\n} \langle
T_{\m\n}\rangle\sim q[(n-2-\r n)]\hbar B,
\ee
for some scalar fields $A$ and $B$ and functions $p[x]$ and $q[x]$ satisfying
$p[0]=q[0]=0$.\\

Furthermore, the allowed counterterms required to absorb the infinities of the
regularization process depend on the value of $\r$. For a generic $\r$, the possible
counterterms will be higher order terms invariant under TDiff\footnote{Similarly,
 in the general analysis
of possible counteterms of \cite{Deser:1976yx}, the possibilities which are
WDiff invariant in
four dimensions but otherwise WTDiff invariant were not considered.}. If the
 symmetry group preserved by the regularization
is enlarged, the possible counterterms will be fewer. For the WTDiff
preserving scheme, following \cite{Buchmuller:1988wx},
we expect those to correspond to
powers of
\be
\label{WTDcoun}
R_{\phantom{\a}\s\b\n}^{\a}[ g^{-1/n}g_{\m\n}].
\ee
If this is so, the arguments of Chapter \ref{chapternl} still apply
and the equations of motion coming from the renormalized Diff or WTDiff
theories are equivalent. In fact, it is not hard to  argue that the
 Diff and WTDiff invariant theories are equivalent at the semiclassical level. To do it,
let us
consider a regularization scheme that yields an energy-momentum tensor satisfying
\be
\nabla^\m \langle T_{\m\n}\rangle=  \pd_\n A ,
  \quad   g^{\m\n} \langle
T_{\m\n}\rangle=0.
\ee
We can define a {\em covariantly conserved} quantity, $\tilde T_{\m\n}=T_{\m\n}-g_{\m\n}A$,
which satisfies
\be
\nabla^\m \langle \tilde T_{\m\n}\rangle=  0 ,
  \quad   g^{\m\n} \langle\tilde
T_{\m\n}\rangle=n A.
\ee
This last energy-momentum tensor, $\tilde T_{\m\n}$, can be used as a source for the Diff
invariant theory and one can easily see that these equations of motion are equivalent
to those of WTDiff invariant theory with the source $T_{\m\n}$ except for an integration constant.
This also implies that the function $A$ corresponds to the conformal anomaly except for
an integration constant\footnote{A more direct way of verifying this equivalence is by
using dimensional regularization and the counterterms
 consistent with WTDiff, (\ref{WTDcoun}). One can readily see how the conformal anomaly is translated
into a Diff anomaly in this case, with still the same physical predictions as the Diff
invariant case \cite{Blas08b}.}
It is interesting to compare this situation with other cases where
two symmetries are not compatible at the quantum level, as for the $V-A$ anomaly (see,{\em e.g.}
 \cite{Bertlmann}). Usually, the regularization schemes which are compatible
 with different symmetries yield different physical predictions and the
only way to tell which of these theories describes Nature
is performing experiments \cite{Bertlmann}.\\

When one adopts Diff preserving regularization
and renormalization schemes, one finds the same result
as in  (\ref{Ward}) for $\r=1$. This yields the celebrated  {\em conformal anomaly}
\cite{Capper:1974ic,Duff:1993wm}. Before closing this section,
we would like to present a simple calculation at linear order in the
perturbations of the metric where (as argued in the
previous paragraphs) this anomaly can be traded by an anomaly
that breaks the Diff to TDiff. To show it,
we will find a local counterterm which, once added to the action,
changes the anomaly
from one current to the other one. Thus, we want $\Delta S_c$ such that
\be
g^{\m\n}\frac{\delta \Delta S_c}{\d g^{\m\n}}=-g^{\m\n}\langle T_{\m\n} \rangle.
\ee
This term will break the Diff to TDiff and also the Weyl symmetry in such a way that we
recover the Weyl invariance. We will make the computation in  $1+1$ dimensions
where \cite{Bertlmann},
\be
\langle T\rangle=\frac{1}{24\pi} R.
\ee
The first thing that we notice is that Einstein's equations are traceless in two dimensions. Thus,
the Einstein-Hilbert action is not an appropriate counterterm. Recall also that, at linear level,
\be
R=\pd^\m \pd^\n h_{\m\n}-\Box h,
\ee
and that the most general TDiff action with two derivatives is (see Chapter \ref{chapterLorentz}),
\be
\Delta S_c^{L}=\frac{1}{4}\pd^\m h_{\a\b}\pd_\m h^{\a\b}-\frac{1}{2}\pd^\m h_{\m\b}\pd^\n h_{\n}^\b
+\frac{a}{2}\pd^\m h_{\m\b}\pd^\b h-\frac{b}{4}\pd_\m h \pd^\m h.
\ee
For arbitrary $a$ and
$b$ one gets,
\be
\frac{\d \Delta S_c^L}{\d h}=\left(b-\frac{(a+1)}{2}\right)\Box h+(1-a)\pd^\m \pd^\n h_{\m\n},
\ee
which means that for $a=0$, $b=-\frac{1}{2}$ we get the desired counterterm. The addition of
this
counterterm to the action breaks Diff to TDiff and the conservation law for the energy-momentum
tensor is modified by
\be
\pd^\m \frac{\d \Delta S_c^L}{\d h^{\m\n}}=\frac{1}{2}\pd_\m\left((1-b)\pd^\r\pd^\s h_{\r\s}
+(b-a)\Box h\right).
\ee
The non-linear extension together with the application
to other dimensions is left for future research \cite{Blas08}(see also \cite{Guadagnini:1988ck}).\\

%%%%%%%%%%%%%%%%%%%%%%%%%%%%%%%%%%%%%%%%%%%%%%%%%%%%%%%%%%%%%%%%%
\section{BRST Invariance}
%%%%%%%%%%%%%%%%%%%%%%%%%%%%%%%%%%%%%%%%%%%%%%%%%%%%%%%

Once the perturbations of the gravitational
field are considered as quantum fields, it is of the uttermost importance
to check the unitarity of the model.
 A first step in this direction for the TDiff invariant theories was taken in \cite{Kreuzer:1989ec,Buchmuller:1988yn,Dragon:1988qf}
 (see also \cite{Alvarez:2006bx})
where the BRST-anti BRST structure of the TDiff invariant theory was studied. The existence
of the nilpotent BRST transformation is assumed as a necessary condition for the theory to be
well-defined\footnote{The BRST transformation could also be nilpotent except for
a gauge transformation but we will not consider this possibility here.}.
  An important
difference between the gauge invariance of GR and the gauge invariance of TDiff and WTDiff is that
for the TDiff and WTDiff cases, the gauge invariance is \emph{reducible} \cite{HenneauxTeit},
\emph{i.e.}, the parameters
of the gauge transformation are not completely free, but satisfy the condition
$$\pd^\m \xi_\m=0.$$
This  makes the covariant quantization of the theory more involved.
First, as we already noticed in Chapter \ref{chapterLorentz}, the covariant
gauge fixing is a bit more complicated for the TDiff case.
Even worse, the action for the Fadeev-Popov ghosts fields
will have a gauge invariance. This new gauge invariance must be gauge fixed, which implies the
introduction of new ghosts for ghosts  whose action can also have
a gauge invariance \cite{HenneauxTeit}. The appearance of these ghosts for ghosts is not so exotic as it may seem
and they also appear in the quantization of forms. In our case, we will see that the BRST
algebra can be constructed with a finite number of ghosts. The next steps would be to
check the unitarity of the theory, calculate
the gauge fixed Lagrangian and perform a one-loop calculation, but work in this direction
is still in progress (see \cite{Gabadadze:2005uq} for some comments in the equivalence
of GR and TDiff at the loop level) \cite{Blas08}.
An interesting possibility would be to see whether the formalism in
\cite{Bern:2002kj} can be extended to the TDiff and WTDiff cases.\\

We refer to the standard books in QFT for an introduction to BRST symmetry
(see {\em e.g.} \cite{HenneauxTeit} for a monograph an \cite{DeJonghe:1993zc}
for a enlightening review).
The BRST structure of the TDiff gauge invariance has been recently reconsidered in \cite{Alvarez:2006bx}
which we will follow closely
(for a BRST-anti-BRST  formulation
see also \cite{Dragon:1988qf,Kreuzer:1989ec} where a gauge fixed action can also be found).
 Concerning the BRST analysis of Diff gauge theories
it can be found in \cite{Stelle:1976gc,Delbourgo:1976xd} (see also \cite{Kugo:1978rj,Latorre:1987nu}).
 The algebraic structure
that we are going to consider at the non-linear level was studied in
Chapter \ref{chapternl}. It is summarized by the transformation,
\be
\delta^{C}_{\xi,\phi} g_{\m\n}=2 \nabla_{(\m}\xi_{\n)}+\frac{2}{n}\phi g_{\m\n},
\ee
which yields a commutator
\be
[\delta^{C}_{\xi_1,\phi_1},\delta^{C}_{\xi_2,\phi_2}]=
\delta^{C}_{[\xi_1,\xi_2], \left(\xi_1^\m \pd_\m \phi_2- \xi_2^\m \pd_\m \phi_1\right)}
\ee
where $\pd_\m \xi_i^\m=0$. Notice that given two
transverse vector modes, its commutator is also transverse.
The transverse condition for the gauge parameter implies that the
ghosts fields related to this symmetry will also be transverse. More explicitly,
the BRST transformation for the metric is\footnote{We will denote
the BRST transformation of a field $\psi$ by $s\psi$.}
\be
s g_{\m\n}=c_W g_{\m\n}+c^\r \pd_\r g_{\m\n}+
g_{\a(\m}\pd_{\n)}c^\a
\ee
where $c_W$ and $c^\n$ are anticommuting variables of ghostnumber equal to one
\be
\{c^\a, c^\b\}=\{c^\g,c_W\}=0,
\ee
 and $c^\m$ satisfies
\be
\label{ghostcond}
\pd_\m c^\m=0.
\ee
In the language of forms, we can write
\be
\d c_1=0,
\ee
where $\d=(-1)^{n(k-1)}* \mathrm d *$ is the adjoint operator of the exterior derivation of a
$k$-form in $n$
dimensions using
the Hodge star associated to the Minkowski metric\footnote{We follow the
conventions of \cite{Ortin}.}
and
$$c_1=c_\m \mathrm d x^\m,$$
is a ghostly form with components $c_\m=\eta_{\m\n}c^\n$. If we want to impose (\ref{ghostcond}) in a local and covariant way,  we can write $c_1$ as
\be
\label{deltac1}
c_1=\d c_2
\ee
where $c_2$ is a ghostly, Grassmann odd 2-form. Notice, however, that
$c_2$ is not determined by the previous condition. In particular, the
addition of a term $\d C_3$ does not change $c_1$. This new
 invariance appears also in the Lagrangian
and more fields are required to completely fix the gauge \cite{HenneauxTeit}.
 Remember that the
BRST transformation must satisfy the following conditions
\be
s^2=0, \quad s(AB)=(s A)B+(-1)^{g_A} A(s B),
\ee
where $g_A$ is the ghost number of $A$, and that it increases the ghost number by one, \emph{i.e.}
$g_{sA}=g_A+1$.
Nilpotency of the operator $s$ acting on the metric implies
\be
s c^\a= c^\r\pd_\r c^\a, \quad s c_W=c^\r \pd_\r c_W,
\ee
which can be written as
\be
s c_1=\frac{(-1)^n}{2}\d\left(c_1 \wedge c_1\right), \quad s c_W=(-1)^n\d \left(  c_1c_W\right),
\ee
where we treat $c_W$ as a ghost function. Recall also that $c^\m$ are Grassmann numbers which
in particular means that $c_\m c_\n$ is antisymmetric. From (\ref{deltac1}),
\be
s c_2=\frac{(-1)^n}{2}\left(c_1 \wedge c_1\right)-\d c_3.
\ee
Imposing again the nilpotency of $s$ on $c_2$ this means that
\be
\label{c3}
s c_3=\frac{(-1)^n}{3!} c_1\wedge c_1\wedge c_1-\d c_4.
\ee
If we can find $c_3$ and $c_4$ within the fields which we have already introduced
such that (\ref{c3}) is satisfied, thus we have constructed a closed BRST system.
The  BRST transformation of the field $c_W$ involves $c_W$
itself, which means that neither it nor its BRST transformation can be used to build expressions involving just
$c_1$. This means that the first term in the r.h.s. of (\ref{c3}) should come from terms
involving just $c_2$, and this is not possible. Thus, we need to add a new field $c_3$ to the
theory which transforms as (\ref{c3}) under BRST transformations.
By requiring nilpotency again, this process continues and we find
\be
s c_m=\frac{(-1)^n}{m!}\underbrace{c_1\wedge...\wedge c_1}_m -\delta c_{m+1},
\ee
for $m< n$. When we arrive to a form of maximum rank, its BRST
transformation will be given by
\be
s c_n=\frac{(-1)^n}{n!}\underbrace{c_1\wedge...\wedge c_1}_n,
\ee
and nilpotency follows directly, as applying again $s$ to $c_n$ we get a $n+1$ form which cancels.
Thus, for arbitrary space-time dimension $n$, we need $2^n-(n+1)$ ghosts to close
the BRST transformations which can be organized as shown in Table \ref{table1}.
\begin{table}[h!]
\centering
\begin{tabular}{|l|c|l|c|}
\hline
F&dim& $g$& G\\
\hline
$c_2$&$\binom{n}{2}$&$1$&$-1$\\
...&...&...&...\\
$c_m$&$\binom{n}{m}$&$m-1$&$(-1)^{m+1}$\\
...&...&...&...\\
$c_n$&$1$&$n-1$&$(-1)^{n+1}$\\
\hline
\end{tabular}
\caption{\small{Ghost fields that appear in the BRST transfomation proposed in the text.
 F stands for the form, dim is the number of independent components, $g$ is the ghost
number and G stands for the Grassmannian character of the fields.}} \label{table1}
\end{table}

Regarding the BRST transformation for the field $c_W$, it is already nilpotent
and we do not need to add more ghosts to the system.
Despite all this apparent complication, if we impose an appropriate non-covariant
gauge fixing condition, these ghost for ghosts can be decoupled, \emph{i.e.}, any reducible theory
can be recast into a irreducible theory by using appropriate independent gauge generator. However,
this can yield the loss of Lorentz covariance or space-time locality.\\

Concerning the antighosts, they are added as trivial pairs of antighosts satisfying
\be
b_1=b_\m \di x^n\equiv \di b_2, \quad b_W
\ee
and
\bea
&&s b_2=B_2, \quad s b_W=B_W,\nonumber\\
&&s B_2=0, \quad s B_W=0.
\eea
Once we have found the previous BRST system, we can look for a gauge fixed action.
For the BRST-anti-BRST system it was already found in \cite{Kreuzer:1989ec}. The
knowledge of this action allows to prove unitarity and to make
calculations at 1-loop level which can differ from the usual calculations of GR.
Fortunately, the ghosts for ghosts do not appear at 1-loop, which means
that the calculation is not so different from that of GR.  We think that
this is a very interesting project but
\emph{Ars lunga, vita brevis}.

%%%%%%%%%%%%%%%%%%%%%%%%%%%%%%%%%%%%%%%%%%%%%%%%%%%%%%%%%%%%%%%%%
\section{Euclidean Quantum Gravity}
%%%%%%%%%%%%%%%%%%%%%%%%%%%%%%%%%%%%%%%%%%%%%%%%%%%%%%%

Finally, some words are in order about another approach to quantum gravity which can be extended
 to the TDiff or WTDiff cases, Euclidean Quantum Gravity (EQG) (see \cite{Hawking:1980gf}
for a review). This formulation is based on the application of the
 path integral approach of field theory
to GR. One of the difficulties it meets is the fact that, in contrast to what happens
for the Standard Model, the action of the Euclidean continuation of the theory is not bounded from below.
The standard
way to prove this is as follows. Given any metric $g_{\m\n}$, we introduce a new metric related by a
 Weyl transformation to the first metric
$$\tilde g_{\m\n}=e^{2\s}g_{\m\n}.$$
For any metric $g_{\m\n}$,
one can prove that the action of the new metric can be made arbitrarily small by the choice of an
appropriate $\s$.

The TDiff generalization allowed for more general Lagrangians which
modify the action of the conformal mode $\s$. In particular,
this mode can be made well behaved for certain TDiff Lagrangians  \cite{vanderBij:1982}.\\

Concerning the WTDiff case, the fact of dealing with a unique Weyl invariant Lagrangian
means that the previous Weyl transformation does not change the action, and thus
the action has a chance to be bounded from below. However, one can show
that also for the WTDiff case there is a transformation which \emph{mutatis mutandis}
has the same effect as the Weyl transformation and renders the action unbounded
from below. To see it, let us  choose a
foliation of the space-time into space and time $M=\mathbb{R}\times \Sigma_t$ which allows
to decompose (at least locally) any metric as
\bea
\di s^2&=&g_{\m\n}\di x^\n \di x^\m=(N^2-N_j N^j)\di t^2-2N_j \di
x^j \di t-\gamma_{ij}\di x^i
\di x^j,
\eea
where
$N^j=\gamma^{ij}N_j. $
Let us choose the  Wick
rotation
$t\mapsto -\I \tau$. To get a real metric, we must also
Wick rotate the shift fields $N_j\mapsto -\I \tilde N_j$, which also ensures
the negative definiteness of the new Euclidean metric,
\be
\di s^2_E=g^E_{\m\n}\di x_E^\n \di x_E^\m=-(N^2+\tilde N_j
\tilde N^j)\di \tau^2-2\tilde
 N_j \di x^j \di \tau-\gamma_{ij}\di x^i
\di x^j .\ee
The Euclidean version of the WTDiff action will
be\footnote{The sign convention is such that
the linearized action around Minkowski has no ghosts.
Note that we are forgetting about the Gibbons-Hawking boundary term,
which from the usual arguments of GR can be found to be
\be {\mathcal S}_{GH}^{WT}=-\frac{1}{\kappa^{n-2}}\int_{\pd M} \di^{n-1} x
\sqrt{\pm h}  K[\hat g_{\m\n}] .\ee}
\be
{\mathcal
S}_E^{WT}=\frac{1}{2\kappa^{n-2}}\int  \di^n x_E (\hat g^E)^{\m\n}
R_E(\hat g_{\a\b}^E)_{\m\n}.
\ee
As shown in (\ref{nlw}), this action can also be written as\footnote{We will drop the index $E$
that indicates that we are dealing with the Euclidean extension.}
\be
\label{explicitWTaction}
{\mathcal
S}_E^{WT}[g]=\frac{1}{2\kappa^{n-2}}\int \di^n x\sqrt{g}g^{\frac{2-n}{2n}}\left(R
+\frac{(n-1)(n-2)}{4n^2} g^{\m\n}\pd_\m \ln g \pd_\n \ln g\right).
\ee
To show that this action is not bounded from below,
let us consider a generic Euclidean metric $g_{\m\n}$ and build
another metric related to it by a Diff. which does not
belong to WTDiff. That is,
\be \tilde  g_{\m\n}= \frac{\pd x^\r}{\pd
y^\m}\frac{\pd x^\s}{\pd y^\n}g_{\r\s},
\ee
with
\be
\tilde g_{\m\n}\neq  \Omega^2 g_{\m\n}
, \quad  J=\det\left(\frac{\pd x^\r}{\pd
y^\m}\right)\neq 1 .\ee
We can consider, for instance, the transformation
\be
x^0\mapsto y^0=f(x^0) \quad x^i \mapsto y^i=x^i,
\ee
which has $J=\pd_0 f$. As this transformation corresponds to
a change of coordinates, the first term in (\ref{explicitWTaction})
will change with a power of $J$ whereas the second term
will involve derivatives
of the Jacobian. More concretely,
\ba
{\mathcal
S}_E^{WT}[\tilde g]=\frac{1}{2\kappa^{n-2}}&&\int \di^n y \sqrt{g} (J^2g)^{\frac{2-n}{2n}}\nonumber\\
&&\left(R+\frac{(n-1)(n-2)}{4n^2} \frac{\pd y^\m}{\pd
x^\r}\frac{\pd y^\n}{\pd x^\s}g^{\r\s}\pd_\m \ln (J^2 g)\pd_\m\ln (J^2 g)\right).\nonumber
\ea
The previous action has a term
\be
\int \di^n x g^{00}\pd_0 J\pd_0 J,
\ee
and thus, for a Jacobian that varies fast enough, the previous action can be made arbitrarily negative
(remember that $g_{\m\n}$ is negative definite).

The cosmological constant is treated
differently in the
EQG formulation of Diff and WTDiff invariant theories \cite{Ng:1990xz}.
Whereas in the Diff invariant case it is a parameter of the action of the
theory,
in the WTDiff invariant theory it is
an integration constant and the path integral formulation should include
{\em all} the possible values for it.  This seems to select a small cosmological
constant \cite{Ng:1990xz} (see also \cite{Unruh:1988in}).

%%%%%%%%%%%%%%%%%%%%%%%%%%%%%%%%%%%%%%%%%%%%%%%%%%%%%%%%%%%%%%%%%%
\chapter{Further Aspects of Unimodular Gravity and Bigravity}\label{AppendixBig}
%%%%%%%%%%%%%%%%%%%%%%%%%%%%%%%%%%%%%%%%%%%%%%%%%%%%%%%%

In this Appendix we will first
 study  some formal aspects
related to the gauge invariance in
TDiff invariant theories and the integration of tensor densities.
Besides, we present some technical work about some bigravity solutions
which appeared in \cite{Blas:2005yk}. Finally, we present some general methods
to generate solutions of bigravity from solutions of GR.

%%%%%%%%%%%%%%%%%%%%%%%%%%%%%%%%%%%%%%%%%%
\section{Comments on Gauge Issues and Fixed Volume Manifolds}\label{subsectiongauge}
%%%%%%%%%%%%%%%%%%%%%%%%%%%%%%%%%%%%%%%%%%

Once we assumed that the gauge invariance of our theory is not the whole
group of Diff but a subgroup of it (namely TDiff), we must reconsider many results which are well
established in GR. We will devote this section to study some of them.

It is also interesting to note that the TDiff and WTDiff theories can be understood as a restriction
of the general metric-affine gauge theories of \cite{Hehl:1994ue} where
the local translations are restricted to be transverse and
the Weyl transformation of the $GL(n,\mathbb{R})$ group acts only in the
vielbein.\\

Let us briefly discuss some global aspects of Diff and TDiff theories.
Recall that  the
EoM for $\hat g_{\m\n}$ of WTDiff coincide with those for $g_{\m\n}$ of GR in the gauge
 $|g|=1$, which is attainable locally in both
theories. Thus,
\emph{any} solution $g_{\m\n}$
of GR is also a solution $\hat g_{\m\n}$ of WTDiff with the same matter content in
this gauge\footnote{For
 globally non-trivial solutions of GR, we can always relate them to WTDiff
invariant solutions. To do this, it is enough to restrict to
a manifold with two patches (the generalization
to other situations is trivial). Let us consider a solution built out of the two
metrics $g_{\m\n}^{1}$, $g_{\m\n}^{2}$ defined in the first and second patch respectively.
We can now perform a Diff such that the new metrics satisfy $|g^{i}|=1$.
Both metrics will be related in the intersection of the patches
by a Diff belonging to TDiff in these coordinates. Thus,
 the globally defined
$\hat g_{\m\n}$ will be a solution of WTDiff (see also \cite{Ng:1990xz}).}.
However, when the field
 $\hat g_{\m\n}$ is transformed under a general Diff it
 is no longer a solution of the transformed  EoM.
 The message that we want to transmit is
that even if the spaces of solutions of GR and WTDiff  coincide in the gauge $|g|=1$,
the different families of gauge equivalent metrics are different. In GR,
two metrics related by a Diff transformation are considered as equivalent
and if one is a solution of the EoM, the other metric is also a solution in the
transformed coordinates
\cite{Will}. In the WTDiff theory, the equivalent solutions  are related by a TDiff
or a Weyl transformation. An immediate
consequence is that the condition for a metric to be equivalent to Minkowski in the WTDiff
theory is no longer
that its Riemann tensor cancels. Instead, a metric will be flat whenever
\be
g_{\m\n}=e^{\phi(x)}\frac{\pd y^\a}{\pd x^\m}\frac{\pd y^\b}{\pd x^\n}\eta_{\a\b},
\ee
with $\det\left[\frac{\pd y^\a}{\pd x^\m}\right]=1$. The determinant of $g_{\m\n}$ will be free
and determined by $\phi(x)$, whereas $\hat g_{\m\n}$ will be related to $\eta_{\m\n}$ by a
TDiff transformation. Thus, the condition for a metric $g_{\m\n}$ to be equivalent to Minkowski
is
$$
R^{\a}_{\phantom{\a}\m\n\r}[\hat g_{\s\t}]=0.$$
The difference between the equivalence classes of solutions of both theories may also
 imply differences
 when one considers the gauge fixing procedure and the definition of
observables in the quantum theory \cite{Unruh:1988in} (see also
Appendix \ref{AppendixQ}).\\

The restriction in the group of symmetry means the possibility of
building quantities which are invariant under the subgroup under
study but not under the original group\footnote{The
invariance under the whole Diff group can always
be recovered after the introduction of an additional spurious field
in the spirit of the St\"uckelberg field \cite{Alvarez:2007nn,ArkaniHamed:2002sp}.}.
In particular, for the TDiff case, the integration of
{\em densities} of any {\em weight}\footnote{We will define a {\em tensor density
of weight w} as an object $T(x)\in T(M)^p\otimes T^*(M)^n$ which under a general diffeomorphism $y(x)$
transforms
as
$$ T'(y)=\left|\det\left[\frac{\pd y^\a}{\pd x^\m}\right]\right|^{w}T(x).$$}
is a well defined operation as we are going
to see in the rest of this section.

The definition of integration of \emph{form  densities of weight
w} in paracompact oriented manifolds proceeds as the usual construction
for {\em forms} (see, {\em e.g.}
\cite{Wald}). Remember that for a $n$-\emph{form} $\alpha$
in a $n$-dimensional orientable paracompact manifold $M$ we choose
an orientation $\epsilon $  and a covering $\{O_i\}$ of $M$ and
define the integral (with respect to the orientation) as \be
\int_M \alpha=\sum_i \int_{O_i} f_i  \alpha, \ee
 where $\{f_i\}$ is
a partition of the unity subordinate to the covering and the
integration in every open is defined as usual. It can be shown
that the result does not depend neither on $\{O_i\}$ nor on
$\{f_i\}$ (but it depends on the orientation). Now, besides the
orientation we will choose also a \emph{transverse class}, that
is, in every open $O_j$ of the covering we choose a class of
frames related by transformations with a unit Jacobian (notice
that this defines an equivalence relation). Given two open sets
$O_i$ and $O_j$, we say that their classes are compatible if in
$O_i\bigcap O_j$ they are related by a transformation of unit
Jacobian. If we can choose \emph{transverse classes}  on $M$ such
that in $O_i\bigcap O_j$  the classes are compatible $\forall i,j$,
we say that $M$
 is a {\em transverse manifold}. Clearly, a non-orientable manifold
is always non-transverse. Besides,
through a continuous coordinate transformation in $O_i$ we can make the Jacobian to
take any value in the intersection $O_i\bigcap O_j$. In particular, this means that
 every \emph{orientable} manifold is
\emph{transverse} and thus both concepts coincide even if not
every {\em atlas} corresponds to a {\em transverse class}.  Given a
{\em transverse class} $t$ and an {\em orientation} we define the integral of
a $n$-\emph{form density} $\alpha$ over the manifold $M$ as \be
\int_{\{M,t\}} \alpha=\sum_i \int_{O_i} f_i  \alpha(t), \ee where
$\{f_i\}$ is again a partition of the unity and \be \int_{O_i} f_i
\alpha(t)=\int_{\phi_i(O_i)} f_i \alpha_{1...n}\di x_t^1\cdots \di
x_t^n ,
\ee
 where $\alpha_{1...n}$ is the component of $\alpha$ with
respect to the basis $\{x_t\}$, which must belong to the
transverse class\footnote{Indeed, this definition of integration
is valid for every object which transforms as a $n$-form within
the transverse class.}. Clearly, this definition not only depends
on the orientation but also on the {\em transverse class}. It is easy to
prove that this definition does not depend neither on the
partition nor on the covering while we stay  in the transverse
class.  We can also define the external calculus in the
usual way \cite{Wald}. Given a $n$-form
density $\alpha$ of weight $w$ \be
\alpha=\alpha_{\m_1\cdots\m_n}\di x^{\m_1}\wedge \cdots\wedge \di
x^{\m_n} ,\ee
 we define its exterior derivative as \be \di
\alpha=(\partial_\rho \alpha_{\m_1\cdots\m_n}\di x^\rho) \wedge \di
x^{\m_1}\wedge \cdots\wedge \di x^{\m_n} .\ee
In other coordinates,
we may write \ba \di \alpha&=&(\partial_{\rho'}
\alpha_{\m_1'\cdots\m_n'}\di x^{\rho'})\wedge \di x^{\m_1'}\wedge
 \cdots\wedge \di x^{\m_n'}=\nonumber  \\
&&\partial_{\rho'}\left(\Big|\frac{\partial x'}{\partial
x}\Big|^{w/2}
\partial_{\m_1'} x^{\m_1}\cdots
\partial_{\m_n'} x^{\m_n}\alpha_{\m_1\cdots\m_n}\right)\di x^{\rho'}\wedge \di x^{\m_1'}\wedge
 \cdots\wedge \di x^{\m_n'}=\nonumber\\
&&\Big|\frac{\partial x'}{\partial x}\Big|^{w/2} (\partial_\rho
\alpha_{\m_1\cdots\m_n}\di x^\rho) \di x^{\m_1}\wedge \cdots\wedge
\di x^{\m_n} +\partial_{\rho'}\left(\Big|\frac{\partial
x'}{\partial x}\Big|^{w/2}\right)\alpha,
 \ea
  and thus, the
operation is well defined only within the transverse classes and
this allows us to define the integration of the exterior
derivative of a form density, always inside a particular class.
Given a manifold $M$ of dimension $n$ and a embedded submanifold
$S$ of dimension $m$,
 once we choose a transverse class $t$ on $M$, by restricting
to $S$ we define a transverse class on $S$. To show it, take two
different systems of coordinates in the same class $\{x^t_\m\}$
and $\{x'^t_\m\}$. Given a embedded oriented submanifold $S$ there
exists a one to one map $\phi: S \rightarrow \phi(S)\subset M$.
Now consider the following diagram
\begin{displaymath}
\xymatrix{ S \ar[r]^\phi \ar[d]_{\{Q_j\}}& M \ar[d]^{\{O_i\}}\\
           \mathbb{R}^m&\mathbb{R}^n}
\end{displaymath}
where $\{Q_j\}$ and $\{O_i\}$ are open coverings of $S$ and $M$
respectively. In the intersection $Q_j \bigcap \phi^{-1}(O_i)$, we
may express the coordinates on $S$ in this open as \be
(y^1(x^1,...,x^n),...,y^m(x^1,...,x^n)), \ee where $\{x^j\}$ are
the coordinates of $M$ in the open $O_i$. If we consider another
open $O_l$ such that $Q_j \bigcap \phi^{-1}(O_i\bigcap O_j)\neq 0$
and that belongs to the same transverse class as $O_i$, the
coordinates on $Q_j$ are defined as \be
(y^1(x^1(x')^1,...,x^n(x')),...,y^m(x^1(x'),...,x^n(x')))=
(y'^1(x'^1,...,x'^n),...,y'^m(x'^1,...,x'^n)). \ee If we now
calculate the Jacobian of the transformation from one coordinates
to the other ones, \be \det \frac{\partial y'^\mu}{\partial
y^\nu}= \det \frac{\partial y'^\mu}{\partial x'^\alpha}
\frac{\partial x'^\alpha}{\partial x^\beta} \frac{\partial
x^\beta}{\partial y^\nu}=\det \frac{\partial y'^\mu}{\partial
x'^\alpha} \det \frac{\partial x^\beta}{\partial y^\nu}=1, \ee
where we have used the fact that \be \det \frac{\partial
x^\beta}{\partial x'^\nu}=1 \quad \Rightarrow\quad \det
\frac{\partial y'^\beta}{\partial x'^\nu}= \det \frac{\partial
y^\beta}{\partial x^\alpha} .\ee
Thus we see that every transverse class on $M$ induces a transverse class on $S$.\\

As the derivation and the integration within each
transverse class coincide with the usual definitions for forms,
 the Stokes' theorem holds also within these classes, \emph{i.e.}
\be \int_{\{M,t\}} \di \alpha=\int_{\{\partial M,\phi(t)\}} \alpha.
\ee
where $\phi(t)$ is the transverse class induced on $\partial M$ by $t$.\\

The exterior derivative we have defined  is only meaningful
within {\em transverse classes}  and only within those does
it defines a $n+1$-density form from a $n$-density form. We may now
add more structure to the manifold in order to define a derivative
operator which after acting on tensor densities yields tensor
densities. To this end, we introduce a connection
$\Gamma_{\phantom{s}\mu\rho}^{\sigma}$ on the manifold.
From the fact that $\Gamma_{\phantom{\a}\mu\alpha}^{\alpha}=\partial_\mu \ln \sqrt{-g}$
transforms as a vector under transformations inside the transverse class
({\em i.e.} under TDiff), we have more freedom to choose the covariant derivative of
tensor densities.

Let us consider two possibilities. First,  we may define the covariant derivative
of tensor densities as the usual covariant derivative independently of
the weights, namely, for $f$, $v$, $\omega$ a scalar, vector and
covector of weights $w_f$, $w_v$ and $w_\omega$ respectively, we
define
\ba \nabla_\mu f=\partial_\mu f, \quad \nabla_\mu
v^{\nu}=\partial_\mu v^\nu+\Gamma_{\phantom{\n}\mu\alpha}^{\nu} v^{\alpha},
 \quad \nabla_\mu \omega_{\nu}=\partial_\mu \omega_\nu-\Gamma_{\phantom{\a}\mu\nu}^{\alpha} \omega_{\alpha},
\ea
and using the Leibnitz property, extend the definition to every tensor
density. This definition, as the exterior derivative, is well
defined only within each {\em transverse class}.

As a second possibility, for  $T$
 a tensor density of weight $w_T$, we can define a derivative\footnote{As
 we said, from the fact that $\Gamma_{\phantom{\a}\mu\alpha}^{\alpha}$
 behaves as a vector for the TDiff subgroup, we could consider and arbitrary
value for $w_T$ in this expression.}
operator \cite{Ortin}
\ba
\label{wcovariant}\nabla_\mu^w T=\nabla_\m T
 +w_T\Gamma_{\phantom{\a}\mu\alpha}^{\alpha}T.
\ea
 Since
 \be
 \label{gammaex}
 \Gamma_{\phantom{\a}\mu\alpha}^{\alpha}=\partial_\mu \ln \sqrt{-g}
 \ee
 and $g$ is a scalar density of weight $-2$, the previous covariant
 derivative preserves the {\em weight} of the tensor $T$ under the whole Diff. The
curvature of both derivations coincide.  Notice that, as
$\nabla_\m$ is not a well defined operator over the tensor densities, both derivations
differ by a term which is not an antisymmetric tensor density
field. In particular, this means that the expression of the
exterior derivative in terms of the derivation $\nabla^\omega$
will be given by
\ba
\di \alpha&=&(\nabla_{\rho}
\alpha_{\m_1\cdots\m_n}\di x^{\rho})\wedge \di x^{\m_1}\wedge
 \cdots\wedge \di x^{\m_n}=\nonumber\\
&& (\nabla^w_{\rho} \alpha_{\m_1\cdots\m_n}\di x^{\rho})\wedge \di
x^{\m_1}\wedge
 \cdots\wedge \di x^{\m_n}-w_\alpha\Gamma_{\phantom{\a}\rho\nu}^{\nu}\alpha
 _{\m_1\cdots\m_n}\di x^{\rho}\wedge \di x^{\m_1}\wedge
 \cdots\wedge \di x^{\m_n}\nonumber.
\ea
Finally, let us express the Stokes' theorem in terms of these
operators. We will use the terminology of \cite{Wald}.
For a vector density $v^\mu$ of
weight $w_v$ in a metric manifold we can construct the form
density of the same weight
\be
\alpha_{\mu_1\cdots\mu_{n-1}}=\epsilon_{\mu\mu_1\cdots\mu_{n-1}}
v^\mu ,
\ee
where $\epsilon_{\mu\mu_1\cdots\mu_{n-1}}$ is the volume
element associated to the metric. We can easily prove that
\be \di
\alpha=\nabla_\mu v^\mu \epsilon=(\nabla^w_\mu v^\mu+w_v
\Gamma_{\phantom{\r}\rho\mu}^{\rho} v^\mu)\epsilon,
\ee
which
from Stokes' theorem means that
\be
\int_{\{M,t\}} \nabla_\mu^w
v^\mu \epsilon=\int_{\{\partial M, \phi(t)\}} n_\mu v^\mu \tilde
\epsilon +w_v\int_{\{M,t\}} \Gamma_{\rho\mu}^{\phantom
{\rho\m}\rho} v^\mu\epsilon,
\ee
where $\tilde
\epsilon$ is the volume element induced in $\delta M$.
Notice that for the metric, as for any tensor, both derivative operators
coincide which in particular means that the metric is compatible with
both operators. This does not happen for arbitrary $w_T$ in
(\ref{wcovariant}). Besides,
\be
\label{compatib}
\nabla^w_\mu
g=\partial_\mu g-2\Gamma_{\phantom{\r}\mu\rho}^{\rho}g=0.
\ee
Finally, let us see the implications of the previous
results for partial
integration. We will proceed in parallel with both derivative
operators. Consider two tensor densities $n$ and $m$ of ranks
$(p_n,q_n)$, $(p_m,q_m)$ and eights $w_n$ and $w_m$. If
$p_n+q_n-1=p_m+q_m=N$ we can saturate indexes of these quantities
and build a scalar vector of weight $w_m+w_n$. For $\nabla$ any derivative operator
$\nabla$, one finds
\ba
&&\int_{\{M,t\}}
m_{\mu_1\cdots\mu_N} \nabla_\alpha
n^{\alpha\mu_1\cdots\mu_N}\epsilon= \nonumber\\
&&\int_{\{M,t\}}
n^{\alpha\mu_1\cdots\mu_N}\nabla_\alpha
m_{\mu_1\cdots\mu_N}\epsilon - \int_{\{M,t\}}
\nabla_\alpha(n^{\alpha\mu_1\cdots\mu_N}
m_{\mu_1\cdots\mu_N}\epsilon) .
\ea
As an example, let us consider
the integral \be \int_{\{M,t\}} f(g) \nabla_\alpha v^\alpha
\epsilon ,\ee where $v^\alpha$ is a vector (\emph{i.e.} it has null
weight) and $f(g)$ is an arbitrary function of the determinant of
the metric of {\em weight w}. The previous equation will be identical to
\begin{displaymath}
=\left\{\begin{array}{l}
\int_{\{M,t\}}\nabla_\alpha(v^\alpha f(g)) \epsilon-\int_{\{M,t\}} v^\alpha\nabla_\alpha f(g) \epsilon\\
\int_{\{M,t\}}\nabla^w_\alpha(v^\alpha f(g))
\epsilon-\int_{\{M,t\}}v^\alpha\nabla^w_\alpha f(g)\epsilon,
\end{array}
\right.
\end{displaymath}
From the compatibility of
the metric (\ref{compatib}) and the Stokes' theorem we find
\begin{displaymath}
=\left\{\begin{array}{l}
\int_{\{\partial M,t\}}n_\alpha v^\alpha f(g)\tilde \epsilon-\int_{\{M,t\}} v^\alpha\partial_\alpha f(g) \epsilon\\
\int_{\{M,t\}}\nabla^w_\alpha(v^\alpha f(g)) \epsilon.
\end{array}
\right.
\end{displaymath}
For the previous particular integral, one can  see that
both expressions coincide. We will choose the {\em usual covariant operator} (without
any reference to the weight) as the
differential operator, which amounts to considering the density
tensors as tensors. The main result is  that $\nabla g\neq 0$
and thus terms of the sort
\be
f(g)\nabla_\mu v^\mu
\ee
are not
pure boundary terms.

The choice of other derivative operators amounts to ``non-minimal" coupling
of the fields to gravity. In any case, from the expression (\ref{gammaex})
we see that only the determinant of the metric enters in this coupling
and the freedom of considering arbitrary functions of the determinant in the TDiff
theory has already been considered in the last section. For the WTDiff case,
the connection compatible with the combination $\hat g_{\m\n}$
satisfies $\Gamma_{\phantom{\b}\r\a}^\r=0$, which means that there is no freedom
in the choice of the covariant derivative.

%%%%%%%%%%%%%%%%%%%%%%%%%%%%%%%%%%%%%%%%%%%%%%
\section{Maximally symmetric metrics and Type II solutions}
%%%%%%%%%%%%%%%%%%%%%%%%%%%%%%%%%%%%%%%%%%%%%%%%%
Here we show that the most general Type II solution for the potential
(\ref{interaction}) where one of the
metrics satisfies
\be
G_{\m\n}^g=\Lambda g_{\m\n},
\label{Einstein}
\ee
 is such that $f_{\mu\nu}= \gamma
g_{\mu\nu}$, where $\gamma$ is a constant whose value is given by
the equations of motion.\\

From (\ref{Einstein}), we have $K T^g_{tt} + J T^g_{rr}=0$, and
plugging expressions (\ref{formg}) and (\ref{formf}) into Eqs.
(\ref{emg}), we have
\ba
\label{Tg1lin} K T^g_{tt} + J T^g_{rr}
&=& \frac{\zeta B}{2 r^4} \left(\frac{\Delta B^2
}{JKr^4}\right)^{v-1} (AJ-CK)(3B-2r^2)=0.
\ea Since we are now
assuming that $B \neq (2/3) r^2$, it follows that \be
\label{casetypeII} AJ-CK = 0. \ee Hence, from (\ref{formg}) and
(\ref{formf}) plugged into (\ref{emf}),
 \ba \label{Tf1lin} A T^f_{tt} + C T^f_{rr} &=& -
\frac{\zeta }{2 B} \left(\frac{JK r^4}{\Delta
B^2}\right)^{u}(AJ-CK)(3B-2r^2) =0, \ea and from the equations of
motion \be A R^f_{tt} + C R^f_{rr} = 0. \ee From this we obtain
(see {\em e.g.} \cite{Isham:1977rj} for the explicit expressions of the
Ricci tensor components), \be -B^{\prime \prime} + \frac{B^{\prime
2}}{2B} + \frac{\Delta^\prime B^\prime}{2 \Delta}=0. \ee A first
integral is given by \be \label{eqdiffB} \frac{B^{\prime 2}}{B}=4
a^2 \Delta \ee where $a$ is the constant of integration.

Let us now consider the linear
combination \ba r^2 T^g_{tt} + J T^g_{\theta \theta} &=&
-\frac{\zeta}{2K r^2} \left(\frac{AB^2
C}{JKr^4}\right)^{v-1}(BJ-Cr^2)(BK-3AB+Ar^2), \ea which again must
vanish if $g$ is a is a solution of (\ref{Einstein}). Thus one
either has \be \label{case2} BK
+ A r^2 = 3 AB, \ee or \be \label{case1} BJ = C r^2.\ee In both cases
\ba B T^f_{tt} + C T^f_{\theta
\theta} &=& \frac{\zeta}{2 AB} \left(\frac{JKr^4}{AB^2C}\right)^u
(BJ-Cr^2)(BK-3AB+Ar^2)=0.\label{cath} \ea
Note that (\ref{Tf1lin}) and (\ref{cath}) imply that $T^f_{\mu\nu}=H(r)f_{\mu\nu}$. The equations of motion require that $T^f$ must be covariantly conserved, which implies that $H$ is a constant. Therefore, $f$ is a solution of Einstein's equations with a cosmological constant.

Consider first the case when (\ref{case2}) is satisfied. From
this equation and (\ref{casetypeII}), we can eliminate $A$ and $C$
as functions of $B$ and $J=K^{-1}$. We get from (\ref{eqdiffB})
\be \label{second} \frac{B^{\prime 2}}{B^3} = \frac{4 a^2}{(3B -
r^2)^2}. \ee
With the change of variable
\be
B(r)=r^2F^2(r),
\ee
the differential equation (\ref{eqdiffB}) is written as
\be
 r F^{\prime}= \frac{a F^2}{(3F^2 -1)}-F,
\ee
which can be easily integrated to give
\be
c r = {1\over F}\Big(\frac{\sqrt{12+a^2}-a+6F}{\sqrt{12+a^2}+a-6F}\Big)^{\frac{a}{\sqrt{12+a^2}}},\label{tess}
\ee
where $c$ is an integration constant. Notice that
\be
F(r)=(\sqrt{12+a^2}+a)/6, \label{constantF}
\ee
is a solution for $a>0,c\to \infty$ and for $a<0, c=0$, which means
\begin{equation}
B \propto r^2.
\label{brsq}
\end{equation}
In fact, as we shall see, Eq. (\ref{brsq}) must hold in general. The equation of motion $B R_{tt}^f+C R^f_
{\theta\theta}=0$ takes the form \cite{Isham:1977rj}
\be
\label{gordy}
BC''-CB''+2\Delta+(CB'-BC'){\Delta'\over 2\Delta}=0.
\ee
From (\ref{casetypeII}) and (\ref{case2}), we have
\be
A={BK\over 3B-r^2}, \quad C={BJ\over 3B-r^2},\label{aacc}
\ee
and hence
\be
\Delta={B^2 \over (3B-r^2)^2}.\label{dd}
\ee
Now, Eqs. (\ref{aacc}) and (\ref{dd}) can be used in (\ref{gordy}) in order
to eliminate $\Delta$ and $C$ in terms of $B$ and its derivatives
(as well as the known function $J$ and its derivatives). The derivatives of $B$
can be eliminated from (\ref{eqdiffB}), and with this Eq. (\ref{gordy}) becomes an
algebraic equation relating $B$ and $r$. Substituting $B=r^2F^2$, and then
 eliminating $r$ from Eq. (\ref{tess}), we find an algebraic equation involving
 only $F$ and the integration constants $a$ and $c$. It turns out that this
 algebraic equation does {\em not} vanish identically. Indeed, the first terms
  in an expansion in powers of $F$ are given by
\ba
&&B R_{tt}^f+C R^f_
{\theta\theta}=
O(F^2)+\frac{J(r)F(r)}{(3F(r)^2-1)^4 r}
\Big[\Big(\frac{\sqrt{12+a^2}-a}{\sqrt{12+a^2}+a}\Big)
^{\frac{3a}{\sqrt{12+a^2}}}c^{-3}\Lambda_g\nonumber \\
&+&\Big\{2a\Big(\frac{\sqrt{12+a^2}-a}{\sqrt{12+a^2}+a}\Big)
^{\frac{a}{\sqrt{12+a^2}}}c^{-1}+
9a\Big(\frac{\sqrt{12+a^2}-a}{\sqrt{12+a^2}+a}\Big)
^{\frac{3a}{\sqrt{12+a^2}}
}c^{-3}\Lambda_g-6M\Big\}F\Big],\nonumber
\ea
where we have used $J=1-2M/r+\Lambda_g r^2/3$. For the zeroth and first order to cancel identically, one needs
\be
\Lambda_g=0, \quad M=\frac{a c^{-1}}{3}
\Big(\frac{\sqrt{12+a^2}-a}{\sqrt{12+a^2}+a}\Big)
^{\frac{a}{\sqrt{12+a^2}}},
\ee
but then going to the next order in $F$ the expression (\ref{gordy}) does not cancel for any value of $a$. Thus, $F$ is fixed to be a constant whose value is determined by (\ref{gordy}). From this (\ref{brsq}) follows.\footnote{Provided, of course, that the algebraic equation has any solution at all. Otherwise there simply aren't any solutions under the assumption (\ref{case2}). Note, in particular, from (\ref{tess}) and the subsequent discussion, that the constancy of $F$ can only be achieved
for very special values of the integration constants, but these turn out to be the only relevant ones.} Now, it is easy to show that whenever $B\propto r^2$ both metrics must be proportional to each other. Indeed, it follows from Eq. (\ref{eqdiffB}) that $\Delta=AC=const.$ and $B=(a^2\Delta) r^2$. Also, using $JK=1$ and (\ref{casetypeII}) we have $A=\Delta^{1/2} K$ and $C=\Delta^{1/2} J$. On the other hand, for constant $\Delta$, Eq. (\ref{gordy}) reads$$
B C''-C B'' +2\Delta=0.
$$
Using $B=(a^2\Delta) r^2$, $C=\Delta^{1/2} J$ and $J=1-2M/r+\Lambda_g r^2/3$, where $M$ and $\Lambda_g$ are constants, it follows immediately that $a^2=\Delta^{-1/2}$, which implies $B=\Delta^{1/2} r^2$. It is then clear that $f_{\mu\nu}=\gamma g_{\mu\nu}$, where $\gamma=\Delta^{1/2}$ is a constant, as we intended to show.

Next, let us consider the case (\ref{case1}). Here, we can use (\ref{casetypeII}) and (\ref{eqdiffB}) to
obtain \be \label{finaleqdiff} \frac{B^{\prime 2}}{B^3}\propto
\frac{1}{r^4}, \ee
and equation (\ref{finaleqdiff}) yields \be B=\frac{\gamma
r^2}{(1+\alpha r)^2}. \ee Since we have assumed that $g$ satisfies
Einstein's equations with a cosmological constant, Eq.
(\ref{Einstein}), $T_{\mu\nu}^g$
should be proportional to $g_{\mu\nu}$ with a constant
proportionality factor. This is achieved only for $\alpha=0$ which
means $C=\gamma J$. This means that both metrics will be
proportional, with \be f_{\mu\nu}=\gamma g_{\mu\nu}. \ee This completes our proof.

As discussed in the text, the remaining equations of motion determine
the constant $\gamma$ in terms of the parameters in the
Lagrangian.
%
%
%

%%%%%%%%%%%%%%%%%%%%%%%%%%%%%%%%%%%%%%%%%%%%%%%%%%%%
\section{Methods to Generate Bigravity Solutions}\label{methodsbi}
%%%%%%%%%%%%%%%%%%%%%%%%%%%%%%%%%%%%%%%%%%%%%%%%%%%

Finally, let us propose a possible method to generate solutions of bigravity
departing from a solution of GR. Consider a family of
solutions of Einstein's equations with or
without a cosmological constant  $f_{\m\n}(\a_i;\Lambda)$ where $\a_i$ are integration
constants.

These metrics transform under GCT and they are still solutions of Einstein's equations
in the new coordinates.
After identifying the new coordinates with the old ones, we find a new family of
solutions of the (vacuum)
 Einstein's equations which we use to define the metric $g_{\m\n}$ (remember
that the gauge invariance of {\em bigravity}
is only the subgroup of diagonal diffeomorphisms which means that $g_{\m\n}$ and
$f_{\m\n}$ are not equivalent). To get a solution
of the bigravity system we also need the traces of the matrix $\M$ to be constant.
More precisely, the second family of solutions will be given by
\be
g_{\m\n}(x)=\pd_\m y^\r(x)\pd_\n y^\s(x)f_{\r\s}(y(x);\a_i',\Lambda'),
\ee
from which
\be
{\mathcal M}^\m_\n\equiv f^{\m\b}g_{\b\n}=
 f^{\m\b}(x;\a_i,\Lambda)\pd_\b y^\r(x)\pd_\n y^\s(x)f_{\r\s}(y(x);\a_i',\Lambda').
\ee
If the first four traces of this matrix are constant then we can find
conditions for these two metrics to be a solution of bigravity.

As an example,
let us consider a generic metric $f_{\m\n}$ and a {\em constant} matrix
$$N_\m^\n=\pd_\m y^\n(x).$$
The traces of ${\mathcal M}$ will not
be constant in general. One possible choice which produces a constant matrix $\M$ is provided by
$$ N=\mathrm{diag}\{\l_1,...,\l_4\},$$
and $\a_i'=\a_i$.
In general this produces a solution of bigravity which breaks the symmetries
of the original metric $f_{\m\n}$. By doing such a transformation and perturbing
the solution we can
get Lorentz-breaking massive terms for the gravitons in Schwarzschild-(A)de Sitter or Kerr space
and this possibility is currently under research \cite{Blas08}. For the Schwarzschild case, this
is  particularly interesting as these Lorentz-breaking perturbations
may constitute a new sort of hair for the black hole \cite{Blas08}.
Besides, the existence of (non-proportional) rotating solutions in {\em bigravity}
is also interesting as
they seem to be problematic in other approaches to massive gravity (see, \emph{e.g}
\cite{Dubovsky:2007zi}).
\\

Another method for  finding solutions of {\em bigravity}
would be to, given a metric  $f_{\m\n}$, identifying a vielbein $e_{\phantom{a}\m}^{a}$ such that
\be
f_{\m\n}=e_{\phantom{a}\m}^{a}e_{\phantom{a}\n}^{b}\eta_{ab}.
\ee
Remember that the vielbein $e_{\phantom{a}\m}^{a}$ is determined
up to local Lorentz-transformations.These local transformations
allow to take any other symmetric
tensor to a diagonal form (with non-constant eigenvalues). For a {\em
bigravity} system, the vielbein where both of the metrics are proportional, being
one of them Minkowski is completely determined, and
 we may call it $\bar e_{\phantom{a}\m}^{a}=L_\m^\n(x)
e_{\phantom{a}\n}^{a}$ for any vielbein $e_{\phantom{a}\n}^{a}$. It satisfies
\be
\label{gmetric}
f_{\m\n}=\bar e_{\phantom{a}\m}^{a}\bar e_{\phantom{a}\n}^{b}\eta_{ab}, \quad
g_{\m\n}=\bar e_{\phantom{a}\m}^{a}\bar e_{\phantom{a}\n}^{b}\l_a(x)\eta_{ab}.
\ee
The previous eigenvalues $\l_a(x)$ will coincide with those of the matrix $\M$
in this frame. Thus, if they have constant values there will be a potential which
will have the previous metrics as a solution. Of course, the metric
$g_{\m\n}$ which we have built is not a solution of Einstein's equations in general. Thus,
the problem of finding a solution of bigravity in this framework
translates into finding a local Lorentz transformation
$L_\m^\n(x)$
(which can depend on new parameters) and four constants $\l_i$ such that the metric
$g_{\m\n}$ of (\ref{gmetric}) is a solution of Einstein's equations with a cosmological constant.
This method has not yet been explored.  A first natural question is whether by using it,
 we can recover the Type I solutions.

When applied, the previous method allows to look for ordinary solutions of GR in bigravity.
 It is also very interesting to look for solutions of bigravity which differ from GR. Recently
 a particular solution of this form has been found in \cite{Berezhiani:2008nr}.

\selectlanguage{english}

\bibliographystyle{blas2}
\bibliography{books,articles}

\end{document}